\documentclass[seceqn,nameyear]{elsart}
\usepackage{amssymb,amsmath,graphicx,mathptm}

\newcommand{\J}{\mathop\mathrm{J_0}}
\newcommand{\I}{\mathop\mathrm{I_0}}
\newcommand{\arcsecf}{\hbox{$.\!\!^{\prime\prime}$}}
\newcommand{\arcminf}{\hbox{$.\!\!^{\prime}$}}
\newcommand{\A}{\mathcal{A}}
\newcommand{\s}{\mathrm{(s)}}

\makeatletter
\let\@internalcite\cite
\def\cite{\@ifstar{\citefull}{\citepars}}
\def\citepars{\def\astroncite##1##2{##1 (##2)}\@internalcite}
\def\citefull{\def\astroncite##1##2{##1 ##2}\@internalcite}
\def\citeyear{\def\astroncite##1##2{##2}\@internalcite}
\def\citename{\def\astroncite##1##2{##1}\@internalcite}

\def\@citex[#1]#2{%
  \if@filesw\immediate\write\@auxout{\string\citation{#2}}\fi
  \def\@citea{}\@cite{\@for\@citeb:=#2\do
    {\@citea\def\@citea{; }\@ifundefined
       {b@\@citeb}{{\bf ?}\@warning
       {Citation `\@citeb' on page \thepage \space undefined}}%
{\csname b@\@citeb\endcsname}}}{#1}}
\def\@cite#1#2{#1\if@tempswa , #2\fi}
\makeatother
\begin{document}

\begin{frontmatter}

\title{Weak Gravitational Lensing}

\author{Matthias Bartelmann} and \author{Peter Schneider}
\address{Max-Planck-Institut f\"ur Astrophysik, P.O. Box 1523,
  D--85740 Garching, Germany}

\begin{abstract}

We review theory and applications of weak gravitational lensing. After
summarising Friedmann-Lema{\^\i}tre cosmological models, we present
the formalism of gravitational lensing and light propagation in
arbitrary space-times. We discuss how weak-lensing effects can be
measured. The formalism is then applied to reconstructions of
galaxy-cluster mass distributions, gravitational lensing by
large-scale matter distributions, QSO-galaxy correlations induced by
weak lensing, lensing of galaxies by galaxies, and weak lensing of the
cosmic microwave background.

\end{abstract}

\end{frontmatter}

\newpage

\tableofcontents

  % -*- LaTeX -*-

\section{\label{sc:1}Introduction}

\subsection{Gravitational Light Deflection}

Light rays are deflected when they propagate through an inhomogeneous
gravitational field. Although several researchers had speculated about
such an effect well before the advent of General Relativity (see
\cite*{sef92} for a historical account), it was Einstein's theory
which elevated the deflection of light by masses from a hypothesis to
a firm prediction. Assuming light behaves like a stream of particles,
its deflection can be calculated within Newton's theory of
gravitation, but General Relativity predicts that the effect is twice
as large. A light ray grazing the surface of the Sun is deflected by
$1.75\,$arc seconds compared to the $0.87\,$arc seconds predicted by
Newton's theory. The confirmation of the larger value in 1919 was
perhaps the most important step towards accepting General Relativity
as the correct theory of gravity (\cite*{ED20.1}).

\begin{figure}[ht]
  \includegraphics[width=\hsize]{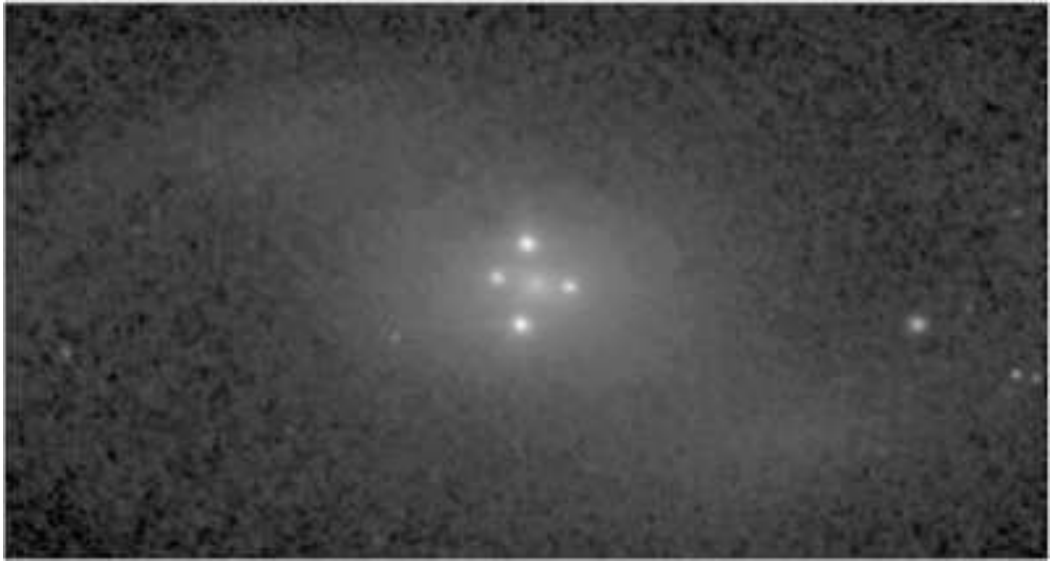}
\caption{The gravitational lens system 2237$+$0305 consists of a
nearby spiral galaxy at redshift $z_\mathrm{d}=0.039$ and four images
of a background quasar with redshift $z_\mathrm{s}=1.69$. It was
discovered by \protect\cite{HU85.1}. The image was taken by the {\em
Hubble Space Telescope\/} and shows only the innermost region of the
lensing galaxy. The central compact source is the bright galaxy core,
the other four compact sources are the quasar images. They differ in
brightness because they are magnified by different amounts. The four
images roughly fall on a circle concentric with the core of the
lensing galaxy. The mass inside this circle can be determined with
very high accuracy (\protect\cite*{RI92.1}). The largest separation
between the images is $1.8''$.}
\label{fig:1.1}
\end{figure}

Cosmic bodies more distant, more massive, or more compact than the Sun
can bend light rays from a single source sufficiently strongly so that
multiple light rays can reach the observer. The observer sees an image
in the direction of each ray arriving at their position, so that the
source appears multiply imaged. In the language of General Relativity,
there may exist more than one null geodesic connecting the world-line
of a source with the observation event. Although predicted long
before, the first multiple-image system was discovered only in 1979
(\cite*{WA79.1}). From then on, the field of {\em gravitational
lensing\/} developed into one of the most active subjects of
astrophysical research. Several dozens of multiply-imaged sources have
since been found. Their quantitative analysis provides accurate masses
of, and in some cases detailed information on, the deflectors. An
example is shown in Fig.~\ref{fig:1.1}.

\begin{figure}[ht]
  \centerline{\includegraphics[width=0.8\hsize]{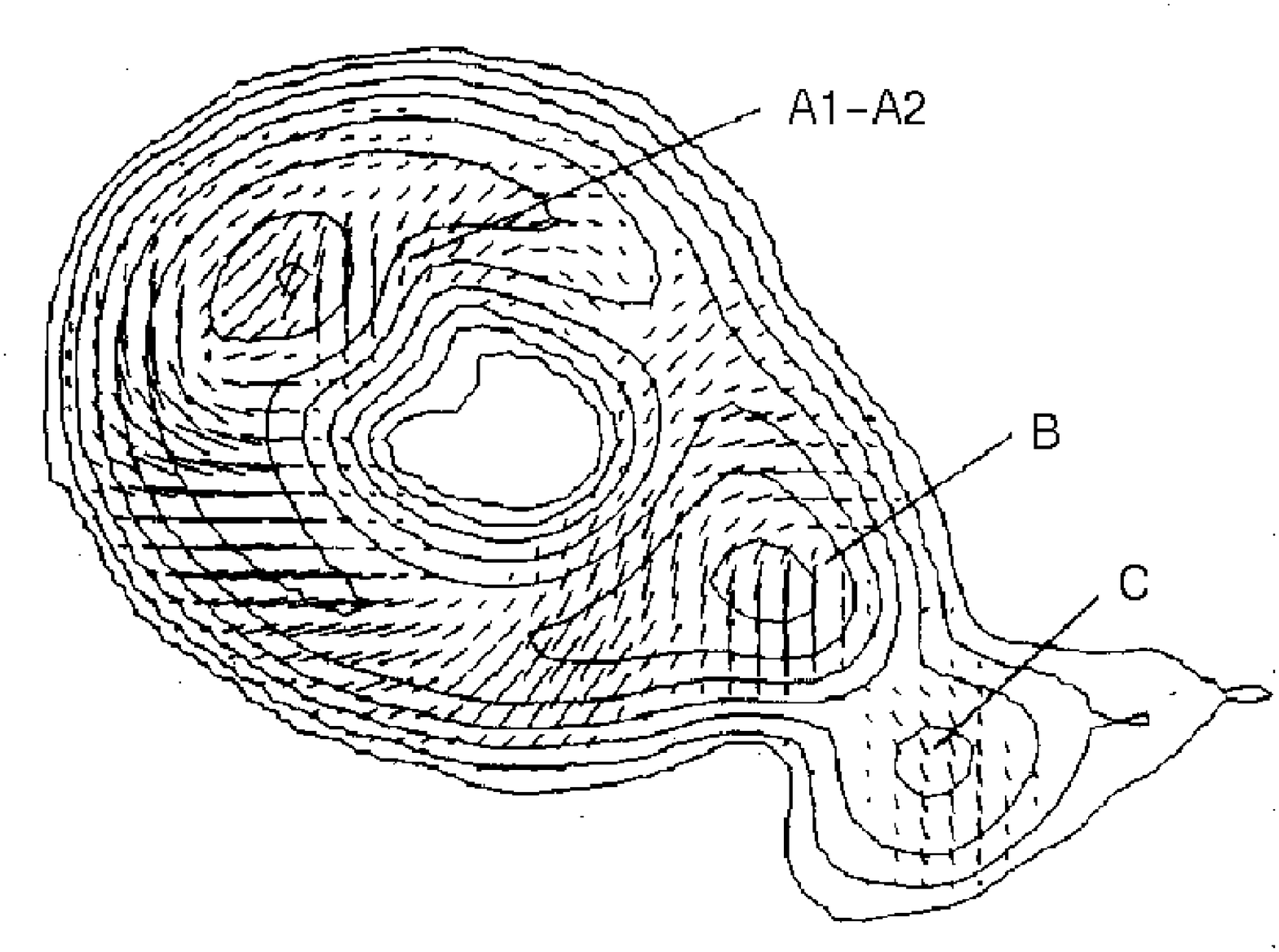}}
\caption{The radio source MG~1131$+$0456 was discovered by
\protect\cite{HE88.1} as the first example of a so-called {\em
Einstein ring\/}. If a source and an axially symmetric lens are
co-aligned with the observer, the symmetry of the system permits the
formation of a ring-like image of the source centred on the lens. If
the symmetry is broken (as expected for all realistic lensing matter
distributions), the ring is deformed or broken up, typically into four
images (see Fig.~\ref{fig:1.1}). However, if the source is
sufficiently extended, ring-like images can be formed even if the
symmetry is imperfect. The 6~cm radio map of MG~1131$+$0456 shows a
closed ring, while the ring breaks up at higher frequencies where the
source is smaller. The ring diameter is $2.1''$.}
\label{fig:1.2}
\end{figure}

\begin{figure}[ht]
  \includegraphics[width=\hsize]{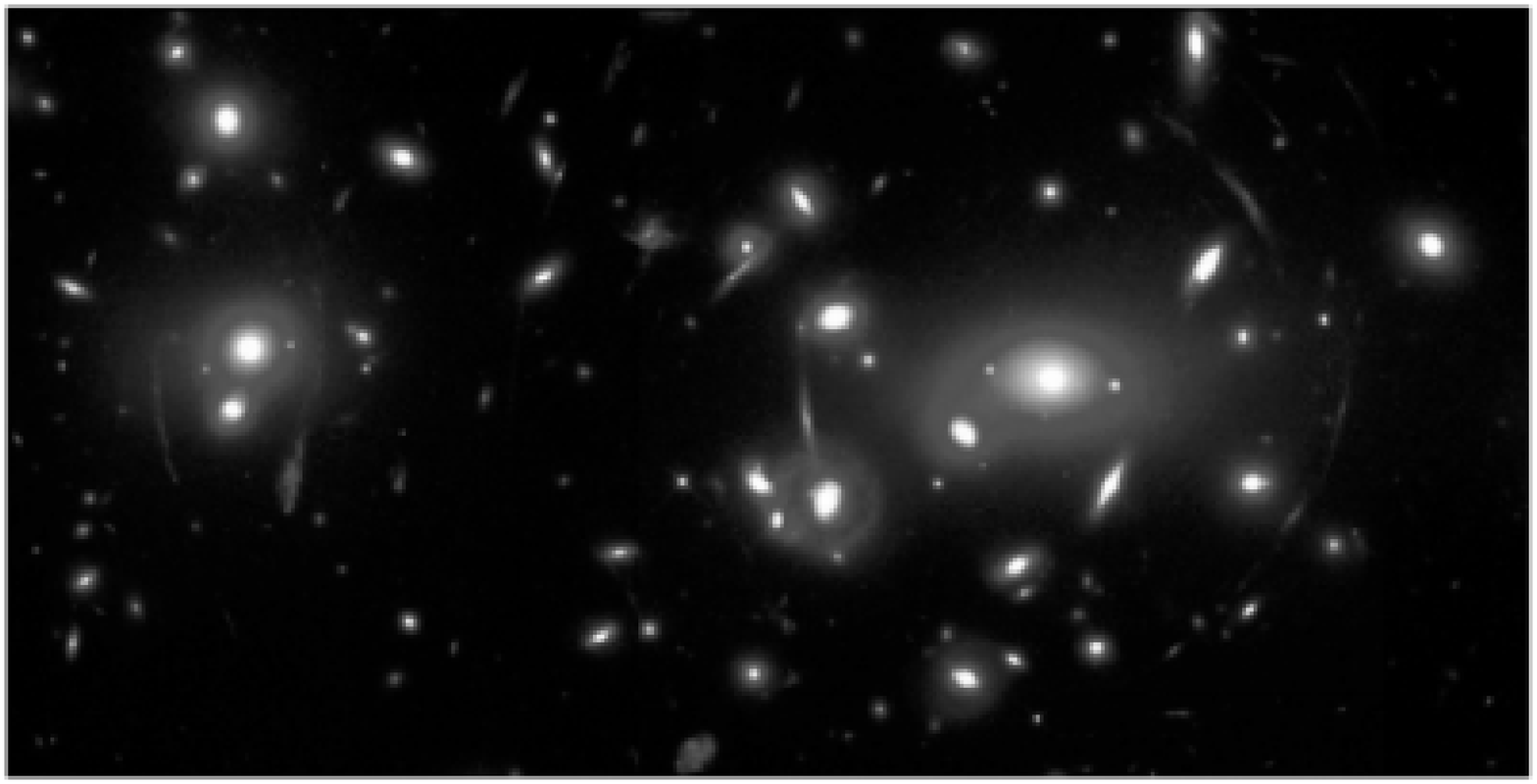}
\caption{The cluster Abell~2218 hosts one of the most impressive
collections of arcs. This {\em HST\/} image of the cluster's central
region shows a pattern of strongly distorted galaxy images
tangentially aligned with respect to the cluster centre, which lies
close to the bright galaxy in the upper part of this image. The frame
measures about $80''\times160''$.}
\label{fig:1.3}
\end{figure}

Tidal gravitational fields lead to differential deflection of light
bundles. The size and shape of their cross sections are therefore
changed. Since photons are neither emitted nor absorbed in the process
of gravitational light deflection, the surface brightness of lensed
sources remains unchanged. Changing the size of the cross section of a
light bundle therefore changes the flux observed from a source. The
different images in multiple-image systems generally have different
fluxes. The images of extended sources, i.e.~sources which can
observationally be resolved, are deformed by the gravitational tidal
field. Since astronomical sources like galaxies are not intrinsically
circular, this deformation is generally very difficult to identify in
individual images. In some cases, however, the distortion is strong
enough to be readily recognised, most noticeably in the case of {\em
Einstein rings\/} (see Fig.~\ref{fig:1.2}) and {\em arcs\/} in galaxy
clusters (Fig.~\ref{fig:1.3}).

If the light bundles from some sources are distorted so strongly that
their images appear as giant luminous arcs, one may expect many more
sources behind a cluster whose images are only weakly
distorted. Although weak distortions in individual images can hardly
be recognised, the net distortion averaged over an ensemble of images
can still be detected. As we shall describe in Sect.~\ref{sc:2.3},
deep optical exposures reveal a dense population of faint galaxies on
the sky. Most of these galaxies are at high redshift, thus distant,
and their image shapes can be utilised to probe the tidal
gravitational field of intervening mass concentrations. Indeed, the
tidal gravitational field can be reconstructed from the coherent
distortion apparent in images of the faint galaxy population, and from
that the density profile of intervening clusters of galaxies can be
inferred (see Sect.~\ref{sc:4}).

\subsection{Weak Gravitational Lensing}

This review deals with {\em weak gravitational lensing\/}. There is no
generally applicable definition of weak lensing despite the fact that
it constitutes a flourishing area of research. The common aspect of
all studies of weak gravitational lensing is that measurements of its
effects are statistical in nature. While a single multiply-imaged
source provides information on the mass distribution of the deflector,
weak lensing effects show up only across ensembles of sources. One
example was given above: The shape distribution of an ensemble of
galaxy images is changed close to a massive galaxy cluster in the
foreground, because the cluster's tidal field polarises the images. We
shall see later that the size distribution of the background galaxy
population is also locally changed in the neighbourhood of a massive
intervening mass concentration.

Magnification and distortion effects due to weak lensing can be used
to probe the statistical properties of the matter distribution between
us and an ensemble of distant sources, provided some assumptions on
the source properties can be made. For example, if a {\em standard
candle\/}\footnote{The term {\em standard candle\/} is used for any
class of astronomical objects whose intrinsic luminosity can be
inferred independently of the observed flux. In the simplest case, all
members of the class have the same luminosity. More typically, the
luminosity depends on some other known and observable parameters, such
that the luminosity can be inferred from them. The luminosity distance
to any standard candle can directly be inferred from the square root
of the ratio of source luminosity and observed flux. Since the
luminosity distance depends on cosmological parameters, the geometry
of the Universe can then directly be investigated. Probably the best
current candidates for standard candles are supernovae of Type Ia.
They can be observed to quite high redshifts, and thus be utilised to
estimate cosmological parameters (e.g.~\cite*{rfc98}).} at high
redshift is identified, its flux can be used to estimate the
magnification along its line-of-sight. It can be assumed that the
orientation of faint distant galaxies is random. Then, any coherent
alignment of images signals the presence of an intervening tidal
gravitational field. As a third example, the positions on the sky of
cosmic objects at vastly different distances from us should be
mutually independent. A statistical association of foreground objects
with background sources can therefore indicate the magnification
caused by the foreground objects on the background sources.

All these effects are quite subtle, or weak, and many of the current
challenges in the field are observational in nature. A coherent
alignment of images of distant galaxies {\em can\/} be due to an
intervening tidal gravitational field, but {\em could\/} also be due
to propagation effects in the Earth's atmosphere or in the
telescope. A variation in the number density of background sources
around a foreground object {\em can\/} be due to a magnification
effect, but {\em could\/} also be due to non-uniform photometry or
obscuration effects. These potential systematic effects have to be
controlled at a level well below the expected weak-lensing effects. We
shall return to this essential point at various places in this review.

\subsection{Applications of Gravitational Lensing}

Gravitational lensing has developed into a versatile tool for
observational cosmology. There are two main reasons:

\begin{enumerate}

\item The deflection angle of a light ray is determined by the
gravitational field of the matter distribution along its
path. According to Einstein's theory of General Relativity, the
gravitational field is in turn determined by the stress-energy tensor
of the matter distribution. For the astrophysically most relevant case
of non-relativistic matter, the latter is characterised by the density
distribution alone. Hence, the gravitational field, and thus the
deflection angle, depend neither on the nature of the matter nor on
its physical state. Light deflection probes the total matter density,
without distinguishing between ordinary (baryonic) matter or dark
matter. In contrast to other dynamical methods for probing
gravitational fields, no assumption needs to be made on the dynamical
state of the matter. For example, the interpretation of radial
velocity measurements in terms of the gravitating mass requires the
applicability of the virial theorem (i.e., the physical system is
assumed to be in virial equilibrium), or knowledge of the orbits (such
as the circular orbits in disk galaxies). However, as will be
discussed in Sect.~\ref{sc:3}, lensing measures only the mass
distribution projected along the line-of-sight, and is therefore
insensitive to the extent of the mass distribution {\em along\/} the
light rays, as long as this extent is small compared to the distances
from the observer and the source to the deflecting mass. Keeping this
in mind, mass determinations by lensing do not depend on any symmetry
assumptions.

\item Once the deflection angle as a function of impact parameter is
given, gravitational lensing reduces to simple geometry. Since most
lens systems involve sources (and lenses) at moderate or high
redshift, lensing can probe the geometry of the Universe. This was
noted by \cite{RE64.2}, who pointed out that lensing can be used to
determine the Hubble constant and the cosmic density
parameter. Although this turned out later to be more difficult than
anticipated at the time, first measurements of the Hubble constant
through lensing have been obtained with detailed models of the matter
distribution in multiple-image lens systems and the difference in
light-travel time along the different light paths corresponding to
different images of the source (e.g., \cite*{KU97.1}; \cite*{SC97.1};
\cite*{big98}). Since the volume element per unit redshift interval
and unit solid angle also depends on the geometry of space-time, so
does the number of lenses therein.  Hence, the lensing probability for
distant sources depends on the cosmological parameters (e.g.,
\cite*{PR73.1}). Unfortunately, in order to derive constraints on the
cosmological model with this method, one needs to know the evolution
of the lens population with redshift. Nevertheless, in some cases,
significant constraints on the cosmological parameters
(\cite*{KO93.2}, \citeyear{KO96.1}; \cite*{MA93.4}; \cite*{BA98.1};
\cite*{FA98.1}), and on the evolution of the lens population
(\cite*{MA94.1}) have been derived from multiple-image and arc
statistics.

\end{enumerate}

The possibility to directly investigate the dark-matter distribution
led to substantial results over recent years. Constraints on the size
of the dark-matter haloes of spiral galaxies were derived (e.g.,
\cite*{BR96.1}), the presence of dark-matter haloes in elliptical
galaxies was demonstrated (e.g., \cite*{MA93.4}; \cite*{GR96.2}), and
the projected total mass distribution in many cluster of galaxies was
mapped (e.g., \cite*{KN96.1}; \cite*{hfk98}; \cite*{kwl98}). These
results directly impact on our understanding of structure formation,
supporting hierarchical structure formation in cold dark matter (CDM)
models. Constraints on the nature of dark matter were also obtained.
Compact dark-matter objects, such as black holes or brown dwarfs,
cannot be very abundant in the Universe, because otherwise they would
lead to observable lensing effects (e.g., \cite*{SC93.4};
\cite*{DA94.1}). Galactic microlensing experiments constrained the
density and typical mass scale of massive compact halo objects in our
Galaxy (see \cite*{pac96}, \cite*{RO97.1} and \cite*{mao99} for
reviews). We refer the reader to the reviews by \cite{BL92.1},
\cite{sch96} and \cite{NA97.1} for a detailed account of the
cosmological applications of gravitational lensing.

We shall concentrate almost entirely on weak gravitational lensing
here. Hence, the flourishing fields of multiple-image systems and
their interpretation, Galactic microlensing and its consequences for
understanding the nature of dark matter in the halo of our Galaxy, and
the detailed investigations of the mass distribution in the inner
parts of galaxy clusters through arcs, arclets, and multiply imaged
background galaxies, will not be covered in this review. In addition
to the references given above, we would like to point the reader to
\cite{RE94.1}, \cite{FO94.1}, and \cite{WU96.2} for more recent
reviews on various aspects of gravitational lensing, to \cite{mel98}
for a very recent review on weak lensing, and to the monograph
(\cite*{sef92}) for a detailed account of the theory and applications
of gravitational lensing.

\subsection{Structure of this Review}

Many aspects of weak gravitational lensing are intimately related to
the cosmological model and to the theory of structure formation in the
Universe. We therefore start the review by giving some cosmological
background in Sect.~\ref{sc:2}. After summarising
Friedmann-Lema{\^\i}tre-Robertson-Walker models, we sketch the theory
of structure formation, introduce astrophysical objects like QSOs,
galaxies, and galaxy clusters, and finish the Section with a general
discussion of correlation functions, power spectra, and their
projections. Gravitational light deflection in general is the subject
of Sect.~\ref{sc:3}, and the specialisation to weak lensing is
described in Sect.~\ref{sc:4}. One of the main aspects there is how
weak lensing effects can be quantified and measured. The following two
sections describe the theory of weak lensing by galaxy clusters
(Sect.~\ref{sc:5}) and cosmological mass distributions
(Sect.~\ref{sc:6}). Apparent correlations between background QSOs and
foreground galaxies due to the magnification bias caused by
large-scale matter distributions are the subject of
Sect.~\ref{sc:7}. Weak lensing effects of foreground galaxies on
background galaxies are reviewed in Sect.~\ref{sc:8}, and
Sect.~\ref{sc:9} finally deals with weak lensing of the most distant
and most extended source possible, i.e.~the Cosmic Microwave
Background. We present a brief summary and an outlook in
Sect.~\ref{sc:10}.

We use standard astronomical units throughout: $1\,M_\odot =
1\,\hbox{solar mass} = 2\times10^{33}\,\mathrm{g}$; $1\,\mathrm{Mpc} =
1\,\hbox{megaparsec} = 3.1\times10^{24}\,\mathrm{cm}$.

  % -*- LaTeX -*-

\section{\label{sc:2}Cosmological Background}

We review in this section those aspects of the standard cosmological
model which are relevant for our further discussion of weak
gravitational lensing. This standard model consists of a description
for the cosmological background which is a homogeneous and isotropic
solution of the field equations of General Relativity, and a theory
for the formation of structure.

The background model is described by the Robertson-Walker metric
(\cite*{RO35.1}; \cite*{WA35.1}), in which hypersurfaces of constant
time are homogeneous and isotropic three-spaces, either flat or
curved, and change with time according to a scale factor which depends
on time only. The dynamics of the scale factor is determined by two
equations which follow from Einstein's field equations given the
highly symmetric form of the metric.

Current theories of structure formation assume that structure grows
via gravitational instability from initial seed perturbations whose
origin is yet unclear. Most common hypotheses lead to the prediction
that the statistics of the seed fluctuations is Gaussian. Their
amplitude is low for most of their evolution so that linear
perturbation theory is sufficient to describe their growth until late
stages. For general references on the cosmological model and on the
theory of structure formation, cf.~\cite{WE72.1}, \cite{MI73.1},
\cite{PE80.1}, \cite{BO88.1}, \cite{PA93.2}, \cite{pee93}, and
\cite{pea99}.

\subsection{\label{sc:2.1}Friedmann-Lema{\^\i}tre Cosmological Models}

\subsubsection{\label{sc:2.1.1}Metric}

Two postulates are fundamental to the standard cosmological model,
which are:

\begin{enumerate}

\item {\em When averaged over sufficiently large scales, there exists
a mean motion of radiation and matter in the Universe with respect to
which all averaged observable properties are isotropic.\/}

\item {\em All fundamental observers, i.e.~imagined observers which
follow this mean motion, experience the same history of the Universe,
i.e.~the same averaged observable properties, provided they set their
clocks suitably.\/} Such a universe is called {\em
observer-homogeneous.\/}

\end{enumerate}

General Relativity describes space-time as a four-dimensional manifold
whose metric tensor $g_{\alpha\beta}$ is considered as a dynamical
field. The dynamics of the metric is governed by Einstein's field
equations, which relate the Einstein tensor to the stress-energy
tensor of the matter contained in space-time. Two events in space-time
with coordinates differing by $\d x^\alpha$ are separated by $\d s$,
with $\d s^2=g_{\alpha\beta}\d x^\alpha\d x^\beta$. The {\em
eigentime\/} (proper time) of an observer who travels by $\d s$
changes by $c^{-1}\d s$. Greek indices run over $0\ldots3$ and Latin
indices run over the spatial indices $1\ldots3$ only.

The two postulates stated above considerably constrain the admissible
form of the metric tensor. Spatial coordinates which are constant for
fundamental observers are called comoving coordinates. In these
coordinates, the mean motion is described by $\d x^i=0$, and hence $\d
s^2=g_{00}\d t^2$. If we require that the {\em eigentime\/} of
fundamental observers equal the cosmic time, this implies
$g_{00}=c^2$.

Isotropy requires that clocks can be synchronised such that the
space-time components of the metric tensor vanish, $g_{0i}=0$. If this
was impossible, the components of $g_{0i}$ identified one particular
direction in space-time, violating isotropy. The metric can therefore
be written
\begin{equation}
  \d s^2 = c^2\d t^2 + g_{ij}\d x^i\d x^j\;,
\label{eq:2.1}
\end{equation}
where $g_{ij}$ is the metric of spatial hypersurfaces. In order not to
violate isotropy, the spatial metric can only isotropically contract
or expand with a scale function $a(t)$ which must be a function of
time only, because otherwise the expansion would be different at
different places, violating homogeneity. Hence the metric further
simplifies to
\begin{equation}
  \d s^2 = c^2\d t^2 - a^2(t)\d l^2\;,
\label{eq:2.2}
\end{equation}
where $\d l$ is the line element of the homogeneous and isotropic
three-space. A special case of the metric (\ref{eq:2.2}) is the
Minkowski metric, for which $\d l$ is the Euclidian line element and
$a(t)$ is a constant. Homogeneity also implies that all quantities
describing the matter content of the Universe, e.g.~density and
pressure, can be functions of time only.

The spatial hypersurfaces whose geometry is described by $\d l^2$ can
either be flat or curved. Isotropy only requires them to be
spherically symmetric, i.e.~spatial surfaces of constant distance
from an arbitrary point need to be two-spheres. Homogeneity permits us
to choose an arbitrary point as coordinate origin. We can then
introduce two angles $\theta,\phi$ which uniquely identify positions
on the unit sphere around the origin, and a radial coordinate $w$. The
most general admissible form for the spatial line element is then
\begin{equation}
  \d l^2 = \d w^2 + 
  f_K^2(w)\left(\d\phi^2+\sin^2\theta\d\theta^2\right)
  \equiv \d w^2 + f_K^2(w)\d\omega^2\;.
\label{eq:2.3}
\end{equation}
Homogeneity requires that the radial function $f_K(w)$ is either a
trigonometric, linear, or hyperbolic function of $w$, depending on
whether the curvature $K$ is positive, zero, or
negative. Specifically,
\begin{equation}
  f_K(w) = \left\{
  \begin{array}{ll}
    K^{-1/2}\sin(K^{1/2}w) & (K>0) \\
    w & (K=0) \\
    (-K)^{-1/2}\sinh[(-K)^{1/2}w] & (K<0) \\
  \end{array}\right.\;.
\label{eq:2.4}
\end{equation}
Note that $f_K(w)$ and thus $|K|^{-1/2}$ have the dimension of a
length. If we define the radius $r$ of the two-spheres by
$f_K(w)\equiv r$, the metric $\d l^2$ takes the alternative form
\begin{equation}
  \d l^2 = \frac{\d r^2}{1-Kr^2} + r^2\d\omega^2\;.
\label{eq:2.5}
\end{equation}

\subsubsection{\label{sc:2.1.2}Redshift}

Due to the expansion of space, photons are redshifted while they
propagate from the source to the observer. Consider a comoving source
emitting a light signal at $t_\mathrm{e}$ which reaches a comoving
observer at the coordinate origin $w=0$ at time $t_\mathrm{o}$. Since
$\d s=0$ for light, a backward-directed radial light ray propagates
according to $|c\d t|=a\d w$, from the metric. The (comoving)
coordinate distance between source and observer is constant by
definition,
\begin{equation}
  w_\mathrm{eo} = \int_\mathrm{o}^\mathrm{e}\d w =
  \int_{t_\mathrm{e}}^{t_\mathrm{o}(t_\mathrm{e})}
  \frac{c\d t}{a} = \hbox{constant}\;,
\label{eq:2.6}
\end{equation}
and thus in particular the derivative of $w_\mathrm{eo}$ with respect
to $t_\mathrm{e}$ is zero. It then follows from eq.~(\ref{eq:2.6})
\begin{equation}
  \frac{\d t_\mathrm{o}}{\d t_\mathrm{e}} = 
  \frac{a(t_\mathrm{o})}{a(t_\mathrm{e})}\;.
\label{eq:2.7}
\end{equation}
Identifying the inverse time intervals $(\d t_\mathrm{e,o})^{-1}$ with
the emitted and observed light frequencies $\nu_\mathrm{e,o}$, we can
write
\begin{equation}
  \frac{\d t_\mathrm{o}}{\d t_\mathrm{e}} =
  \frac{\nu_\mathrm{e}}{\nu_\mathrm{o}} =
  \frac{\lambda_\mathrm{o}}{\lambda_\mathrm{e}}\;.
\label{eq:2.8}
\end{equation}
Since the redshift $z$ is defined as the relative change in
wavelength, or $1+z=\lambda_\mathrm{o}\lambda_\mathrm{e}^{-1}$, we
find
\begin{equation}
  1+z = \frac{a(t_\mathrm{o})}{a(t_\mathrm{e})}\;.
\label{eq:2.9}
\end{equation}
This shows that light is redshifted by the amount by which the
Universe has expanded between emission and observation.

\subsubsection{\label{sc:2.1.3}Expansion}

To complete the description of space-time, we need to know how the
scale function $a(t)$ depends on time, and how the curvature $K$
depends on the matter which fills space-time. That is, we ask for the
dynamics of the space-time. Einstein's field equations relate the
Einstein tensor $G_{\alpha\beta}$ to the stress-energy tensor
$T_{\alpha\beta}$ of the matter,
\begin{equation}
  G_{\alpha\beta} = \frac{8\pi G}{c^2}\,T_{\alpha\beta} +
  \Lambda\,g_{\alpha\beta}\;.
\label{eq:2.10}
\end{equation}
The second term proportional to the metric tensor $g_{\alpha\beta}$ is
a generalisation introduced by Einstein to allow static cosmological
solutions of the field equations. $\Lambda$ is called the cosmological
constant. For the highly symmetric form of the metric given by
(\ref{eq:2.2}) and (\ref{eq:2.3}), Einstein's equations imply that
$T_{\alpha\beta}$ has to have the form of the stress-energy tensor of
a homogeneous perfect fluid, which is characterised by its density
$\rho(t)$ and its pressure $p(t)$. Matter density and pressure can
only depend on time because of homogeneity. The field equations then
simplify to the two independent equations
\begin{equation}
  \left(\frac{\dot{a}}{a}\right)^2 = \frac{8\pi G}{3}\rho -
  \frac{Kc^2}{a^2} + \frac{\Lambda}{3}
\label{eq:2.11}
\end{equation}
and
\begin{equation}
  \frac{\ddot{a}}{a} = -\frac{4}{3}\pi G
  \left(\rho+\frac{3p}{c^2}\right) + \frac{\Lambda}{3} \;.
\label{eq:2.12}
\end{equation}
The scale factor $a(t)$ is determined once its value at one instant of
time is fixed. We choose $a=1$ at the present epoch $t_0$. Equation
(\ref{eq:2.11}) is called {\em Friedmann's equation\/} (\cite*{fri22},
\citeyear{fri24}). The two equations (\ref{eq:2.11}) and
(\ref{eq:2.12}) can be combined to yield the {\em adiabatic
equation\/}
\begin{equation}
  \frac{\d}{\d t}\left[a^3(t)\rho(t)c^2\right] + 
  p(t)\frac{\d a^3(t)}{\d t} = 0\;,
\label{eq:2.13}
\end{equation}
which has an intuitive interpretation. The first term $a^3\rho$ is
proportional to the energy contained in a fixed comoving volume, and
hence the equation states that the change in `internal' energy equals
the pressure times the change in proper volume. Hence eq.~(\ref{eq:2.13})
is the first law of thermodynamics in the cosmological context.

A metric of the form given by eqs.~(\ref{eq:2.2}), (\ref{eq:2.3}), and
(\ref{eq:2.4}) is called the Robertson-Walker metric. If its scale
factor $a(t)$ obeys Friedmann's equation (\ref{eq:2.11}) and the
adiabatic equation (\ref{eq:2.13}), it is called the
Friedmann-Lema{\^\i}tre-Robertson-Walker metric, or the
Friedmann-Lema{\^\i}tre metric for short. Note that
eq.~(\ref{eq:2.12}) can also be derived from Newtonian gravity except
for the pressure term in (\ref{eq:2.12}) and the cosmological
constant. Unlike in Newtonian theory, pressure acts as a source of
gravity in General Relativity.

\subsubsection{\label{sc:2.1.4}Parameters}

The relative expansion rate $\dot{a}a^{-1}\equiv H$ is called the {\em
Hubble parameter\/}, and its value at the present epoch $t=t_0$ is the
{\em Hubble constant\/}, $H(t_0)\equiv H_0$. It has the dimension of
an inverse time. The value of $H_0$ is still uncertain. Current
measurements roughly fall into the range $H_0=(50-80)\,$km s$^{-1}$
Mpc$^{-1}$ (see \cite*{fre96} for a review), and the uncertainty in
$H_0$ is commonly expressed as $H_0=100\,h\,$km s$^{-1}$ Mpc$^{-1}$,
with $h=(0.5-0.8)$. Hence
\begin{equation}
  H_0 \approx 3.2\times10^{-18}\,h\,\mathrm{s}^{-1}
  \approx 1.0\times10^{-10}\,h\,\mathrm{yr}^{-1}\;.
\label{eq:2.14}
\end{equation}
The time scale for the expansion of the Universe is the inverse Hubble
constant, or $H_0^{-1}\approx10^{10}\,h^{-1}\,$years.

The combination
\begin{equation}
  \frac{3H_0^2}{8\pi G} \equiv \rho_\mathrm{cr} \approx
  1.9\times10^{-29}\,h^2\,\mathrm{g}\,\mathrm{cm}^{-3}
\label{eq:2.15}
\end{equation}
is the {\em critical density\/} of the Universe, and the density
$\rho_0$ in units of $\rho_\mathrm{cr}$ is the {\em density
parameter\/} $\Omega_0$,
\begin{equation}
  \Omega_0 = \frac{\rho_0}{\rho_\mathrm{cr}}\;.
\label{eq:2.16}
\end{equation}
If the matter density in the universe is critical,
$\rho_0=\rho_\mathrm{cr}$ or $\Omega_0=1$, and if the cosmological
constant vanishes, $\Lambda=0$, spatial hypersurfaces are flat, $K=0$,
which follows from (\ref{eq:2.11}) and will become explicit in
eq.~(\ref{eq:2.29}) below. We further define
\begin{equation}
  \Omega_\Lambda \equiv \frac{\Lambda}{3H_0^2}\;.
\label{eq:2.17}
\end{equation}
The {\em deceleration parameter\/} $q_0$ is defined by
\begin{equation}
  q_0 = -\frac{\ddot{a}a}{\dot{a}^2}
\label{eq:2.17a}
\end{equation}
at $t=t_0$.

\subsubsection{\label{sc:2.1.5}Matter Models}

For a complete description of the expansion of the Universe, we need
an equation of state $p=p(\rho)$, relating the pressure to the energy
density of the matter. Ordinary matter, which is frequently called
{\em dust\/} in this context, has $p\ll\rho c^2$, while $p=\rho c^2/3$
for radiation or other forms of relativistic matter. Inserting these
expressions into eq.~(\ref{eq:2.13}), we find
\begin{equation}
  \rho(t) = a^{-n}(t)\,\rho_0\;,
\label{eq:2.18}
\end{equation}
with
\begin{equation}
  n = \left\{\begin{array}{ll}
             3 & \hbox{for dust,}\; p=0 \\
             4 & \hbox{for relativistic matter,}\; p=\rho\,c^2/3 \\
	     \end{array}\right.\;.
\label{eq:2.19}
\end{equation}
The energy density of relativistic matter therefore drops more rapidly
with time than that of ordinary matter.

\subsubsection{\label{sc:2.1.6}Relativistic Matter Components}

There are two obvious candidates for relativistic matter today,
photons and neutrinos. The energy density contained in photons today
is determined by the temperature of the Cosmic Microwave Background,
$T_\mathrm{CMB}=2.73\,$K (\cite*{fcg96}). Since the CMB has an
excellent black-body spectrum, its energy density is given by the
Stefan-Boltzmann law,
\begin{equation}
  \rho_\mathrm{CMB} = \frac{1}{c^2}\,\frac{\pi^2}{15}\,
  \frac{(kT_\mathrm{CMB})^4}{(\hbar c)^3}
  \approx 4.5\times10^{-34}\,\mathrm{g}\,\mathrm{cm}^{-3}\;.
\label{eq:2.20}
\end{equation}
In terms of the cosmic density parameter $\Omega_0$
[eq.~(\ref{eq:2.16})], the cosmic density contributed by the photon
background is
\begin{equation}
  \Omega_\mathrm{CMB,0} = 2.4\times10^{-5}\,h^{-2}\;.
\label{eq:2.21}
\end{equation}

Like photons, neutrinos were produced in thermal equilibrium in the
hot early phase of the Universe. Interacting weakly, they decoupled
from the cosmic plasma when the temperature of the Universe was
$kT\approx1\,\mathrm{MeV}$ because later the time-scale of their
leptonic interactions became larger than the expansion time-scale of
the Universe, so that equilibrium could no longer be maintained. When
the temperature of the Universe dropped to
$kT\approx0.5\,\mathrm{MeV}$, electron-positron pairs annihilated to
produce $\gamma$ rays. The annihilation heated up the photons but not
the neutrinos which had decoupled earlier. Hence the neutrino
temperature is lower than the photon temperature by an amount
determined by entropy conservation. The entropy $S_\mathrm{e}$ of the
electron-positron pairs was dumped completely into the entropy of the
photon background $S_\gamma$. Hence,
\begin{equation}
  (S_\mathrm{e}+S_\gamma)_\mathrm{before} =
  (S_\gamma)_\mathrm{after}\;,
\label{eq:2.22}
\end{equation}
where ``before'' and ``after'' refer to the annihilation
time. Ignoring constant factors, the entropy per particle species is
$S\propto g\,T^3$, where $g$ is the statistical weight of the
species. For bosons $g=1$, and for fermions $g=7/8$ per spin
state. Before annihilation, we thus have
$g_\mathrm{before}=4\cdot7/8+2=11/2$, while after the annihilation
$g=2$ because only photons remain. From eq.~(\ref{eq:2.22}),
\begin{equation}
  \left(\frac{T_\mathrm{after}}{T_\mathrm{before}}\right)^3 =
  \frac{11}{4}\;.
\label{eq:2.23}
\end{equation}
After the annihilation, the neutrino temperature is therefore lower
than the photon temperature by the factor $(11/4)^{1/3}$. In
particular, the neutrino temperature today is
\begin{equation}
  T_{\nu,0} = \left(\frac{4}{11}\right)^{1/3}\,T_\mathrm{CMB}
  = 1.95\,\mathrm{K}\;.
\label{eq:2.24}
\end{equation}
Although neutrinos have long been out of thermal equilibrium, their
distribution function remained unchanged since they decoupled, except
that their temperature gradually dropped in the course of cosmic
expansion. Their energy density can thus be computed from a
Fermi-Dirac distribution with temperature $T_\nu$, and be converted to
the equivalent cosmic density parameter as for the photons. The result
is
\begin{equation}
  \Omega_{\nu,0} = 2.8\times10^{-6}\,h^{-2}
\label{eq:2.25}
\end{equation}
per neutrino species.

Assuming three relativistic neutrino species, the total density
parameter in relativistic matter today is
\begin{equation}
  \Omega_\mathrm{R,0} = \Omega_\mathrm{CMB,0} + 
  3\times\Omega_{\nu,0} = 3.2\times10^{-5}\,h^{-2}\;.
\label{eq:2.26}
\end{equation}
Since the energy density in relativistic matter is almost five orders
of magnitude less than the energy density of ordinary matter today if
$\Omega_0$ is of order unity, the expansion of the Universe today is
matter-dominated, or $\rho=a^{-3}(t)\rho_0$. The energy densities of
ordinary and relativistic matter were equal when the scale factor
$a(t)$ was
\begin{equation}
  a_\mathrm{eq} = \frac{\Omega_\mathrm{R,0}}{\Omega_0} =
  3.2\times10^{-5}\,\Omega_0^{-1}\,h^{-2}\;,
\label{eq:2.27}
\end{equation}
and the expansion was radiation-dominated at yet earlier times,
$\rho=a^{-4}\rho_0$. The epoch of equality of matter and radiation
density will turn out to be important for the evolution of structure
in the Universe discussed below.

\subsubsection{\label{sc:2.1.7}Spatial Curvature and Expansion}

With the parameters defined previously, Friedmann's equation
(\ref{eq:2.11}) can be written
\begin{equation}
  H^2(t) = H_0^2\,\left[
  a^{-4}(t)\Omega_\mathrm{R,0} + a^{-3}(t)\Omega_0 -
  a^{-2}(t)\frac{Kc^2}{H_0^2} + \Omega_\Lambda
  \right]\;.
\label{eq:2.28}
\end{equation}
Since $H(t_0)\equiv H_0$, and $\Omega_\mathrm{R,0}\ll\Omega_0$,
eq.~(\ref{eq:2.28}) implies
\begin{equation}
  K = \left(\frac{H_0}{c}\right)^2\,(\Omega_0+\Omega_\Lambda-1)\;,
\label{eq:2.29}
\end{equation}
and eq.~(\ref{eq:2.28}) becomes
\begin{equation}
  H^2(t) = H_0^2\,\left[
  a^{-4}(t)\Omega_\mathrm{R,0} +
  a^{-3}(t)\Omega_0 + 
  a^{-2}(t)(1-\Omega_0-\Omega_\Lambda) +
  \Omega_\Lambda\right]\;.
\label{eq:2.30}
\end{equation}
The curvature of spatial hypersurfaces is therefore determined by the
sum of the density contributions from matter, $\Omega_0$, and from the
cosmological constant, $\Omega_\Lambda$. If
$\Omega_0+\Omega_\Lambda=1$, space is flat, and it is closed or
hyperbolic if $\Omega_0+\Omega_\Lambda$ is larger or smaller than
unity, respectively. The spatial hypersurfaces of a low-density
universe are therefore hyperbolic, while those of a high-density
universe are closed [cf.~eq.~(\ref{eq:2.4})]. A
Friedmann-Lema{\^\i}tre model universe is thus characterised by four
parameters: the expansion rate at present (or Hubble constant) $H_0$,
and the density parameters in matter, radiation, and the cosmological
constant.

Dividing eq.~(\ref{eq:2.12}) by eq.~(\ref{eq:2.11}), using
eq.~(\ref{eq:2.29}), and setting $p=0$, we obtain for the deceleration
parameter $q_0$
\begin{equation}
  q_0 = \frac{\Omega_0}{2} - \Omega_\Lambda\;.
\label{eq:2.31}
\end{equation}

The age of the universe can be determined from eq.~(\ref{eq:2.30}).
Since $\d t=\d a\,\dot{a}^{-1}=\d a(aH)^{-1}$, we have, ignoring
$\Omega_\mathrm{R,0}$,
\begin{equation}
  t_0 = \frac{1}{H_0}\,\int_0^1\,\d a\,\left[
  a^{-1}\Omega_0 + (1-\Omega_0-\Omega_\Lambda) + a^2\Omega_\Lambda
\right]^{-1/2}\;.
\label{eq:2.32}
\end{equation}
It was assumed in this equation that $p=0$ holds for all times $t$,
while pressure is not negligible at early times. The corresponding
error, however, is very small because the universe spends only a very
short time in the radiation-dominated phase where $p>0$.

\begin{figure}[ht]
  \includegraphics[width=\hsize]{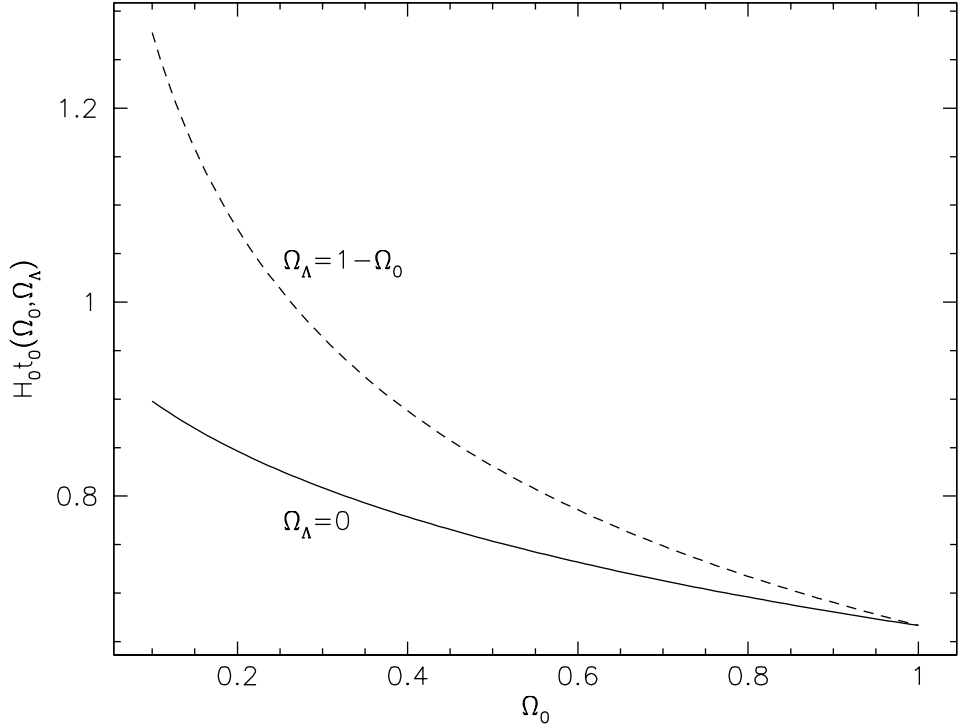}
\caption{Cosmic age $t_0$ in units of $H_0^{-1}$ as a function of
$\Omega_0$, for $\Omega_\Lambda=0$ (solid curve) and
$\Omega_\Lambda=1-\Omega_0$ (dashed curve).}
\label{fig:2.1}
\end{figure}

Figure \ref{fig:2.1} shows $t_0$ in units of $H_0^{-1}$ as a function
of $\Omega_0$, for $\Omega_\Lambda=0$ (solid curve) and
$\Omega_\Lambda=1-\Omega_0$ (dashed curve). The model universe is
older for lower $\Omega_0$ and higher $\Omega_\Lambda$ because the
deceleration decreases with decreasing $\Omega_0$ and the acceleration
increases with increasing $\Omega_\Lambda$.

In principle, $\Omega_\Lambda$ can have either sign. We have
restricted ourselves in Fig.~\ref{fig:2.1} to non-negative
$\Omega_\Lambda$ because the cosmological constant is usually
interpreted as the energy density of the vacuum, which is positive
semi-definite.

The time evolution (\ref{eq:2.30}) of the Hubble function $H(t)$
allows one to determine the dependence of $\Omega$ and
$\Omega_\Lambda$ on the scale function $a$. For a matter-dominated
universe, we find
\begin{eqnarray}
  \Omega(a) &=& \frac{8\pi G}{3H^2(a)}\,\rho_0\,a^{-3} =
  \frac{\Omega_0}{a+\Omega_0(1-a)+\Omega_\Lambda(a^3-a)}\;,
  \nonumber\\
  \Omega_\Lambda(a) &=& \frac{\Lambda}{3H^2(a)} =
  \frac{\Omega_\Lambda\,a^3}
       {a+\Omega_0(1-a)+\Omega_\Lambda(a^3-a)}\;.
\label{eq:2.33}
\end{eqnarray}
These equations show that, whatever the values of $\Omega_0$ and
$\Omega_\Lambda$ are at the present epoch, $\Omega(a)\to1$ and
$\Omega_\Lambda\to0$ for $a\to0$. This implies that for sufficiently
early times, all matter-dominated Friedmann-Lema{\^\i}tre model
universes can be described by Einstein-de Sitter models, for which
$K=0$ and $\Omega_\Lambda=0$. For $a\ll1$, the right-hand side of
Friedmann's equation (\ref{eq:2.30}) is therefore dominated by the
matter and radiation terms because they contain the strongest
dependences on $a^{-1}$. The Hubble function $H(t)$ can then be
approximated by
\begin{equation}
  H(t) = H_0\,\left[\Omega_\mathrm{R,0}\,a^{-4}(t) + 
  \Omega_0\,a^{-3}(t)\right]^{1/2}\;.
\label{eq:2.34}
\end{equation}
Using the definition of $a_\mathrm{eq}$,
$a_\mathrm{eq}^{-4}\Omega_\mathrm{R,0}=a_\mathrm{eq}^{-3}\Omega_0$
[cf.~eq.~(\ref{eq:2.27})], eq.~(\ref{eq:2.34}) can be written
\begin{equation}
  H(t) = H_0\,\Omega_0^{1/2}\,a^{-3/2}\,
  \left(1+\frac{a_\mathrm{eq}}{a}\right)^{1/2}\;.
\label{eq:2.35}
\end{equation}
Hence,
\begin{equation}
  H(t) = H_0\,\Omega_0^{1/2}\,\left\{\begin{array}{ll}
    a_\mathrm{eq}^{1/2}\,a^{-2} & (a\ll a_\mathrm{eq}) \\
    a^{-3/2} & (a_\mathrm{eq}\ll a\ll 1) \\
  \end{array}\right.\;.
\label{eq:2.36}
\end{equation}
Likewise, the expression for the cosmic time reduces to
\begin{equation}
  t(a) = \frac{2}{3H_0}\,\Omega_0^{-1/2}\left[
  a^{3/2}\,\left(1-2\frac{a_\mathrm{eq}}{a}\right)\,
  \left(1+\frac{a_\mathrm{eq}}{a}\right)^{1/2} +
  2\,a_\mathrm{eq}^{3/2}\right]\;,
\label{eq:2.37}
\end{equation}
or
\begin{equation}
  t(a) = \frac{1}{H_0}\,\Omega_0^{-1/2}\,\left\{\begin{array}{ll}
    \frac{1}{2}\,a_\mathrm{eq}^{-1/2}\,a^2 & (a\ll a_\mathrm{eq}) \\
    \frac{2}{3}\,a^{3/2} & (a_\mathrm{eq}\ll a\ll 1) \\
  \end{array}\right.\;.
\label{eq:2.38}
\end{equation}
Equation (\ref{eq:2.35}) is called the Einstein-de Sitter limit of
Friedmann's equation. Where not mentioned otherwise, we consider in
the following only cosmic epochs at times much later than
$t_\mathrm{eq}$, i.e., when $a\gg a_\mathrm{eq}$, where the Universe
is dominated by dust, so that the pressure can be neglected, $p=0$.

\subsubsection{Necessity of a Big Bang}

Starting from $a=1$ at the present epoch and integrating Friedmann's
equation (\ref{eq:2.11}) back in time shows that there are
combinations of the cosmic parameters such that $a>0$ at all
times. Such models would have no Big Bang. The necessity of a Big Bang
is usually inferred from the existence of the cosmic microwave
background, which is most naturally explained by an early, hot phase
of the Universe. \cite{boe88} showed that two simple observational
facts suffice to show that the Universe must have gone through a Big
Bang, if it is properly described by the class of
Friedmann-Lema{\^\i}tre models. Indeed, the facts that there are
cosmological objects at redshifts $z>4$, and that the cosmic density
parameter of non-relativistic matter, as inferred from observed
galaxies and clusters of galaxies is $\Omega_0>0.02$, exclude models
which have $a(t)>0$ at all times. Therefore, if we describe the
Universe at large by Friedmann-Lema{\^\i}tre models, we must assume a
Big Bang, or $a=0$ at some time in the past.

\subsubsection{\label{sc:2.1.8}Distances}

The meaning of ``distance'' is no longer unique in a curved
space-time. In contrast to the situation in Euclidian space, distance
definitions in terms of different measurement prescriptions lead to
different distances. Distance measures are therefore defined in
analogy to relations between measurable quantities in Euclidian
space. We define here four different distance scales, the proper
distance, the comoving distance, the angular-diameter distance, and
the luminosity distance.

Distance measures relate an emission event and an observation event on
two separate geodesic lines which fall on a common light cone, either
the forward light cone of the source or the backward light cone of the
observer. They are therefore characterised by the times $t_2$ and
$t_1$ of emission and observation respectively, and by the structure
of the light cone. These times can uniquely be expressed by the values
$a_2=a(t_2)$ and $a_1=a(t_1)$ of the scale factor, or by the redshifts
$z_2$ and $z_1$ corresponding to $a_2$ and $a_1$. We choose the latter
parameterisation because redshifts are directly observable. We also
assume that the observer is at the origin of the coordinate system.

The {\em proper distance\/} $D_\mathrm{prop}(z_1,z_2)$ is the distance
measured by the travel time of a light ray which propagates from a
source at $z_2$ to an observer at $z_1<z_2$. It is defined by $\d
D_\mathrm{prop}=-c\d t$, hence $\d D_\mathrm{prop}=-c\d
a\dot{a}^{-1}=-c\d a(aH)^{-1}$. The minus sign arises because, due to
the choice of coordinates centred on the observer, distances increase
away from the observer, while the time $t$ and the scale factor $a$
increase towards the observer. We get
\begin{equation}
  D_\mathrm{prop}(z_1,z_2) =
  \frac{c}{H_0}\,\int_{a(z_2)}^{a(z_1)}\left[
  a^{-1}\Omega_0 + (1-\Omega_0-\Omega_\Lambda) + a^2\Omega_\Lambda
  \right]^{-1/2}\d a\;.
\label{eq:2.39}
\end{equation}

The {\em comoving distance\/} $D_\mathrm{com}(z_1,z_2)$ is the
distance on the spatial hypersurface $t=t_0$ between the worldlines of
a source and an observer comoving with the cosmic flow. Due to the
choice of coordinates, it is the coordinate distance between a source
at $z_2$ and an observer at $z_1$, $\d D_\mathrm{com}=\d w$. Since
light rays propagate with $\d s=0$, we have $c\d t=-a\d w$ from the
metric, and therefore $\d D_\mathrm{com}=-a^{-1}c\d t=-c\d
a(a\dot{a})^{-1}=-c\d a(a^2H)^{-1}$. Thus
\begin{eqnarray}
  D_\mathrm{com}(z_1,z_2) &=&
  \frac{c}{H_0}\,\int_{a(z_2)}^{a(z_1)}\left[
  a\Omega_0 + a^2(1-\Omega_0-\Omega_\Lambda) + a^4\Omega_\Lambda
  \right]^{-1/2}\d a \nonumber \\
  &=& w(z_1,z_2) \; .
\label{eq:2.40}
\end{eqnarray}

The {\em angular-diameter distance\/} $D_\mathrm{ang}(z_1,z_2)$ is
defined in analogy to the relation in Euclidian space between the
physical cross section $\delta A$ of an object at $z_2$ and the solid
angle $\delta\omega$ that it subtends for an observer at $z_1$,
$\delta\omega D_\mathrm{ang}^2=\delta A$. Hence,
\begin{equation}
  \frac{\delta A}{4\pi a^2(z_2)\,f_K^2[w(z_1,z_2)]} =
  \frac{\delta\omega}{4\pi}\;,
\label{eq:2.41}
\end{equation}
where $a(z_2)$ is the scale factor at emission time and
$f_K[w(z_1,z_2)]$ is the radial coordinate distance between the
observer and the source. It follows
\begin{equation}
  D_\mathrm{ang}(z_1,z_2) = 
  \left(\frac{\delta A}{\delta\omega}\right)^{1/2} = 
  a(z_2)\,f_K[w(z_1,z_2)]\;.
\label{eq:2.42}
\end{equation}
According to the definition of the comoving distance, the
angular-diameter distance therefore is
\begin{equation}
  D_\mathrm{ang}(z_1,z_2) = a(z_2)\,
  f_K[D_\mathrm{com}(z_1,z_2)]\;.
\label{eq:2.43}
\end{equation}

The {\em luminosity distance\/} $D_\mathrm{lum}(a_1,a_2)$ is defined
by the relation in Euclidian space between the luminosity $L$ of an
object at $z_2$ and the flux $S$ received by an observer at $z_1$. It
is related to the angular-diameter distance through
\begin{equation}
  D_\mathrm{lum}(z_1,z_2) =
  \left(\frac{a(z_1)}{a(z_2)}\right)^2\,
  D_\mathrm{ang}(z_1,z_2) =
  \frac{a(z_1)^2}{a(z_2)}\,f_K[D_\mathrm{com}(z_1,z_2)]\;.
\label{eq:2.44}
\end{equation}
The first equality in (\ref{eq:2.44}), which is due to \cite{ET33.1},
is valid in arbitrary space-times. It is physically intuitive because
photons are redshifted by $a(z_1)a(z_2)^{-1}$, their arrival times are
delayed by another factor $a(z_1)a(z_2)^{-1}$, and the area of the
observer's sphere on which the photons are distributed grows between
emission and absorption in proportion to $[a(z_1)a(z_2)^{-1}]^2$. This
accounts for a total factor of $[a(z_1)a(z_2)^{-1}]^4$ in the flux,
and hence for a factor of $[a(z_1)a(z_2)^{-1}]^2$ in the distance
relative to the angular-diameter distance.

\begin{figure}[ht]
  \includegraphics[width=\hsize]{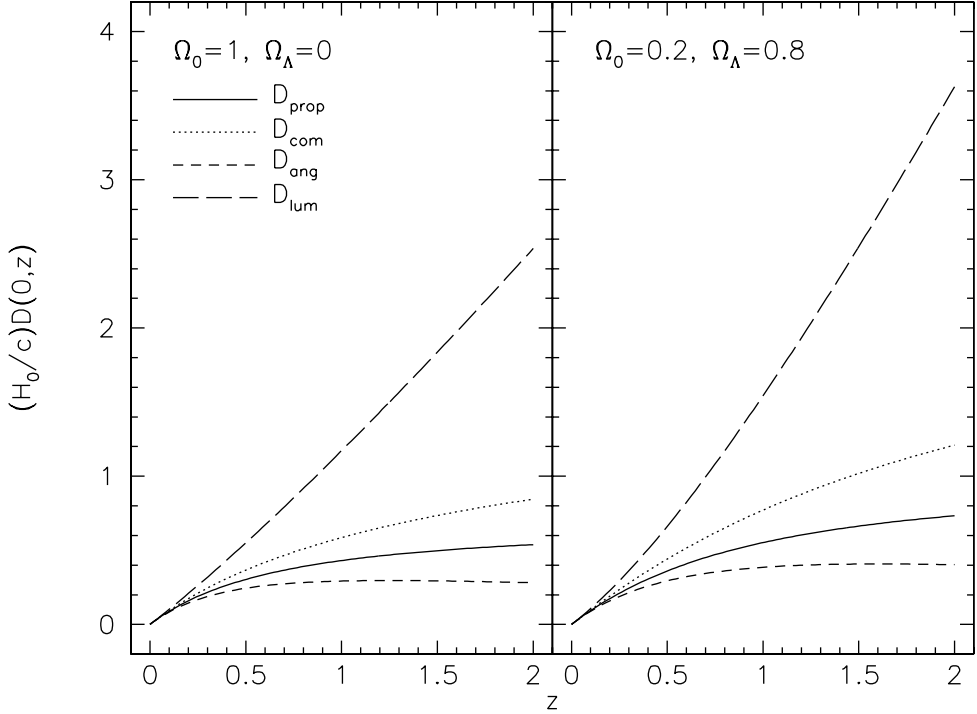}
\caption{Four distance measures are plotted as a function of source
redshift for two cosmological models and an observer at redshift
zero. These are the proper distance $D_\mathrm{prop}$ (1, solid line),
the comoving distance $D_\mathrm{com}$ (2, dotted line), the
angular-diameter distance $D_\mathrm{ang}$ (3, short-dashed line), and
the luminosity distance $D_\mathrm{lum}$ (4, long-dashed line).}
\label{fig:2.2}
\end{figure}

We plot the four distances $D_\mathrm{prop}$, $D_\mathrm{com}$,
$D_\mathrm{ang}$, and $D_\mathrm{lum}$ for $z_1=0$ as a function of
$z$ in Fig.~\ref{fig:2.2}.

The distances are larger for lower cosmic density and higher
cosmological constant. Evidently, they differ by a large amount at
high redshift. For small redshifts, $z\ll1$, they all follow the
Hubble law,
\begin{equation}
  \hbox{distance} = \frac{cz}{H_0} + O(z^2) \;.
\label{eq:2.45}
\end{equation}

\subsubsection{\label{sc:2.1.9}The Einstein-de Sitter Model}

In order to illustrate some of the results obtained above, let us now
specialise to a model universe with a critical density of dust,
$\Omega_0=1$ and $p=0$, and with zero cosmological constant,
$\Omega_\Lambda=0$. Friedmann's equation then reduces to
$H(t)=H_0\,a^{-3/2}$, and the age of the Universe becomes
$t_0=2(3H_0)^{-1}$. The distance measures are
\begin{eqnarray}
  D_\mathrm{prop}(z_1,z_2) &=&
  \frac{2c}{3H_0}\left[(1+z_1)^{-3/2}-(1+z_2)^{-3/2}\right]
  \\
  D_\mathrm{com}(z_1,z_2) &=&
  \frac{2c}{H_0}\left[(1+z_1)^{-1/2}-(1+z_2)^{-1/2}\right]
  \nonumber\\
  D_\mathrm{ang}(z_1,z_2) &=&
  \frac{2c}{H_0}\,\frac{1}{1+z_2}
  \left[(1+z_1)^{-1/2}-(1+z_2)^{-1/2}\right]
  \nonumber\\
  D_\mathrm{lum}(z_1,z_2) &=&
  \frac{2c}{H_0}\,\frac{1+z_2}{(1+z_1)^2}
  \left[(1+z_1)^{-1/2}-(1+z_2)^{-1/2}\right]\;.\nonumber
\label{eq:2.46}
\end{eqnarray}

\subsection{\label{sc:2.2}Density Perturbations}

The standard model for the formation of structure in the Universe
assumes that there were small fluctuations at some very early initial
time, which grew by gravitational instability. Although the origin of
the seed fluctuations is yet unclear, they possibly originated from
quantum fluctuations in the very early Universe, which were blown up
during a later inflationary phase. The fluctuations in this case are
uncorrelated and the distribution of their amplitudes is
Gaussian. Gravitational instability leads to a growth of the
amplitudes of the relative density fluctuations. As long as the
relative density contrast of the matter fluctuations is much smaller
than unity, they can be considered as small perturbations of the
otherwise homogeneous and isotropic background density, and linear
perturbation theory suffices for their description.

The linear theory of density perturbations in an expanding universe is
generally a complicated issue because it needs to be relativistic
(e.g.~\cite*{lif46}; \cite*{bar80}). The reason is that perturbations
on any length scale are comparable to or larger than the size of the
horizon\footnote{\label{fn:2.1}In this context, the size of the
horizon is the distance $ct$ by which light can travel in the time $t$
since the big bang.} at sufficiently early times, and then Newtonian
theory ceases to be applicable. In other words, since the horizon
scale is comparable to the curvature radius of space-time, Newtonian
theory fails for larger-scale perturbations due to non-zero spacetime
curvature. The main features can nevertheless be understood by fairly
simple reasoning. We shall not present a rigourous mathematical
treatment here, but only quote the results which are relevant for our
later purposes. For a detailed qualitative and quantitative
discussion, we refer the reader to the excellent discussion in
chapter~4 of the book by \cite{PA93.2}.

\subsubsection{\label{sc:2.2.1}Horizon Size}

The size of causally connected regions in the Universe is called the
{\em horizon size\/}. It is given by the distance by which a photon
can travel in the time $t$ since the Big Bang. Since the appropriate
time scale is provided by the inverse Hubble parameter $H^{-1}(a)$,
the horizon size is $d'_\mathrm{H}=c\,H^{-1}(a)$, and the {\em
comoving\/} horizon size is
\begin{equation}
  d_\mathrm{H} = \frac{c}{a\,H(a)} = \frac{c}{H_0}\,\Omega_0^{-1/2}\,
  a^{1/2}\,\left(1 + \frac{a_\mathrm{eq}}{a}\right)^{-1/2}\;,
\label{eq:2.47}
\end{equation}
where we have inserted the Einstein-de Sitter limit (\ref{eq:2.35}) of
Friedmann's equation. The length $c\,H_0^{-1}=3\,h^{-1}\,\mathrm{Gpc}$
is called the {\em Hubble radius\/}. We shall see later that the
horizon size at $a_\mathrm{eq}$ plays a very important r\^ole for
structure formation. Inserting $a=a_\mathrm{eq}$ into
eq.~(\ref{eq:2.47}), yields
\begin{equation}
  d_\mathrm{H}(a_\mathrm{eq}) = \frac{c}{\sqrt{2}\,H_0}\,
  \Omega_0^{-1/2}\,a_\mathrm{eq}^{1/2} \approx
  12\,(\Omega_0\,h^2)^{-1}\,\mathrm{Mpc}\;,
\label{eq:2.48}
\end{equation}
where $a_\mathrm{eq}$ from eq.~(\ref{eq:2.27}) has been inserted.

\subsubsection{\label{sc:2.2.2}Linear Growth of Density Perturbations}

\begin{figure}[ht]
  \includegraphics[width=\hsize]{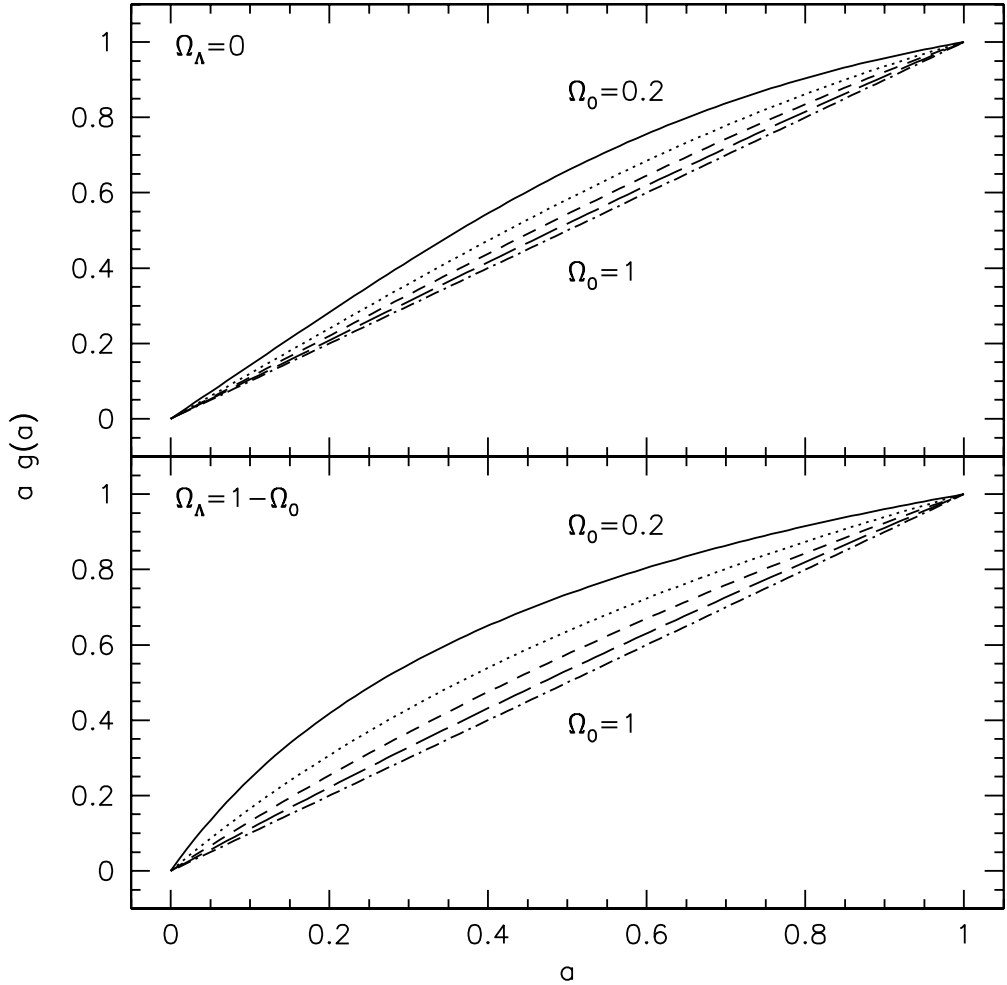}
\caption{The growth function $a\,g(a)\equiv a\,g'(a)/g'(1)$ given in
eqs.~(\ref{eq:2.51}) and (\ref{eq:2.52}) for $\Omega_0$ between $0.2$
and $1.0$ in steps of $0.2$. Top panel: $\Omega_\Lambda=0$; bottom
panel: $\Omega_\Lambda=1-\Omega_0$. The growth rate is constant for
the Einstein-de Sitter model ($\Omega_0=1$, $\Omega_\Lambda=0$), while
it is higher for $a\ll1$ and lower for $a\approx1$ for low-$\Omega_0$
models. Consequently, structure forms earlier in low- than in
high-$\Omega_0$ models.}
\label{fig:2.3}
\end{figure}

We adopt the commonly held view that the density of the Universe is
dominated by weakly interacting dark matter at the relatively late
times which are relevant for weak gravitational lensing, $a\gg
a_\mathrm{eq}$. Dark-matter perturbations are characterised by the
density contrast
\begin{equation}
  \delta(\vec x,a) =
  \frac{\rho(\vec x,a)-\bar{\rho}(a)}{\bar{\rho}(a)}\;,
\label{eq:2.49}
\end{equation}
where $\bar{\rho}=\rho_0\,a^{-3}$ is the average cosmic
density. Relativistic and non-relativistic perturbation theory shows
that linear density fluctuations, i.e.~perturbations with
$\delta\ll1$, grow like
\begin{equation}
  \delta(a) \propto a^{n-2} = \left\{\begin{array}{ll}
    a^2 & \mathrm{before}\; a_\mathrm{eq} \\
    a   & \mathrm{after} \; a_\mathrm{eq} \\
  \end{array}\right.
\label{eq:2.50}
\end{equation}
as long as the Einstein-de Sitter limit holds. For later times, $a\gg
a_\mathrm{eq}$, when the Einstein-de Sitter limit no longer applies if
$\Omega_0\ne1$ or $\Omega_\Lambda\ne0$, the linear growth of density
perturbations is changed according to
\begin{equation}
  \delta(a) = \delta_0\,a\,\frac{g'(a)}{g'(1)} \equiv
  \delta_0\,a\,g(a)\;,
\label{eq:2.51}
\end{equation}
where $\delta_0$ is the density contrast linearly extrapolated to the
present epoch, and the density-dependent growth function $g'(a)$ is
accurately fit by (\cite*{CA92.1})
\begin{equation}
  g'(a;\Omega_0,\Omega_\Lambda) = \frac{5}{2}\,\Omega(a)\,
  \left[\Omega^{4/7}(a) - \Omega_\Lambda(a) +
  \left(1+\frac{\Omega(a)}{2}\right)
  \left(1+\frac{\Omega_\Lambda(a)}{70}\right)\right]^{-1}\;.
\label{eq:2.52}
\end{equation}
The dependence of $\Omega$ and $\Omega_\Lambda$ on the scale factor
$a$ is given in eqs.~(\ref{eq:2.33}). The growth function
$a\,g(a;\Omega_0,\Omega_\Lambda)$ is shown in Fig.~\ref{fig:2.3} for a
variety of parameters $\Omega_0$ and $\Omega_\Lambda$.

The cosmic microwave background reveals relative temperature
fluctuations of order $10^{-5}$ on large scales. By the Sachs-Wolfe
effect (\cite*{SA67.1}), these temperature fluctuations reflect
density fluctuations of the same order of magnitude. The cosmic
microwave background originated at $a\approx10^{-3}\gg a_\mathrm{eq}$,
well after the Universe became
matter-dominated. Equation~(\ref{eq:2.50}) then implies that the
density fluctuations today, expected from the temperature fluctuations
at $a\approx10^{-3}$, should only reach a level of $10^{-2}$. Instead,
structures (e.g.~galaxies) with $\delta\gg1$ are observed. How can
this discrepancy be resolved? The cosmic microwave background displays
fluctuations in the baryonic matter component only. If there is an
additional matter component that only couples through weak
interactions, fluctuations in that component could grow as soon as it
decoupled from the cosmic plasma, well before photons decoupled from
baryons to set the cosmic microwave background free. Such fluctuations
could therefore easily reach the amplitudes observed today, and
thereby resolve the apparent mismatch between the amplitudes of the
temperature fluctuations in the cosmic microwave background and the
present cosmic structures. This is one of the strongest arguments for
the existence of a dark matter component in the Universe.

\subsubsection{\label{sc:2.2.3}Suppression of Growth}

It is convenient to decompose the density contrast $\delta$ into
Fourier modes. In linear perturbation theory, individual Fourier
components evolve independently. A perturbation of (comoving)
wavelength $\lambda$ is said to ``enter the horizon'' when
$\lambda=d_\mathrm{H}(a)$. If $\lambda<d_\mathrm{H}(a_\mathrm{eq})$,
the perturbation enters the horizon while radiation is still
dominating the expansion. Until $a_\mathrm{eq}$, the expansion
time-scale, $t_\mathrm{exp}=H^{-1}$, is determined by the radiation
density $\rho_\mathrm{R}$, which is shorter than the collapse
time-scale of the dark matter, $t_\mathrm{DM}$:
\begin{equation}
  t_\mathrm{exp} \sim (G\rho_\mathrm{R})^{-1/2} <
  (G\rho_\mathrm{DM})^{-1/2} \sim t_\mathrm{DM}\;.
\label{eq:2.53}
\end{equation}
In other words, the fast radiation-driven expansion prevents
dark-matter perturbations from collapsing. Light can only cross
regions that are smaller than the horizon size. The suppression of
growth due to radiation is therefore restricted to scales smaller than
the horizon, and larger-scale perturbations remain unaffected. This
explains why the horizon size at $a_\mathrm{eq}$,
$d_\mathrm{H}(a_\mathrm{eq})$, sets an important scale for structure
growth.

\begin{figure}[ht]
  \includegraphics[width=\hsize]{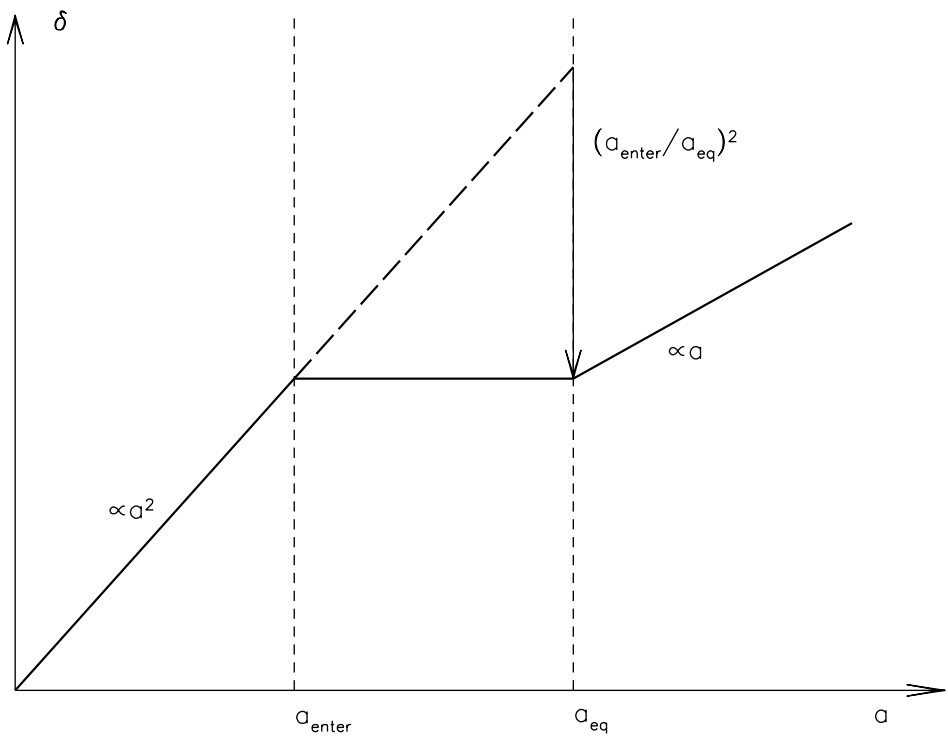}
\caption{Sketch illustrating the suppression of structure growth
during the radiation-dominated phase. The perturbation grows $\propto
a^2$ before $a_\mathrm{eq}$, and $\propto a$ thereafter. If the
perturbation is smaller than the horizon at $a_\mathrm{eq}$, it enters
the horizon at $a_\mathrm{enter}<a_\mathrm{eq}$ while radiation is
still dominating. The rapid radiation-driven expansion prevents the
perturbation from growing further. Hence it stalls until
$a_\mathrm{eq}$. By then, its amplitude is smaller by
$f_\mathrm{sup}=(a_\mathrm{enter}/a_\mathrm{eq})^2$ than it would be
without suppression.}
\label{fig:2.4}
\end{figure}

Figure~\ref{fig:2.4} illustrates the growth of a perturbation with
$\lambda<d_\mathrm{H}(a_\mathrm{eq})$, that is small enough to enter
the horizon at $a_\mathrm{enter}<a_\mathrm{eq}$. It can be read off
from the figure that such perturbations are suppressed by the factor
\begin{equation}
  f_\mathrm{sup} = \left(
  \frac{a_\mathrm{enter}}{a_\mathrm{eq}}\right)^2\;.
\label{eq:2.54}
\end{equation}

It remains to be evaluated at what time $a_\mathrm{enter}$ a density
perturbation with comoving wavelength $\lambda$ enters the
horizon. The condition is
\begin{equation}
  \lambda = d_\mathrm{H}(a_\mathrm{enter}) =
  \frac{c}{a_\mathrm{enter}\,H(a_\mathrm{enter})}\;.
\label{eq:2.55}
\end{equation}
Well in the Einstein-de Sitter regime, the Hubble parameter is given
by eq.~(\ref{eq:2.36}). Inserting that expression into (\ref{eq:2.55})
yields
\begin{equation}
  \lambda \propto \left\{\begin{array}{ll}
    a_\mathrm{enter} & (a_\mathrm{enter}\ll a_\mathrm{eq}) \\
    a^{1/2}_\mathrm{enter} &
    (a_\mathrm{eq}\ll a_\mathrm{enter}\ll 1) \\
  \end{array}\right.\;.
\label{eq:2.56}
\end{equation}
Let now $k=\lambda^{-1}$ be the wave number of the perturbation, and
$k_0=d_\mathrm{H}^{-1}(a_\mathrm{eq})$ the wave number corresponding
to the horizon size at $a_\mathrm{eq}$. The suppression factor
(\ref{eq:2.54}) can then be written
\begin{equation}
  f_\mathrm{sup} = \left(\frac{k_0}{k}\right)^2\;.
\label{eq:2.57}
\end{equation}
From eq.~(\ref{eq:2.48}),
\begin{equation}
  k_0 \approx 0.083\,(\Omega_0\,h^2)\,\mathrm{Mpc}^{-1} \approx
  250\,(\Omega_0\,h)\,\hbox{(Hubble radii)}^{-1}\;.
\label{eq:2.58}
\end{equation}

\subsubsection{\label{sc:2.2.4}Density Power Spectrum}

The assumed Gaussian density fluctuations $\delta(\vec x)$ at the
comoving position $\vec x$ can completely be characterised by their
power spectrum $P_\delta(k)$, which can be defined by (see
Sect.~\ref{sc:2.4})
\begin{equation}
  \left\langle\hat{\delta}(\vec k)
              \hat{\delta}^*(\vec k')\right\rangle =
  (2\pi)^3\,\delta_\mathrm{D}(\vec k-\vec k')\,P_\delta(k)\;,
\label{eq:2.59}
\end{equation}
where $\hat\delta(\vec k)$ is the Fourier transform of $\delta$, and
the asterisk denotes complex conjugation. Strictly speaking, the
Fourier decomposition is valid only in flat space. However, at early
times space is flat in any cosmological model, and at late times the
interesting scales $k^{-1}$ of the density perturbations are much
smaller than the curvature radius of the Universe. Hence, we can apply
Fourier decomposition here.

Consider now the primordial perturbation spectrum at some very early
time, $P_\mathrm{i}(k)=|\delta^2_\mathrm{i}(k)|$. Since the density
contrast grows as $\delta\propto a^{n-2}$ [eq.~(\ref{eq:2.50})], the
spectrum grows as $P_\delta(k)\propto a^{2(n-2)}$. At
$a_\mathrm{enter}$, the spectrum has therefore changed to
\begin{equation}
  P_\mathrm{enter}(k)\propto a_\mathrm{enter}^{2(n-2)}\,
  P_\mathrm{i}(k)\propto k^{-4}\,P_\mathrm{i}(k)\;.
\label{eq:2.60}
\end{equation}
where eq.~(\ref{eq:2.56}) was used for $k\gg k_0$.

It is commonly assumed that the total power of the density
fluctuations at $a_\mathrm{enter}$ should be scale-invariant. This
implies $k^3\,P_\mathrm{enter}(k)=\mathrm{const.}$, or
$P_\mathrm{enter}(k)\propto k^{-3}$. Accordingly, the primordial
spectrum has to scale with $k$ as $P_\mathrm{i}(k)\propto k$. This
{\em scale-invariant\/} spectrum is called the {\em
Harrison-Zel'dovich\/} spectrum (\cite*{har70}; \cite*{pey70};
\cite*{zel72}). Combining that with the suppression of small-scale
modes (\ref{eq:2.57}), we arrive at
\begin{equation}
  P_\delta(k) \propto \left\{\begin{array}{ll}
    k & \hbox{for $k\ll k_0$}\\
    k^{-3} & \hbox{for $k\gg k_0$}\\
  \end{array}\right.\;.
\label{eq:2.61}
\end{equation}

An additional complication arises when the dark matter consists of
particles moving with a velocity comparable to the speed of light. In
order to keep them gravitationally bound, density perturbations then
have to have a certain minimum mass, or equivalently a certain minimum
size. All perturbations smaller than that size are damped away by free
streaming of particles. Consequently, the density perturbation
spectrum of such particles has an exponential cut-off at large
$k$. This clarifies the distinction between {\em hot\/} and {\em
cold\/} dark matter: Hot dark matter (HDM) consists of fast particles
that damp away small-scale perturbations, while cold dark matter (CDM)
particles are slow enough to cause no significant damping.

\subsubsection{\label{sc:2.2.5}Normalisation of the Power Spectrum}

Apart from the shape of the power spectrum, its normalisation has to
be fixed. Several methods are available which usually yield different
answers:

\begin{enumerate}

\item Normalisation by microwave-background anisotropies: The COBE
satellite has measured fluctuations in the temperature of the
microwave sky at the {\em rms\/} level of $\Delta
T/T\sim1.3\times10^{-5}$ at an angular scale of $\sim7^\circ$
(\cite*{bgb97}). Adopting a shape for the power spectrum, these
fluctuations can be translated into an amplitude for $P_\delta(k)$.
Due to the large angular scale of the measurement, this kind of
amplitude determination specifies the amplitude on large physical
scales (small $k$) only. In addition, microwave-background
fluctuations measure the amplitude of scalar {\em and\/} tensor
perturbation modes, while the growth of density fluctuations is
determined by the fluctuation amplitude of scalar modes only.

\item Normalisation by the local variance of galaxy counts, pioneered
by \cite{dap83}: Galaxies are supposed to be biased tracers of
underlying dark-matter fluctuations (\cite*{kai84}; \cite*{bbk86};
\cite*{wde87}). By measuring the local variance of galaxy counts
within certain volumes, and assuming an expression for the bias, the
amplitude of dark-matter fluctuations can be inferred. Conventionally,
the variance of galaxy counts $\sigma_{8,\mathrm{galaxies}}$ is
measured within spheres of radius $8\,h^{-1}\,\mathrm{Mpc}$, and the
result is $\sigma_{8,\mathrm{galaxies}}\approx1$. The problem of
finding the corresponding variance $\sigma_8$ of matter-density
fluctuations is that the exact bias mechanism of galaxy formation is
still under debate (e.g.~\cite*{kns97}).

\item Normalisation by the local abundance of galaxy clusters
(\cite*{wef93}; \cite*{EK96.1}; \cite*{vil96}): Galaxy clusters form
by gravitational instability from dark-matter density
perturbations. Their spatial number density reflects the amplitude of
appropriate dark-matter fluctuations in a very sensitive manner. It is
therefore possible to determine the amplitude of the power spectrum by
demanding that the local spatial number density of galaxy clusters be
reproduced. Typical scales for dark-matter fluctuations collapsing to
galaxy clusters are of order $10\,h^{-1}\,\mathrm{Mpc}$, hence cluster
normalisation determines the amplitude of the power spectrum on just
that scale.

\end{enumerate}

\begin{figure}[ht]
  \includegraphics[width=\hsize]{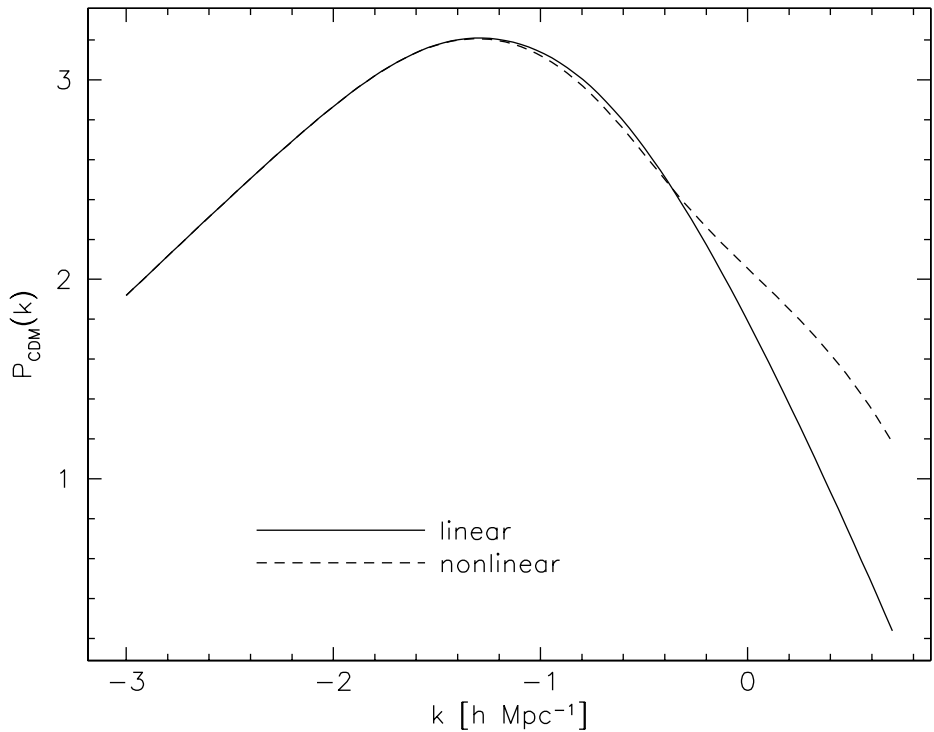}
\caption{CDM power spectrum, normalised to the local abundance of
galaxy clusters, for an Einstein-de Sitter universe with $h=0.5$. Two
curves are displayed. The solid curve shows the linear, the dashed
curve the non-linear power spectrum. While the linear power spectrum
asymptotically falls off $\propto k^{-3}$, the non-linear power
spectrum, according to \protect\cite{PE96.3}, illustrates the
increased power on small scales due to non-linear effects, at the
expense of larger-scale structures.}
\label{fig:2.5}
\end{figure}

Since gravitational lensing by large-scale structures is generally
sensitive to scales comparable to
$k_0^{-1}\sim12\,(\Omega_0\,h^2)\,\mathrm{Mpc}$, cluster normalisation
appears to be the most appropriate normalisation method for the
present purposes. The solid curve in Fig.~\ref{fig:2.5} shows the CDM
power spectrum, linearly and non-linearly evolved to $z=0$ (or $a=1$)
in an Einstein-de Sitter universe with $h=0.5$, normalised to the
local cluster abundance.

\subsubsection{\label{sc:2.2.6}Non-Linear Evolution}

At late stages of the evolution and on small scales, the growth of
density fluctuations begins to depart from the linear behaviour of
eq.~(\ref{eq:2.51}). Density fluctuations grow non-linear, and
fluctuations of different size interact. Generally, the evolution of
$P(k)$ then becomes complicated and needs to be evaluated
numerically. However, starting from the bold {\em ansatz\/} that the
two-point correlation functions in the linear and non-linear regimes
are related by a general scaling relation (\cite*{hmk91}), which turns
out to hold remarkably well, analytic formulae describing the
non-linear behaviour of $P(k)$ have been derived (\cite*{jmw95};
\cite*{PE96.3}). It will turn out in subsequent chapters that the
non-linear evolution of the density fluctuations is crucial for
accurately calculating weak-lensing effects by large-scale
structures. As an example, we show as the dashed curve in
Fig.~\ref{fig:2.5} the CDM power spectrum in an Einstein-de Sitter
universe with $h=0.5$, normalised to the local cluster abundance,
non-linearly evolved to $z=0$. The non-linear effects are immediately
apparent: While the spectrum remains unchanged for large scales ($k\ll
k_0$), the amplitude on small scales ($k\gg k_0$) is substantially
increased at the expense of scales just above the peak. It should be
noted that non-linearly evolved density fluctuations are no longer
fully characterised by the power spectrum only, because then
non-Gaussian features develop.

\subsubsection{\label{sc:2.2.7}Poisson's Equation}

Localised density perturbations which are much smaller than the
horizon and whose peculiar velocities relative to the mean motion in
the Universe are much smaller than the speed of light, can be
described by Newtonian gravity. Their gravitational potential obeys
Poisson's equation,
\begin{equation}
  \nabla_r^2\Phi' = 4\pi G\rho\;,
\label{eq:2.62}
\end{equation}
where $\rho=(1+\delta)\bar\rho$ is the total matter density, and
$\Phi'$ is the sum of the potentials of the smooth background
$\bar\Phi$ and the potential of the perturbation $\Phi$. The gradient
$\nabla_r$ operates with respect to the physical, or proper,
coordinates. Since Poisson's equation is linear, we can subtract the
background contribution $\nabla_r^2\bar\Phi=4\pi
G\bar\rho$. Introducing the gradient with respect to comoving
coordinates $\nabla_x=a\nabla_r$, we can write eq.~(\ref{eq:2.62}) in
the form
\begin{equation}
  \nabla_x^2\Phi = 4\pi G\,a^2\,\bar\rho\,\delta\;.
\label{eq:2.63}
\end{equation}
In the matter-dominated epoch, $\bar\rho=a^{-3}\bar\rho_0$. With the
critical density (\ref{eq:2.15}), Poisson's equation can be re-written
as
\begin{equation}
  \nabla_x^2\Phi = \frac{3H_0^2}{2a}\Omega_0\delta\;.
\label{eq:2.64}
\end{equation}

\subsection{\label{sc:2.3}Relevant Properties of Lenses and Sources}

Individual reviews have been written on galaxies (e.g.~\cite*{fag79};
\cite*{bst88}; \cite*{gih91}; \cite*{kok92}; \cite*{ell97}), clusters
of galaxies (e.g.~\cite*{bah77}; \cite*{roo81}; \cite*{foj82};
\cite*{bah88}; \cite*{sar86}), and active galactic nuclei
(e.g.~\cite*{ree84}; \cite*{WE86.2}; \cite*{bnw90}; \cite*{has90};
\cite*{wah90}; \cite*{ant93}; \cite*{pet97}). A detailed presentation
of these objects is not the purpose of this review. It suffices here
to summarise those properties of these objects that are relevant for
understanding the following discussion. Properties and peculiarities
of individual objects are not necessary to know; rather, we need to
specify the objects statistically. This section will therefore focus
on a statistical description, leaving subtleties aside.

\subsubsection{\label{sc:2.3.1}Galaxies}

For the purposes of this review, we need to characterise the
statistical properties of galaxies as a class. Galaxies can broadly be
grouped into two populations, dubbed {\em early-type\/} and {\em
late-type\/} galaxies, or {\em ellipticals\/} and {\em spirals\/},
respectively. While spiral galaxies include disks structured by more
or less pronounced spiral arms, and approximately spherical bulges
centred on the disk centre, elliptical galaxies exhibit amorphous
projected light distributions with roughly elliptical isophotes. There
are, of course, more elaborate morphological classification schemes
(e.g.~\cite*{ddc91}; \cite*{bmd94}; \cite*{nlb95}; \cite*{nls95}), but
the broad distinction between ellipticals and spirals suffices for
this review.

Outside galaxy clusters, the galaxy population consists of about $3/4$
spiral galaxies and $1/4$ elliptical galaxies, while the fraction of
ellipticals increases towards cluster centres. Elliptical galaxies are
typically more massive than spirals. They contain little gas, and
their stellar population is older, and thus `redder', than in spiral
galaxies. In spirals, there is a substantial amount of gas in the
disk, providing the material for ongoing formation of new
stars. Likewise, there is little dust in ellipticals, but possibly
large amounts of dust are associated with the gas in spirals.

Massive galaxies have of order $10^{11}$~solar masses, or
$2\times10^{44}\,\mathrm{g}$ within their visible radius. Such
galaxies have luminosities of order $10^{10}$~times the solar
luminosity. The kinematics of the stars, gas and molecular clouds in
galaxies, as revealed by spectroscopy, indicate that there is a
relation between the characteristic velocities inside galaxies and
their luminosity (\cite*{faj76}; \cite*{tuf77}); brighter galaxies
tend to have larger masses.

The differential luminosity distribution of galaxies can very well be
described by the functional form
\begin{equation}
  \Phi(L)\,\frac{\d L}{L_*} = 
  \Phi_0\,\left(\frac{L}{L_*}\right)^{-\nu}\,
  \exp\left(-\frac{L}{L_*}\right)\,\frac{\d L}{L_*}\;,
\label{eq:2.65}
\end{equation}
proposed by \cite{sch76}. The parameters have been measured to be
\begin{equation}
  \nu\approx1.1\;,\quad L_*\approx1.1\times10^{10}\,L_\odot\;,\quad
  \Phi_*\approx1.5\times10^{-2}\,h^3\,\mathrm{Mpc}^{-3}
\label{eq:2.66}
\end{equation}
(e.g.~\cite*{eep88}; \cite*{mgh94}; \cite*{mhg94}). This distribution
means that there is essentially a sharp cut-off in the galaxy
population above luminosities of $\sim L_*$, and the mean separation
between $L_*$-galaxies is of order
$\sim\Phi_*^{-1/3}\approx4\,h^{-1}\,\mathrm{Mpc}$.

The stars in elliptical galaxies have randomly oriented orbits, while
by far the most stars in spirals have orbits roughly coplanar with the
galactic disks. Stellar velocities are therefore characterised by a
velocity dispersion $\sigma_\mathrm{v}$ in ellipticals, and by an
asymptotic circular velocity $v_\mathrm{c}$ in
spirals.\footnote{\label{fn:2.2a}The circular velocity of stars and
gas in spiral galaxies turns out to be fairly independent of radius,
except close to their centre. These flat rotations curves cannot be
caused by the observable matter in these galaxies, but provide strong
evidence for the presence of a dark halo, with density profile
$\rho\propto r^{-2}$ at large radii.} These characteristic velocities
are related to galaxy luminosities by laws of the form
\begin{equation}
  \frac{\sigma_\mathrm{v}}{\sigma_{\mathrm{v},*}} =
  \left(\frac{L}{L_*}\right)^{1/\alpha} =
  \frac{v_\mathrm{c}}{v_{\mathrm{c},*}}\;,
\label{eq:2.67}
\end{equation}
where $\alpha$ ranges around $3-4$. For spirals, eq.~(\ref{eq:2.67})
is called Tully-Fisher (\cite*{tuf77}) relation, for ellipticals
Faber-Jackson (\cite*{faj76}) relation. Both velocity scales
$\sigma_{\mathrm{v},*}$ and $v_{\mathrm{c},*}$ are of order
$220\,\mathrm{km\,s}^{-1}$. Since $v_\mathrm{c}=\sqrt{2}\sigma_v$,
ellipticals with the same luminosity are more massive than spirals.

Most relevant for weak gravitational lensing is a population of faint
galaxies emitting bluer light than local galaxies, the so-called {\em
faint blue galaxies\/} (\cite*{TY88.1}; see \cite*{ell97} for a
review). There are of order $30-50$ such galaxies per square arc
minute on the sky which can be mapped with current ground-based
optical telescopes, i.e.~there are $\approx20,000-40,000$ such
galaxies on the area of the full moon. The picture that the sky is
covered with a `wall paper' of those faint and presumably distant blue
galaxies is therefore justified. It is this fine-grained pattern on
the sky that makes many weak-lensing studies possible in the first
place, because it allows the detection of the coherent distortions
imprinted by gravitational lensing on the images of the faint blue
galaxy population.

Due to their faintness, redshifts of the faint blue galaxies are hard
to measure spectroscopically. The following picture, however, seems to
be reasonably secure. It has emerged from increasingly deep and
detailed observations (see, e.g.~\cite*{bes88}; \cite*{ces91};
\cite*{ceb93}; \cite*{lcg91}; \cite*{lil93}; \cite*{cll95}; and also
the reviews by \cite*{kok92} and \cite*{ell97}). The redshift
distribution of faint galaxies has been found to agree fairly well
with that expected for a non-evolving comoving number density. While
the galaxy number counts in blue light are substantially above an
extrapolation of the local counts down to increasingly faint
magnitudes, those in the red spectral bands agree fairly well with
extrapolations from local number densities. Further, while there is
significant evolution of the luminosity function in the blue, in that
the luminosity scale $L_*$ of a Schechter-type fit increases with
redshift, the luminosity function of the galaxies in the red shows
little sign of evolution. Highly resolved images of faint blue
galaxies obtained with the {\em Hubble Space Telescope\/} are now
becoming available. In red light, they reveal mostly ordinary spiral
galaxies, while their substantial emission in blue light is more
concentrated to either spiral arms or bulges. Spectra exhibit emission
lines characteristic of star formation.

These findings support the view that the galaxy evolution towards
higher redshifts apparent in blue light results from enhanced
star-formation activity taking place in a population of galaxies
which, apart from that, may remain unchanged even out to redshifts of
$z\gtrsim1$. The redshift distribution of the faint blue galaxies is
then sufficiently well described by
\begin{equation}
  p(z)\d z = \frac{\beta}{z_0^3\,\Gamma(3/\beta)}\,z^2\,
  \exp\left[-\left(\frac{z}{z_0}\right)^\beta\right]\,\d z\;.
\label{eq:2.68}
\end{equation}
This expression is normalised to $0\le z<\infty$ and provides a good
fit to the observed redshift distribution (e.g.~\cite*{shy95}). The
mean redshift $\langle z\rangle$ is proportional to $z_0$, and the
parameter $\beta$ describes how steeply the distribution falls off
beyond $z_0$. For $\beta=1.5$, $\langle z\rangle\approx1.5\,z_0$. The
parameter $z_0$ depends on the magnitude cutoff and the colour
selection of the galaxy sample.

Background galaxies would be ideal tracers of distortions caused by
gravitational lensing if they were intrinsically circular. Then, any
measured ellipticity would directly reflect the action of the
gravitational tidal field of the lenses. Unfortunately, this is not
the case. To first approximation, galaxies have intrinsically
elliptical shapes, but the ellipses are randomly oriented. The
intrinsic ellipticities introduce noise into the inference of the
tidal field from observed ellipticities, and it is important for the
quantification of the noise to know the intrinsic ellipticity
distribution. Let $|\epsilon|$ be the ellipticity of a galaxy image,
defined such that for an ellipse with axes $a$ and $b<a$,
\begin{equation}
  |\epsilon| \equiv \frac{a-b}{a+b}\;.
\label{eq:2.69}
\end{equation}
Ellipses have an orientation, hence the ellipticity has two components
$\epsilon_{1,2}$, with $|\epsilon|=(\epsilon_1^2+\epsilon_2^2)^{1/2}$.
It turns out empirically that a Gaussian is a good description for the
ellipticity distribution,
\begin{equation}
  p_\epsilon(\epsilon_1,\epsilon_2)\d\epsilon_1\d\epsilon_2 =
  \frac{\exp(-|\epsilon|^2/\sigma_\epsilon^2)}
  {\pi\sigma_\epsilon^2\left[
    1-\exp(-1/\sigma_\epsilon^2)
  \right]}\,\d\epsilon_1\d\epsilon_2\;,
\label{eq:2.70}
\end{equation}
with a characteristic width of $\sigma_\epsilon\approx0.2$
(e.g.~\cite*{MI91.1}; \cite*{tys88}; \cite*{BR96.1}). We will later
(Sect.~\ref{sc:4.2}) define galaxy ellipticities for the general
situation where the isophotes are not ellipses. This completes our
summary of galaxy properties as required here.

\subsubsection{\label{sc:2.3.2}Groups and Clusters of Galaxies}

\begin{figure}[ht]
  \centerline{\includegraphics[width=0.6\hsize]{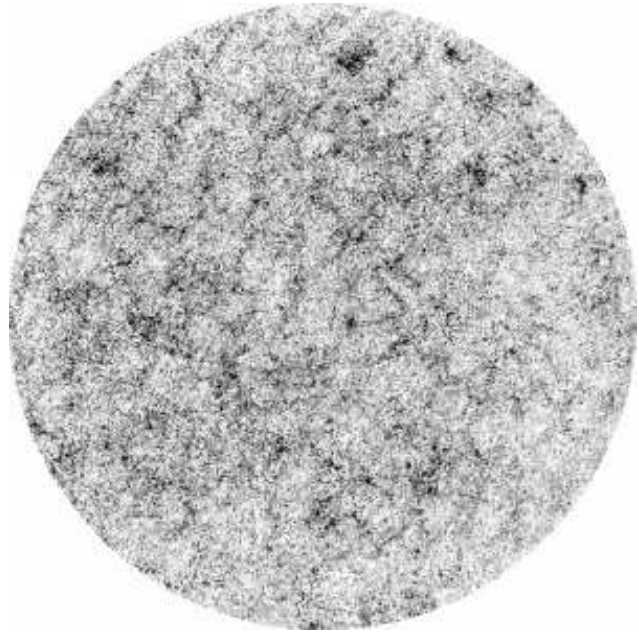}}
\caption{The Lick galaxy counts within $50^\circ$ radius around the
North Galactic pole (\protect\cite*{ssg77}). The galaxy number density
is highest at the black and lowest at the white regions on the
map. The picture illustrates structure in the distribution of fairly
nearby galaxies, viz.~under-dense regions, long extended filaments,
and clusters of galaxies.}
\label{fig:2.6}
\end{figure}

Galaxies are not randomly distributed in the sky. Their positions are
correlated, and there are areas in the sky where the galaxy density is
noticeably higher or lower than average (cf.~the galaxy count map in
Fig.~\ref{fig:2.6}). There are groups consisting of a few galaxies,
and there are {\em clusters of galaxies\/} in which some hundred up to
a thousand galaxies appear very close together.

The most prominent galaxy cluster in the sky covers a huge area
centred on the Virgo constellation. Its central region has a diameter
of about $7^\circ$, and its main body extends over roughly
$15^\circ\times40^\circ$. It was already noted by Sir William Herschel
in the 18th century that the entire Virgo cluster covers about $1/8$th
of the sky, while containing about $1/3$rd of the galaxies observable
at that time.

\citename{zwi33} noted in \citeyear{zwi33} that the galaxies in the
Coma cluster and other rich clusters move so fast that the clusters
required about ten to 100 times more mass to keep the galaxies bound
than could be accounted for by the luminous galaxies themselves. This
was the earliest indication that there is invisible mass, or dark
matter, in at least some objects in the Universe.

Several thousands of galaxy clusters are known
today. \citename{abe58}'s (\citeyear{abe58}) cluster catalog lists
2712 clusters north of $-20^\circ$ declination and away from the
Galactic plane. Employing a less restrictive definition of galaxy
clusters, the catalog by \cite{zhw68} identifies 9134 clusters north
of $-3^\circ$ declination. Cluster masses can exceed
$10^{48}\,\mathrm{g}$ or $5\times10^{14}\,M_\odot$, and they have
typical radii of $\approx5\times10^{24}\,\mathrm{cm}$ or
$\approx1.5\,\mathrm{Mpc}$.

\begin{figure}[ht]
  \centerline{\includegraphics[width=0.6\hsize]{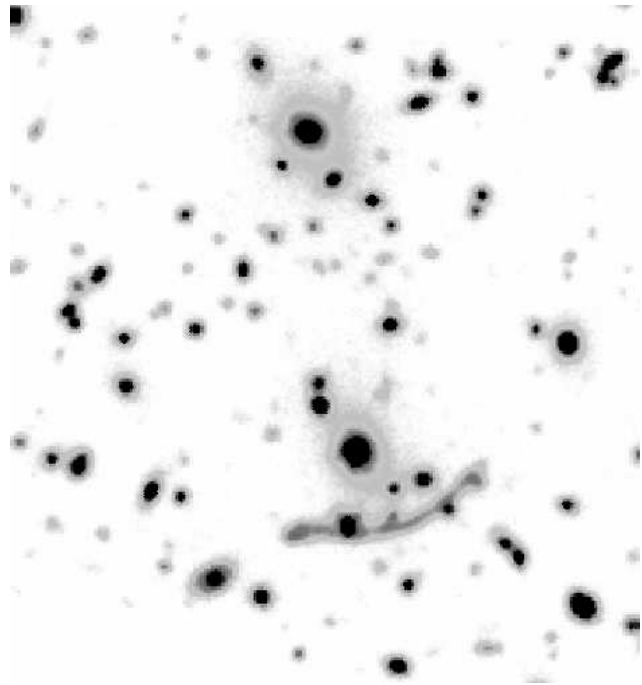}}
\caption{The galaxy cluster Abell~370, in which the first
gravitationally lensed arc was detected (\protect\cite*{LY86.1};
\protect\cite*{SO87.1}, \protect\citeyear{SO87.2}). Most of the bright
galaxies seen are cluster members at $z=0.37$, whereas the arc,
i.e.~the highly elongated feature, is the image of a galaxy at
redshift $z=0.724$ (\protect\cite*{SO88.1}).}
\label{fig:2.7}
\end{figure}

When X--ray telescopes became available after 1966, it was discovered
that clusters are powerful X--ray emitters. Their X--ray luminosities
fall within $(10^{43}-10^{45})\,\mathrm{erg}\,\mathrm{s}^{-1}$,
rendering galaxy clusters the most luminous X--ray sources in the
sky. Improved X--ray telescopes revealed that the source of X--ray
emission in clusters is extended rather than point-like, and that the
X--ray spectra are best explained by thermal {\em bremsstrahlung\/}
(free-free radiation) from a hot, dilute plasma with temperatures in
the range $(10^7-10^8)\,\mathrm{K}$ and densities of $\sim10^{-3}$
particles per cm$^3$. Based on the assumption that this intra-cluster
gas is in hydrostatic equilibrium with a spherically symmetric
gravitational potential of the total cluster matter, the X--ray
temperature and flux can be used to estimate the cluster mass. Typical
results {\em approximately\/} (i.e.~up to a factor of $\sim2$) agree
with the mass estimates from the kinematics of cluster galaxies
employing the virial theorem. The mass of the intra-cluster gas
amounts to about $10\%$ of the total cluster mass. The X--ray emission
thus independently confirms the existence of dark matter in galaxy
clusters. \cite{sar86} reviews clusters of galaxies focusing on their
X--ray emission.

Later, luminous arc-like features were discovered in two galaxy
clusters (\cite*{LY86.1}; \cite*{SO87.1}, \citeyear{SO87.2}; see
Fig.~\ref{fig:2.7}). Their light is typically bluer than that from the
cluster galaxies, and their length is comparable to the size of the
central cluster region. \cite{PA87.1} suggested that these {\em
arcs\/} are images of galaxies in the background of the clusters which
are strongly distorted by the gravitational tidal field close to the
cluster centres. This explanation was generally accepted when
spectroscopy revealed that the sources of the arcs are much more
distant than the clusters in which they appear (\cite*{SO88.1}).

Large arcs require special alignment of the arc source with the
lensing cluster. At larger distance from the cluster centre, images of
background galaxies are only weakly deformed, and they are referred to
as {\em arclets\/} (\cite*{FO88.1}; \cite*{TY90.1}). The high number
density of faint arclets allows one to measure the coherent distortion
caused by the tidal gravitational field of the cluster out to fairly
large radii. One of the main applications of weak gravitational
lensing is to reconstruct the (projected) mass distribution of galaxy
clusters from their measurable tidal fields. Consequently, the
corresponding theory constitutes one of the largest sections of this
review.

Such strong and weak gravitational lens effects offer the possibility
to detect and measure the entire cluster mass, dark and luminous,
without referring to any equilibrium or symmetry assumptions like
those required for the mass estimates from galactic kinematics or
X--ray emission. For a review on arcs and arclets in galaxy clusters
see \cite{FO94.1}.

Apart from being spectacular objects in their own right, clusters are
also of particular interest for cosmology. Being the largest
gravitationally bound entities in the cosmos, they represent the
high-mass end of collapsed structures. Their number density, their
individual properties, and their spatial distribution constrain the
power spectrum of the density fluctuations from which the structure in
the universe is believed to have originated (e.g.~\cite*{vil96};
\cite*{EK96.1}). Their formation history is sensitive to the
parameters that determine the geometry of the universe as a whole. If
the matter density in the universe is high, clusters tend to form
later in cosmic history than if the matter density is low (first noted
by \cite*{rlt92}). This is due to the behaviour of the growth factor
shown in Fig.~\ref{fig:2.3}, combined with the Gaussian nature of the
initial density fluctuations. Consequently, the compactness and the
morphology of clusters reflect the cosmic matter density, and this has
various observable implications. One method to normalise the
density-perturbation power spectrum fixes its overall amplitude such
that the local spatial number density of galaxy clusters is
reproduced. This method, called {\em cluster normalisation\/} and
pioneered by \cite{wef93}, will frequently be used in this review.

In summary, clusters are not only regions of higher galaxy number
density in the sky, but they are gravitationally bound bodies whose
member galaxies contribute only a small fraction of their mass. About
80\% of their mass is dark, and roughly 10\% is in the form of the
diffuse, X--ray emitting gas spread throughout the cluster. Mass
estimates inferred from galaxy kinematics, X--ray emission, and
gravitational-lensing effects generally agree to within about a factor
of two, typically arriving at masses of order $5\times10^{14}$~solar
masses, or $10^{48}\,\mathrm{g}$. Typical sizes of galaxy clusters are
of order several megaparsecs, or $5\times10^{24}\,\mathrm{cm}$. In
addition, there are smaller objects, called {\em galaxy groups\/},
which contain fewer galaxies and have typical masses of order
$10^{13}$~solar masses.

\subsubsection{\label{sc:2.3.3}Active Galactic Nuclei}

The term `active galactic nuclei' (AGNs) is applied to galaxies which
show signs of non-stellar radiation in their centres. Whereas the
emission from `normal' galaxies like our own is completely dominated
by radiation from stars and their remnants, the emission from AGNs is
a combination of stellar light and non-thermal emission from their
nuclei. In fact, the most prominent class of AGNs, the quasi-stellar
radio sources, or quasars, have their names derived from the fact that
their optical appearance is point-like. The nuclear emission almost
completely outshines the extended stellar light of its host galaxy.

AGNs do not form a homogeneous class of objects. Instead, they are
grouped into several types. The main classes are: quasars,
quasi-stellar objects (QSOs), Seyfert galaxies, BL~Lacertae objects
(BL Lacs), and radio galaxies. What unifies them is the non-thermal
emission from their nucleus, which manifests itself in various ways:
(1) radio emission which, owing to its spectrum and polarisation, is
interpreted as synchrotron radiation from a power-law distribution of
relativistic electrons; (2) strong ultraviolet and optical emission
lines from highly ionised species, which in some cases can be
extremely broad, corresponding to Doppler velocities up to
$\sim20,000\,\mathrm{km\,s}^{-1}$, thus indicating the presence of
semi-relativistic velocities in the emission region; (3) a flat
ultraviolet-to-optical continuum spectrum, often accompanied by
polarisation of the optical light, which cannot naturally be explained
by a superposition of stellar (Planck) spectra; (4) strong X--ray
emission with a hard power-law spectrum, which can be interpreted as
inverse Compton radiation by a population of relativistic electrons
with a power-law energy distribution; (5) strong gamma-ray emission;
(6) variability at all wavelengths, from the radio to the gamma-ray
regime. Not all these phenomena occur at the same level in all the
classes of AGNs. QSOs, for example, can roughly be grouped into
radio-quiet QSOs and quasars, the latter emitting strongly at radio
wavelengths.

Since substantial variability cannot occur on timescales shorter than
the light-travel time across the emitting region, the variability
provides a rigourous constraint on the compactness of the region
emitting the bulk of the nuclear radiation. In fact, this causality
argument based on light-travel time can mildly be violated if
relativistic velocities are present in the emitting region. Direct
evidence for this comes from the observation of the so-called
superluminal motion, where radio-source components exhibit apparent
velocities in excess of $c$ (e.g.~\cite*{ZE87.1}). This can be
understood as a projection effect, combining velocities close to (but
of course smaller than) the velocity of light with a velocity
direction close to the line-of-sight to the observer. Observations of
superluminal motion indicate that bulk velocities of the
radio-emitting plasma components can have Lorentz factors of order 10,
i.e., they move at $\sim0.99\,c$.

The standard picture for the origin of this nuclear activity is that a
supermassive black hole (or order $10^8\,M_\odot$), situated in the
centre of the host galaxy, accretes gas from the host. In this
process, gravitational binding energy is released, part of which can
be transformed into radiation. The appearance of an AGN then depends
on the black-hole mass and angular momentum, the accretion rate, the
efficiency of the transformation of binding energy into radiation, and
on the orientation relative to the line-of-sight. The understanding of
the physical mechanisms in AGNs, and how they are related to their
phenomenology, is still rather incomplete. We refer the reader to the
books and articles by \cite{bbr84}, \cite{WE86.2}, \cite{bnw90},
\cite{pet97}, and \cite{kro99}, and references therein, for an
overview of the phenomena in AGNs, and of our current ideas on their
interpretation. For the current review, we only make use of one
particular property of AGNs:

QSOs can be extremely luminous. Their optical luminosity can reach a
factor of thousand or more times the luminosity of normal
galaxies. Therefore, their nuclear activity completely outshines that
of the host galaxy, and the nuclear sources appear point-like on
optical images. Furthermore, the high luminosity implies that QSOs can
be seen to very large distances, and in fact, until a few years ago
QSOs held the redshift record. In addition, the comoving number
density of QSOs evolves rapidly with redshift. It was larger than
today by a factor of $\sim100$ at redshifts between 2 and 3. Taken
together, these two facts imply that a flux-limited sample of QSOs has
a very broad redshift distribution, in particular, very distant
objects are abundant in such a sample.

However, it is quite difficult to obtain a `complete' flux-limited
sample of QSOs. Of all point-like objects at optical wavelengths, QSOs
constitute only a tiny fraction, most being stars. Hence, morphology
alone does not suffice to obtain a candidate QSO sample which can be
verified spectroscopically. However, QSOs are found to have very blue
optical colours, by which they can efficiently be selected. Colour
selection typically yields equal numbers of white dwarfs and QSOs with
redshifts below $\sim2.3$. For higher-redshift QSOs, the strong
Ly$\alpha$ emission line moves from the U-band filter into the B-band,
yielding redder U$-$B colours. For these higher-redshift QSOs,
multi-colour or emission-line selection criteria must be used
(cf.~\cite*{fan99}). In contrast to optical selection, AGNs are quite
efficiently selected in radio surveys. The majority of sources
selected at centimeter wavelengths are AGNs. A flux-limited sample of
radio-selected AGNs also has a very broad redshift distribution. The
large fraction of distant objects in these samples make AGNs
particularly promising sources for the gravitational lensing effect,
as the probability of finding an intervening mass concentration close
to the line-of-sight increases with the source distance. In fact, most
of the known multiple-image gravitational lens systems have AGN
sources.

In addition to their high redshifts, the number counts of AGNs are
important for lensing. For bright QSOs with apparent B-band magnitudes
$B\lesssim19$, the differential source counts can be approximated by a
power law, $n(S)\propto S^{-(\alpha+1)}$, where $n(S)\,\d S$ is the
number density of QSOs per unit solid angle with flux within $\d S$ of
$S$, and $\alpha\approx2.6$. At fainter magnitudes, the differential
source counts can also be approximated by a power law in flux, but
with a much flatter index of $\alpha\sim0.5$. The source counts at
radio wavelengths are also quite steep for the highest fluxes, and
flatten as the flux decreases. The steepness of the source counts will
be the decisive property of AGNs for the magnification bias, which
will be discussed in Sect.~\ref{sc:6}.

\subsection{\label{sc:2.4}Correlation Functions, Power Spectra, and
  their Projections}

\subsubsection{Definitions; Homogeneous and Isotropic Random Fields}

In this subsection, we define the correlation function and the power
spectrum of a random field, which will be used extensively in later
sections. One example already occurred above, namely the power
spectrum $P_\delta$ of the density fluctuation field $\delta$.

Consider a random field $g(\vec x)$ whose expectation value is zero
everywhere. This means that an average over many realisations of the
random field should vanish, $\langle g(\vec x)\rangle=0$, for all
$\vec x$. This is not an important restriction, for if that was not
the case, we could consider the field $g(\vec x)-\langle g(\vec
x)\rangle$ instead, which would have the desired property. Spatial
positions $\vec x$ have $n$ dimensions, and the field can be either
real or complex.

A random field $g(\vec x)$ is called {\em homogeneous\/} if it cannot
statistically be distinguished from the field $g(\vec x+\vec y)$, where
$\vec y$ is an arbitrary translation vector. Similarly, a random field
$g(\vec x)$ is called {\em isotropic\/} if it has the same statistical
properties as the random field $g(\mathcal{R}\vec x)$, where
$\mathcal{R}$ is an arbitrary rotation matrix in $n$
dimensions. Restricting our attention to homogeneous and isotropic
random fields, we note that the {\em two-point correlation function\/}
\begin{equation}
  \langle g(\vec x)\,g^*(\vec y)\rangle = C_{gg}(|\vec x-\vec y|)
\label{eq:2.71}
\end{equation}
can only depend on the absolute value of the difference vector between
the two points $\vec x$ and $\vec y$. Note that $C_{gg}$ is real, even
if $g$ is complex. This can be seen by taking the complex conjugate of
(\ref{eq:2.71}), which is equivalent to interchanging $\vec x$ and
$\vec y$, leaving the right-hand-side unaffected.

We define the Fourier-transform pair of $g$ as
\begin{equation}
  \hat g(\vec k) = \int_{\Rset^n}\d^n x\,g(\vec x)\,
  \mathrm{e}^{\mathrm{i}\vec x\cdot\vec k}\;;\quad
  g(\vec x) = \int_{\Rset^n}\frac{\d^n k}{(2\pi)^n}\hat g(\vec k)\,
  \mathrm{e}^{-\mathrm{i}\vec x\cdot\vec k}\;.
\label{eq:2.72}
\end{equation}
We now calculate the correlation function in Fourier space, 
\begin{equation}
  \langle\hat g(\vec k)\hat g^*(\vec k')\rangle =
  \int_{\Rset^n}\d^n x\,\mathrm{e}^{\mathrm{i}\vec x\cdot\vec k}
  \int_{\Rset^n}\d^n x'\,\mathrm{e}^{-\mathrm{i}\vec x'\cdot\vec k'}
  \langle g(\vec x)\,g^*(\vec x')\rangle\;.
\label{eq:2.73}
\end{equation}
Using (\ref{eq:2.71}) and substituting $\vec x'=\vec x+\vec y$, this
becomes
\begin{eqnarray}
  \langle\hat g(\vec k)\hat g^*(\vec k')\rangle &=&
  \int_{\Rset^n}\d^n x\,\mathrm{e}^{\mathrm{i}\vec x\cdot\vec k}
  \int_{\Rset^n}\d^n y\,\mathrm{e}^{-\mathrm{i}
  (\vec x+\vec y)\cdot\vec k'}\,C_{gg}(|\vec y|)\nonumber\\
  &=&
  (2\pi)^n\delta_\mathrm{D}(\vec k-\vec k')\,
  \int_{\Rset^n}\d^n y\,\mathrm{e}^{-\mathrm{i}\vec y\cdot\vec k}\,
  C_{gg}(|\vec y|)\nonumber\\
  &\equiv&
  (2\pi)^n\delta_\mathrm{D}(\vec k-\vec k')\,P_g(|\vec k|)\;.
\label{eq:2.74}
\end{eqnarray}
In the final step, we defined the {\em power spectrum\/} of the
homogeneous and isotropic random field $g$,
\begin{equation}
  P_g(|\vec k|) = \int_{\Rset^n}\d^n y\,
  \mathrm{e}^{-\mathrm{i}\vec y\cdot\vec k}\,C_{gg}(|\vec y|)\;,
\label{eq:2.75}
\end{equation}
which is the Fourier transform of the two-point correlation
function. Isotropy of the random field implies that $P_g$ can only
depend on the modulus of $\vec k$.

{\em Gaussian random fields\/} are characterised by the property that
the probability distribution of any linear combination of the random
field $g(\vec x)$ is Gaussian. More generally, the joint probability
distribution of a number $M$ of linear combinations of the random
variable $g(\vec x_i)$ is a multivariate Gaussian. This is equivalent
to requiring that the Fourier components $\hat g(\vec k)$ are mutually
statistically independent, and that the probability densities for the
$\hat g(\vec k)$ are Gaussian with dispersion $P_g(|\vec k|)$. Thus, a
Gaussian random field is fully characterised by its power spectrum.

\subsubsection{Projections; Limber's Equation}

We now derive a relation between the power spectrum (or the
correlation function) of a homogeneous isotropic random field in three
dimensions, and its projection onto two dimensions. Specifically, for
the three-dimensional field, we consider the density contrast
$\delta[f_K(w)\vec\theta,w]$, where $\vec\theta$ is a two-dimensional
vector, which could be an angular position on the sky. Hence,
$f_K(w)\vec \theta$ and $w$ form a local comoving isotropic Cartesian
coordinate system. We define two different projections of $\delta$
along the backward-directed light cone of the observer at $w=0$,
$t=t_0$,
\begin{equation}
  g_i(\vec\theta) = \int\d w\,q_i(w)\,\delta[f_K(w)\vec\theta,w]\;,
\label{eq:2.76}
\end{equation}
for $i=1,2$. The $q_i(w)$ are weight functions, and the integral
extends from $w=0$ to the horizon $w=w_\mathrm{H}$. Since $\delta$ is
a homogeneous and isotropic random field, so is its projection.
Consider now the correlation function
\begin{eqnarray}
  C_{12} &=& \langle g_1(\vec\theta)g_2(\vec\theta')\rangle
  \nonumber\\ &=&
  \int\d w\,q_1(w)\int \d w'\,q_2(w')\,\langle
    \delta[f_K(w)\vec\theta,w]\,\delta[f_K(w')\vec\theta',w']
  \rangle\;.
\label{eq:2.77}
\end{eqnarray}
We assume that there is no power in the density fluctuations on scales
larger than a coherence scale $L_\mathrm{coh}$. This is justified
because the power spectrum $P_\delta$ declines $\propto k$ as $k\to
0$; see (\ref{eq:2.61}). This implies that the correlation function on
the right-hand side of eq.~(\ref{eq:2.77}) vanishes for
$w_\mathrm{H}\gg|w-w'|\gtrsim L_\mathrm{coh}$. Although $\delta$
evolves cosmologically, it can be considered constant over a time
scale on which light travels across a comoving distance
$L_\mathrm{coh}$. We note that the second argument of $\delta$
simultaneously denotes the third local spatial dimension and the
cosmological epoch, related through the light-cone condition $|c\d
t|=a\d w$. Furthermore, we assume that the weight functions $q_i(w)$
do not vary appreciably over a scale $\Delta w\le
L_\mathrm{coh}$. Consequently, $|w-w'|\lesssim L_\mathrm{coh}$ over
the scale where $C_{\delta\delta}$ is non-zero, and we can set
$f_K(w')\approx f_K(w)$ and $q_2(w')=q_2(w)$ to obtain
\begin{equation}
  C_{12}(\theta) = \int\d w\,q_1(w)q_2(w)\int\d(\Delta w)\,
  C_{\delta\delta}\left(
    \sqrt{f_K^2(w)\theta^2+(\Delta w)^2},w
  \right)\;.
\label{eq:2.78}
\end{equation}
The second argument of $C_{\delta\delta}$ now denotes the dependence
of the correlation function on cosmic epoch. Equation~(\ref{eq:2.78})
is one form of Limber's (\citeyear{lim53}) equation, which relates the
two-point correlation of the {\em projected\/} field to that of the
{\em three-dimensional\/} field.

Another very useful form of this equation relates the projected
two-point correlation function to the power spectrum of the
three-dimensional field. The easiest way to derive this relation is by
replacing the $\delta$'s in (\ref{eq:2.77}) by their Fourier
transforms, where upon
\begin{eqnarray}
  C_{12} &=& \int\d w\,q_1(w)\int\d w'\,
  q_2(w')\int\frac{\d^3k}{(2\pi)^3}
         \int\frac{\d^3k'}{(2\pi)^3} \nonumber\\
  &\times&
  \langle\hat\delta(\vec k,w)\,\hat\delta^*(\vec k',w')\rangle\,
  \mathrm{e}^{-\mathrm{i}f_K(w)\vec k_\perp\cdot\vec\theta}\,
  \mathrm{e}^{\mathrm{i}f_K(w')\vec k'_\perp\cdot\vec\theta'}\,
  \mathrm{e}^{-\mathrm{i}k_3 w} 
  \mathrm{e}^{\mathrm{i}k'_3 w'}\;.
\label{eq:2.79}
\end{eqnarray}
$\vec k_\perp$ is the two-dimensional wave vector perpendicular to the
line-of-sight. The correlator can be replaced by the power spectrum
$P_\delta$ using (\ref{eq:2.74}). This introduces a Dirac delta
function $\delta_\mathrm{D}(\vec k-\vec k')$, which allows us to carry
out the $\vec k'$-integration. Under the same assumptions on the
spatial variation of $q_i(w)$ and $f_K(w)$ as before, we find
\begin{eqnarray}
  C_{12} &=& \int\d w\,q_1(w)q_2(w)\,
  \int\frac{\d^3k}{(2\pi)^3}\,P_\delta(|\vec k|,w)\,
  \mathrm{e}^{-\mathrm{i}f_K(w)\vec k_\perp\cdot
    (\vec\theta-\vec\theta')}\,
  \mathrm{e}^{-\mathrm{i}k_3 w}\nonumber\\
  &\times&
  \int\d w'\,\mathrm{e}^{\mathrm{i}k_3w'}\;.
\label{eq:2.80}
\end{eqnarray}
The final integral yields $2\pi\delta_\mathrm{D}(k_3)$, indicating
that only such modes contribute to the projected correlation function
whose wave-vectors lie in the plane of the sky
(\cite*{BL91.1}). Finally, carrying out the trivial $k_3$-integration
yields
\begin{eqnarray}
  C_{12}(\theta) &=& \int\d w\,q_1(w)q_2(w)\,
  \int\frac{\d^2k_\perp}{(2\pi)^2}\,P_\delta(|\vec k_\perp|,w)\,
  \mathrm{e}^{-\mathrm{i}f_K(w)\vec k_\perp\cdot\vec\theta}
  \label{eq:2.81}\\
  &=&
  \int\d w\,q_1(w)q_2(w)\,\int\frac{k\d k}{2\pi}\,
  P(k,w)\,\mathrm{J}_0[f_K(w)\,\theta\,k]\;.
\label{eq:2.82}
\end{eqnarray}
The definition (\ref{eq:2.72}) of the Fourier transform, and the
relation (\ref{eq:2.75}) between power spectrum and correlation
function allow us to write the (cross) power spectrum $P_{12}(l)$ as
\begin{eqnarray}
  P_{12}(l) &=& \int\d^2\theta\,C_{12}(\theta)
  \mathrm{e}^{\mathrm{i}\vec l\cdot\vec\theta}\nonumber\\
  &=&
  \int\d w\,q_1(w)q_2(w)\,\int\frac{\d^2k_\perp}{(2\pi)^2}\,
  P_\delta(|\vec k_\perp|,w)\,
  (2\pi)^2\,\delta_\mathrm{D}[\vec l-f_k(w)\vec k_\perp] \nonumber\\
  &=&
  \int\d w\,\frac{q_1(w)q_2(w)}{f_K^2(w)}\,
  P_\delta \left( \frac{l}{f_K(w)}, w\right)\;,
\label{eq:2.83}
\end{eqnarray}
which is Limber's equation in Fourier space (\cite*{KA92.1},
\citeyear{KA98.1}). We shall make extensive use of these relations in
later sections.

  % -*- LaTeX -*-
 
\section{\label{sc:3}Gravitational Light Deflection}

In this section, we summarise the theoretical basis for the
description of light deflection by gravitational fields. Granted the
validity of Einstein's Theory of General Relativity, light propagates
on the null geodesics of the space-time metric. However, most
astrophysically relevant situations permit a much simpler approximate
description of light rays, which is called gravitational lens theory;
we first describe this theory in Sect.~\ref{sc:3.1}. It is sufficient
for the treatment of lensing by galaxy clusters in Sect.~\ref{sc:5},
where the deflecting mass is localised in a region small compared to
the distance between source and deflector, and between deflector and
observer. In contrast, mass distributions on a cosmic scale cause
small light deflections all along the path from the source to the
observer. The magnification and shear effects resulting therefrom
require a more general description, which we shall develop in
Sect.~\ref{sc:3.2}. In particular, we outline how the gravitational
lens approximation derives from this more general description.

\subsection{\label{sc:3.1}Gravitational Lens Theory}

\begin{figure}[ht]
  \centerline{\includegraphics[width=0.6\hsize]{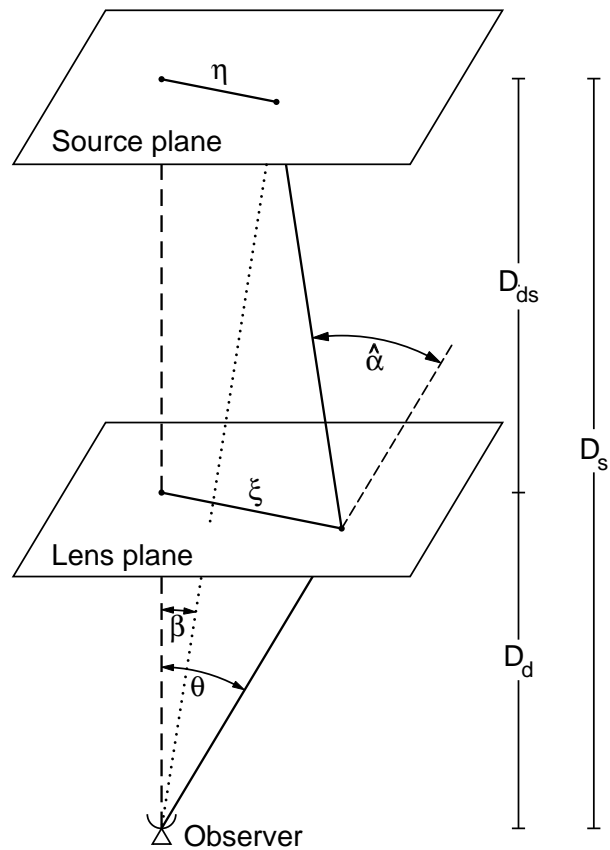}}
\caption{Sketch of a typical gravitational lens system.}
\label{fig:3.1}
\end{figure}

A typical situation considered in gravitational lensing is sketched in
Fig.~\ref{fig:3.1}, where a mass concentration at redshift
$z_\mathrm{d}$ (or angular diameter distance $D_\mathrm{d}$) deflects
the light rays from a source at redshift $z_\mathrm{s}$ (or angular
diameter distance $D_\mathrm{s}$). If there are no other deflectors
close to the line-of-sight, and if the extent of the deflecting mass
along the line-of-sight is very much smaller than both $D_\mathrm{d}$
and the angular diameter distance $D_\mathrm{ds}$ from the deflector
to the source,\footnote{\label{fn:3.1}This condition is very well
satisfied in most astrophysical situations. A cluster of galaxies, for
instance, has a typical size of a few Mpc, whereas the distances
$D_\mathrm{d}$, $D_\mathrm{s}$, and $D_\mathrm{ds}$ are fair fractions
of the Hubble length $cH_0^{-1}=3\,h^{-1}\times10^3\,\mathrm{Mpc}$.}
the actual light rays which are smoothly curved in the neighbourhood
of the deflector can be replaced by two straight rays with a kink near
the deflector. The magnitude and direction of this kink is described
by the {\em deflection angle\/} $\hat{\vec\alpha}$, which depends on
the mass distribution of the deflector and the impact vector of the
light ray.

\subsubsection{\label{sc:3.1.1}The Deflection Angle}

Consider first the deflection by a point mass $M$. If the light ray
does not propagate through the strong gravitational field close to the
horizon, that is, if its impact parameter $\xi$ is much larger than
the Schwarzschild radius of the lens, $\xi\gg
R_\mathrm{S}\equiv2GM\,c^{-2}$, then General Relativity predicts that
the deflection angle $\hat\alpha$ is
\begin{equation}
  \hat\alpha = \frac{4GM}{c^2\,\xi}\;.
\label{eq:3.1}
\end{equation}
This is just twice the value obtained in Newtonian gravity (see the
historical remarks in \cite*{sef92}). According to the condition
$\xi\gg R_\mathrm{S}$, the deflection angle is small,
$\hat\alpha\ll1$.

The field equations of General Relativity can be linearised if the
gravitational field is weak. The deflection angle of an ensemble of
point masses is then the (vectorial) sum of the deflections due to
individual lenses. Consider now a three-dimensional mass distribution
with volume density $\rho(\vec r)$. We can divide it into cells of
size $\d V$ and mass $\d m=\rho(\vec r)\,\d V$. Let a light ray pass
this mass distribution, and describe its spatial trajectory by
$\left(\xi_1(\lambda),\xi_2(\lambda),r_3(\lambda)\right)$, where the
coordinates are chosen such that the incoming light ray (i.e.~far from
the deflecting mass distribution) propagates along $r_3$. The actual
light ray is deflected, but if the deflection angle is small, it can
be approximated as a straight line in the neighbourhood of the
deflecting mass. This corresponds to the Born approximation in atomic
and nuclear physics. Then, $\vec\xi(\lambda)\equiv\vec\xi$,
independent of the affine parameter $\lambda$. Note that
$\vec\xi=(\xi_1,\xi_2)$ is a two-dimensional vector. The impact vector
of the light ray relative to the mass element $\d m$ at $\vec
r=(\xi_1',\xi_2',r'_3)$ is then $\vec\xi-\vec\xi'$, independent of
$r_3'$, and the total deflection angle is
\begin{eqnarray}
  \hat{\vec\alpha}(\vec\xi) &=& \frac{4G}{c^2}\,
  \sum\d m(\xi_1',\xi_2',r_3')\,
  \frac{\vec\xi-\vec\xi'}{|\vec\xi-\vec\xi'|^2}
  \nonumber\\ &=&
  \frac{4G}{c^2}\int\d^2\xi'\int\d r_3'\,\rho(\xi_1',\xi_2',r_3')\,
  \frac{\vec\xi-\vec\xi'}{|\vec\xi-\vec\xi'|^2}\;,
\label{eq:3.2}
\end{eqnarray}
which is also a two-dimensional vector. Since the last factor in
eq.~(\ref{eq:3.2}) is independent of $r_3'$, the $r_3'$-integration
can be carried out by defining the {\em surface mass density\/}
\begin{equation}
  \Sigma(\vec\xi) \equiv \int\d r_3\,\rho(\xi_1,\xi_2,r_3)\;,
\label{eq:3.3}
\end{equation}
which is the mass density projected onto a plane perpendicular to the
incoming light ray. Then, the deflection angle finally becomes
\begin{equation}
  \hat{\vec\alpha}(\vec\xi) = \frac{4G}{c^2}\,
  \int\d^2\xi'\,\Sigma(\vec\xi')\,
  \frac{\vec\xi-\vec\xi'}{|\vec\xi-\vec\xi'|^2}\;.
\label{eq:3.4}
\end{equation}
This expression is valid as long as the deviation of the actual light
ray from a straight (undeflected) line within the mass distribution is
small compared to the scale on which the mass distribution changes
significantly. This condition is satisfied in virtually all
astrophysically relevant situations (i.e.~lensing by galaxies and
clusters of galaxies), unless the deflecting mass extends all the way
from the source to the observer (a case which will be dealt with in
Sect.~\ref{sc:6}). It should also be noted that in a lensing situation
such as displayed in Fig.~\ref{fig:3.1}, the incoming light rays are
not mutually parallel, but fall within a beam with opening angle
approximately equal to the angle which the mass distribution subtends
on the sky. This angle, however, is typically {\em very\/} small (in
the case of cluster lensing, the relevant angular scales are of order
1~arc min~$\approx2.9\times10^{-4}$).

\subsubsection{\label{sc:3.1.2}The Lens Equation}

We now require an equation which relates the true position of the
source to its observed position on the sky. As sketched in
Fig.~\ref{fig:3.1}, the source and lens planes are defined as planes
perpendicular to a straight line (the optical axis) from the observer
to the lens at the distance of the source and of the lens,
respectively. The exact definition of the optical axis does not matter
because of the smallness of angles involved in a typical lens
situation, and the distance to the lens is well defined for a
geometrically-thin matter distribution. Let $\vec\eta$ denote the
two-dimensional position of the source on the source plane. Recalling
the definition of the angular-diameter distance, we can read off
Fig.~\ref{fig:3.1}
\begin{equation}
  \vec\eta = \frac{D_\mathrm{s}}{D_\mathrm{d}}\,
  \vec\xi - D_\mathrm{ds}\hat{\vec\alpha}(\vec\xi)\;.
\label{eq:3.5}
\end{equation}
Introducing angular coordinates by $\vec\eta=D_\mathrm{s}\vec\beta$
and $\vec\xi=D_\mathrm{d}\vec\theta$, we can transform
eq.~(\ref{eq:3.5}) to
\begin{equation}
  \vec\beta = \vec\theta-\frac{D_\mathrm{ds}}{D_\mathrm{s}}\,
  \hat{\vec\alpha}(D_\mathrm{d}\vec\theta) \equiv
  \vec\theta-\vec\alpha(\vec\theta)\;,
\label{eq:3.6}
\end{equation}
where we defined the scaled deflection angle $\vec\alpha(\vec\theta)$
in the last step. The interpretation of the lens equation
(\ref{eq:3.6}) is that a source with true position $\vec\beta$ can be
seen by an observer at angular positions $\vec\theta$ satisfying
(\ref{eq:3.6}). If (\ref{eq:3.6}) has more than one solution for fixed
$\vec\beta$, a source at $\vec\beta$ has images at several positions
on the sky, i.e.~the lens produces multiple images. For this to
happen, the lens must be `strong'. This can be quantified by the
dimension-less surface mass density
\begin{equation}
  \kappa(\vec\theta) = 
  \frac{\Sigma(D_\mathrm{d}\vec\theta)}{\Sigma_\mathrm{cr}}
  \quad\hbox{with}\quad
  \Sigma_\mathrm{cr} = \frac{c^2}{4\pi G}\,
  \frac{D_\mathrm{s}}{D_\mathrm{d}\,D_\mathrm{ds}}\;,
\label{eq:3.7}
\end{equation}
where $\Sigma_\mathrm{cr}$ is called the critical surface mass density
(which depends on the redshifts of source and lens). A mass
distribution which has $\kappa\ge1$ somewhere,
i.e.~$\Sigma\ge\Sigma_\mathrm{cr}$, produces multiple images for some
source positions $\vec\beta$ (see \cite*{sef92}, Sect.~5.4.3). Hence,
$\Sigma_\mathrm{cr}$ is a characteristic value for the surface mass
density which distinguishes between `weak' and `strong' lenses. Note
that $\kappa\ge1$ is sufficient but not necessary for producing
multiple images. In terms of $\kappa$, the scaled deflection angle
reads
\begin{equation}
  \vec\alpha(\vec\theta) = \frac{1}{\pi}\,
  \int_{\Rset^2}\d^2\theta'\,\kappa(\vec\theta')\,
  \frac{\vec\theta-\vec\theta'}{|\vec\theta-\vec\theta'|^2}\;.
\label{eq:3.8}
\end{equation}

Equation~(\ref{eq:3.8}) implies that the deflection angle can be
written as the gradient of the {\em deflection potential\/},
\begin{equation}
  \psi(\vec\theta) = \frac{1}{\pi}\int_{\Rset^2}\d^2\theta'\,
  \kappa(\vec\theta')\,\ln|\vec\theta-\vec\theta'|\;,
\label{eq:3.9}
\end{equation}
as $\vec\alpha=\nabla\psi$. The potential $\psi(\vec\theta)$ is the
two-dimensional analogue of the Newtonian gravitational potential and
satisfies the Poisson equation
$\nabla^2\psi(\vec\theta)=2\kappa(\vec\theta)$.

\subsubsection{\label{sc:3.1.3}Magnification and Distortion}

The solutions $\vec\theta$ of the lens equation yield the angular
positions of the images of a source at $\vec\beta$. The shapes of the
images will differ from the shape of the source because light bundles
are deflected differentially. The most visible consequence of this
distortion is the occurrence of giant luminous arcs in galaxy
clusters. In general, the shape of the images must be determined by
solving the lens equation for all points within an extended
source. Liouville's theorem and the absence of emission and absorption
of photons in gravitational light deflection imply that lensing
conserves surface brightness (or specific intensity). Hence, if
$I^{(s)}(\vec\beta)$ is the surface brightness distribution in the
source plane, the observed surface brightness distribution in the lens
plane is
\begin{equation}
  I(\vec\theta) = I^{(s)}[\vec\beta(\vec\theta)]\;.
\label{eq:3.10}
\end{equation}
If a source is much smaller than the angular scale on which the lens
properties change, the lens mapping can locally be linearised. The
distortion of images is then described by the Jacobian matrix
\begin{equation}
  \mathcal{A}(\vec\theta) =
  \frac{\partial\vec\beta}{\partial\vec\theta} =
  \left(\delta_{ij} -
    \frac{\partial^2\psi(\vec\theta)}{\partial\theta_i\partial\theta_j}
  \right) = \left(
    \begin{array}{cc}
      1-\kappa-\gamma_1 & -\gamma_2 \\ 
      -\gamma_2 & 1-\kappa+\gamma_1 \\
    \end{array}
  \right)\;,
\label{eq:3.11}
\end{equation}
where we have introduced the components of the shear
$\gamma\equiv\gamma_1+\mathrm{i}\gamma_2 =
|\gamma|\mathrm{e}^{2\mathrm{i}\varphi}$,
\begin{equation}
  \gamma_1 = \frac{1}{2}(\psi_{,11}-\psi_{,22})\;,\quad
  \gamma_2 = \psi_{,12}\;,
\label{eq:3.12}
\end{equation}
and $\kappa$ is related to $\psi$ through Poisson's equation. Hence,
if $\vec\theta_0$ is a point within an image, corresponding to the
point $\vec\beta_0=\vec\beta(\vec\theta_0)$ within the source, we find
from (\ref{eq:3.10}) using the locally linearised lens equation
\begin{equation}
  I(\vec\theta) = I^{(s)}\left[
    \vec\beta_0+\mathcal{A}(\vec\theta_0)
    \cdot(\vec\theta-\vec\theta_0)
  \right]\;.
\label{eq:3.13}
\end{equation}
According to this equation, the images of a circular source are
ellipses. The ratios of the semi-axes of such an ellipse to the radius
of the source are given by the inverse of the eigenvalues of
$\mathcal{A}(\vec\theta_0)$, which are $1-\kappa\pm|\gamma|$, and the
ratio of the solid angles subtended by an image and the unlensed
source is the inverse of the determinant of $\mathcal{A}$. The fluxes
observed from the image and from the unlensed source are given as
integrals over the brightness distributions $I(\vec\theta)$ and
$I^{(s)}(\vec\beta)$, respectively, and their ratio is the {\em
magnification\/} $\mu(\vec\theta_0)$. From (\ref{eq:3.13}), we find
\begin{equation}
  \mu = \frac{1}{\det\mathcal{A}} = 
  \frac{1}{(1-\kappa)^2-|\gamma|^2}\;.
\label{eq:3.14}
\end{equation}
The images are thus distorted in shape and size. The shape distortion
is due to the tidal gravitational field, described by the shear
$\gamma$, whereas the magnification is caused by both isotropic
focusing caused by the local matter density $\kappa$ and anisotropic
focusing caused by shear.

Since the shear is defined by the trace-free part of the symmetric
Jacobian matrix $\mathcal{A}$, it has two independent
components. There exists a one-to-one mapping from symmetric,
trace-free $2\times2$ matrices onto complex numbers, and we shall
extensively use complex notation. Note that the shear transforms as
$\mathrm{e}^{2\mathrm{i}\varphi}$ under rotations of the coordinate
frame, and is therefore not a vector. Equations~(\ref{eq:3.9}) and
(\ref{eq:3.12}) imply that the complex shear can be written
\begin{eqnarray}
  \gamma(\vec\theta) &=& \frac{1}{\pi}\int_{\Rset^2}\d^2\theta'\,
  \mathcal{D}(\vec\theta-\vec\theta')\,
  \kappa(\vec\theta')\;,\nonumber\\
  \hbox{with}&&
  \mathcal{D}(\vec\theta) \equiv
  \frac{\theta_2^2-\theta_1^2-2\mathrm{i}\theta_1\theta_2}
  {|\vec\theta|^4}
  = \frac{-1}{(\theta_1-\mathrm{i}\theta_2)^2}\;.
\label{eq:3.15}
\end{eqnarray}

\subsubsection{\label{sc:3.1.4}Critical Curves and Caustics}

Points in the lens plane where the Jacobian $\mathcal{A}$ is singular,
i.e.~where $\det\mathcal{A}=0$, form closed curves, the {\em critical
curves\/}. Their image curves in the source plane are called {\em
caustics\/}. Equation~(\ref{eq:3.14}) predicts that sources on
caustics are infinitely magnified; however, infinite magnification
does not occur in reality, for two reasons. First, each astrophysical
source is extended, and its magnification (given by the surface
brightness-weighted point-source magnification across its solid angle)
remains finite. Second, even point sources would be magnified by a
finite value since for them, the geometrical-optics approximation
fails near critical curves, and a wave-optics description leads to a
finite magnification (e.g.~\cite*{OH83.1}; \cite*{sef92},
Chap.~7). For the purposes of this review, the first effect always
dominates. Nevertheless, images near critical curves can be magnified
and distorted substantially, as is demonstrated by the giant luminous
arcs which are formed from source galaxies close to caustics. (Point)
sources which move across a caustic have their number of images
changed by $\pm2$, and the two additional images appear or disappear
at the corresponding critical curve in the lens plane. Hence, only
sources inside a caustic are multiply imaged.

\subsubsection{\label{sc:3.1.5}An Illustrative Example: Isothermal
  Spheres}

The rotation curves of spiral galaxies are observed to be
approximately flat out to the largest radii where they can be
measured. If the mass distribution in a spiral galaxy followed the
light distribution, the rotation curves would have to decrease at
large radii in roughly Keplerian fashion. Flat rotation curves thus
provide the clearest evidence for dark matter on galactic scales. They
can be understood if galactic disks are embedded in a dark halo with
density profile $\rho\propto r^{-2}$ for large $r$. The projected mass
density then behaves like $\theta^{-1}$. Such density profiles are
obtained by assuming that the velocity dispersion of the dark matter
particles is spatially constant. They are therefore also called
isothermal profiles. We shall describe some simple properties of a
gravitational lens with an isothermal mass profile, which shall later
serve as a reference.

The projected surface mass density of a {\em singular isothermal
sphere\/} is
\begin{equation}
  \Sigma(\xi) = \frac{\sigma_v^2}{2G\xi}\;,
\label{eq:3.16}
\end{equation}
where $\sigma_v$ is the line-of-sight velocity dispersion of the
`particles' (e.g.~stars in galaxies, or galaxies in clusters of
galaxies) in the gravitational potential of the mass distribution,
assuming that they are in virial equilibrium. The corresponding
dimensionless surface mass density is
\begin{equation}
  \kappa(\theta) = \frac{\theta_\mathrm{E}}{2\theta}\;,
  \quad\hbox{where}\quad
  \theta_\mathrm{E} = 4\pi\,\left(\frac{\sigma_v}{c}\right)^2
  \frac{D_\mathrm{ds}}{D_\mathrm{s}}
\label{eq:3.17}
\end{equation}
is called the {\em Einstein deflection angle\/}. As can easily be
verified from (\ref{eq:3.8}), the magnitude of the scaled deflection
angle is constant for this mass profile,
$|\vec\alpha|=\theta_\mathrm{E}$, and the deflection potential is
$\psi=\theta_\mathrm{E}|\vec\theta|$. From that, the shear is obtained
using (\ref{eq:3.12})\footnote{\label{fn:3.2}For axially-symmetric
projected mass profiles, the magnitude of the shear can be calculated
from $|\gamma|(\theta)= \bar\kappa(\theta)-\kappa(\theta)$, where
$\bar\kappa(\theta)$ is the mean surface mass density inside a circle
of radius $\theta$ from the lens centre. Accordingly, the magnitude of
the deflection angle is $|\vec\alpha|=\theta\bar\kappa(\theta)$.},
\begin{equation}
  \gamma(\vec\theta) = -\frac{\theta_\mathrm{E}}
  {2|\vec\theta|}\mathrm{e}^{2\mathrm{i}\varphi}\;,
\label{eq:3.18}
\end{equation}
and the magnification is
\begin{equation}
  \mu(\vec\theta) =
  \frac{|\vec\theta|}{|\vec\theta|-\theta_\mathrm{E}}\;.
\label{eq:3.19}
\end{equation}
This shows that $|\vec\theta|=\theta_\mathrm{E}$ defines a critical
curve, which is called the {\em Einstein circle\/}. The corresponding
caustic, obtained by mapping the Einstein circle back into the source
plane under the lens equation, degenerates to a single point at
$\vec\beta=\vec0$. Such degenerate caustics require highly symmetric
lenses. Any perturbation of the mass distribution breaks the
degeneracy and expands the singular caustic point into a caustic curve
(see Chapter 6 in \cite*{sef92} for a detailed treatment of critical
curves and caustics). The lens (\ref{eq:3.17}) produces two images
with angular separation $2\theta_\mathrm{E}$ for a source with
$|\vec\beta|<1$, and one image otherwise.

The mass distribution (\ref{eq:3.17}) has two unsatisfactory
properties. The surface mass density diverges for $|\vec\theta|\to0$,
and the total mass of the lens is infinite. Clearly, both of these
properties will not match real mass distributions. Despite this fact,
the singular isothermal sphere fits many of the observed lens systems
fairly well. In order to construct a somewhat more realistic lens
model, one can cut off the distribution at small and large distances,
e.g.~by
\begin{equation}
  \kappa(\vec\theta) = \frac{\theta_\mathrm{E}}
  {2\sqrt{|\vec\theta|^2+\theta_\mathrm{c}^2}} -
  \frac{\theta_\mathrm{E}}
  {2\sqrt{|\vec\theta|^2+\theta_\mathrm{t}^2}}\;,
\label{eq:3.20}
\end{equation}
which has a core radius $\theta_\mathrm{c}$, and a truncation radius
$\theta_\mathrm{t}$. For
$\theta_\mathrm{c}\ll|\vec\theta|\ll\theta_\mathrm{t}$, this mass
distribution behaves like $\theta^{-1}$. This lens can produce three
images, but only if $\theta_\mathrm{c}\theta_\mathrm{t}\,
(\theta_\mathrm{c}+\theta_\mathrm{t})^{-1}<\theta_\mathrm{E}/2$. One
of the three images occurs near the centre of the lens and is strongly
de-magnified if $\theta_\mathrm{c}\ll\theta_\mathrm{E}$. In most of
the multiple-image QSO lens systems, there is no indication for a
third central image, imposing strict upper bounds on
$\theta_\mathrm{c}$, whereas for some arc systems in clusters, a
finite core size is required when a lens model like (\ref{eq:3.20}) is
assumed.

\subsection{\label{sc:3.2}Light Propagation in Arbitrary Spacetimes}

We now turn to a more rigourous description of the propagation of
light rays, based on the theory of geometrical optics in General
Relativity. We then specialise the resulting propagation equations to
the case of weak gravitational fields and metric perturbations to the
background of an expanding universe. These equations contain the
gravitational lens equation discussed previously as a special case. We
shall keep the discussion brief and follow closely the work of
\citename{sef92} (\citeyear{sef92}, Chaps.~3 \& 4), and \cite{SE94.5},
where further references can be found.

\subsubsection{\label{sc:3.2.1}Propagation of Light Bundles}

In Sect.~\ref{sc:3.1.2}, we have derived the lens equation
(\ref{eq:3.5}) in a heuristic way. A rigourous derivation in an
arbitrary spacetime must account for the fact that distance vectors
between null geodesics are four-vectors. Nevertheless, by choosing an
appropriate coordinate system, the separation transverse to the
line-of-sight between two neighbouring light rays can effectively be
described by a two-dimensional vector $\vec\xi$. We outline this
operation in the following two paragraphs.

We first consider the propagation of infinitesimally thin light beams
in an arbitrary space-time, characterised by the metric tensor
$g_{\mu\nu}$. The propagation of a fiducial ray $\gamma_0$ of the
bundle is determined by the geodesic equation (e.g.~\cite*{MI73.1};
\cite*{WE72.1}). We are interested here in the evolution of the shape
of the bundle as a function of the affine parameter along the fiducial
ray. Consider an observer O with four-velocity $U^\mu_\mathrm{o}$,
satisfying $U^\mu_\mathrm{o}U_{\mathrm{o}\mu}=1$. The physical wave
vector $k^\mu$ of a photon depends on the light frequency. We define
$\tilde k^\mu\equiv-c^{-1}\omega_\mathrm{o}\,k^\mu$ as a past-directed
dimensionless wave vector which is independent of the frequency
$\omega_\mathrm{o}$ measured by the observer. We choose an affine
parameter $\lambda$ of the rays passing through O such that (1)
$\lambda=0$ at the observer, (2) $\lambda$ increases along the
backward light cone of O, and (3) $U^\mu_\mathrm{o}\tilde k_\mu=-1$ at
O. Then, with the definition of $\tilde k^\mu$, it follows that
$\tilde k^\mu=\d x^\mu/\d\lambda$, and that $\lambda$ measures the
proper distance along light rays for events close to O.

Let $\gamma^\mu(\vec\theta,\lambda)$ characterise the rays of a light
beam with vertex at O, such that $\vec\theta$ is the angle between a
ray and the fiducial ray with
$\gamma_0^\mu(\lambda)\equiv\gamma^\mu(\vec 0,\lambda)$. Further, let
$Y^\mu(\vec\theta,\lambda)=
\gamma^\mu(\vec\theta,\lambda)-\gamma^\mu(\vec 0,\lambda)=
[\partial\gamma^\mu(\vec\theta,\lambda)/\partial\theta_k]\theta_k$
denote the vector connecting the ray characterised by $\vec\theta$
with the fiducial ray at the same affine parameter $\lambda$, where we
assumed sufficiently small $|\vec\theta|$ so that $Y^\mu$ can be
linearised in $\vec\theta$. We can then decompose $Y^\mu$ as
follows. At O, the vectors $U_\mathrm{o}^\mu$ and $\tilde k^\mu$
define a two-dimensional plane perpendicular to both
$U_\mathrm{o}^\mu$ and $\tilde k^\mu$. This plane is tangent to the
sphere of directions seen by the observer. Now choose orthonormal unit
vectors $E_1$ and $E_2$ to span that plane. Hence, $E_1^\mu E_{2
\mu}=0$, $E_k^\mu E_{k\mu}=-1$, $E_k^\mu\tilde k_\mu=E_k^\mu
U_{\mathrm{o}\mu}=0$, for $k=1,2$. Transporting the four vectors
$\tilde k^\mu$, $U_\mathrm{o}^\mu$, $E_1^\mu$, and $E_2^\mu$ parallel
along the fiducial ray defines a {\em vierbein\/} at each event along
the fiducial ray. The deviation vector can then be decomposed into
\begin{equation}
  Y^\mu(\vec\theta,\lambda) =
  -\xi_1(\vec\theta,\lambda)\,E_1^\mu
  -\xi_2(\vec\theta,\lambda)\,E_2^\mu
  -\xi_0(\vec\theta,\lambda)\,\tilde k^\mu\;.
\label{eq:3.21}
\end{equation}
Thus, the two-dimensional vector $\vec\xi(\vec\theta,\lambda)$ with
components $\xi_{1,2}(\vec\theta,\lambda)$ describes the transverse
separation of two light rays at affine parameter $\lambda$, whereas
$\xi_0$ allows for a deviation component along the beam direction. Due
to the linearisation introduced above, $\vec\xi$ depends linearly on
$\vec\theta$, and the choice of $\lambda$ assures that
$\d\vec\xi/\d\lambda(\lambda=0)=\vec\theta$. Hence, we can write the
linear propagation equation
\begin{equation}
  \vec\xi(\lambda)=\mathcal{D}(\lambda)\,\vec\theta\;.
\label{eq:3.22}
\end{equation}
The $2\times2$ matrix $\mathcal{D}$ satisfies the Jacobi differential
equation
\begin{equation}
  \frac{\d^2\mathcal{D}(\lambda)}{\d\lambda^2} =
  \mathcal{T}(\lambda)\,\mathcal{D}(\lambda)\;,
\label{eq:3.23}
\end{equation}
with initial conditions
\begin{equation}
  \mathcal{D}(0) = \mathcal{O}
  \quad\hbox{and}\quad
  \frac{\d\mathcal{D}}{\d\lambda}(0) = \mathcal{I}\;.
\label{eq:3.24}
\end{equation}
The {\em optical tidal matrix\/} $\mathcal{T}(\lambda)$ is symmetric,
\begin{equation}
  \mathcal{T}(\lambda) = \left(\begin{array}{cc}
    \mathcal{R}(\lambda)+\Re[\mathcal{F}(\lambda)] & 
    \Im[\mathcal{F}(\lambda)] \\
    \Im[\mathcal{F}(\lambda)] &
    \mathcal{R}(\lambda)-\Re[\mathcal{F}(\lambda)] \\
  \end{array}\right)\;,
\label{eq:3.25}
\end{equation}
and its components depend on the curvature of the metric. $\Re(z)$ and
$\Im(z)$ denote the real and imaginary parts of the complex number
$z$. Specifically,
\begin{equation}
  \mathcal{R}(\lambda) = -\frac{1}{2}R_{\mu\nu}(\lambda)
  \tilde k^\mu(\lambda)\tilde k^\nu(\lambda)\;,
\label{eq:3.26}
\end{equation}
where $R_{\mu\nu}(\lambda)$ is the Ricci tensor at
$\gamma_0^\mu(\lambda)$. The complex quantity $\mathcal{F}(\lambda)$
is more complicated and depends on the Weyl curvature tensor at
$\gamma_0^\mu\lambda)$. The {\em source of convergence\/}
$\mathcal{R}(\lambda)$ leads to an isotropic focusing of light
bundles, in that a circular light beam continues to have a circular
cross section. In contrast, a non-zero {\em source of shear\/}
$\mathcal{F}(\lambda)$ causes an anisotropic focusing, changing the
shape of the light bundle. For a similar set of equations, see,
e.g.~\cite{BL91.1} and \cite{pee93}.

To summarise this subsection, the transverse separation vector
$\vec\xi$ of two infinitesimally close light rays, enclosing an angle
$\vec\theta$ at the observer, depends linearly on $\vec\theta$. The
matrix which describes this linear mapping is obtained from the Jacobi
differential equation (\ref{eq:3.23}). The optical tidal matrix
$\mathcal{T}$ can be calculated from the metric. This exact result
from General Relativity is of course not easily applied to practical
calculations in general space-times, as one first has to calculate the
null geodesic $\gamma_0^\mu(\lambda)$, and from that the components of
the tidal matrix have to be determined. However, as we shall show
next, the equations attain rather simple forms in the case of weak
gravitational fields.

\subsubsection{\label{sc:3.2.2}Specialisation to Weak Gravitational
  Fields}

We shall now specialise the transport equation (\ref{eq:3.23}) to the
situation of a homogeneous and isotropic universe, and to weak
gravitational fields. In a metric of the Robertson-Walker type
(\ref{eq:2.2}, page~\pageref{eq:2.2}), the source of shear
$\mathcal{F}$ must vanish identically because of isotropy; otherwise
preferred directions would exist. Initially circular light bundles
therefore remain circular. Hence, the optical tidal matrix
$\mathcal{T}$ is proportional to the unit matrix,
$\mathcal{T}(\lambda)=\mathcal{R}(\lambda)\,\mathcal{I}$, and the
solution of (\ref{eq:3.23}) must be of the form
$\mathcal{D}(\lambda)=D(\lambda)\,\mathcal{I}$. According to
(\ref{eq:3.22}), the function $D(\lambda)$ is the angular-diameter
distance as a function of the affine parameter. As we shall
demonstrate next, this function indeed agrees with the angular
diameter distance as defined in (\ref{eq:2.42},
page~\pageref{eq:2.42}).

To do so, we first have to find $\mathcal{R}(\lambda)$. The Ricci
tensor deviates from the Einstein tensor by two terms proportional to
the metric tensor $g_{\mu\nu}$, one involving the Ricci scalar, the
other containing the cosmological constant. These two terms do not
contribute to (\ref{eq:3.26}), since $\tilde k^\mu$ is a null
vector. We can thus replace the Ricci tensor in (\ref{eq:3.26}) by the
energy-momentum tensor according to Einstein's field equation. Since
$k^0=c^{-1}\omega=(1+z)c^{-1}\omega_\mathrm{o}$, we have $\tilde
k^0=-(1+z)$, and the spatial components of $\tilde k^\mu$ are
described by a direction and the constraint that $\tilde k^\mu$ is a
null vector. Then, using the energy-momentum tensor of a perfect fluid
with density $\rho$ and pressure $p$, (\ref{eq:3.26}) becomes
\begin{equation}
  \mathcal{R}(\lambda) = -\frac{4\pi G}{c^2}
  \left(\rho+\frac{p}{c^2}\right)(1+z)^2\;.
\label{eq:3.27}
\end{equation}
Specialising to a universe filled with dust, i.e.~$p=0$, we find from
(\ref{eq:2.16}, page~\pageref{eq:2.16}) and (\ref{eq:2.18},
page~\pageref{eq:2.18})
\begin{equation}
  \mathcal{R}(\lambda) = -\frac{3}{2}\,
  \left(\frac{H_0}{c}\right)^2\,\Omega_0\,(1+z)^5\;.
\label{eq:3.28}
\end{equation}
The transport equation (\ref{eq:3.23}) then transforms to
\begin{equation}
  \frac{\d^2D}{\d\lambda^2} = -\frac{3}{2}\,
  \left(\frac{H_0}{c}\right)^2\,\Omega_0\,(1+z)^5\,D\;.
\label{eq:3.29}
\end{equation}
In order to show that the solution of (\ref{eq:3.29}) with initial
conditions $D=0$ and $\d D=\d\lambda$ at $\lambda=0$ is equivalent to
(\ref{eq:2.42}, page~\pageref{eq:2.42}), we proceed as follows. First
we note that (\ref{eq:2.42}) for $z_1=0$ can be written as an
initial-value problem,
\begin{equation}
  \frac{\d^2}{\d w^2}\left(\frac{D_\mathrm{ang}}{a}\right)
  = -K\,\left(\frac{D_\mathrm{ang}}{a}\right)\;,
\label{eq:3.30}
\end{equation}
with $D_\mathrm{ang}(0)=0$ and $\d D_\mathrm{ang}=\d w$ at $w=0$,
because of the properties of the function $f_K$; cf.~(\ref{eq:2.4},
page~\pageref{eq:2.4}). Next, we need a relation between $\lambda$ and
$w$. The null component of the photon geodesic is
$x^0=c(t_0-t)$. Then, from $\d x^\mu=\tilde k^\mu\d\lambda$, we obtain
$\d\lambda=-ac\d t$. Using $\d t=\dot a^{-1}\d a$, we find
\begin{equation}
  \d a = -\frac{\dot a}{c\,a}\d\lambda\;,
  \quad\hbox{or}\quad
  \d z = \frac{\dot a}{c\,a^3}\d\lambda\;.
\label{eq:3.31}
\end{equation}
Since $c\,\d t=-a\,\d w$ for null rays, we have $\dot a^{-1}\d a=\d t=
-ac^{-1}\d w$, which can be combined with (\ref{eq:3.31}) to yield
\begin{equation}
  \d\lambda = a^2\,\d w\;.
\label{eq:3.32}
\end{equation}
We can now calculate the analogous expression of (\ref{eq:3.30}) for
$D$,
\begin{equation}
  \frac{\d^2}{\d w^2}\left(\frac{D}{a}\right)
  = a^2\frac{\d}{\d\lambda}
    \left[a^2\,\frac{\d}{\d\lambda}\left(\frac{D}{a}\right)\right]
  = a^3\,D''-a^2\,a''\,D\;,
\label{eq:3.33}
\end{equation}
where a prime denotes differentiation with respect to $\lambda$. From
(\ref{eq:3.31}), $a'=-(ac)^{-1}\dot a$, and
\begin{equation}
  a'' = \frac{1}{2}\frac{\d(a')^2}{\d a} = \frac{1}{2c^2}
  \frac{\d}{\d a}\left(\frac{\dot a^2}{a^2}\right) = 
  \frac{1}{2c^2}\frac{\d H^2}{\d a}\;,
\label{eq:3.34}
\end{equation}
with $H$ given in (\ref{eq:2.30},
page~\pageref{eq:2.30}). Substituting (\ref{eq:3.29}) into the first
term on the right-hand side of (\ref{eq:3.33}), and (\ref{eq:3.34})
into the second term, we immediately see that $D$ satisfies the
differential equation (\ref{eq:3.30}). Since $D$ has the same initial
conditions as $D_\mathrm{ang}$, they indeed agree.

For computational convenience, we can also transform (\ref{eq:3.29})
into a differential equation for $D(z)$. Using (\ref{eq:3.31}) and
(\ref{eq:2.30}), one finds
\begin{eqnarray}
  && (1+z)\,\left[
    (1+\Omega_0 z)-\Omega_\Lambda\left(1-\frac{1}{(1+z)^2}\right)
  \right]
  \frac{\d^2D}{\d z^2} \nonumber\\
  &+& \left[\frac{7}{2}\Omega_0 z+\frac{\Omega_0}{2}+3
  - \Omega_\Lambda\left(3-\frac{2}{(1+z)^2}\right)\right]
  \frac{\d D}{\d z} + \frac{3}{2}\Omega_0D=0\;.
\label{eq:3.35}
\end{eqnarray}

We next turn to the case of a weak isolated mass inhomogeneity with a
spatial extent small compared to the Hubble distance $cH_0^{-1}$, like
galaxies or clusters of galaxies. In that case, the metric can locally
be approximated by the post-Minkowskian line element
\begin{equation}
  \d s^2 = \left(1+\frac{2\Phi}{c^2}\right)\,c^2\d t^2 
  - \left(1-\frac{2\Phi}{c^2}\right)\d x^2\;,
\label{eq:3.36}
\end{equation}
where $\d x^2$ is the line element of Euclidian three-space, and
$\Phi$ is the Newtonian gravitational potential which is assumed to be
weak, $\Phi\ll c^2$. Calculating the curvature tensor of the metric
(\ref{eq:3.36}), and using Poisson's equation for $\Phi$, we find that
for a light ray which propagates into the three-direction, the sources
of convergence and shear are
\begin{equation}
  \mathcal{R} = -\frac{4\pi G}{c^2}\rho\;,
  \quad\hbox{and}\quad
  \mathcal{F} = -\frac{1}{c^2}\left(
    \Phi_{,11}-\Phi_{,22} + 2\mathrm{i}\Phi_{,12}
  \right)\;.
\label{eq:3.37}
\end{equation}

Now the question is raised as to how an isolated inhomogeneity can be
combined with the background model of an expanding universe. There is
no exact solution of Einstein's field equations which describes a
universe with density fluctuations, with the exception of a few very
special cases such as the Swiss-Cheese model (\cite*{eis45}). We
therefore have to resort to approximation methods which start from
identifying `small' parameters of the problem, and expanding the
relevant quantities into a Taylor series in these parameters. If the
length scales of density inhomogeneities are much smaller than the
Hubble length $cH_0^{-1}$, the associated Newtonian gravitational
potential $\Phi\ll c^2$ (note that this does not imply that the
relative density fluctuations are small!), and the peculiar velocities
$v\ll c$, then an approximate metric is
\begin{equation}
  \d s^2 = a^2(\tau)\left[
    \left(1+\frac{2\Phi}{c^2}\right)\,c^2\d\tau^2-
    \left(1-\frac{2\Phi}{c^2}\right)
    \left(\d w^2+f_K^2(w)\d\omega^2\right)
  \right]\;,
\label{eq:3.38}
\end{equation}
where $\d\tau=a^{-1}\d t$ is the conformal time element, and $\Phi$
satisfies Poisson's equation with source $\Delta\rho$, the density
enhancement or reduction relative to the mean cosmic density
(\cite*{FU89.3}; \cite*{FU89.2}; \cite*{JA93.1}).

In the case of weak metric perturbations, the sources of convergence
and shear of the background metric and the perturbations can be
added. Recalling that both $\mathcal{R}$ and $\mathcal{F}$ are
quadratic in $\tilde k^\mu\propto(1+z)$, so that the expressions in
(\ref{eq:3.37}) have to be multiplied by $(1+z)^2$, we find for the
optical tidal matrix
\begin{equation}
  \mathcal{T}_{ij}(\lambda) = -\frac{3}{2}\,
  \left(\frac{H_0}{c}\right)^2\,\Omega_0\,(1+z)^5\,\delta_{ij}-
  \frac{(1+z)^2}{c^2}
  \left(2\Phi_{,ij}+\delta_{ij}\Phi_{,33}\right)\;,
\label{eq:3.39}
\end{equation}
where we have assumed that the local Cartesian coordinates are chosen
such that the light ray propagates in $x_3$-direction. The same result
is obtained from the metric (\ref{eq:3.38}).

The lens equation as discussed in Sect.~\ref{sc:3.1} can now be
derived from the previous relations. To do so, one has to assume a
geometrically thin matter distribution, i.e.~one approximates the
density perturbation $\Delta\rho$ by a distribution which is
infinitely thin in the direction of photon propagation. It is then
characterised by its surface mass density $\Sigma(\vec\xi)$. The
corresponding Newtonian potential $\Phi$ can then be inserted into
(\ref{eq:3.39}). The integration over $\Phi_{,33}$ along the light ray
vanishes, and (\ref{eq:3.23}) can be employed to calculate the change
of $\d\mathcal{D}/\d\lambda$ across the thin matter sheet (the lens
plane), whereas the components of $\mathcal{D}$ far from the lens
plane are given by a linear combination of solutions of the transport
equation~(\ref{eq:3.29}). Continuity and the change of derivative at
$\lambda_\mathrm{d}$, corresponding to the lens redshift
$z_\mathrm{d}$, then uniquely fix the solution. If
$\mathcal{D}(\vec\theta,\lambda_\mathrm{s})$ denotes the solution at
redshift $z_\mathrm{s}$, then
$\mathcal{D}(\vec\theta,\lambda_\mathrm{s})=
\partial\vec\eta/\partial\vec\theta$ in the notation of
Sect.~\ref{sc:3.1}. Line integration of this relation then leads to
the lens equation (\ref{eq:3.2}). See \cite{SE94.5} for details, and
\cite{PY96.1} for an alternative derivation.

  % -*- LaTeX -*-

\section{\label{sc:4}Principles of Weak Gravitational Lensing}

\subsection{\label{sc:4.1}Introduction}

If the faint, and presumably distant, galaxy population is observed
through the gravitational field of a deflector, the appearance of the
galaxies is changed. The tidal component of the gravitational field
distorts the {\em shapes\/} of galaxy images, and the magnification
associated with gravitational light deflection changes their apparent
{\em brightness\/}. If all galaxies were intrinsically circular, any
galaxy image would immediately provide information on the local tidal
gravitational field. With galaxies being intrinsically elliptical, the
extraction of significant information from individual images is
impossible, except for {\em giant luminous arcs\/} (see
Fig.~\ref{fig:2.7}, page~\pageref{fig:2.7}, for an example) whose
distortion is so extreme that it can easily be determined.

However, assuming that the galaxies are intrinsically randomly
oriented\footnote{\label{fn:4.1}This assumption is not seriously
challenged. Whereas galaxies in a cluster may have non-random
orientations relative to the cluster centre, or pairs of galaxies may
be aligned due to mutual tidal interaction, the faint galaxies used
for lensing studies are distributed over a large volume enclosed by a
narrow cone with opening angle selected by the angular resolution of
the mass reconstruction (see below) and length comparable to the
Hubble radius, since the redshift distribution of faint galaxies is
fairly broad. Thus, the faint galaxies typically have large spatial
separations, which is also reflected by their weak two-point angular
auto-correlation (\cite*{bsm95}; \cite*{vil97}).}, the strength of the
tidal gravitational field can be inferred from a sample of galaxy
images, provided its net ellipticity surmounts the Poisson noise
caused by the finite number of galaxy images in the sample and by the
intrinsic ellipticity distribution.

Since lensing conserves surface brightness, magnification increases
the size of galaxy images at a fixed surface-brightness level. The
resulting flux enhancement enables galaxies to be seen down to fainter
intrinsic magnitudes, and consequently the local number density of
galaxy images above a certain flux threshold can be altered by
lensing.

In this section, we introduce the principles of weak gravitational
lensing. In Sect.~\ref{sc:4.2}, we present the laws of the
transformation between source and image ellipticities and sizes, and
in particular we introduce a convenient definition of the {\em
ellipticity\/} of irregularly-shaped objects. Sect.~\ref{sc:4.3}
focuses on the determination of the local tidal gravitational field
from an ensemble of galaxy images. We derive practical estimators for
the shear and compare their relative merits. The effects of
magnification on the observed galaxy images are discussed in
Sect.~\ref{sc:4.4}. We derive an estimate for the detectability of a
deflector from its weak-lensing imprint on galaxy-image ellipticities
in Sect.~\ref{sc:4.5}, and the final subsection \ref{sc:4.6} is
concerned with practical aspects of the measurement of galaxy
ellipticities.

\subsection{\label{sc:4.2}Galaxy Shapes and Sizes, and their
  Transformation}

If a galaxy had elliptical isophotes, its shape and size could simply
be defined in terms of axis ratio and area enclosed by a boundary
isophote. However, the shapes of faint galaxies can be quite irregular
and not well approximated by ellipses. In addition, observed galaxy
images are given in terms of pixel brightness on CCDs. We therefore
require a definition of size and shape which accounts for the
irregularity of images, and which is well adapted to observational
data.

Let $I(\vec\theta)$ be the surface brightness of a galaxy image at
angular position $\vec\theta$. We first assume that the galaxy image
is isolated, so that $I$ can be measured to large angular separations
from the centre $\bar{\vec\theta}$ of the image,
\begin{equation}
  \bar{\vec\theta} \equiv
  \frac{\int\!\d^2\theta\,q_I[I(\vec\theta)]\,\vec\theta}
       {\int\!\d^2\theta\,q_I[I(\vec\theta)]}\;,
\label{eq:4.1}
\end{equation}
where $q_I(I)$ is a suitably chosen weight function. For instance, if
$q_I(I)=\mathrm{H}(I-I_\mathrm{th})$ is the Heaviside step function,
$\bar{\vec\theta}$ is the centre of the area enclosed by a limiting
isophote $I=I_\mathrm{th}$. Alternatively, if $q_I(I)=I$,
$\bar{\vec\theta}$ is the centre of light. As a third example, if
$q_I(I)=I\,\mathrm{H}(I-I_\mathrm{th})$, $\bar{\vec\theta}$ is the
centre of light within the limiting isophote $I=I_{\rm th}$. Having
chosen $q_I(I)$, we define the tensor of second brightness moments,
\begin{equation}
  Q_{ij} = \frac{
    \int\!\d^2\theta\,q_I[I(\vec\theta)]\,
    (\theta_i-\bar\theta_i)\,(\theta_j-\bar\theta_j)
  }{
    \int\!\d^2\theta\,q_I[I(\vec\theta)]
  }\;,\quad i,j\in \{1,2\}\;,
\label{eq:4.2}
\end{equation}
(e.g.~\cite*{BL91.1}). In writing (\ref{eq:4.1}) and (\ref{eq:4.2}),
we implicitly assumed that $q_I(I)$ is chosen such that the integrals
converge. We can now define the {\em size\/} of an image in terms of
the two invariants of the symmetric tensor $Q$. For example, we can
define the size by
\begin{equation}
  \omega = \left(Q_{11}Q_{22}-Q_{12}^2\right)^{1/2}\;,
\label{eq:4.3}
\end{equation}
so that it is proportional to the solid angle enclosed by the limiting
isophote if $q(I)$ is a step function. We quantify the {\em shape\/}
of the image by the {\em complex ellipticity\/}
\begin{equation}
  \chi \equiv \frac{Q_{11}-Q_{22}+2\mathrm{i}Q_{12}}
  {Q_{11}+Q_{22}}\;.
\label{eq:4.4}
\end{equation}
If the image has elliptical isophotes with axis ratio $r\le1$, then
$\chi=(1-r^2)(1+r^2)^{-1}\exp(2\mathrm{i}\vartheta)$, where the phase
of $\chi$ is twice the position angle $\vartheta$ of the major
axis. This definition assures that the complex ellipticity is
unchanged if the galaxy image is rotated by $\pi$, for this rotation
leaves an ellipse unchanged.

If we define the centre of the source $\bar{\vec\beta}$ and the tensor
of second brightness moments $Q^{(s)}_{ij}$ of the source in complete
analogy to that of the image, i.e.~with $I(\vec\theta)$ replaced by
$I^{(s)}(\vec\beta)$ in eqs.~(\ref{eq:4.1}) and (\ref{eq:4.2}), and
employ the conservation of surface brightness (\ref{eq:3.10},
page~\pageref{eq:3.10}) and the linearised lens equation
(\ref{eq:3.13}, page~\pageref{eq:3.13}), we find that the tensors of
second brightness moments of source and image are related through
\begin{equation}
  Q^{(s)} = \mathcal{A}\,Q\,\mathcal{A}^T = 
  \mathcal{A}\,Q\,\mathcal{A}\;,
\label{eq:4.5}
\end{equation}
where $\mathcal{A}\equiv\mathcal{A}(\bar{\vec\theta})$ is the Jacobian
matrix of the lens equation at position $\bar{\vec\theta}$. Defining
further the complex ellipticity of the source $\chi^{(s)}$ in analogy
to (\ref{eq:4.4}) in terms of $Q^{(s)}$, ellipticities transform
according to
\begin{equation}
  \chi^{(s)} = \frac{\chi-2g+g^2\chi^*}{1+|g|^2-2\Re(g\chi^*)}
\label{eq:4.6}
\end{equation}
(\cite*{SC95.1}; similar transformation formulae were previously
derived by \cite*{KO90.4} and \cite*{MI91.1}), where the asterisk
denotes complex conjugation, and $g$ is the {\em reduced shear\/}
\begin{equation}
  g(\vec\theta) \equiv
  \frac{\gamma(\vec\theta)}{1-\kappa(\vec\theta)}\;.
\label{eq:4.7}
\end{equation}
The inverse transformation is obtained by interchanging $\chi$ and
$\chi^{(s)}$ and replacing $g$ by $-g$ in (\ref{eq:4.6}). Equation
(\ref{eq:4.6}) shows that the transformation of image ellipticities
depends only on the reduced shear, and not on the shear and the
surface mass density individually. Hence, the reduced shear or
functions thereof are the only quantities accessible through
measurements of image ellipticities. This can also immediately be seen
by writing $\mathcal{A}$ as
\begin{equation}
  \mathcal{A}=(1-\kappa)
  \left(
    \begin{array}{cc}
      1-g_1 & -g_2 \\ 
      -g_2 & 1+g_1 \\
    \end{array}
  \right)\;.
\label{eq:4.8}
\end{equation}
The pre-factor $(1-\kappa)$ only affects the size, but not the shape of
the images. From (\ref{eq:4.5}) and (\ref{eq:4.3}), we immediately see
that the sizes of source and image are related through
\begin{equation}
  \omega = \mu(\vec\theta)\,\omega^{(s)}\;.
\label{eq:4.9}
\end{equation}

We point out that the dimension-less surface mass density $\kappa$,
and therefore also the shear $\gamma$, depend not only on the redshift
of the lens, but also on the redshift of the sources, because the
critical surface mass density (\ref{eq:3.7}, page~\pageref{eq:3.7})
involves the source redshift. More precisely, for fixed lens redshift
$z_\mathrm{d}$, the lens strength is proportional to the distance
ratio $D_\mathrm{ds}/D_\mathrm{s}$. This implies that the
transformation (\ref{eq:4.6}) generally also depends on source
redshift. We shall return to these redshift effects in
Sect.~\ref{sc:4.3}, and assume for now that the lens redshift
$z_\mathrm{d}$ is sufficiently small so that the ratio
$D_\mathrm{ds}/D_\mathrm{s}$ is approximately the same for all faint
galaxy images.

Instead of $\chi$, we can define different ellipticity parameters (see
\cite*{BO95.1}). One of these definitions turns out to be quite
useful, namely
\begin{equation}
  \epsilon \equiv \frac{Q_{11}-Q_{22}+2\mathrm{i}Q_{12}}
  {Q_{11}+Q_{22}+2(Q_{11}Q_{22}-Q_{12}^2)^{1/2}}\;,
\label{eq:4.10}
\end{equation}
which we shall also call {\em complex ellipticity\/}. (Since we shall
use the notation $\chi$ and $\epsilon$ consistently throughout this
article, there should be no confusion from using the same name for two
different quantities.) $\epsilon$ has the same phase as $\chi$, and
for elliptical isophotes with axis ratio $r\le1$,
$|\epsilon|=(1-r)(1+r)^{-1}$. $\epsilon$ and $\chi$ are related
through
\begin{equation}
  \epsilon = \frac{\chi}{1+(1-|\chi|^2)^{1/2}}\;,\quad
  \chi = \frac{2\epsilon}{1+|\epsilon|^2}\;.
\label{eq:4.11}
\end{equation}
The transformation between source and image ellipticity in terms of
$\epsilon$ is given by
\begin{equation}
  \epsilon^{(s)} = \left\{\begin{array}{ll}
    \displaystyle\frac{\epsilon-g}{1-g^*\epsilon} &
    \quad\hbox{for}\quad |g|\le1 \\
    \\
    \displaystyle\frac{1-g\epsilon^*}{\epsilon^*-g^*} & 
    \quad\hbox{for}\quad |g|>1 \\
  \end{array}\right.
\label{eq:4.12}
\end{equation}
(\cite*{SE97.1}), and the inverse transformation is obtained by
interchanging $\epsilon$ and $\epsilon^{(s)}$ and replacing $g$ by
$-g$ in (\ref{eq:4.12}). Although the transformation of $\epsilon$
appears more complicated because of the case distinction, we shall see
in the next subsection that it is often useful to work in terms of
$\epsilon$ rather than $\chi$; cf.~eq.~(\ref{eq:4.17}) below.

For the case of weak lensing, which we define for the purpose of this
section by $\kappa\ll1$, $|\gamma|\ll1$, and thus $|g|\ll1$,
(\ref{eq:4.12}) becomes $\epsilon\approx\epsilon^{(s)}+g$, provided
$|\epsilon|\approx|\epsilon^{(s)}|\lesssim 1/2$. Likewise,
eq.~(\ref{eq:4.6}) simplifies to $\chi\approx\chi^{(s)}+2g$ in this
case.

\subsection{\label{sc:4.3}Local Determination of the Distortion}

As mentioned earlier, the observed ellipticity of a single galaxy
image provides only little information about the local tidal
gravitational field of the deflector, for the intrinsic ellipticity of
the source is unknown. However, based on the assumption that the
sources are randomly oriented, information on the local tidal field
can be inferred from a local ensemble of images. Consider for example
galaxy images at positions $\vec\theta_i$ close enough to a fiducial
point $\vec\theta$ so that the local lens properties $\kappa$ and
$\gamma$ do not change appreciably over the region encompassing these
galaxies. The expectation value of their corresponding source
ellipticities is assumed to vanish,
\begin{equation}
  \mathrm{E}(\chi^{(s)}) = 0 = \mathrm{E}(\epsilon^{(s)})\;.
\label{eq:4.13}
\end{equation}

\subsubsection{\label{sc:4.3.1}All Sources at the Same Redshift}

We first consider the case that all sources are at the same
redshift. Then, as mentioned following eq.~(\ref{eq:3.13},
page~\pageref{eq:3.13}), the ellipticity of a circular source
determines the ratio of the local eigenvalues of the Jacobian matrix
$\mathcal{A}$. This also holds for the net image ellipticity of an
ensemble of sources with vanishing net ellipticity. From
(\ref{eq:3.11}, page~\pageref{eq:3.11}), we find for the ratio of the
eigenvalues of $\mathcal{A}$ in terms of the reduced shear $g$
\begin{equation}
  r = \frac{1\mp|g|}{1\pm|g|}\;.
\label{eq:4.14}
\end{equation}
Interestingly, if we replace $g$ by $1/g^*$, $r$ switches sign, but
$|r|$ and the phase of $\epsilon$ remain unchanged. The sign of $r$
cannot be determined observationally, and hence measurements cannot
distinguish between $g$ and $1/g^*$. This is called {\em local
degeneracy\/}. Writing $\det\mathcal{A}=(1-\kappa)^2(1-|g|^2)$, we see
that the degeneracy between $g$ and $1/g^*$ means that we cannot
distinguish between observed images inside a critical curve (so that
$\det\mathcal{A}<0$ and $|g|>1$) or outside. Therefore, only functions
of $g$ which are invariant under $g\to1/g^*$ are accessible to (local)
measurements, as for instance the {\em complex distortion\/}
\begin{equation}
  \delta \equiv \frac{2g}{1+|g|^2}\;.
\label{eq:4.15}
\end{equation}

Replacing the expectation value in (\ref{eq:4.13}) by the average over
a local ensemble of image ellipticities,
$\langle\chi^{(s)}\rangle\approx\mathrm{E}(\chi^{(s)})=0$,
\cite{SC95.1} showed that $\langle\chi^{(s)}\rangle=0$ is equivalent
to
\begin{equation}
  \sum_i\,u_i\,\frac{\chi_i-\delta}{1-\Re(\delta\chi_i^*)} = 0\;,
\label{eq:4.16}
\end{equation}
where the $u_i$ are weight factors depending on
$|\vec\theta_i-\vec\theta|$ which can give larger weight to galaxies
closer to the fiducial point. Additionally, the $u_i$ can be chosen
such as to account for measurement uncertainties in the image
ellipticities by giving less weight to images with larger measurement
error. Equation~(\ref{eq:4.16}) has a unique solution $\delta$, so
that the distortion can locally be determined. It is readily solved by
a quickly converging iteration starting from
$\delta=\langle\chi\rangle$.

The $\delta$ obtained from (\ref{eq:4.16}) is an unbiased estimate of
the distortion. Its dispersion about the true value depends on the
dispersion $\sigma_\chi$ of the intrinsic ellipticity distribution,
and on the number of galaxy images. A fairly accurate estimate of the
{\em rms\/} error of $\delta$ is
$\sigma_\delta\approx\sigma_\chi\,N^{-1/2}$, where $N$ is the
effective number of galaxies used for the local average, $N=\left(\sum
u_i\right)^2\left(\sum u_i^2\right)^{-1}$. This overestimates the
error for large values of $|\delta|$ (\cite*{SC95.1}). It is important
to note that the expectation value of $\chi$ is {\em not\/} $\delta$,
but differs from it by a factor which depends both on $|\delta|$ and
the intrinsic ellipticity distribution of the sources. In contrast to
that, it follows from (\ref{eq:4.13}) and (\ref{eq:4.12}) that the
expectation value of the complex ellipticity $\epsilon$ of the images
{\em is\/} the reduced shear or its inverse, $\mathrm{E}(\epsilon)=g$
if $|g|<1$ and $\mathrm{E}(\epsilon)=1/g^*$ if $|g|>1$
(\cite*{SC95.5}; \cite*{SE97.1}). Hence,
\begin{equation}
  \langle\epsilon\rangle = \frac{\sum_i u_i\epsilon_i}{\sum_i u_i}
\label{eq:4.17}
\end{equation}
is an unbiased local estimate for $g$ or $1/g^*$. The ellipticity
parameter $\epsilon$ is useful exactly because of this property. If
one deals with sub-critical lenses (i.e.~lenses which are not dense
enough to have critical curves, so that
$\det\mathcal{A}(\vec\theta)>0$ everywhere), or with the region
outside the critical curves in critical lenses, the degeneracy between
$g$ and $1/g^*$ does not occur, and $\langle\epsilon\rangle$ is a
convenient estimate for the local reduced shear. The {\em rms\/} error
of this estimate is approximately
$\sigma_g\approx\sigma_\epsilon\,(1-|g|^2)\,N^{-1/2}$ (\cite*{ske99}),
where $\sigma_\epsilon$ is the dispersion of the intrinsic source
ellipticity $\epsilon^{(s)}$. As we shall see in a moment, $\epsilon$
is the more convenient ellipticity parameter when the sources are
distributed in redshift.

The estimates for $\delta$ and $g$ discussed above can be derived
without knowing the intrinsic ellipticity distribution. If, however,
the intrinsic ellipticity distribution is known (e.g.~from deep {\em
Hubble Space Telescope\/} images), we can exploit this additional
information and determine $\delta$ (or $g$) through a
maximum-likelihood method (\cite*{gou95}; \cite*{LO98.1}). Depending
on the shape of the intrinsic ellipticity distribution, this approach
can yield estimates of the distortion which have a smaller {\em rms\/}
error than the estimates discussed above. However, if the intrinsic
ellipticity distribution is approximately Gaussian, the {\em rms\/}
errors of both methods are identical. It should be noted that the
intrinsic ellipticity distribution is likely to depend on the apparent
magnitude of the galaxies, possibly on their redshifts, and on the
wavelength at which they are observed, so that this distribution is
not easily determined observationally. Knowledge of the intrinsic
ellipticity distribution can also be used to determine $\delta$ from
the orientation of the images (that is, the phase of $\chi$) only
(\cite*{KO90.4}; \cite*{SC95.1}; \cite*{dei96}, unpublished). This may
provide a useful alternative to the method above since the orientation
of images is much less affected by seeing than the modulus of
$\chi$. We return to the practical estimate of the image ellipticities
and the corresponding distortion in Sect.~\ref{sc:4.5}.

In the case of weak lensing, defined by $\kappa\ll1$ and
$|\gamma|\ll1$, implying $|g|\ll1$, we find from
(\ref{eq:4.11}--\ref{eq:4.16}) that
\begin{equation}
  \gamma \approx g \approx \frac{\delta}{2} \approx
  \langle\epsilon\rangle \approx
  \frac{\langle\chi\rangle}{2}\;.
\label{eq:4.18}
\end{equation}

\subsubsection{\label{sc:4.3.2}Sources Distributed in Redshift}

So far, we assumed that all source galaxies are at the same redshift,
or more precisely, that the ratio $D_\mathrm{ds}/D_\mathrm{s}$ between
the lens-source and observer-source distances is the same for all
sources. This ratio enters into the scaling (\ref{eq:3.7},
page~\pageref{eq:3.7}) of the physical surface mass density $\Sigma$
to the dimension-less convergence $\kappa$. The deflection angle, the
deflection potential, and the shear are all linear in $\kappa$, so
that the distance ratio $D_\mathrm{ds}/D_\mathrm{s}$ is sufficient to
specify the lens strength as a function of source redshift. Provided
$z_\mathrm{d}\lesssim0.2$, this ratio is fairly constant for sources
with redshift $z_\mathrm{s}\gtrsim0.8$, so that the approximation used
so far applies to relatively low-redshift deflectors. However, for
higher-redshift lenses, the redshift distribution of the sources must
explicitly be taken into account.

\begin{figure}[ht]
  \includegraphics[width=\hsize]{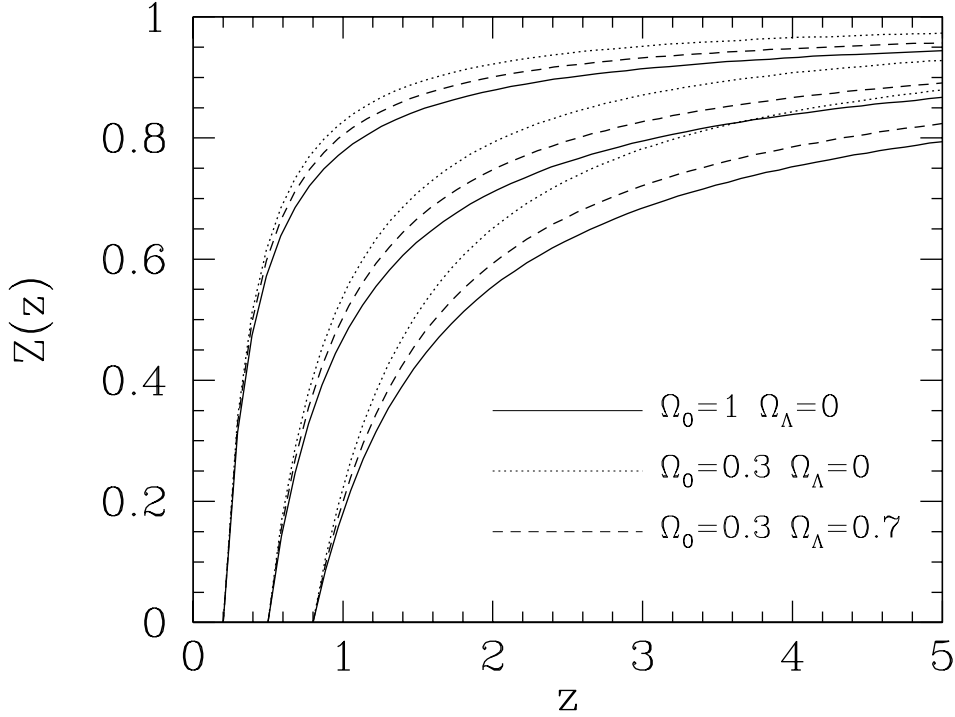}
\caption{The function $Z(z)$ defined in eq.~(\ref{eq:4.19}) describes
the relative lens strength as a function of source redshift $z$. We
show $Z(z)$ for three cosmological models as indicated in the figure,
and for three values for the lens redshift, $z_\mathrm{d}=0.2, 0.5,
0.8$. By definition, $Z(z)\to0$ as $z\to z_\mathrm{d}$, and $Z(z)\to1$
as $z\to\infty$. For sources close to the deflector, $Z(z)$ varies
strongly in a way depending relatively weakly on cosmology.}
\label{fig:4.1}
\end{figure}

For a fixed lens redshift $z_\mathrm{d}$, the dimension-less surface
mass density and the shear depend on the source redshift. We define
\begin{eqnarray}
  Z(z) & \equiv &
  \frac{\lim_{z\to\infty}\Sigma_\mathrm{cr}(z_\mathrm{d},z)}
       {\Sigma_\mathrm{cr}(z_\mathrm{d},z)}\;
  \mathrm{H}(z-z_\mathrm{d}) \nonumber \\
  & = &
  \frac{f_K[w(z_\mathrm{d},z)]}{f_K[w(0,z)]}\;
  \frac{f_K[w(0,\infty)]}{f_K[w(z_\mathrm{d},\infty)]}\;
  \mathrm{H}(z-z_\mathrm{d}) \;,
\label{eq:4.19}
\end{eqnarray}
using the notation of Sect.~\ref{sc:2.1} (page~\pageref{sc:2.1}). The
Heaviside step function accounts for the fact that sources closer than
the deflector are not lensed. Then,
$\kappa(\vec\theta,z)=Z(z)\kappa(\vec\theta)$, and
$\gamma(\vec\theta,z)=Z(z)\gamma(\vec\theta)$ for a source at $z$, and
$\kappa$ and $\gamma$ refer to a fictitious source at redshift
infinity. The function $Z(z)$ is readily evaluated for any
cosmological model using (\ref{eq:2.40}, page~\pageref{eq:2.40}) and
(\ref{eq:2.4}, page~\pageref{eq:2.4}). We plot $Z(z)$ for various
cosmologies and lens redshifts in Fig.~\ref{fig:4.1}.

The expectation value for the ellipticity of images with redshift $z$
now becomes
\begin{equation}
  \mathrm{E}[\epsilon(z)] = \left\{\begin{array}{ll}
    \displaystyle\frac{Z(z)\,\gamma}{1-Z(z)\,\kappa}
    & \quad\hbox{for}\quad \mu(z)\ge0 \\
    \\
    \displaystyle\frac{1-Z(z)\,\kappa}{Z(z)\,\gamma^*} 
    & \quad\hbox{for}\quad \mu(z)<0 \\
  \end{array}\right.\;,
\label{eq:4.20}
\end{equation}
where $\mu(z)$ is the magnification as a function of source redshift,
\begin{equation}
  \mu(z) = \left\{[1-Z(z)\kappa]^2-Z^2(z)|\gamma|^2\right\}^{-1}\;.
\label{eq:4.21}
\end{equation}
We refer to {\em sub-critical\/} lensing if $\mu(z)>0$ for all
redshifts, which is equivalent to $1-\kappa-|\gamma|>0$.

Without redshift information, only the mean ellipticity averaged over
all redshifts can be observed. We first consider this case, for which
the source redshift distribution is assumed to be known. We define the
probability $p_z(z)\d z$ that a galaxy image (in the selected
magnitude range) has a redshift within $\d z$ of $z$. The image
redshift distribution will in general be different from the source
redshift distribution since magnified sources can be seen to higher
redshifts than unlensed ones. Therefore, the redshift distribution
will depend on the local lens parameters $\kappa$ and $\gamma$ through
the magnification (\ref{eq:4.21}). If, however, the magnification is
small, or if the redshift distribution depends only weakly on the
flux, the simplification of identifying the two redshift distributions
is justified. We shall drop it later. Given $p_z(z)$, the expectation
value of the image ellipticity becomes the weighted average
\begin{equation}
  \mathrm{E}(\epsilon) = 
  \int\d z\,p_z(z)\,\mathrm{E}[\epsilon(z)] =
  \gamma\left[ X(\kappa,\gamma)+|\gamma|^{-2}
  Y(\kappa,\gamma)\right]\;,
\label{eq:4.22}
\end{equation}
with
\begin{eqnarray}
  X(\kappa,\gamma) &=& \int_{\mu(z)\ge0} \d z\;p_z(z)\,
  \frac{Z(z)}{1-Z(z)\kappa},\nonumber\\
  Y(\kappa,\gamma) &=& \int_{\mu(z)<0} \d z\;p_z(z)\,
  \frac{1-Z(z)\kappa}{Z(z)}\;,
\label{eq:4.23}
\end{eqnarray}
and the integration boundaries depend on the values of $\kappa$ and
$|\gamma|$ through the magnification.

If the lens is sub-critical, $\mu(z)>0$ for all $z$. Then $Y=0$, and
only the first term in (\ref{eq:4.22}) remains. Also, $X$ no longer
depends on $\gamma$ in this case, and
$\mathrm{E}(\epsilon)=\gamma\,X(\kappa)$. An accurate approximation
for $X(\kappa)$, valid for $\kappa\lesssim0.6$, has been derived in
\cite{SE97.1},
\begin{equation}
  \gamma = \frac{\mathrm{E}(\epsilon)}{\langle Z\rangle}\,
  \left(
    1-\frac{\langle Z^2\rangle}{\langle Z\rangle}\,\kappa
  \right)\;,
\label{eq:4.24}
\end{equation}
where $\langle Z^n\rangle\equiv\int\d z\,p_z(z)\,Z^n$. 

Specialising further to the weak-lensing regime, the expectation
value of the image ellipticity is simply
\begin{equation}
  \mathrm{E}(\epsilon) \approx \langle Z\rangle\gamma\;.
\label{eq:4.25}
\end{equation} 
Thus, in the weak-lensing case, a source redshift distribution can be
collapsed on a single redshift $z_\mathrm{s}$ satisfying
$Z(z_\mathrm{s})=\langle Z\rangle$.

We now drop the simplification introduced above and define $n_0(S,z)\d
S\d z$ as the number of galaxy images per unit solid angle with flux
within $\d S$ of $S$ and redshift within $\d z$ of $z$ in the absence
of lensing. At a point $\vec\theta$ with surface mass density $\kappa$
and shear $\gamma$, the number density can be changed by
magnification. Images of a fixed set of sources are distributed over a
larger solid angle, reducing the number density by a factor
$\mu^{-1}(z)$. On the other hand, the magnification allows the
observation of fainter sources. In total, the expected number density
becomes
\begin{equation}
  n(S,z) = \frac{1}{\mu^2(z)}\,
  n_0\left(\frac{S}{\mu(z)},z\right)\;,
\label{eq:4.26}
\end{equation}
with $\mu(z)$ given in (\ref{eq:4.21}). This yields the redshift
distribution
\begin{equation}
  p(z;S,\kappa,\gamma) = \frac{n_0\left[\mu^{-1}(z)S,z\right]}
  {\mu^2(z)\int\d z'\,\mu^{-2}(z')\,
  n_0\left[\mu^{-1}(z)S,z'\right]}\;, 
\label{eq:4.27}
\end{equation}
which depends on the flux $S$ and the local lens parameters $\kappa$
and $\gamma$ through the magnification. This function can now be
substituted for $p_z(z)$ in eq.~(\ref{eq:4.22}).

\subsubsection{\label{sc:4.3.3}Practical Estimates of the Shear}

We saw before that $\langle\epsilon\rangle=\sum_i u_i
\epsilon_i/\sum_i u_i$ is an unbiased estimate of the local reduced
shear $g$ if all sources are at the same redshift. We now generalise
this result for sources distributed in redshift. Then, the expectation
value of $\epsilon$ is no longer a simple function of $\kappa$ and
$\gamma$, and therefore estimates of $\gamma$ for an assumed value for
$\kappa$ will be derived.

We first assume that redshifts for individual galaxies are
unavailable, but that only the normalised redshift distribution
$p_z(z)$ is known, or the distribution in
eq.~(\ref{eq:4.27}). Replacing the expectation value of the image
ellipticity by the mean, eq.~(\ref{eq:4.22}) implies that the solution
$\gamma^{(1)}$ of
\begin{equation}
  \gamma = \left[X(\kappa,\gamma)+|\gamma|^{-2}
  Y(\kappa,\gamma)\right]^{-1}\,\langle\epsilon\rangle
\label{eq:4.28}
\end{equation}
provides an unbiased estimator for the shear $\gamma$.  This is not a
particularly explicit expression for the shear estimate, but it is
still extremely useful, as we shall see in the next section. The shear
estimate considerably simplifies if we assume a sub-critical
lens. Then,
\begin{equation}
  \gamma^{(1,\mathrm{sc})} =
  \langle\epsilon\rangle X^{-1}(\kappa) \approx
  \frac{\langle\epsilon\rangle}{\langle Z\rangle}\,
  \left(
    1-\frac{\langle Z^2\rangle}{\langle Z\rangle}\,\kappa
  \right)\;,
\label{eq:4.29}
\end{equation}
where we used eq.~(\ref{eq:4.24}) in the second step. Specialising
further to weak lensing, the shear estimate simplifies to
\begin{equation}
  \gamma^{(1,\mathrm{wl})} = \langle\epsilon\rangle\,
  \langle Z\rangle^{-1}\;.
\label{eq:4.30}
\end{equation}

Next, we assume that the redshifts of all galaxy images are known. At
first sight, this appears entirely unrealistic, because the galaxy
images are so faint that a complete spectroscopic survey at the
interesting magnitude limits seems to be out of reach. However, it has
become clear in recent years that accurate redshift estimates, the
so-called photometric redshifts, can be obtained from multi-colour
photometry alone (see, e.g., \cite*{ccs95}). The accuracy of
photometric redshifts depends on the number of wave bands for which
photometry is available, the photometric accuracy, and the galaxy
type; typical errors are $\Delta z\sim0.1$ for faint, high-redshift
galaxies. This uncertainty is small compared to the range over which
the function $Z(z)$ varies appreciably, so that photometric redshifts
are (almost) as good as precise spectroscopic redshifts for our
purposes.

If the redshifts $z_i$ of the galaxies are known, more precise shear
estimates than before can be derived. Consider the weighted sum
$F\equiv\sum_i u_i\,|\epsilon_i-\mathrm{E}(\epsilon_i)|^2$, where the
expectation value is given by eq.~(\ref{eq:4.20}), and $Z=Z_i\equiv
Z(z_i)$. For an assumed value of $\kappa$, an unbiased estimate of
$\gamma$ is given by the $\gamma^{(2)}$ minimising $F$. Due to the
case distinction in eq.~(\ref{eq:4.20}), this estimator is complicated
to write down analytically, but can easily be calculated numerically.

This case distinction is no longer necessary in the sub-critical case,
for which the resulting estimator reads
\begin{equation}
  \gamma^{(2,\mathrm{sc})} =
  \frac{\sum_i u_i\,Z_i\,\epsilon_i\,(1-Z_i \kappa)^{-1}}
       {\sum_i u_i\,Z_i^2\,(1-Z_i\kappa)^{-2}}\;.
\label{eq:4.31}
\end{equation}
In the case of weak lensing, this becomes
\begin{equation}
  \gamma^{(2,\mathrm{wl})} =
  \frac{\sum_i u_i\,Z_i\,\epsilon_i}
       {\sum_i u_i\,Z_i^2}\;.
\label{eq:4.32}
\end{equation}

\begin{figure}[ht]
  \includegraphics[width=\hsize]{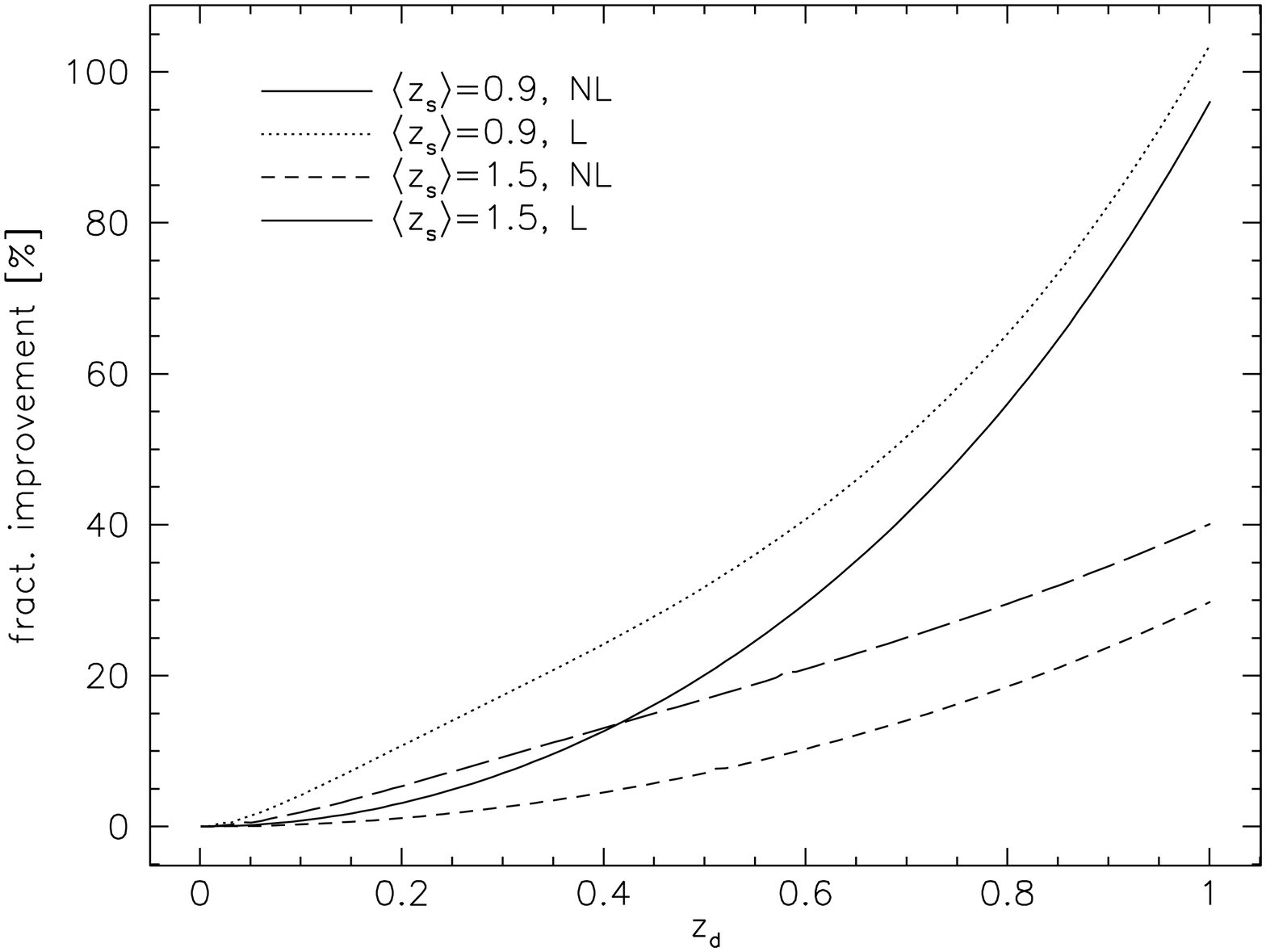}
\caption{The fractional accuracy gain in the shear estimate due to the
knowledge of the source redshifts is plotted, more precisely the
deviation of the square root of (\ref{eq:4.37}) from unity in per
cent. The four curves shown correspond to two different values of the
mean source redshift, and to the cases without lensing
($\kappa=0=\gamma$), and with lensing ($\kappa=0.3=|\gamma|$),
labelled NL and L, respectively. We assumed the redshift distribution
(\ref{eq:2.68}) with $\beta=3/2$, and an Einstein-de Sitter
cosmology. As expected, the higher the lens redshift $z_\mathrm{d}$,
the more substantially is the shear estimate improved by redshift
information, since for low values of $z_\mathrm{d}$, the function
$Z(z)$ is nearly constant. Furthermore, the lower the mean redshift of
the source distribution, the more important the knowledge of
individual redshifts becomes, for example to distinguish between
foreground and background galaxies. Finally, redshift information is
relatively more important for larger lens strength.}
\label{fig:4.2}
\end{figure}

We now compare the accuracy of the shear estimates with and without
redshift information of the individual galaxies. For simplicity, we
assume sub-critical lensing and set all weight factors to unity,
$u_i=1$. The dispersion of the estimate
$\gamma^{(1,\mathrm{sc})}=(N\,X)^{-1}\sum_i \epsilon_i$ for $N$ galaxy
images is
\begin{equation}
  \sigma^2\left(\gamma^{(1,\mathrm{sc})}\right) =
  \mathrm{E}\left(|\gamma^{(1,\mathrm{sc})}|^2\right)-|\gamma|^2 =
  \left[N\,X(\kappa)\right]^{-2}\,
  \mathrm{E}\left(\sum_{ij}\epsilon_i\,\epsilon^*_j\right)-|\gamma|^2\;.
\label{eq:4.33}
\end{equation}
The expectation value in the final expression can be estimated noting
that the image ellipticity is to first order given by
$\epsilon_i=\epsilon_i^{(s)}+\gamma$, and that the intrinsic
ellipticities are uncorrelated. If we further assume that the
redshifts of any two galaxies are uncorrelated, we find
\begin{eqnarray}
  \mathrm{E}\left(\epsilon_i\,\epsilon^*_j\right) &\approx&
  \left\langle
    \frac{Z_iZ_j}{(1-Z_i\kappa)(1-Z_j\kappa)}
  \right\rangle
  |\gamma|^2+\delta_{ij}\sigma_\epsilon^2\nonumber\\
  &=&
  X^2(\kappa)|\gamma|^2 +
  \delta_{ij}\left(\sigma_X^2|\gamma|^2+\sigma_\epsilon^2\right)\;,
\label{eq:4.34}
\end{eqnarray}
where we used the definition (\ref{eq:4.23}) of $X(\kappa)$, and
defined $\sigma_X^2(\kappa)\equiv\langle
Z^2(1-Z\kappa)^{-2}\rangle-X^2$. Angular brackets denote averages over
the redshift distribution $p_z$. Inserting (\ref{eq:4.34}) into
(\ref{eq:4.33}) yields
\begin{equation}
  \sigma^2\left(\gamma^{(1,\mathrm{sc})}\right) =
  \frac{\sigma_X^2|\gamma|^2+\sigma_\epsilon^2}{N\,X^2}\;.
\label{eq:4.35}
\end{equation}
Likewise, the dispersion of the estimate $\gamma^{(2,\mathrm{sc})}$ is
\begin{eqnarray}
  \sigma^2\left(\gamma^{(2,\mathrm{sc})}\right) &=&
  \frac{\sum_{ij}Z_iZ_j(1-Z_i\kappa)^{-1}(1-Z_j\kappa)^{-1}
        \mathrm{E}\left(\epsilon_i\,\epsilon^*_j\right)}
       {\left[ \sum_i Z_i^2 (1-Z_i\kappa)^{-2}\right]^2}
  -|\gamma|^2\nonumber\\ &=&
  \frac{\sigma_\epsilon^2}{\sum_iZ_i^2(1-Z_i\kappa)^{-2}} \approx
  \frac{\sigma_\epsilon^2}{N[X^2(\kappa)+\sigma_X^2(\kappa)]}\;.
\label{eq:4.36}
\end{eqnarray}
We used eq.~(\ref{eq:4.34}), but noted that $Z$ is now no longer a
statistical variable, so that we can put $\sigma_X^2=0$ in
(\ref{eq:4.34}). In the final step, we have replaced the denominator
by its expectation value under ensemble averaging. We then find the
ratio of the dispersions,
\begin{equation}
  \frac{\sigma^2\left(\gamma^{(1,\mathrm{sc})}\right)}
       {\sigma^2\left(\gamma^{(2,\mathrm{sc})}\right)} =
  \left(1+|\gamma|^2\frac{\sigma_X^2}{\sigma_\epsilon^2}\right)
  \left(1+\frac{\sigma_X^2}{X^2}\right)\;.
\label{eq:4.37}
\end{equation}
We thus see that the relative accuracy of these two estimates depends
on the fractional width of the distribution of $Z/(1-Z\kappa)$, and on
the ratio between the dispersion of this quantity and the ellipticity
dispersion. Through its explicit dependence on $|\gamma|^2$, and
through the dependence of $\sigma_X$ and $X$ on $\kappa$, the relative
accuracy also depends on the lens parameters. Quantitative estimates
of (\ref{eq:4.37}) are given in Fig.~\ref{fig:4.2}.

The figure shows that the accuracy of the shear estimate is noticeably
improved, in particular once the lens redshift becomes a fair fraction
of the mean source redshift. The dependence of the lens strength on
the deflector redshift implies that the lens signal will become
smaller for increasing deflector redshift, so that the accuracy gained
by redshift information becomes significant. In addition, the
assumptions used to derive (\ref{eq:4.35}) were quite optimistic,
since we have assumed in (\ref{eq:4.34}) that the sample of galaxies
over which the average is taken is a fair representation of the galaxy
redshift distribution $p_z(z)$. Given that these galaxies come from a
small area (small enough to assume that $\kappa$ and $\gamma$ are
constant across this area), and that the redshift distribution of
observed galaxies in pencil beams shows strong correlations (see,
e.g., \cite*{bek90}, \cite*{sad98}, \cite*{cbh99}), this assumption is
not very realistic. Indeed, the strong clustering of galaxy redshifts
means that the effective $\sigma_X$ will be considerably larger than
the analytical estimate used above. In any case, redshift information
on the source galaxies will substantially improve the accuracy of weak
lensing results.

\subsection{\label{sc:4.4}Magnification Effects}

In addition to the distortion of image {\em shapes\/}, by which the
(reduced) shear can be measured locally, gravitational light
deflection also magnifies the images, leaving the surface brightness
invariant. The magnification changes the size, and therefore the flux,
of individual galaxy images. Moreover, for a fixed set of sources, the
number density of images decreases by a factor $\mu$ as the sky is
locally stretched. Combining the latter effect with the flux
magnification, the lensed and unlensed source counts are changed
according to (\ref{eq:4.26}). Two strategies to measure the
magnification effect have been suggested in the literature, namely
either through the change in the local source counts, perhaps combined
with the associated change (\ref{eq:4.27}) in the redshift
distribution (\cite*{BR95.1}), or through the change of image sizes at
fixed surface brightness (\cite*{BA95.5}).

\subsubsection{\label{sc:4.4.1}Number Density Effect}

Let $n_0(>S,z)\d z$ be the unlensed number density of galaxies with
redshift within $\d z$ of $z$ and with flux larger than $S$. Then, at
an angular position $\vec\theta$ where the magnification is
$\mu(\vec\theta,z)$, the number counts are changed according to
(\ref{eq:4.26}),
\begin{equation}
  n(>S,z) = \frac{1}{\mu(\vec\theta,z)}\,
  n_0\left(>\frac{S}{\mu(\vec\theta,z)},z\right)\;.
\label{eq:4.38}
\end{equation}
Accordingly, magnification can either increase or decrease the local
number counts, depending on the shape of the unlensed number-count
function. This change of number counts is called {\em magnification
bias\/}, and is a very important effect for gravitational lensing of
QSOs (see \cite*{sef92} for references).\footnote{\label{fn:4.2}Bright
QSOs have a very steep number-count function, and so the flux
enhancement of the sources outweighs the number reduction due to the
stretching of the sky by a large margin. Whereas the lensing
probability even for a high-redshift QSO is probably too small to
affect the overall sources counts significantly, the fraction of
multiply-imaged QSOs in flux-limited samples is increased through the
magnification bias by a substantial factor over the probability that
any individual QSO is multiply imaged (see, e.g.~\cite*{TU84.1};
\cite*{naw93} and references therein).}

Magnification allows the observation of fainter sources. Since the
flux from the sources is correlated with their redshift, the redshift
distribution is changed accordingly,
\begin{equation}
  p(z;>S,\kappa,\gamma) = 
  \frac{n_0\left[>\mu^{-1}(z)S,z\right]}
       {\mu(z)\int\d z'\,\mu^{-1}(z')\,
        n_0\left[>\mu^{-1}(z')S,z'\right]}\;,
\label{eq:4.39}
\end{equation}
in analogy to the redshift distribution (\ref{eq:4.27}) at fixed flux
$S$. Since the objects of interest here are very faint, spectroscopic
redshift information is in general difficult to obtain, and so one can
only observe the redshift-integrated counts
\begin{equation}
  n(>S) = \int\d z\,\frac{1}{\mu(z)}\,
  n_0\left(>\mu^{-1}(z)S,z\right)\;.
\label{eq:4.40}
\end{equation}
The number counts of faint galaxies are observed to very closely
follow a power law over a wide range of fluxes, and so we write the
unlensed counts as
\begin{equation}
  n_0(>S,z) = a\,S^{-\alpha}\,p_0(z;S)\;,
\label{eq:4.41}
\end{equation}
where the exponent $\alpha$ depends on the wave band of the
observation (e.g.~\cite*{SM95.1}), and $p_0(z;S)$ is the redshift
probability distribution of galaxies with flux $>S$. Whereas this
redshift distribution is fairly well known for brighter galaxies which
are accessible to current spectroscopy, little is known about the
faint galaxies of interest here. The ratio of the lensed and unlensed
source counts is then found by inserting (\ref{eq:4.41}) into
(\ref{eq:4.40}),
\begin{equation}
  \frac{n(>S)}{n_0(>S)} = \int\d z\,\mu^{\alpha-1}(z)\,
  p_0\left(z;\mu^{-1}(z)S\right)\;.
\label{eq:4.42}
\end{equation}
We should note that the lensed counts do not strictly follow a power
law in $S$, for $p_0$ depends on $z$. Since the redshift distribution
$p_0(z,S)$ is currently unknown, the change of the number counts due
to the magnification cannot be predicted. For very faint flux
thresholds, however, the redshift distribution is likely to be
dominated by galaxies at relatively high redshift. For lenses at
fairly small redshift (say $z_\mathrm{d}\lesssim0.3$), we can
approximate the redshift-dependent magnification $\mu(z)$ by the
magnification $\mu$ of a fiducial source at infinity, in which case
\begin{equation}
  \frac{n(>S)}{n_0(>S)} = \mu^{\alpha-1}\;.
\label{eq:4.43}
\end{equation}
Thus, a local estimate of the magnification can be obtained through
(\ref{eq:4.43}) and from a measurement of the local change of the
number density of images. If the slope of the source counts is unity,
$\alpha=1$, there will be no magnification bias, while it will cause a
decrease of the local number density for flatter slopes. \cite{bro95}
pointed out that one can immediately obtain (for sub-critical lensing,
i.e.~$\det\mathcal{A}>0$) an estimate for the local surface mass
density from a measurement of the local magnification and the local
reduced shear $g$, $\kappa=1-[\mu(1-|g|^2)]^{-1/2}$. In the absence of
shape information, (\ref{eq:4.43}) can be used in the weak lensing
limit [where $\kappa\ll1$, $|\gamma|\ll1$, so that
$\mu\approx(1+2\kappa)$] to obtain an estimate of the surface mass
density,
\begin{equation}
  \kappa \approx \frac{n(>S)-n_0(>S)}{n_0(>S)}\,
  \frac{1}{2(\alpha -1)}\;.
\label{eq:4.44}
\end{equation}

\subsubsection{\label{sc:4.4.2}Size Effect}

Since lensing conserves surface brightness, the magnification can be
obtained from the change in galaxy-image sizes at fixed surface
brightness. Let $I$ be some convenient measure of the surface
brightness. For example, if $\omega$ is the solid angle of an image,
defined by the determinant of the tensor of second brightness moments
as in (\ref{eq:4.3}), one can set $I=S/\omega$.

Denoting by $n(\omega,I,z)\d\omega$ the number density of images with
surface brightness $I$, redshift $z$, and solid angle within
$\d\omega$ of $\omega$, the relation between the lensed and the
unlensed number density can be written
\begin{equation}
  n(\omega,I,z) = \frac{1}{\mu^2}\,
  n_0\left(\frac{\omega}{\mu},I,z\right)\;.
\label{eq:4.45}
\end{equation}
For simplicity, we only consider the case of a moderately small lens
redshift, so that the magnification can be assumed to be locally
constant for all images, irrespective of galaxy redshift. We can then
drop the variable $z$ here. The mean image size
$\langle\omega\rangle(I)$ at fixed surface brightness $I$ is then
related to the mean image size $\langle\omega\rangle_0(I)$ in the
absence of lensing through
\begin{equation}
  \langle\omega\rangle(I) = \mu\langle\omega\rangle_0(I)\;.
\label{eq:4.46}
\end{equation}
If the mean image size in the absence of lensing can be measured
(e.g.~by deep {\em HST\/} exposures of blank fields), the local value
$\mu$ of the magnification can therefore be determined by comparing
the observed image sizes to those in the blank fields. This method has
been discussed in detail in \cite{BA95.5}. For instance, if we assume
that the logarithm of the image size is distributed as a Gaussian with
mean $\langle\ln\omega\rangle_0(I)$ and dispersion $\sigma(I)$, we
obtain an estimate for the local magnification from a set of $N$
galaxy images,
\begin{equation}
  \ln\mu = \sum_{i=1}^N\,
  \frac{\ln\omega_i-\langle\ln\omega\rangle_0(I_i)}{\sigma^2(I_i)}\,
  \left(
    \sum_{i=1}^N\frac{1}{\sigma^2(I_i)}
  \right)^{-1}\;.
\label{eq:4.47}
\end{equation}
A typical value for the dispersion is $\sigma(I)\approx0.5$
(\cite*{BA95.5}).

\subsubsection{\label{sc:4.4.3}Relative Merits of Shear and
  Magnification Effect}

It is interesting to compare the prospects of measuring shear and
magnification caused by a deflector. We consider a small patch of the
sky containing an expected number $N$ of galaxy images (in the absence
of lensing), which is sufficiently small so that the lens parameters
$\kappa$ and $\gamma$ can be assumed to be constant. We also restrict
the discussion to weak lensing case.

The dispersion of a shear estimate from averaging over galaxy
ellipticities is $\sigma_\epsilon^2/N$, so that the signal-to-noise
ratio is
\begin{equation}
  \left(\frac{\mathrm{S}}{\mathrm{N}}\right)_\mathrm{shear} =
  \frac{|\gamma|}{\sigma_\epsilon}\,\sqrt{N}\;.
\label{eq:4.48}
\end{equation}
According to (\ref{eq:4.44}), the expected change in galaxy number
counts is $|\Delta N|=2\kappa|\alpha-1|N$. Assuming Poissonian noise,
the signal-to-noise ratio in this case is
\begin{equation}
  \left(\frac{\mathrm{S}}{\mathrm{N}}\right)_\mathrm{counts} =
  2\kappa|\alpha-1|\sqrt{N}\;.
\label{eq:4.49}
\end{equation}
Finally, the signal-to-noise ratio for the magnification estimate
(\ref{eq:4.47}) is
\begin{equation}
  \left(\frac{\mathrm{S}}{\mathrm{N}}\right)_\mathrm{size} =
  \frac{2\kappa}{\sigma(I)}\,\sqrt{N}\;,
\label{eq:4.50}
\end{equation}
assuming all $\sigma(I)$ are equal.

Comparing the three methods, we find
\begin{equation}
  \frac{(\mathrm{S}/\mathrm{N})_\mathrm{shear}}
       {(\mathrm{S}/\mathrm{N})_\mathrm{counts}} =
  \frac{|\gamma|}{\kappa}\,
  \frac{1}{2\sigma_\epsilon|\alpha-1|}\;,\quad
  \frac{(\mathrm{S}/\mathrm{N})_\mathrm{counts}}
       {(\mathrm{S}/\mathrm{N})_\mathrm{size}} =
  2\sigma(I)|\alpha-1|\;.
\label{eq:4.51}
\end{equation}
If the lens situation is such that $\kappa\approx|\gamma|$ as for
isothermal spheres, the first of eqs.~(\ref{eq:4.51}) implies that the
signal-to-noise of the shear measurement is considerably larger than
that of the magnification. Even for number-count slopes as flat as
$\alpha\sim0.5$, this ratio is larger than five, with
$\sigma_\epsilon\sim0.2$. The second of eqs.~(\ref{eq:4.51}) shows
that the size effect yields a somewhat larger signal-to-noise ratio
than the number-density effect. We therefore conclude from these
considerations that shear measurements should yield more significant
results than magnification measurements.

This, however, is not the end of the story. Several additional
considerations come into play when these three methods of measuring
lensing effects are compared. First, the shear measurement is the only
one for which we know precisely what to expect in the absence of
lensing, whereas the other two methods need to compare the
measurements with calibration fields void of lensing. These
comparisons require very accurate photometry. Second,
eq.~(\ref{eq:4.49}) overestimates the signal-to-noise ratio since we
assumed Poissonian errors, while real galaxies are known to cluster
even at very faint magnitudes (e.g., \cite*{vil97}), and so the error
is substantially underestimated. Third, as we shall discuss in
Sect.~\ref{sc:4.6}, observational effects such as atmospheric seeing
affect the observable ellipticities and sizes of galaxy images,
whereas the observed flux of galaxies is much less affected. Hence,
the shear and size measurements require better seeing conditions than
the number-count method. Both the number counts and the size
measurements (at fixed surface brightness) require accurate
photometry, which is not very important for the shear measurements. As
we shall see in the course of this article, most weak-lensing
measurements have indeed been obtained from galaxy ellipticities.

\subsection{\label{sc:4.5}Minimum Lens Strength for its Weak Lensing
  Detection}

After our detailed discussion of shear estimates and signal-to-noise
ratios for local lensing measurements, it is interesting to ask how
strong a deflecting mass distribution needs to be for a weak lensing
measurement to recognise it. Our simplified consideration here
suffices to gain insight into the dependence on the lens mass of the
signal-to-noise ratio for a lens detection, and on the redshifts of
lens and sources.

We model the deflector as a singular isothermal sphere (see
Sect.~\ref{sc:3.1.5}, page~\pageref{sc:3.1.5}). Let there be $N$
galaxy images with ellipticities $\epsilon_i$ in an annulus centred on
the lens and bounded by angular radii
$\theta_\mathrm{in}\le\theta_i\le\theta_\mathrm{out}$. For simplicity,
we restrict ourselves to weak lensing, so that
$\mathrm{E}(\epsilon)\approx\gamma$. For an axially-symmetric mass
distribution, the shear is always tangentially oriented relative to
the direction towards the mass centre, which is expressed by
eq.~(\ref{eq:3.18}) on page~\pageref{eq:3.18}. We therefore consider
the ellipticity component projected onto the tangential direction. It
is formally defined by $\epsilon_\mathrm{t}\equiv-\Re(\epsilon\,
\mathrm{e}^{-2\mathrm{i}\varphi})$, where $\varphi$ is the polar angle
of the galaxy position relative to the lens centre [see
(\ref{eq:3.18}), page~\pageref{eq:3.18}]. We now define an estimator
for the lens strength by
\begin{equation}
  X \equiv \sum_{i=1}^N a_i\,\epsilon_{\mathrm{t}i}\;.
\label{eq:4.52}
\end{equation}
The factors $a_i=a(\theta_i)$ are arbitrary at this point, and will be
chosen later such as to maximise the signal-to-noise ratio of the
estimator (\ref{eq:4.52}). Note that the expectation value of $X$ is
zero in the absence of lensing, so that a significant non-zero value
of $X$ signifies the presence of a lens. The expectation value for an
isothermal sphere is
$\mathrm{E}(X)=\theta_\mathrm{E}\sum_ia_i/(2\theta_i)$, where we used
(\ref{eq:3.18}, page~\pageref{eq:3.18}), and
\begin{equation}
  \mathrm{E}(X^2) = \sum_{i,j=1}^Na_ia_j\,
  \mathrm{E}(\epsilon_{\mathrm{t}i}\epsilon_{\mathrm{t}j}) =
  [\mathrm{E}(X)]^2+\frac{\sigma_\epsilon^2}{2}\sum_{i=1}^Na_i^2\;.
\label{eq:4.53}
\end{equation}
We employed $\mathrm{E}(\epsilon_{\mathrm{t}i}\epsilon_{\mathrm{t}j})
= \gamma_{\mathrm{t}}(\theta_i)\gamma_{\mathrm{t}}(\theta_j)
+\delta_{ij}\sigma_\epsilon^2/2$ here, and the factor two is due to
the fact that the ellipticity dispersion only refers to one component
of the ellipticity, while $\sigma_\epsilon$ is defined as the
dispersion of the two-component ellipticity. Therefore, the
signal-to-noise ratio for a detection of the lens is
\begin{equation}
  \frac{\mathrm{S}}{\mathrm{N}} = 
  \frac{\theta_\mathrm{E}}{\sqrt{2}\sigma_\epsilon}\,
  \frac{\sum_ia_i\,\theta_i^{-1}}{\sqrt{\sum_ia_i^2}}\;.
\label{eq:4.54}
\end{equation}
Differentiating (S/N) with respect to $a_j$, we find that (S/N) is
maximised if the $a_i$ are chosen $\propto\theta_i^{-1}$. Inserting
this choice into (\ref{eq:4.54}) yields
$\mathrm{S/N}=2^{-1/2}\theta_\mathrm{E}\sigma_\epsilon^{-1}
\allowbreak\left(\sum_i \theta_i^{-2}\right)^{1/2}$. We now replace
the sum by its ensemble average over the annulus,
$\left\langle\sum_i\theta_i^{-2}\right\rangle =
N\left\langle\theta^{-2}\right\rangle = 2n\pi
\ln(\theta_\mathrm{out}/\theta_\mathrm{in})$, where we used $N=\pi n
(\theta_\mathrm{out}^2-\theta_\mathrm{in}^2)$, with the number density
of galaxy images $n$. Substituting this result into (\ref{eq:4.54}),
and using the definition of the Einstein radius (\ref{eq:3.17},
page~\pageref{eq:3.17}), the signal-to-noise ratio becomes
\begin{eqnarray}
 \frac{\mathrm{S}}{\mathrm{N}} &=&
 \frac{\theta_\mathrm{E}}{\sigma_\epsilon}\,\sqrt{\pi n}\,
 \sqrt{\ln(\theta_\mathrm{out}/\theta_\mathrm{in})}
 \label{eq:4.55} \\ &=&
 12.7\,
 \left(\frac{n}{30\,\mathrm{arc~min}^{-2}}\right)^{1/2}
 \left(\frac{\sigma_\epsilon}{0.2}\right)^{-1}
 \left(\frac{\sigma_v}{600\,\mathrm{km\,s}^{-1}}\right)^2
 \nonumber\\ &\times& \phantom{12.7\,}
 \left(
   \frac{\ln(\theta_\mathrm{out}/\theta_\mathrm{in})}{\ln10}
 \right)^{1/2} 
 \left\langle\frac{D_\mathrm{ds}}{D_\mathrm{s}}\right\rangle\;.
 \nonumber 
\end{eqnarray}

As expected, the signal-to-noise ratio is proportional the square root
of the number density of galaxies and the inverse of the intrinsic
ellipticity dispersion. Furthermore, it is proportional to the square
of the velocity dispersion $\sigma_v$. Assuming the fiducial values
given in eq.~(\ref{eq:4.55}) and a typical value of
$(D_\mathrm{ds}/D_\mathrm{s})\sim 0.5$, lenses with velocity
dispersion in excess of $\sim600\,\mathrm{km\,s}^{-1}$ can be detected
with a signal-to-noise $\gtrsim6$. This shows that galaxy clusters
will yield a significant weak lensing signal, and explains why
clusters have been the main target for weak-lensing research up to
now. Individual galaxies with $\sigma_v\sim200\,\mathrm{km\,s}^{-1}$
cannot be detected with weak-lensing techniques. If one is interested
in the statistical properties of the mass distribution of galaxies,
the lensing effects of $N_\mathrm{gal}$ galaxies need to be
statistically superposed, increasing (S/N) by a factor of
$\sqrt{N_\mathrm{gal}}$. Thus, it is necessary to superpose several
hundred galaxies to obtain a significant galaxy-galaxy lensing
signal. We shall return to this topic in Sect.~\ref{sc:7} on
page~\pageref{sc:7}.

We finally note that (\ref{eq:4.55}) also demonstrates that the
detection of lenses will become increasingly difficult with increasing
lens redshift, as the last factor is a sensitive function of
$z_\mathrm{d}$. Therefore, most lenses so far investigated with
weak-lensing techniques have redshifts below $0.5$. High-redshift
clusters have only recently become the target of detailed lensing
studies.

\subsection{\label{sc:4.6}Practical Consideration for Measuring Image
  Shapes}

\subsubsection{\label{sc:4.6.1}General Discussion} 

Real astronomical data used for weak lensing are supplied by CCD
images. The steps from a CCD image to a set of galaxy images with
measured ellipticities are highly non-trivial and cannot be explained
in any detail in the frame of this review. Nevertheless, we want to
mention some of the problems together with the solutions which were
suggested and applied.

The steps from CCD frames to image ellipticities can broadly be
grouped into four categories; data reduction, image detection, shape
determination, and corrections for the point-spread function. The
data-reduction process is more or less standard, involving de-biasing,
flat-fielding, and removal of cosmic rays and bad pixels. For the
latter purpose, it is essential to have several frames of the same
field, slightly shifted in position. This also allows the the flat
field to be determined from the images themselves (a nice description
of these steps is given in \cite*{MO94.1}). To account for telescope
and instrumental distortions, the individual frames have to be
re-mapped before being combined into a final image. In order to do
this, the geometric distortion has to be either known or stable. In
the latter case, it can be determined by measuring the positions and
shapes of stellar images (e.g., from a globular cluster). In
\cite{MO94.1}, the classical optical aberrations were determined and
found to be in good agreement with the system's specifications
obtained from ray-tracing analysis.

With the individual frames stacked together in the combined image, the
next step is to detect galaxies and to measure their shapes. This may
appear simple, but is in fact not quite as straightforward, for
several reasons. Galaxy images are not necessarily isolated on the
image, but they can overlap, e.g.~with other galaxies. Since
weak-lensing observations require a large number density of galaxy
images, such merged images are not rare. The question then arises
whether a detected object is a single galaxy, or a merged pair, and
depending on the choice made, the measured ellipticities will be much
different. Second, the image is noisy because of the finite number of
photons per pixel and the noise intrinsic to the CCD
electronics. Thus, a local enhancement of counts needs to be
classified as a statistically significant source detection, and a
conservative signal-to-noise threshold reduces the number of galaxy
images. Third, galaxy images have to be distinguished from stars. This
is not a severe problem, in particular if the field studied is far
from the Galactic plane where the number density of stars is small.

Several data-analysis software packages exist, such as FOCAS
(\cite*{jat81}) and SExtractor (\cite*{bea96}). They provide routines,
based on algorithms developed from experience and simulated data, for
objective selection of objects and measuring their centroids, their
multipole moments, their magnitudes, and classify them as stars or
extended objects. \cite{KA95.4} developed their own object detection
algorithm. It is based on convolving the CCD image with
two-dimensional Mexican hat-shaped filter functions of variable width
$\theta_\mathrm{s}$. For each value of $\theta_\mathrm{s}$, the maxima
of the smoothed intensity map are localised. Varying
$\theta_\mathrm{s}$, these maxima form curves in the three-dimensional
space spanned by $\vec\theta$ and $\theta_\mathrm{s}$. Along each such
curve, the significance of a source detection is calculated, and the
maximum of the significance is defined as the location $\vec\theta$ of
an object with corresponding size $\theta_\mathrm{s}$.

Once an object is found, the quadrupole moments can in principle be
obtained from (\ref{eq:4.2}). In practice, however, this is not
necessarily the most practical definition of the moment tensor. The
function $q_I(I)$ in (\ref{eq:4.2}) should be chosen such that it
vanishes for surface brightnesses close to and smaller than the sky
brightness; otherwise, one would sample too much noise. On the other
hand, if $q_I$ is cut off at too bright values of $I$, the area within
which the quadrupole moments are measured becomes too small, and the
effects of seeing (see below) become overwhelming. Also, with a too
conservative cut-off, many galaxy images would be missed. Assume, for
instance, that $q_I(I)=I\,\mathrm{H}(I-I_\mathrm{th})$. One would then
choose $I_\mathrm{th}$ such that it is close to, but a few
$\sigma_\mathrm{noise}$ above the sky background, and the quadrupole
moments would then be measured inside the resulting limiting
isophote. Since this isophote is close to the sky background, its
shape is affected by sky noise. This implies that the measured
quadrupole moments will depend highly non-linearly on the brightness
on the CCD; in particular, the effect of noise will enter the measured
ellipticities in a non-linear fashion. A more robust measurement of
the quadrupole moments is obtained by replacing the weight function
$q_I[I(\vec\theta)]$ in (\ref{eq:4.2}) by $I\,W(\vec\theta)$, where
$W(\vec\theta)$ explicitly depends on $\vec\theta$. \cite{KA95.4} use
a Gaussian of size $\theta_\mathrm{s}$ as their weight function $W$,
i.e., the size of their $W$ is the scale on which the object was
detected at highest significance. It should be noted that the
quadrupole moments obtained with a weight function $W(\vec\theta)$ do
not obey the transformation law (\ref{eq:4.5}), and therefore, the
expectation value of the ellipticity, $\mathrm{E}(\epsilon)$, will be
different from the reduced shear $g$. We return to this issue further
below.

Another severe difficulty for the determination of the local shear is
atmospheric seeing. Due to atmospheric turbulence, a point-like source
will be seen from the ground as an extended image; the source is
smeared-out. Mathematically, this can be described as a convolution.
If $I(\vec\theta)$ is the surface brightness before passing the
Earth's atmosphere, the observed brightness distribution
$I^{(\mathrm{obs})}(\vec\theta)$ is
\begin{equation}
  I^{(\mathrm{obs})}(\vec\theta) = \int\d^2\vartheta\,
  I(\vec\vartheta)\,P(\vec\theta-\vec\vartheta)\;,
\label{eq:4.56}
\end{equation}
where $P(\vec\theta)$ is the {\em point-spread function\/} (PSF) which
describes the brightness distribution of a point source on the
CCD. $P(\vec\theta)$ is normalised to unity and centred on $\vec
0$. The characteristic width of the PSF is called the size of the
seeing disc. The smaller it is, the less smeared the images are. A
seeing well below $1\,$arc second is required for weak-lensing
observations, and there are only a handful of telescope sites where
such seeing conditions are regularly met. The reason for this strong
requirement on the data quality lies in the fact that weak-lensing
studies require a high number density of galaxy images, i.e., the
observations have to be extended to faint magnitudes. But the
characteristic angular size of faint galaxies is below $1\,$arc
second. If the seeing is larger than that, the shape information is
diluted or erased.

The PSF includes not only the effects of the Earth's atmosphere, but
also pointing errors of the telescope (e.g., caused by wind
shake). Therefore, the PSF will in general be slightly
anisotropic. Thus, seeing has two important effects on the observed
image ellipticities: Small elliptical images become rounder, and the
anisotropy of the PSF introduces a systematic, spurious image
ellipticity. The PSF can be determined directly from the CCD once a
number of isolated stellar images are identified. The shape of the
stars (which serve as point sources) reflects the PSF. Note that the
PSF is not necessarily constant across the CCD. If the number density
of stellar images is sufficiently large, one can empirically describe
the PSF variation across the field by a low-order polynomial. An
additional potential difficulty is the chromaticity of the PSF,
i.e.~the dependence of the PSF on the spectral energy distribution of
the radiation. The PSF as measured from stellar images is not
necessarily the same as the PSF which applies to galaxies, due to
their different spectra. The difference of the PSFs is larger for
broader filters. However, it is assumed that the PSF measured from
stellar images adequately represents the PSF for galaxies.

In the idealised case, in which the quadrupole moments are defined
with the weight function $q_I(I)=I$, the effect of the PSF on the
observed image ellipticities can easily be described. If $P_{ij}$
denotes the quadrupole tensor of the PSF, defined in complete analogy
to (\ref{eq:4.2}), then the observed quadrupole tensor
$Q^{(\mathrm{obs})}_{ij}$ is related to the true one by
$Q^{(\mathrm{obs})}_{ij}=P_{ij}+Q_{ij}$ (see \cite*{VA83.1}). The
ellipticity $\chi$ then transforms like
\begin{equation}
  \chi^{(\mathrm{obs})} =
  \frac{\chi+T\chi^{(\mathrm{PSF})}}{1+T}\;,
\label{eq:4.57}
\end{equation}
where 
\begin{equation}
  T = \frac{P_{11}+P_{22}}{Q_{11}+Q_{22}}\;;\quad
  \chi^{(\mathrm{PSF})} = \frac{P_{11}-P_{22}+2\mathrm{i}P_{12}}
  {P_{11}+P_{22}}\;.
\label{eq:4.58}
\end{equation}
Thus, $T$ expresses the ratio of the PSF size to the image size before
convolution, and $\chi^{(\mathrm{PSF})}$ is the PSF ellipticity. It is
evident from (\ref{eq:4.57}) that the smaller $T$, the less
$\chi^{(\mathrm{obs})}$ deviates from $\chi$. In the limit of very
large $T$, $\chi^{(\mathrm{obs})}$ approaches
$\chi^{(\mathrm{PSF})}$. In principle, the relation (\ref{eq:4.57})
could be inverted to obtain $\chi$ from $\chi^{(\mathrm{obs})}$.
However, this inversion is unstable unless $T$ is sufficiently small,
in the sense that noise affecting the measurement of
$\chi^{(\mathrm{obs})}$ is amplified by the inversion
process. Unfortunately, these simple transformation laws only apply
for the specific choice of the weight function. For weighting schemes
that can be applied to real data, the resulting transformation becomes
much more complicated.

If a galaxy image features a bright compact core which emits a
significant fraction of the galaxy's light, this core will be smeared
out by the PSF. In that case, $\chi^{(\mathrm{obs})}$ may be dominated
by the core and thus contain little information about the galaxy
ellipticity. This fact motivated \cite{BO95.1} to define the
quadrupole moments with a weight function $W(\vec\theta)$ which not
only cuts off at large angular separations, but which is also small
near $\vec\theta=\vec 0$. Hence, their weight function $q$ is
significantly non-zero in an annulus with radius and width both being
of the order of the size of the PSF.

The difficulties mentioned above prohibit the determination of the
local reduced shear by straight averaging over the directly measured
image ellipticities. This average is affected by the use of a
angle-dependent weight function $W$ in the practical definition of the
quadrupole moments, by the finite size of the PSF and its anisotropy,
and by noise. \cite{BO95.1} have performed detailed simulations of CCD
frames which resemble real observations as close as possible,
including an anisotropic PSF. With these simulations, the efficiency
of object detection, the accuracy of their centre positions, and the
relation between true and measured image ellipticities can be
investigated in detail, and so the relation between mean ellipticity
and (reduced) shear can approximately be calibrated. \cite{WI96.2}
followed a very similar approach, except that the analysis of their
simulated CCD frames was performed with FOCAS. Assuming an isotropic
PSF, the mean image ellipticity is proportional to the reduced shear,
$g\approx f\langle\epsilon\rangle$, with a correction factor $f$
depending on the limiting galaxy magnitude, the photometric depth of
the image, and the size of the seeing disk. For a seeing of
$0\arcsecf8$, \citename{BO95.1} obtained a correction factor $f\sim6$,
whereas the correction factor in \citename{WI96.2} for the same seeing
is $f\sim1.5$. This large difference is not a discrepancy, but due to
the different definitions of the quadrupole tensor. Although the
correction factor is much larger for the \citename{BO95.1} method,
they show that their measured (and calibrated) shear estimate is more
accurate than that obtained with FOCAS. \cite{KA95.4} used CCD frames
taken with WFPC2 on board {\em HST\/} which are unaffected by
atmospheric seeing, sheared them, and degraded the resulting images by
a PSF typical for ground-based images and by adding noise. In this
way, they calibrated their shear measurement and tested their removal
of an anisotropic contribution of the PSF.

However, calibrations relying on simulated images are not fully
satisfactory since the results will depend on the assumptions
underlying the simulations. \cite{KA95.4} and \cite{LU97.1} presented
a perturbative approach for correcting the observed image
ellipticities for PSF effects, with additional modifications made by
\cite{hfk98} and \cite{hgd98}. Since the measurement of ellipticities
lies at the heart of weak lensing studies, we shall present this
approach in the next subsection, despite its being highly technical.

\subsubsection{\label{sc:4.6.2}The KSB Method} 

Closely following the work by \cite{KA95.4}, this subsection provides
a relation between the observed image ellipticity and a source
ellipticity known to be isotropically distributed. The relation
corrects for PSF smearing and its anisotropy, and it also takes into
account that the transformation (\ref{eq:4.5}) no longer applies if
the weight factor explicitly depends on $\vec\theta$.

We consider the quadrupole tensor
\begin{equation}
  Q_{ij} = \int\d^2\theta\,
  (\theta_i-\bar\theta_i)(\theta_j-\bar\theta_j)\,
  I(\vec\theta)\,W\left(
    |\vec\theta-\bar{\vec\theta}|^2/\sigma^2
  \right)\;,
\label{eq:4.59}
\end{equation}
where $W$ contains a typical scale $\sigma$, and $\bar{\vec\theta}$ is
defined as in (\ref{eq:4.1}), but with the new weight function. Note
that, in contrast to the definition (\ref{eq:4.2}), this tensor is no
longer normalised by the flux, but this does not affect the definition
(\ref{eq:4.4}) of the complex ellipticity.

The relation between the observed surface brightness
$I^\mathrm{obs}(\vec\theta)$ and the true surface brightness $I$ is
given by (\ref{eq:4.56}). We assume in the following that $P$ is
nearly isotropic, so that the anisotropic part of $P$ is small. Then,
we define the isotropic part $P^\mathrm{iso}$ of $P$ as the azimuthal
average over $P$, and decompose $P$ into an isotropic and an
anisotropic part as
\begin{equation}
  P(\vec\vartheta) = \int\d^2\varphi\,q(\vec\varphi)\,
  P^\mathrm{iso}(\vec\vartheta-\vec\varphi)\;,
\label{eq:4.60}
\end{equation}
which defines $q$ uniquely. In general, $q(\vec\varphi)$ will be an
almost singular function, but we shall show later that it has
well-behaved moments. Both $P^\mathrm{iso}$ and $q$ are normalised to
unity and have vanishing first moments. With $P^\mathrm{iso}$, we
define the brightness profiles
\begin{eqnarray}
  I^\mathrm{iso}(\vec\theta) &=& \int\d^2\varphi\,
  I(\vec\varphi)\,P^\mathrm{iso}(\vec\theta-\vec\varphi)
  \nonumber\\
  I^0(\vec\theta) &=& \int\d^2\varphi\,I^\mathrm{s}(\vec\varphi)\,
  P^\mathrm{iso}(\vec\theta-\vec\varphi)\;.
\label{eq:4.61}
\end{eqnarray}
The first of these would be observed if the true image was smeared
only with an isotropic PSF, and the second is the unlensed source
smeared with $P^\mathrm{iso}$. Both of these brightness profiles are
unobservable, but convenient for the following discussion. For each of
them, we can define a quadrupole tensor as in (\ref{eq:4.59}). From
each quadrupole tensor, we define the complex ellipticity
$\chi=\chi_1+\mathrm{i}\chi_2$, in analogy to (\ref{eq:4.4}).

If we define the centres of images including a spatial weight
function, the property that the centre of the image is mapped onto the
centre of the source through the lens equation is no longer strictly
true. However, the deviations are expected to be very small in general
and will be neglected in the following. Hence, we choose coordinates
such that $\bar{\vec\theta}=\vec0$, and approximate the other centres
to be at the origin as well.

According to our fundamental assumption that the intrinsic
ellipticities are randomly oriented, this property is shared by the
ellipticities $\chi^0$ defined in terms of $I^0$ [see
(\ref{eq:4.61})], because it is unaffected by an isotropic
PSF. Therefore, we can replace (\ref{eq:4.13}) by
$\mathrm{E}(\chi^0)=0$ in the determination of $g$. The task is then
to relate the observed image ellipticity $\chi^\mathrm{obs}$ to
$\chi^0$. We break it into several steps.

\paragraph*{From $\chi^\mathrm{iso}$ to $\chi^\mathrm{obs}$.}

We first look into the effect of an anisotropic PSF on the observed
ellipticity. According to (\ref{eq:4.60}) and (\ref{eq:4.61}),
\begin{equation}
  I^\mathrm{obs}(\vec\theta) = \int\d^2\varphi\,
  q(\vec\theta-\vec\varphi)\,I^\mathrm{iso}(\vec\varphi)\;. 
\label{eq:4.62}
\end{equation}
Let $f(\vec\theta)$ be an arbitrary function, and consider
\begin{eqnarray}
  \lefteqn{%
    \int\d^2\theta\,f(\vec\theta)\,I^\mathrm{obs}(\vec\theta) =
    \int\d^2\varphi\,I^\mathrm{iso}(\vec\varphi)
    \int\d^2\vartheta\,f(\vec\varphi+\vec\vartheta)\,
    q(\vec\vartheta)
  } \nonumber\\ &=&
  \int\d^2\varphi\,I^\mathrm{iso}(\vec\varphi)\,f(\vec\varphi) +
  \frac{1}{2}\,q_{kl}\int\d^2\varphi\,
  I^\mathrm{iso}(\vec\varphi)
  \frac{\partial^2f}{\partial\varphi_k\partial\varphi_l} +
  \mathcal{O}(q^2)\;.
\label{eq:4.63}
\end{eqnarray}
We used the fact that $q$ is normalised and has zero mean, and defined
\begin{equation}
  q_{ij} = \int\d^2\varphi\,q(\vec\varphi)\,
  \varphi_i\varphi_j\;,\quad
  q_1 \equiv q_{11}-q_{22}\;,\quad q_2 \equiv 2q_{12}\;. 
\label{eq:4.64}
\end{equation}
The tensor $q_{ij}$ is trace-less, $q_{11}=-q_{22}$, following from
(\ref{eq:4.60}). We consider in the following only terms up to linear
order in $q$. To that order, we can replace $I^\mathrm{iso}$ by
$I^\mathrm{obs}$ in the final term in (\ref{eq:4.63}), since the
difference would yield a term $\propto\mathcal{O}(q^2)$. Hence,
\begin{equation}
  \int\d^2\varphi\,I^\mathrm{iso}(\vec\varphi)\,
  f(\vec\varphi) \approx
  \int\d^2\theta\,f(\vec\theta)\,I^\mathrm{obs}(\vec\theta) -
  \frac{1}{2}\,q_{kl}\int\d^2\varphi\,I^\mathrm{obs}(\vec\varphi)
  \frac{\partial^2f}{\partial\varphi_k\partial\varphi_l}\;.
\label{eq:4.65}
\end{equation}
Setting $\sigma_\mathrm{iso}=\sigma_\mathrm{obs}\equiv\sigma$ in the
definition of the quadrupole tensors $Q^\mathrm{iso}$ and
$Q^\mathrm{obs}$, and choosing $f(\vec\theta)=\theta_i\theta_j
W(|\vec\theta|^2/\sigma^2)$, yields
\begin{equation}
  Q^\mathrm{iso}_{ij} = Q^\mathrm{obs}_{ij} -
  \frac{1}{2}Z_{ijkl}q_{kl}\;,
\label{eq:4.66}
\end{equation}
where the Einstein summation convention was adopted, and where
\begin{equation}
  Z_{ijkl} = \int\d^2\varphi\,I^\mathrm{obs}(\vec\varphi)\, 
  \frac{\partial^2}{\partial\varphi_k\partial\varphi_l}
  \left[\varphi_i\varphi_j\,W\left(
    \frac{|\vec\varphi|^2}{\sigma_\mathrm{iso}^2}
  \right)\right]\;.
\label{eq:4.67}
\end{equation}
This then yields
\begin{eqnarray}
  \mathrm{tr}(Q^\mathrm{iso}) &=&
  \mathrm{tr}(Q^\mathrm{obs}) - x_\alpha q_\alpha\;,
  \nonumber\\
  (Q_{11}^\mathrm{iso}-Q_{22}^\mathrm{iso}) &=&
  (Q_{11}^\mathrm{obs}-Q_{22}^\mathrm{obs}) - X_{1\alpha}q_\alpha\;,
  \quad\hbox{and}\nonumber\\
  2Q_{12}^\mathrm{iso} &=&
  2Q_{12}^\mathrm{obs} - X_{2\alpha}q_\alpha\;,
\label{eq:4.68}
\end{eqnarray}
where the sums run over $\alpha=1,2$.\footnote{\label{fn:4.3}We use
Greek instead of Latin indices $\alpha,\beta=1,2$ to denote that they
are not tensor indices. In particular, the components of $\chi$ do not
transform like a vector, but like the trace-less part of a symmetric
tensor.} Up to linear order in $q_\alpha$,
\begin{equation}
  \chi^\mathrm{iso}_\alpha = \chi^\mathrm{obs}_\alpha -
  P_{\alpha\beta}^\mathrm{sm}\,q_\beta\;,
\label{eq:4.69}
\end{equation}
with the definitions
\begin{eqnarray}
  P_{\alpha\beta}^\mathrm{sm} &=&
  \frac{1}{\mathrm{tr} Q^\mathrm{obs}}\left(
    X_{\alpha\beta}-\chi_\alpha^\mathrm{obs}\, x_\beta
  \right)\;\nonumber\\
  X_{\alpha\beta} &=&
    \int\d^2\varphi\,I^\mathrm{obs}(\vec\varphi)\,\left[\left(
      W+2|\vec\varphi|^2\frac{W'}{\sigma_\mathrm{iso}^2}
    \right)\delta_{\alpha\beta} +
    \eta_\alpha(\vec\varphi)\eta_\beta(\vec\varphi)
    \frac{W''}{\sigma_\mathrm{iso}^4}
  \right]\;,\nonumber\\
  x_\alpha &=&
    \int\d^2\varphi\,I^\mathrm{obs}(\vec\varphi)\,
    \eta_\alpha(\vec\varphi)\left(
      \frac{2 W'}{\sigma_\mathrm{iso}^2}+|\vec\varphi|^2
      \frac{W''}{\sigma_\mathrm{iso}^4}
    \right)\;,
\label{eq:4.70}
\end{eqnarray}
and
\begin{equation}
  \eta_1(\vec\theta) = \theta_1^2-\theta_2^2\;;\quad
  \eta_2(\vec\theta) = 2\theta_1\theta_2\;.
\label{eq:4.71}
\end{equation}
$P^\mathrm{sm}_{\alpha\beta}$ was dubbed {\em smear polarisability\/}
in \cite{KA95.4}. It describes the (linear) response of the
ellipticity to a PSF anisotropy. Note that
$P^\mathrm{sm}_{\alpha\beta}$ depends on the observed brightness
profile. In particular, its size decreases for larger images, as
expected: The ellipticities of larger images are less affected by a
PSF anisotropy than those of smaller images.

\paragraph*{The determination of $q_\alpha$.}

Equation~(\ref{eq:4.69}) provides a relation between the ellipticities
of an observed image and a hypothetical image smeared by an isotropic
PSF. In order to apply this relation, the anisotropy term $q_\alpha$
needs to be known. It can be determined from the shape of stellar
images. 

Since stars are point-like and unaffected by lensing, their
isotropically smeared images have zero ellipticity,
$\chi^{*,\mathrm{iso}}=0$. Hence, from (\ref{eq:4.69}),
\begin{equation}
  q_\alpha=(P^{*,\mathrm{sm}})^{-1}_{\alpha\beta}\,
  \chi^{*,\mathrm{obs}}_\beta\;.
\label{eq:4.72}
\end{equation}
In general, the PSF varies with the position of an image. If this
variation is sufficiently smooth, $q$ can be measured for a set of
stars, and approximated by a low-order polynomial across the data
field. As pointed out by \cite{hfk98}, the scale size $\sigma$ in the
measurement of $q$ is best chosen to be the same as that of the galaxy
image under consideration. Hence, for each value of $\sigma$, such a
polynomial fit is constructed. This approach works well and provides
an estimate of $q$ at the position of all galaxies, which can then be
used in the transformation (\ref{eq:4.69}).

\paragraph*{From $\chi^0$ to $\chi^\mathrm{iso}$.}

We now relate $\chi^\mathrm{iso}$ to the ellipticity $\chi^0$ of a
hypothetical image obtained from isotropic smearing of the source. To
do so, we use (\ref{eq:4.61}) and (\ref{eq:3.10}) in the form
$I(\vec\theta)=I^s(\mathcal{A}\vec\theta)$, and consider
\begin{eqnarray}
  I^\mathrm{iso}(\vec\theta) &=& \int\d^2\varphi\,
  I^\mathrm{s}(\mathcal{A}\vec\varphi)\,
  P^\mathrm{iso}(\vec\theta-\vec\varphi) \nonumber\\
  &=&
  \frac{1}{\det\mathcal{A}}\int\d^2\zeta\,I^\mathrm{s}(\vec\zeta)\,
  P^\mathrm{iso}(\vec\theta-\mathcal{A}^{-1}\vec\zeta) \equiv
  \hat I(\mathcal{A}\vec\theta)\;.
\label{eq:4.74}
\end{eqnarray}
The second step is merely a transformation of the integration
variable, and in the final step we defined the brightness moment
\begin{equation}
  \hat I(\vec\theta) = \int\d^2\varphi\,I^\mathrm{s}(\vec\varphi)\,
  \hat P(\vec\theta-\vec\varphi)\quad\hbox{with}\quad
  \hat P(\vec\theta) \equiv \frac{1}{\det\mathcal{A}} 
  P^\mathrm{iso}(\mathcal{A}^{-1}\vec\theta)\;.
\label{eq:4.75}
\end{equation}
The function $\hat P$ is normalised and has zero mean. It can be
interpreted as a PSF relating $\hat I$ to $I^\mathrm{s}$. The presence
of shear renders $\hat P$ anisotropic.

We next seek to find a relation between the ellipticities of
$I^\mathrm{iso}$ and $\hat I$:
\begin{eqnarray}
  \hat Q_{ij} &=& \int\d^2\beta\,\beta_i\beta_j\,
  \hat I(\vec\beta)\,W\left(
    \frac{|\vec\beta|^2}{\hat\sigma^2}
  \right) \nonumber\\
  &=&
  \det\mathcal{A}\,\mathcal{A}_{ik}\mathcal{A}_{jl}
  \int\d^2\theta\,\theta_k\theta_l\,I^\mathrm{iso}(\vec\theta)\,
  W\left(
    \frac{|\vec\theta|^2-\delta_\alpha\,\eta_\alpha(\vec\theta)}
    {\sigma^2}
  \right)\;.
\label{eq:4.76}
\end{eqnarray}
The relation between the two filter scales is given by
$\hat\sigma^2=(1-\kappa)^2(1+|g|^2)\sigma^2$, and $\delta$ is the
distortion (\ref{eq:4.15}). For small $\delta$, we can employ a
first-order Taylor expansion of the weight function $W$ in the
previous equation. This results in the following relation between
$\hat\chi$ and $\chi^\mathrm{iso}$:
\begin{equation}
  \chi_\alpha^\mathrm{iso}-\hat\chi_\alpha =
  C_{\alpha\beta}g_\beta\;,
\label{eq:4.77}
\end{equation}
where
\begin{eqnarray}
  C_{\alpha\beta} &=& 2\delta_{\alpha\beta} -
  2\chi_\alpha^\mathrm{iso}\chi_\beta^\mathrm{iso} +
  \frac{2}{\mathrm{tr}(Q^\mathrm{iso})}\,
  \chi_\alpha^\mathrm{iso}\,L_\beta -
  \frac{2}{\mathrm{tr}(Q^\mathrm{iso})}\,
  B_{\alpha\beta}\;,\nonumber\\
  B_{\alpha\beta} &=& -\int\d^2\theta\,
  I^\mathrm{iso}(\vec\theta)\,
  W'\left(\frac{|\vec\theta|^2}{\sigma^2}\right)
  \frac{1}{\sigma^2}\,\eta_\alpha(\vec\theta)\,
  \eta_\beta(\vec\theta)\nonumber\\
  L_\alpha &=& -\int\d^2\theta\,|\vec\theta|^2\,
  I^\mathrm{iso}(\vec\theta)\,
  W'\left(\frac{|\vec\theta|^2}{\sigma^2}\right)
  \frac{1}{\sigma^2}\,\eta_\alpha(\vec\theta)\;.
\label{eq:4.78}
\end{eqnarray}
$C$ is the {\em shear polarisability\/} of \cite{KA95.4}. Whereas $C$
is defined in terms of $I^\mathrm{iso}$, owing to the assumed
smallness of $q$, the difference of $C$ calculated with
$I^\mathrm{iso}$ and $I^\mathrm{obs}$ would cause a second-order
change in (\ref{eq:4.77}) and is neglected, so that we can calculate
$C$ directly from the observed brightness profile.

In analogy to (\ref{eq:4.60}), we can decompose $\hat P$ into an
isotropic and an anisotropic part, the latter one being small due to
the assumed smallness of the shear,
\begin{equation}
  \hat P(\vec\theta) = \int\d^2\varphi\,
  \hat P^\mathrm{iso}(\vec\varphi)\,\hat q(\vec\theta-\vec\varphi)\;.
\label{eq:4.79}
\end{equation}
Defining the brightness profile which would be obtained from smearing
the source with the isotropic PSF $\hat P^\mathrm{iso}$, $\hat
I^0(\vec\theta)=\int\d^2\varphi\,I^s(\vec\varphi)\,\hat
P^\mathrm{iso}(\vec\theta-\vec\varphi)$, one finds
\begin{equation}
  \hat I(\vec\theta) = \int\d^2\varphi\,\hat I^0(\vec\varphi)\,
  \hat q(\vec\theta-\vec\varphi)\;.
\label{eq:4.80}
\end{equation}
Thus, the relation between $\hat I$ and $\hat I^0$ is the same as that
between $I^\mathrm{obs}$ and $I^\mathrm{iso}$, and we can write
\begin{equation}
  \hat\chi^0_\alpha = \hat\chi_\alpha -
  P_{\alpha\beta}^\mathrm{sm}\,\hat q_\beta\;.
\label{eq:4.81}
\end{equation}
Note that $P^\mathrm{sm}$ should in principle be calculated by using
$\hat I$ instead of $I^\mathrm{obs}$ in (\ref{eq:4.69}). However, due
to the assumed smallness of $g$ and $q$, the differences between
$I^\mathrm{obs}$, $I^\mathrm{iso}$, and $\hat I$ are small, namely of
first order in $g$ and $q$. Since $\hat q$ is of order $g$ [as is
obvious from its definition, and will be shown explicitly in
(\ref{eq:4.83})], this difference in the calculation of
$P^\mathrm{sm}$ would be of second order in (\ref{eq:4.81}) and is
neglected here.

Eliminating $\hat\chi$ from (\ref{eq:4.77}) and (\ref{eq:4.81}), we
obtain
\begin{equation}
  \chi^\mathrm{iso}_\alpha = \hat\chi^0_\alpha +
  C_{\alpha\beta}g_\beta +
  P^\mathrm{sm}_{\alpha\beta} \hat q_\beta\;.
\label{eq:4.82}
\end{equation}
Now, for stellar objects, both $\hat\chi^0$ and $\chi^\mathrm{iso}$
vanish, which implies a relation between $\hat q$ and $g$,
\begin{equation}
  \hat q_\alpha = -(P^{\mathrm{sm}*})^{-1}_{\alpha\beta}\,
  C^*_{\beta\gamma}g_\gamma\;,
\label{eq:4.83}
\end{equation}
where the asterisk indicates that $P^\mathrm{sm}$ and $C$ are to be
calculated from stellar images. Whereas the result should in principle
not depend on the choice of the scale length in the weight function,
it does so in practice. As argued in \cite{hfk98}, one should use the
same scale length in $P^{\mathrm{sm}*}$ and $C^*$ as for the galaxy
object for which the ellipticities are measured. Defining now
\begin{equation}
  P^g_{\alpha\beta} = C_{\alpha\beta} -
  P^\mathrm{sm}_{\alpha\gamma}
  (P^{\mathrm{sm}*})^{-1}_{\gamma\delta}C^*_{\delta\beta}\;,
\label{eq:4.84}
\end{equation}
and combining (\ref{eq:4.69}) and (\ref{eq:4.82}), we finally obtain
\begin{equation}
  \hat{\chi}^0 = \chi^\mathrm{obs}-P^\mathrm{sm}_{\alpha\beta}
  q_\beta-P^g_{\alpha\beta}g_\beta\;.
\label{eq:4.85}
\end{equation}
This equation relates the observed ellipticity to that of the source
smeared by an isotropic PSF, using the PSF anisotropy and the reduced
shear $g$. Since the expectation value of $\hat\chi^0$ is zero,
(\ref{eq:4.85}) yields an estimate of $g$. The two tensors
$P^\mathrm{sm}$ and $P^g$ can be calculated from the brightness
profile of the images. Whereas the treatment has been confined to
first order in the PSF anisotropy and the shear, the simulations in
\cite{KA95.4} and \cite{hfk98} show that the resulting equations can
be applied even for moderately large shear. A numerical implementation
of these relations, the \texttt{imcat} software, is provided by
N.~Kaiser (see \texttt{http://www.ifa.hawaii.edu/$\sim$kaiser}). We
also note that modifications of this scheme were recently suggested
(\cite*{rrg99}, \cite*{kai99}), as well as a completely different
approach to shear measurements (\cite*{kui99}).

  % -*- LaTeX -*-

\section{\label{sc:5}Weak Lensing by Galaxy Clusters}

\subsection{\label{sc:5.1}Introduction}

So far, weak gravitational lensing has chiefly been applied to
determine the mass distribution of medium-redshift galaxy
clusters. The main reason for this can be seen from
eq.~(\ref{eq:4.55}): Clusters are massive enough to be individually
detected by weak lensing. More traditional methods to infer the matter
distribution in clusters are (a) dynamical methods, in which the
observed line-of-sight velocity distribution of cluster galaxies is
used in conjunction with the virial theorem, and (b) the investigation
of the diffuse X--ray emission from the hot ($\sim10^7\,$K)
intra-cluster gas residing in the cluster potential well (see, e.g.,
\cite*{sar86}).

Both of these methods are based on rather strong assumptions. For the
dynamical method to be reliable, the cluster must be in or near virial
equilibrium, which is not guaranteed because the typical dynamical
time scale of a cluster is not much shorter than the Hubble time
$H_0^{-1}$, and the substructure abundantly observed in clusters
indicates that an appreciable fraction of them is still in the process
of formation. Projection effects and the anisotropy of galaxy orbits
in clusters further affect the mass determination by dynamical
methods. On the other hand, X--ray analyses rely on the assumption
that the intra-cluster gas is in hydrostatic equilibrium. Owing to the
finite spatial and energy resolution of existing X--ray instruments,
one often has to conjecture the temperature profile of the gas. Here,
too, the influence of projection effects is difficult to assess.

Whereas these traditional methods have provided invaluable information
on the physics of galaxy clusters, and will continue to do so,
gravitational lensing offers a welcome alternative approach, for it
determines the projected mass distribution of a cluster independent of
the physical state and nature of the matter. In particular, it can be
used to calibrate the other two methods, especially for clusters
showing evidence of recent merger events, for which the equilibrium
assumptions are likely to fail. Finally, as we shall show below, the
determination of cluster mass profiles by lensing is theoretically
simple, and recent results show that the observational challenges can
also be met with modern telescopes and instruments.

Both shear and magnification effects have been observed in a number of
galaxy clusters. In this chapter, we discuss the methods by which the
projected mass distribution in clusters can be determined from the
observed lensing effects, and show some results of mass
reconstructions, together with a brief discussion of their
astrophysical relevance. Sect.~\ref{sc:5.2} presents the principles of
cluster mass reconstruction from estimates of the (reduced) shear
obtained from image ellipticities. In contrast to the two-dimensional
mass maps generated by these reconstructions, the aperture mass
methods discussed in Sect.~\ref{sc:5.3} determine a single number to
characterise the bulk properties of the cluster mass. Observational
results are presented in Sect.~\ref{sc:5.4}. We outline further
developments in the final section, including the combined analysis of
shear and magnification effects, maximum-likelihood methods for the
mass reconstruction, and a method for measuring local lens parameters
from the extragalactic background noise.

\subsection{\label{sc:5.2}Cluster Mass Reconstruction from Image
  Distortions}

We discussed in detail in Sect.~\ref{sc:4} how the distortion of image
shapes can be used to determine the local tidal gravitational field of
a cluster. In this section, we describe how this information can be
used to construct two-dimensional mass maps of clusters.

Shortly after the discovery of giant luminous arcs (\cite*{SO87.1};
\cite*{LY89.1}), \cite{FO88.1} detected a number of distorted galaxy
images in the cluster A~370. They also interpreted these {\em
arclets\/} as distorted background galaxy images, but on a weaker
level than the giant luminous arc in the same cluster. The redshift
determination of one arclet by \cite{ME91.1} provided early support
for this interpretation. \cite{TY90.1} discovered a coherent
distortion of faint galaxy images in the clusters A~1689 and
Cl~1409+52, and constrained their (dark) mass profiles from the
observed `shear'. \cite{KO90.4} and \cite{MI91.1} studied in detail
how parameterised mass models for clusters can be constrained from
such distortion measurements.

The field began to flourish after \cite{KA93.2} found that the
distortions can be used for parameter-free reconstructions of cluster
surface mass densities. Their method, and several variants of it, will
be described in this section. It has so far been applied to about 15
clusters, and this number is currently limited by the number of
available dark nights with good observing conditions at the large
telescopes which are required for observations of weak lensing.

\subsubsection{\label{sc:5.2.1}Linear Inversion of Shear Maps}

Equation~(\ref{eq:3.15}, page~\pageref{eq:3.15}) shows that the shear
$\gamma$ is a convolution of the surface mass density $\kappa$ with
the kernel $\mathcal{D}$. This relation is easily inverted in Fourier
space to return the surface mass density in terms of a linear
functional of the shear. Hence, if the shear can be observed from
image distortions, the surface mass density can directly be
obtained. Let the Fourier transform of $\kappa(\vec\theta)$ be
\begin{equation}
  \hat\kappa(\vec l) = \int_{\Rset^2}\d^2 \theta\,\kappa(\vec \theta)\,
  \exp(\mathrm{i}\vec\theta\cdot\vec l)\;.
\label{eq:5.1}
\end{equation}
The Fourier transform of the complex kernel $\mathcal{D}$ defined in
(\ref{eq:3.15}, page~\pageref{eq:3.15}) is
\begin{equation}
  \hat\mathcal{D}(\vec l) = \pi
  \frac{\left(l_1^2-l_2^2+2\mathrm{i}l_1l_2\right)}
  {|\vec l|^2}\;.
\label{eq:5.2}
\end{equation}
Using the convolution theorem, eq.~(\ref{eq:3.15},
page~\pageref{eq:3.15}) can be written $\hat\gamma(\vec
l)=\pi^{-1}\hat\mathcal{D}(\vec l)\,\hat\kappa(\vec l)$ for $\vec
l\ne\vec 0$. Multiplying both sides of this equation with
$\tilde\mathcal{D}^*$ and using
$\tilde\mathcal{D}\,\tilde\mathcal{D}^*=\pi^2$ gives
\begin{equation}
  \hat\kappa(\vec l) = \pi^{-1}\hat\gamma(\vec l)\,
  \hat\mathcal{D}^*(\vec l)
  \quad\hbox{for}\quad
  \vec l\ne\vec 0\;,
\label{eq:5.3}
\end{equation}
and the convolution theorem leads to the final result
\begin{eqnarray}
  \kappa(\vec\theta) - \kappa_0 &=& 
  \frac{1}{\pi}\int_{\Rset^2}\d^2\theta'\,
  \mathcal{D}^*(\vec\theta-\vec\theta')\,
  \gamma(\vec\theta')\nonumber\\
  &=& \frac{1}{\pi}\int_{\Rset^2}\d^2\theta'\,
  \Re\left[\mathcal{D}^*(\vec\theta-\vec\theta')\,
  \gamma(\vec\theta')\right]
\label{eq:5.4}
\end{eqnarray}
(\cite*{KA93.2}). The constant $\kappa_0$ in (\ref{eq:5.4}) appears
because a constant surface mass density does not cause any shear and
is thus unconstrained by $\gamma$. The two expressions in
(\ref{eq:5.4}) are equivalent because
$\Im(\hat\mathcal{D}^*\,\hat\gamma)\equiv0$, as can be shown from the
Fourier transforms of equations (\ref{eq:3.12},
page~\pageref{eq:3.12}). In applications, the second form of
(\ref{eq:5.4}) should be used to ensure that $\kappa$ is
real. Relation (\ref{eq:5.4}) can either be applied to a case where
all the sources are at the same redshift, in which case $\kappa$ and
$\gamma$ are defined as in eqs.~(\ref{eq:3.7}) and (\ref{eq:3.12}), or
where the sources are distributed in redshift, because $\kappa$ and
$\gamma$ are interpreted as convergence and shear for a hypothetical
source at infinite redshift, as discussed in Sect.~\ref{sc:4.3.2}.

In the case of a weak lens ($\kappa\ll1$, $|\gamma|\ll1$), the shear
map is directly obtained from observations, cf.~(\ref{eq:5.17}). When
inserted into (\ref{eq:5.4}), this map provides a parameter-free
reconstruction of the surface mass density, apart from an overall
additive constant. The importance of this result is obvious, as it
provides us with a novel and simple method to infer the mass
distribution in galaxy clusters.

There are two basic ways to apply (\ref{eq:5.4}) to observational
data. Either, one can derive a shear map from averaging over galaxy
images by calculating the local shear on a grid in $\vec\theta$-space,
as described in Sect.~\ref{sc:4.3}; or, one can replace the integral
in (\ref{eq:5.4}) by a sum over galaxy images at positions
$\vec\theta_i$,
\begin{equation}
  \kappa(\vec\theta)=\frac{1}{n\pi}\,
  \sum_i\Re\left[\mathcal{D}^*(\vec\theta-\vec\theta_i)\,
  \epsilon_i\right]\;.
\label{eq:5.5}
\end{equation}

Unfortunately, this estimate of $\kappa$ has infinite noise
(\cite*{KA93.2}) because of the noisy sampling of the
shear at the discrete background galaxy positions. Smoothing is
therefore necessary to obtain estimators of $\kappa$ with finite
noise. The form of eq.~(\ref{eq:5.5}) is preserved by smoothing, but
the kernel $\mathcal{D}$ is modified to another kernel
$\tilde\mathcal{D}$. In particular, Gaussian smoothing with smoothing
length $\theta_\mathrm{s}$ leads to
\begin{equation}
  \tilde\mathcal{D}(\vec\theta)=\left[
    1-\left(1+\frac{|\vec\theta|^2}{\theta_\mathrm{s}^2}\right)\,
    \exp\left(\frac{|\vec\theta|^2}{\theta_\mathrm{s}^2}\right)
  \right]\,\mathcal{D}(\vec\theta)
\label{eq:5.5a}
\end{equation}
(\cite*{SE95.1}). The {\em rms\/} error of the resulting $\kappa$ map
is of order $\sigma_\epsilon\,N^{-1/2}$, where $N$ is the number of
galaxy images per smoothing window, $N\sim
n\pi\theta_\mathrm{s}^2$. However, the errors will be strongly
spatially correlated. Whereas the estimate (\ref{eq:5.5}) with
$\mathcal{D}$ replaced by $\tilde\mathcal{D}$ uses the observational
data more directly than by first constructing a smoothed shear map and
applying (\ref{eq:5.4}) to it, it turns out that the latter method
yields a mass map which is less noisy than the estimate obtained from
(\ref{eq:5.5}), because (\ref{eq:5.5}) contains the `shot noise' from
the random angular position of the galaxy images (\cite*{SE95.1}).

A lower bound to the smoothing length $\theta_\mathrm{s}$ follows from
the spatial number density of background galaxies, i.e.~their mean
separation. More realistically, a smoothing window needs to encompass
several galaxies. In regions of strong shear signals, $N\sim10$ may
suffice, whereas mass maps in the outskirts of clusters where the
shear is small may be dominated by noise unless $N\sim100$. These
remarks illustrate that a single smoothing scale across a whole
cluster may be a poor choice. We shall return to this issue in
Sect.~\ref{sc:5.5.1}, where improvements will be discussed.

Before applying the mass reconstruction formula (\ref{eq:5.4}) to
real data, one should be aware of the following difficulties:

\begin{enumerate}

\item The integral in (\ref{eq:5.4}) extends over $\Rset^2$, while
real data fields are relatively small (most of the applications shown
in Sect.~\ref{sc:5.4} are based on CCDs with side lengths of about
7~arc min). Since there is no information on the shear outside the
data field, the integration has to be restricted to the field, which
is equivalent to setting $\gamma=0$ outside. This is done explicitly
in (\ref{eq:5.5}). This cut-off in the integration leads to boundary
artefacts in the mass reconstruction. Depending on the strength of the
lens, its angular size relative to that of the data field, and its
location within the data field, these boundary artefacts can be more
or less severe. They are less important if the cluster is weak, small
compared to the data field, and centred on it.

\item The shear is an approximate observable only in the limit of weak
lensing. The surface mass density obtained by (\ref{eq:5.4}) is biased
low in the central region of the cluster where the weak lensing
assumption may not hold (and does not hold in those clusters which
show giant arcs). Thus, if the inversion method is to be applied also
to the inner parts of a cluster, the relation between $\gamma$ and the
observable $\delta$ has to be taken into account.

\item The surface mass density is determined by (\ref{eq:5.4}) only up
to an additive constant. We demonstrate in the next subsection that
there exists a slightly different general invariance transformation
which is present in all mass reconstructions based solely on image
shapes. However, this invariance transformation can be broken by
including the magnification effect.

\end{enumerate}

In the next three subsections, we shall consider points (1) and
(2). In particular, we show that the first two problems can easily be
cured. The magnification effects will be treated in
Sect.~\ref{sc:5.4}.

\subsubsection{\label{sc:5.2.2}Non--Linear Generalisation of the
  Inversion, and an Invariance Transformation}

In this section, we generalise the inversion equation (\ref{eq:5.4})
to also account for strong lensing, i.e.~we shall drop the assumption
$\kappa\ll1$ and $|\gamma|\ll1$. In this case, the shear $\gamma$ is
no longer a direct observable, but at best the reduced shear $g$, or
in general the distortion $\delta$. In this case, the relation between
$\kappa$ and the observable becomes non-linear. Furthermore, we shall
assume here that all sources are at the same redshift, so that the
reduced shear is well-defined.

Consider first the case that the cluster is sub-critical everywhere,
i.e.~$\det\mathcal{A}>0$ for all $\vec\theta$, which implies
$|g(\vec\theta)|<1$. Then, the mean image ellipticity $\epsilon$ is an
unbiased estimate of the local reduced shear, so that
\begin{equation}
  \gamma(\vec\theta) = \left[1-\kappa(\vec\theta)\right]\,
  \langle\epsilon\rangle(\vec\theta)\;,
\label{eq:5.6}
\end{equation}
where the field $\langle\epsilon\rangle(\vec\theta)$ is determined by
the local averaging procedure described in
Sect.~\ref{sc:4.3.1}. Inserting this into (\ref{eq:5.4}) leads to an
integral equation for $\kappa(\vec\theta)$,
\begin{equation}
  \kappa(\vec\theta) - \kappa_0 = 
  \frac{1}{\pi}\,\int_{\Rset^2}\d^2\theta'\,
  \left[1-\kappa(\vec\theta')\right]\,\Re\left[
    \mathcal{D}^*(\vec\theta-\vec\theta')\,
    \langle\epsilon\rangle(\vec\theta')
  \right]\;,
\label{eq:5.7}
\end{equation}
(\cite*{SE95.1}), which is readily solved by iteration. Starting from
$\kappa\equiv0$, a first estimate of $\kappa(\vec\theta)$ is obtained
from (\ref{eq:5.7}), which after insertion into the right-hand side of
(\ref{eq:5.6}) yields an update of $\gamma(\vec\theta)$, etc. This
iteration process converges quickly to the unique solution.

The situation becomes only slightly more complicated if critical
clusters are included. We only need to keep track of $\det\mathcal{A}$
while iterating, because $\gamma$ must be derived from
$1/\langle\epsilon\rangle^*$ rather than from $\langle\epsilon\rangle$
where $\det\mathcal{A}<0$. Hence, the local invariance between $g$ and
$1/g^*$ is broken due to non-local effects: A local jump from $g$ to
$1/g^*$ cannot be generated by any smooth surface mass density.

After a minor modification\footnote{\label{fn:5.1}At points where
$\kappa=1$, $1/g^*=0$ and $E(\epsilon)=0$, while $\gamma$ remains
finite. During the iteration, there will be points $\vec\theta$ where
the field $\kappa$ is very close to unity, but where
$\langle\epsilon\rangle$ is not necessarily small. This leads to large
values of $\gamma$, which render the iteration unstable. However, this
instability can easily be removed if a damping factor like
$\left(1+|\gamma^2(\vec\theta')|\right)
\,\exp\left(-|\gamma^2(\vec\theta')|\right)$ is included in
(\ref{eq:5.4}). This modification leads to fast convergence and
affects the result of the iteration only very slightly.}, this
iteration process converges quickly. See \cite{SE95.1} for more
details on this method and for numerical tests done with a cluster
mass distribution produced by a cosmological $N$-body simulation. It
should have become clear that the non-linear inversion process poses
hardly any additional problem to the mass reconstruction compared to
the linear inversion (\ref{eq:5.4}).

This non-linear inversion still contains the constant $\kappa_0$, and
so the result will depend on this unconstrained constant. However, in
contrast to the linear (weak lensing) case, this constant does not
correspond to adding a sheet of constant surface mass density. In
fact, as can be seen from (\ref{eq:5.7}), the transformation
\begin{eqnarray}
  \kappa(\vec\theta)\to\kappa'(\vec\theta) &=&
  \lambda\kappa(\vec\theta)+(1-\lambda)
  \quad\hbox{or}\nonumber\\
  \left[1-\kappa'(\vec\theta)\right] &=&
  \lambda\left[1-\kappa(\vec\theta)\right]
\label{eq:5.8}
\end{eqnarray}
leads to another solution of the inverse problem for any value of
$\lambda\ne0$. Another and more general way to see this is that the
transformation $\kappa\to\kappa'$ changes $\gamma$ to
$\gamma'(\vec\theta)=\lambda\gamma(\vec\theta)$, cf.~(\ref{eq:3.15},
page~\pageref{eq:3.15}). Hence, the reduced shear
$g=\gamma(1-\kappa)^{-1}$ is invariant under the transformation
(\ref{eq:5.8}), so that the relation between intrinsic and observed
ellipticity is unchanged under the {\em invariance transformation\/}
(\ref{eq:5.8}). This is the mass-sheet degeneracy pointed out by
\cite{fgs85} in a different context. We thus conclude that the
degeneracy due to the invariance transformation (\ref{eq:5.8}) cannot
be lifted if only image shapes are used. However, the magnification
transforms like
\begin{equation}
  \mu'(\vec\theta) = \lambda^{-2}\mu(\vec\theta)\;,
\label{eq:5.9}
\end{equation}
so that the degeneracy can be lifted if magnification effects are
taken into account (see Sect.~\ref{sc:4.4}).

The invariance transformation leaves the critical curves of the lens
mapping invariant. Therefore, even the location of giant luminous arcs
which roughly trace the critical curves does not determine the scaling
constant $\lambda$. In addition, the curve $\kappa=1$ is invariant
under (\ref{eq:5.8}). However, there are at least two ways to
constrain $\lambda$. First, it is reasonable to expect that on the
whole the surface mass density in clusters decreases with increasing
separation from the cluster `centre', so that $\lambda>0$. Second,
since the surface mass density $\kappa$ is non-negative, upper limits
on $\lambda$ are obtained by enforcing this condition.

\subsubsection{\label{sc:5.2.3}Finite--Field Inversion Techniques}

We shall now turn to the problem that the inversion (\ref{eq:5.4}) in
principle requires data on the whole sky, whereas the available data
field is finite. A simple solution of this problem has been attempted
by \cite{SE95.1}. They extrapolated the measured shear field on the
finite region $\mathcal{U}$ outside the data field, using a
parameterised form for the radial decrease of the shear. From a sample
of numerically generated cluster mass profiles, \cite{BA95.6} showed
that this extrapolation yields fairly accurate mass
distributions. However, in these studies the cluster was always
assumed to be isolated and placed close to the centre of the data
field. If these two conditions are not met, the extrapolation can
produce results which are significantly off. In order to remove the
boundary artefacts inherent in applying (\ref{eq:5.4}) to a finite
field, one should therefore aim at constructing an unbiased
finite-field inversion method.

The basis of most finite-field inversions is a result first derived by
\cite{KA95.1}. Equation~(\ref{eq:3.12}, page~\pageref{eq:3.12}) shows
that shear and surface mass density are both given as second partial
derivatives of the deflection potential $\psi$. After partially
differentiating (\ref{eq:3.12}, page~\pageref{eq:3.12}) and combining
suitable terms we find
\begin{equation}
  \nabla\kappa = \left(\begin{array}{c}
    \gamma_{1,1}+\gamma_{2,2} \\
    \gamma_{2,1}-\gamma_{1,2} \\
  \end{array}\right) \equiv \vec u_\gamma(\vec\theta)\;.
\label{eq:5.10}
\end{equation}
The gradient of the surface mass density can thus be expressed by the
first derivatives of the shear, hence $\kappa(\vec\theta)$ can be
determined, up to an additive constant, by integrating (\ref{eq:5.10})
along appropriately selected curves. This can be done in the weak
lensing case where the observed smoothed ellipticity field
$\langle\epsilon\rangle(\vec\theta)$ can be identified with $\gamma$,
and $\vec u_\gamma(\vec\theta)$ can be constructed by finite
differencing. If we insert $\gamma=(1-\kappa)\,g$ into
(\ref{eq:5.10}), we find after some manipulations
\begin{eqnarray}
  \nabla K(\vec\theta) &=& \frac{-1}{1-g_1^2-g_2^2}\,
  \left(\begin{array}{cc}
    1-g_1 & -g_2 \\
    -g_2 & 1+g_1 \\
  \end{array}\right)\,
  \left(\begin{array}{c}
    g_{1,1}+g_{2,2} \\
    g_{2,1}-g_{1,2} \\
  \end{array}\right) \nonumber\\&\equiv& \vec u_g(\vec\theta)\;,
\label{eq:5.11}
\end{eqnarray}
where
\begin{equation}
  K(\vec\theta) \equiv \ln[1-\kappa(\vec\theta)]\;.
\label{eq:5.12}
\end{equation}
Hence, using the smoothed ellipticity field
$\langle\epsilon\rangle(\vec\theta)$ as an unbiased estimator for
$g(\vec\theta)$, and assuming a sub-critical cluster, one can obtain
the vector field $\vec u_g(\vec\theta)$ by finite differencing, and
thus determine $K(\vec\theta)$ up to an additive constant from line
integration, or, equivalently, $1-\kappa(\vec\theta)$ up to an overall
multiplicative constant. This is again the invariance transformation
(\ref{eq:5.8}).

In principle, it is now straightforward to obtain $\kappa(\vec\theta)$
from the vector field $\vec u_\gamma(\vec\theta)$, or $K(\vec\theta)$
from $\vec u_g(\vec\theta)$, simply by a line integration of the type
\begin{equation}
  \kappa(\vec\theta,\vec\theta_0) = \kappa(\vec\theta_0) + 
  \int_{\vec\theta_0}^{\vec\theta}\d\vec l\cdot\vec u_\gamma(\vec l)\;,
\label{eq:5.13}
\end{equation}
where $\vec l$ is a smooth curve connecting $\vec \theta$ with
$\vec\theta_0$. If $\vec u_\gamma$ is a gradient field, as it ideally
is, the resulting surface mass density is independent of the choice of
the curves $\vec l$. However, since $\vec u_\gamma$ is obtained from
noisy data (at least the noise resulting from the intrinsic
ellipticity distribution), it will in general not be a gradient field,
so that (\ref{eq:5.10}) has no solution. Therefore, the various line
integration schemes proposed (\cite*{SC95.4}, \cite*{kai94},
\cite*{BA95.6}) yield different results.

Realising that eq.~(\ref{eq:5.10}) has no exact solution for an
observed field $\vec u_\gamma$, we wish to find a mass distribution
$\kappa(\vec \theta)$ which satisfies (\ref{eq:5.10}) `best'. In
general, $\vec u_\gamma$ can be split into a gradient field and a curl
component, but this decomposition is not unique. However, as pointed
out in \cite{SE96.3}, since the curl component is due to noise, its
mean over the data field is expected to vanish. Imposing this
condition, which determines the decomposition uniquely, they showed
that
\begin{equation}
  \kappa(\vec\theta) - \bar \kappa = \int_\mathcal{U}\d^2\theta'\,
  \vec H(\vec\theta',\vec\theta)\cdot\vec u_\gamma(\vec\theta')\;,
\label{eq:5.16}
\end{equation}
where $\bar \kappa$ is the average of $\kappa(\vec\theta)$ over the
data field $\mathcal{U}$, and the kernel $\vec H$ is the gradient of a
scalar function which is determined through a von Neumann boundary
value problem, with singular source term. This problem can be solved
analytically for circular and rectangular data fields, as detailed in
the Appendix of \cite{SE96.3}. If the data field has a more
complicated geometry, an analytic solution is no longer possible, and
the boundary value problem with a singular source term cannot be
solved numerically.

An alternative method starts with taking the divergence of
(\ref{eq:5.10}) and leads to the new boundary value problem,
\begin{equation}
  \nabla^2 \kappa =\nabla \cdot \vec u_\gamma
  \quad\hbox{with}\quad
  \vec n\cdot \nabla \kappa = \vec n\cdot \vec u_\gamma
  \quad\hbox{on}\quad
  \partial\mathcal{U}\;,
\label{eq:5.17}
\end{equation}
where $\vec n$ is the outward-directed normal on the boundary of
$\mathcal{U}$. As shown in \cite{ses98}, eqs.~(\ref{eq:5.16}) and
(\ref{eq:5.17}) are equivalent. An alternative and very elegant way to
derive (\ref{eq:5.17}) has been found by \cite{LO98.2}. They noticed
that the `best' approximation to a solution of (\ref{eq:5.10})
minimises the `action'
\begin{equation}
  \int_\mathcal{U}\d^2\theta\; |\nabla\kappa(\vec\theta)-\vec
   u_\gamma(\vec\theta)|^2 \;.
\label{eq:5.17a}
\end{equation}
Euler's equations of the variational principle immediately reproduce
(\ref{eq:5.17}). This von Neumann boundary problem is readily solved
numerically, using standard numerically techniques (see Sect.~19.5 of
\cite*{PR86.1}).

A comparison between these different finite-field inversion equations
was performed in \cite{SE96.3} and in \cite{SQ96.3} by numerical
simulations. Of all the inversions tested, the inversion
(\ref{eq:5.17}) performs best on all scales (\cite*{SE96.3}; Fig.~6 of
\cite*{SQ96.3}). Indeed, \cite{LO98.2} showed analytically that the
solution of eq.~(\ref{eq:5.17}) provides the best unbiased estimate of
the surface mass density. The relations (\ref{eq:5.13}) through
(\ref{eq:5.17a}) can be generalised to the non-weak case by replacing
$\kappa$ with $K$ and $\vec u_\gamma$ with $\vec u_g$.

\subsubsection{\label{sc:5.2.4}Accounting for a redshift distribution
of the sources}

We now describe how the preceding mass reconstructions must be
modified if the sources have a broad redshift distribution. In fact,
only minor modifications are needed. The relation
$\langle\epsilon\rangle=g$ for a single source redshift is replaced by
eq.~(\ref{eq:4.28}), which gives an estimate for the shear in terms of
the mean image ellipticities and the surface mass density. This
relation can be applied iteratively:

Begin with $\kappa^{(0)}=0$; then, eq.~(\ref{eq:4.28}) yields a first
guess for the shear $\gamma^{(1)}(\vec\theta)$ by setting $\gamma=0$
on the right-hand side. From (\ref{eq:5.16}), or equivalently by
solving (\ref{eq:5.17}), the corresponding surface mass density
$\kappa^{(1)}(\vec\theta)$ is obtained. Inserting $\kappa^{(1)}$ and
$\gamma^{(1)}$ on the right-hand side of eq.~(\ref{eq:4.28}), a new
estimate $\gamma^{(2)}(\vec\theta)$ for the shear is obtained, and so
forth.

This iteration process quickly converges. Indeed, the difficulty
mentioned in footnote~\ref{fn:5.1} (page~\pageref{fn:5.1}) no longer
occurs since the critical curves and the curve(s) $\kappa=1$ are
effectively smeared out by the redshift distribution, and so the
iteration converges even faster than in the case of a single source
redshift.

Since $\kappa^{(n)}$ is determined only up to an additive constant for
any $\gamma^{(n)}$, the solution of the iteration depends on the
choice of this constant. Hence, one can obtain a one-parameter family
of mass reconstructions, like in (\ref{eq:5.8}).  However, the
resulting mass-sheet degeneracy can no longer be expressed
analytically due to the complex dependence of (\ref{eq:4.28}) on
$\kappa$ and $\gamma$.  In the case of weak lensing, it corresponds to
adding a constant, as before. An approximate invariance transformation
can also be obtained explicitly for mildly non-linear clusters with
$\kappa\lesssim0.7$ and $\det\mathcal{A}>0$ everywhere. In that case,
eq.~(\ref{eq:4.29}) holds approximately, and can be used to show
(\cite*{SE97.1}) that the invariance transformation takes the form
\begin{equation}
  \kappa(\vec\theta) \to \kappa'(\vec\theta) =
  \lambda\,\kappa(\vec\theta) + 
  \frac{(1-\lambda)\langle w\rangle}{\langle w^2\rangle}\;.
\label{eq:5.18}
\end{equation}
In case of a single redshift $z_\mathrm{s}$, such that
$w(z_\mathrm{s})=\langle w\rangle$, this transformation reduces to
(\ref{eq:5.8}) for $\langle w\rangle\kappa$.

We point out that the invariance transformation (\ref{eq:5.18}) in the
case of a redshift distribution of sources is of different nature than
that for a single source redshift. In the latter case, the reduced
shear $g(\vec\theta)$ is invariant under the transformation
(\ref{eq:5.8}). Therefore, the probability distribution of the
observed galaxy ellipticities is invariant, since it involves only the
intrinsic ellipticity distribution and $g$. For a redshift
distribution, the invariance transformation keeps the mean image
ellipticities invariant, but the probability distributions are
changed. Several strategies were explored in \cite{SE97.1} to utilise
this fact for breaking the invariance transformation. While possible
in principle, the corresponding effect on the observed ellipticity
distribution is too small for this approach to be feasible with
existing data.

\subsubsection{\label{sc:5.2.5}Breaking the Mass-Sheet Degeneracy}

Equation~(\ref{eq:5.9}) shows that the invariance transformation
(\ref{eq:5.8}) affects the magnification. Hence, the degeneracy can be
lifted with magnification information. As discussed in
Sect.~\ref{sc:4.4}, two methods to obtain magnification information
have been proposed. Detections of the number-density effect have so
far been reported for two clusters (Cl~0024+16, \cite*{FO97.1};
Abell~1689, \cite*{TA98.1}). Whereas the information provided by the
number density effect is less efficient than shear measurements (see
Sect.~\ref{sc:4.4.3}), these two clusters appear to be massive enough
to allow a significant detection. In fact, \cite{TA98.1} obtained a
two-dimensional mass reconstruction of the cluster A~1689 from
magnification data.

In the case of weak lensing, and thus small magnifications, the
magnification can locally be translated into a surface mass density --
see (\ref{eq:4.44}). In general, the relation between $\mu$ and
$\kappa$ is non-local, since $\mu$ also depends on the shear. Various
attempts to account for this non-locality have been published
(\cite*{vka98}, \cite*{dye98}). However, it must be noted that the
surface mass density cannot be obtained from magnification alone since
the magnification also depends on the shear caused by matter outside
the data field. In practice, if the data field is sufficiently large
and no mass concentration lies close to but outside the data field,
the mass reconstruction obtained from magnification can be quite
accurate.

In order to break the mass-sheet degeneracy, it suffices in principle
to measure one value of the magnification: Either the magnification at
one location in the cluster, or the average magnification over a
region. We shall see later in Sect.~\ref{sc:5.5.1} how local
magnification information can be combined with shear
measurements. Doing it the naive way, expressing $\kappa$ in terms of
$\mu$ and $\gamma$, is a big waste of information: Since there is only
one independent scalar field (namely the deflection potential $\psi$)
describing the lens, one can make much better use of the measurements
of $\gamma$ and $\mu$ than just combining them locally; the relation
between them should be used to reduce the error on $\kappa$.

\subsection{\label{sc:5.3}Aperture Mass and Multipole Measures}

Having reconstructed the mass distribution, we can estimate the local
dispersion of $\kappa$ (e.g., \cite*{LO98.2}). However, the errors at
different points will be strongly correlated, and so it makes little
sense to attach an error bar to each point of the mass map. Although
mass maps contain valuable information, it is sometimes preferable to
reduce them to a small set of numbers such as the mass-to-light ratio,
or the correlation coefficient between the mass map and the light
distribution. One of the quantities of interest is the total mass
inside a given region. As became clear in the last section, this
quantity by itself cannot be determined from observed image
ellipticities due to the invariance transformation. But a quantity
related to it,
\begin{equation}
  \zeta(\vec\theta;\vartheta_1,\vartheta_2) \equiv
  \bar\kappa(\vec\theta;\vartheta_1)
  -\bar\kappa(\vec\theta;\vartheta_1,\vartheta_2)\;,
\label{eq:5.21}
\end{equation}
the difference between the mean surface mass densities in a circle of
radius $\vartheta_1$ around $\vec\theta$ and in an annulus of inner
and outer radii $\vartheta_1$ and $\vartheta_2$, respectively, can be
determined in the weak-lensing case, since then the invariance
transformation corresponds to an additive constant in $\kappa$ which
drops out of (\ref{eq:5.21}). We show in this section that quantities
like (\ref{eq:5.21}) can directly be obtained from the image
ellipticities without the need for a two-dimensional mass map. In
Sect.~\ref{sc:5.3.1}, we derive a generalised version of
(\ref{eq:5.21}), whereas we consider the determination of mass
multipoles in Sect.~\ref{sc:5.3.2}. The prime advantage of all these
aperture measures is that the error analysis is relatively
straightforward.

\subsubsection{\label{sc:5.3.1}Aperture Mass Measures}

Generally, aperture mass measures are weighted integrals of the local
surface mass density,
\begin{equation}
  M_\mathrm{ap}(\vec\theta_0) = \int\d^2\theta\,\kappa(\vec\theta)\,
  U(\vec\theta-\vec\theta_0)\;,
\label{eq:5.22}
\end{equation}
with weight function $U(\vec\theta)$. Assume now that the weight
function is constant on self-similar concentric curves. For example,
the $\zeta$-statistics (\ref{eq:5.21}), introduced by \cite{KA95.1},
is of the form (\ref{eq:5.22}), with a weight function that is
constant on circles, $U(\vartheta)=(\pi \vartheta_1^2)^{-1}$ for
$0\le\vartheta\le \vartheta_1$, $U(\vartheta)=[\pi (\vartheta_2^2-
\vartheta_1^2]^{-1}$ for $\vartheta_1 < \vartheta\le \vartheta_2$, and
zero otherwise.

Let the shape of the aperture be described by a closed curve $\vec
c(\lambda)$, $\lambda\in I$, where $I$ is a finite interval, such that
$\vec c\times\dot{\vec c}\equiv c_1\dot c_2-c_2\dot c_1>0$ for all
$\lambda\in I$. We can then uniquely define a new coordinate system
$(b,\lambda)$ by choosing a centre $\vec\theta_0$ and defining
$\vec\theta=\vec\theta_0+b\vec c(\lambda)$. The weight function should
be constant on the curves $\vec c(\lambda)$ so that it depends only on
$b$. In the new coordinate system, (\ref{eq:5.22}) reads
\begin{equation}
  M_\mathrm{ap}(\vec\theta_0) = \int_0^\infty\d b\;
  b\,U(b)\oint_I\d\lambda\;\vec c\times\dot{\vec c}\,
  \kappa[\vec\theta_0+b\vec c(\lambda)]\;,
\label{eq:5.23}
\end{equation}
where the factor $b\,\vec c\times\dot{\vec c}$ is the Jacobian
determinant of the coordinate transformation.
Equation~(\ref{eq:5.23}) can now be transformed in three steps; first,
by a partial integration with respect to $b$; second, by replacing
partial derivatives of $\kappa$ with partial derivatives of $\gamma$
using eq.~(\ref{eq:5.10}); and third by removing partial derivatives
of $\gamma$ in another partial integration. In carrying out these
steps, we assume that the weight function is compensated,
\begin{equation}
  \int\d b\,b\,U(b) = 0\;.
\label{eq:5.24}
\end{equation}
Introducing
\begin{equation}
  Q(b) \equiv \frac{2}{b^2}\int_0^b\d b'\,b'\,U(b')- U(b)
\label{eq:5.25}
\end{equation}
and writing the curve $\vec c$ in complex notation,
$C(\lambda)=c_1(\lambda)+\mathrm{i}\,c_2(\lambda)$, leads to the final
result (\cite*{scb97})
\begin{equation}
  M_\mathrm{ap}(\vec\theta_0) = \int\d^2\theta\,
  {Q[b(\vec\theta)]}\,
  \frac{\Im[\gamma(\vec\theta)C^*\dot C^*]}{\Im[C^*\dot C]}\;,
\label{eq:5.26}
\end{equation}
where the argument $\lambda$ of $C$ is to be evaluated at position
$\vec\theta=\vec\theta_0+b\vec
c(\lambda).$\footnote{\label{fn:5.3}There are of course other ways to
derive (\ref{eq:5.26}), e.g.~by inserting (\ref{eq:5.4}) into
(\ref{eq:5.22}). See \cite{SQ96.3} for a different approach using
Gauss's law.} The numerator in the final term of (\ref{eq:5.26})
projects out a particular component of the shear, whereas the
denominator is part of the Jacobian of the coordinate
transformation. The constraint (\ref{eq:5.24}) assures that an
additive constant in $\kappa$ does not affect $M_\mathrm{ap}$. The
expression (\ref{eq:5.26}) has several nice properties which render it
useful:

\begin{enumerate}

\item If the function $U(b)$ is chosen such that it vanishes for
$b>b_2$, then from (\ref{eq:5.24}) and (\ref{eq:5.25}), $Q(b)=0$ for
$b>b_2$. Thus, the aperture mass can be derived from the shear in a
finite region.

\item If $U(b)=\,\mathrm{const}$ for $0\le b\le b_1$, then $Q(b)=0$ in
that interval. This means that the aperture mass can be determined
solely from the shear in an annulus $b_1<b<b_2$. This has two
advantages which are relevant in practice. First, if the aperture is
centred on a cluster, the bright central cluster galaxies may prevent
the detection of a large number of faint background galaxies there, so
that the shear in the central part of the cluster may be difficult to
measure. In that case it is still possible to determine the total mass
inside the cluster core using (\ref{eq:5.26}) with an appropriately
chosen weight function $U$. Second, although in general the shear
cannot be determined directly from the image ellipticities [but only
the reduced shear $\gamma(1-\kappa)^{-1}$], we can choose the size
$b_1$ of the inner boundary of the annulus sufficiently large that
$\kappa\ll1$ in the annulus, and then $\gamma\approx g$ is an accurate
approximation. Hence, in that case the mean image ellipticity directly
yields an estimate of the shear. Then, the integral (\ref{eq:5.26})
can be transformed into a sum over galaxy images lying in the annulus,
yielding $M_\mathrm{ap}$ directly in terms of the observables. This in
turn has the great advantage that an error analysis of $M_\mathrm{ap}$
is fairly simple.

\end{enumerate}

We consider circular apertures as an example, for which
$(b,\lambda)=(\theta,\varphi)$ and
$C(\varphi)=\exp(\mathrm{i}\varphi)$. Then, $\Im(C^*\dot C)=1$, and
\begin{equation}
  \Im(\gamma C^*\dot C^*) =
  \gamma_\mathrm{t}(\vec\theta;\vec\theta_0) :=
  -\left[\gamma_1\cos(2\varphi)+\gamma_2\sin(2\varphi)\right]
  =-\Re[\gamma(\vec\theta+\vec\theta_0)
   \mathrm{e}^{-2\mathrm{i}\varphi}] \;,
\label{eq:5.27}
\end{equation}
where we have defined the {\em tangential component\/}
$\gamma_\mathrm{t}$ of the shear relative to the point
$\vec\theta_0$. Hence, for circular apertures (\ref{eq:5.26}) becomes
\begin{equation}
  M_\mathrm{ap}(\vec\theta_0) = \int\d^2\theta\,{Q(|\vec\theta|)}\,
  \gamma_\mathrm{t}(\vec\theta;\vec\theta_0)
\label{eq:5.28}
\end{equation}
(\cite*{kai94}; \cite*{SC96.3}). The $\zeta$-statistics
(\ref{eq:5.21}) is obtained from (\ref{eq:5.28}) by setting
$Q(\theta)=\vartheta_2^2\,\theta^{-2}\left[\pi
(\vartheta_2^2-\vartheta_1^2)\right]^{-1}$ for
$\vartheta_1\le\theta\le\vartheta_2$ and $Q(\theta)=0$ otherwise, so
that
\begin{equation}
  \zeta(\vec\theta_0;\vartheta_1,\vartheta_2) = 
  \frac{\vartheta_2^2}{\left[\pi(\vartheta_2^2-\vartheta_1^2)\right]}
  \int\d^2\theta\,\frac{\gamma_\mathrm{t}(\vec\theta;\vec\theta_0)}
  {|\vec\theta|^2}\;,
\label{eq:5.29}
\end{equation}
where the integral is taken over the annulus
$\vartheta_1\le\theta\le\vartheta_2$.

For practical purposes, the integral in (\ref{eq:5.28}) is transformed
into a sum over galaxy images. Recalling that $\epsilon$ is an
estimator for $\gamma$ in the weak-lensing case, and that the weight
function can be chosen to avoid the strong-lensing regime, we can
write
\begin{equation}
  M_\mathrm{ap}(\vec\theta_0) = \frac{1}{n}\,
  \sum_i {Q(|\vec\theta_i-\vec\theta_0|)}\,
  \epsilon_{\mathrm{t}i}(\vec\theta_0)\;,
\label{eq:5.30}
\end{equation}
where we have defined, in analogy to $\gamma_\mathrm{t}$, the
tangential component $\epsilon_{\mathrm{t}i}$ of the ellipticity of an
image at $\vec\theta_i$ relative to the point $\vec\theta_0$ by
\begin{equation}
  \epsilon_{\mathrm{t}i} = 
  -\Re(\epsilon\,\mathrm{e}^{-2\mathrm{i}\varphi})\;,
\label{eq:5.31}
\end{equation}
$\varphi$ is the polar angle of $\vec\theta-\vec\theta_0$, and $n$ is
the number density of galaxy images. The {\em rms\/} dispersion
$\sigma(M_\mathrm{ap})$ of $M_\mathrm{ap}$ in the case of no lensing
is found from the (two-dimensional) dispersion $\sigma_\epsilon$ of
the intrinsic ellipticity of galaxies,
\begin{equation}
  \sigma(M_\mathrm{ap}) = \frac{\sigma_\epsilon}{2^{1/2}n}
  \left[\sum_i{Q^2(|\vec\theta_i-\vec\theta_0|)}\right]^{1/2}\;.
\label{eq:5.32}
\end{equation}
The {\em rms\/} dispersion in the presence of lensing will deviate
only weakly from $\sigma(M_\mathrm{ap})$ as long as the assumption of
weak lensing in the annulus is satisfied. Hence,
$\sigma(M_\mathrm{ap})$ can be used as an error estimate for the
aperture mass and as an estimate for the signal-to-noise ratio of a
mass measurement.

This opens the interesting possibility to search for (dark) mass
concentrations using the aperture mass (\cite*{SC96.3}). Consider a
weight function $U$ with the shape of a Mexican hat, and a data field
$\mathcal{U}$ on which apertures of angular size $\theta$ can be
placed. For each aperture position, one can calculate $M_\mathrm{ap}$
and the dispersion. The dispersion can be obtained either from the
analytical formula (\ref{eq:5.32}), or it can be obtained directly
from the data, by randomising the position angles of all galaxy images
within the aperture. The dispersion can be obtained from many
realisations of this randomisation process. Large values of
$M_\mathrm{ap}$ will be obtained for mass concentrations whose
characteristic size and shape is close to that of the chosen filter
function $U$. Thus, by varying the size $\theta$ of the filter,
different mass concentrations will preferentially be selected. The
aperture mass is insensitive to mass concentrations of much smaller
and much larger angular scales than the filter size.

We have considered in Sect.~\ref{sc:4.5} the signal-to-noise ratio for
the detection of a singular isothermal sphere from its weak lensing
effect. The estimate (\ref{eq:4.54}) was obtained by an optimal
weighting scheme for this particular mass distribution. Since real
mass concentrations will deviate from this profile, and also from the
assumed symmetry, the filter function $U$ should have a more generic
shape. In that case, the S/N will have the same functional behaviour
as in (\ref{eq:4.54}), but the prefactor depends on the exact shape of
$U$. For the filter function used in \cite{SC96.3}, S/N is about 25\%
smaller than in (\ref{eq:4.54}). Nevertheless, one expects that the
aperture-mass method will be sensitive to search for
intermediate-redshift haloes with characteristic velocity dispersions
above $\sim600\,\mathrm{km\,s}^{-1}$.

This expectation has been verified by numerical simulations, which
also contained larger and smaller scale mass perturbations. In
addition, a detailed strong-lensing investigation of the cluster
MS~1512$+$62 has shown that its velocity dispersion is very close to
$\sim600\,\mathrm{km\,s}^{-1}$, and it can be seen from the
weak-lensing image distortion alone with very high significance
(\cite*{ssb98}), supporting the foregoing quantitative
prediction. Thus, this method appears to be a very promising way to
obtain a {\em mass-selected\/} sample of haloes which would be of great
cosmological interest (cf.~\cite*{reb99}). We shall return to this
issue in Sect.~\ref{sc:6.7.2}.

\subsubsection{\label{sc:5.3.2}Aperture Multipole Moments}

Since it is possible to express the weighted mass within an aperture
as an integral over the shear, with the advantage that in the weak
lensing regime this integral can be replaced by a sum over galaxy
ellipticities, it is natural to ask whether a similar result holds for
multipole moments of the mass. As shown in \cite{scb97}, this is
indeed possible, and we shall briefly outline the method and the
result.

Consider a circular aperture\footnote{\label{fn:5.4}The method is not
restricted to circular apertures, but this case will be most relevant
for measuring multipole moments.} centred on a point
$\vec\theta_0$. Let $U(|\vec\theta|)$ be a radial weight function, and
define the $n$-th multipole moment by
\begin{equation}
  Q^{(n)} \equiv \int_0^\infty \d \theta\,\theta^{n+1}\,U(\theta)
  \int_0^{2\pi}\d\varphi\,\mathrm{e}^{n\mathrm{i}\varphi}\,
  \kappa(\vec\theta_0+\vec\theta) \;.
\label{eq:5.33}
\end{equation}
This can be replaced by an integral over $\gamma$ in two ways:
(\ref{eq:5.33}) can be integrated by parts with respect to $\varphi$
(for $n\ne 0$), or with respect to $\theta$, again utilising
(\ref{eq:5.10}). The resulting expressions are assumed to contain no
boundary terms, which restricts the choice for the weight function
$U(\theta)$. The remaining integrals then contain partial derivatives
of $\kappa$ with respect to $\varphi$ and $\theta$,
respectively. Writing (\ref{eq:5.10}) in polar coordinates, these
partial derivatives can be replaced by partial derivatives of the
shear components with respect to $\varphi$ and $\theta$. Integrating
those by parts with respect to the appropriate coordinate, and
enforcing vanishing boundary terms, we find two different expressions
for the $Q^{(n)}$:
\begin{equation}
  Q^{(n)}_{\varphi,\theta} = \int\d^2\theta\,
  q^{(n)}_{\varphi,\theta}(\vec\theta)\,
  \gamma(\vec\theta_0+\vec\theta) \;.
\label{eq:5.34}
\end{equation}
The two expressions for $q^{(n)}$ are formally very different,
although it can be shown that the resulting two expressions for
$Q^{(n)}$ are equivalent. The two very different equations for the
same result are due to the fact that the two components of the shear
$\gamma$ are not mutually independent, which was not used in the
derivation of (\ref{eq:5.34}).

We now have substantial freedom to choose the weight function and to
select one of the two expressions for $Q^{(n)}$, or even to take a
linear combination of them. We note the following interesting
examples:

\begin{enumerate}

\item The weight function $U(\theta)$ can be chosen to vanish outside
an annulus, to be piece-wise differentiable, and to be zero on the
inner and outer boundary of the annulus. The $Q^{(n)}$ for $n\ne 0$
can then be expressed as integrals of the shear over the annulus, with
no further restrictions on $U$. In particular, $U(\theta)$ does not
need to be a compensated weight function.

\item $U(\theta)$ can be a piece-wise differentiable weight function
which is constant for $\theta\le\theta_1$, and decreases smoothly to
zero at $\theta=\theta_2>\theta_1$. Again, $Q^{(n)}$ for $n\ne 0$ can
be expressed as an integral of the shear in the annulus
$\theta_1\le\theta\le\theta_2$. Hence, as for the aperture mass,
multipole moments in the inner circle can be probed with the shear in
the surrounding annulus.

\item One can choose, for $n>2$, a piece-wise differentiable weight
function $U(\theta)$ which behaves like $\theta^{-2n}$ for
$\theta>\theta_2$ and decreases to zero at
$\theta=\theta_1<\theta_2$. In that case, the multipole moments of the
matter outside an annulus can be probed with data inside the annulus.

\end{enumerate}

For practical applications, the integral in (\ref{eq:5.34}) is
replaced by a sum over galaxy ellipticities. The dispersion of this
sum is easily obtained in the absence of lensing, with an expression
analogous to (\ref{eq:5.32}). Therefore, the signal-to-noise ratio for
the multipole moments is easily defined, and thus also the
significance of a multipole-moment detection.

\subsection{\label{sc:5.4}Application to Observed Clusters}

Soon after the parameter-free two-dimensional mass reconstruction was
suggested by \cite{KA93.2}, their method was applied to the cluster
MS~1224 (\cite*{FA94.1}). Since then, several groups have used it to
infer the mass profiles of clusters. In parallel to this, several
methods have been developed to measure the shear from CCD data,
accounting for PSF smearing and anisotropy, image distortion by the
telescope, noise, blending etc. -- see the discussion in Sect.~4.6. We
will now summarise and discuss several of these observational results.

\cite{TY90.1} made the first attempt to constrain the mass
distribution of a cluster from a weak-lensing analysis. They
discovered a statistically significant tangential alignment of faint
galaxy images relative to the centre of the clusters A~1689 and
Cl~1409$+$52. Their ``lens distortion map'' obtained from the image
ellipticities yields an estimate of the mass distribution in these
clusters. A detailed analysis of their method is given in
\cite{KA93.2}. From a comparison with numerical simulations,
\citename{TY90.1} showed that the best isothermal sphere model for the
clusters has a typical velocity dispersion of $\sigma_v\sim1300\pm
200\,\mathrm{km\,s}^{-1}$ for both clusters. In particular, their
analysis showed that diffuse dark matter in the cluster centres is
needed to account for the observed image distortions.

The inversion method developed by \cite{KA93.2} provided a systematic
approach to reconstruct the mass distribution in clusters. It was
first applied to the cluster MS~1224$+$20 (\cite*{FA94.1}) at redshift
$z_\mathrm{d}=0.33$, which had been selected for its high X-ray
luminosity. Their square data field with side-length $\sim14'$ was
composed of several exposures, most of them with excellent
seeing. They estimated the shear from image ellipticities, corrected
for the PSF anisotropy, and applied a correction factor $f$ as defined
in Sect.~\ref{sc:4.6.1}. They found $f\sim 1.5$ in simulations, in
very good agreement with \cite{WI96.2}. The resulting shear pattern,
obtained from 2147 galaxy images, clearly shows a circular pattern
around the cluster centre as defined by the centroid of the optical
and X-ray light. Using the \citename{KA93.2} reconstruction method
(\ref{eq:5.6}), \citename{FA94.1} produced maps of the dimension-less
surface mass density $\kappa(\vec\theta)$, both by taking all galaxy
images into account, and after splitting the galaxy sample into a
`brighter' and `fainter' sample of roughly equal size. Although
differing in detail, the resulting mass show an overall similarity.
In particular, the position of the mass centre is very similar in all
maps.

\citename{FA94.1} applied the aperture mass method to determine the
cluster mass -- see (\ref{eq:5.22}) and (\ref{eq:5.30}) -- in an
annulus centred on the cluster centre with inner radius
$\vartheta_1=2\arcminf76$ and an outer radius such that the annulus
nearly fits into their data field. The lower limit to the mean surface
mass density in the annulus is
$\bar\kappa(2\arcminf76)\ge\zeta=0.06\pm0.013$. To convert this into
an estimate of the physical surface mass density and the total mass
inside the aperture, the mean distance ratio
$D_\mathrm{ds}/D_\mathrm{s}$ for the galaxy population has to be
estimated, or equivalently the mean value of $w$ as defined in
(\ref{eq:5.36}).

While the redshift distribution is known statistically for the
brighter sub-sample from redshift surveys, the use of the fainter
galaxies requires an extrapolation of the galaxy redshifts. From that,
\citename{FA94.1} estimated the mass within a cylinder of radius
$\vartheta_1=2\arcminf 76$, corresponding to $0.48 h^{-1}$~Mpc for an
Einstein-de Sitter cosmology, to be
$\sim3.5\times10^{14}\,h^{-1}\,M_\odot$. This corresponds to a
mass-to-light ratio (in solar units) of $M/L\sim800\,h$. \cite{CA94.4}
obtained 75 redshifts of galaxies in the cluster field, of which 30
are cluster members. From their line-of-sight velocity dispersion, the
cluster mass can be estimated by a virial analysis. The resulting mass
is lower by a factor $\sim 3$ than the weak-lensing estimate. The
mass-to-light ratio from the virial analysis is much closer to typical
values in lower-redshift clusters like Coma, which has
$M/L\approx270\,h^{-1}$. The high mass estimate of this cluster was
recently confirmed in a completely independent study by \cite{fis99}.

The origin of this large apparent discrepancy is not well understood
yet, and several possibilities are discussed in \cite{kai94}. It
should be pointed out that lensing measures the total mass inside a
cone, weighted by the redshift-dependent factor
$D_\mathrm{d}D_\mathrm{ds}/D_\mathrm{s}$, and hence the lensing mass
estimate possibly includes substantial foreground and background
material. While this may cause an overestimate of the mass, it is
quite unlikely to cause an overestimate of the mass-to-light ratio of
the total material inside the cone. Foreground material will
contribute much more strongly to the light than to the measured mass,
and additional matter behind the cluster will not be very efficient as
a lens. The uncertainty in the redshift distribution of the faint
galaxies translates into an uncertainty in the mass. However, all
background galaxies would have to be put at a redshift $\sim4$ to
explain the mass discrepancy, while redshift surveys show that the
brighter sub-sample of \citename{FA94.1} has a mean redshift below
unity. The mass estimate is only weakly dependent on the assumed
cosmological model. On the other hand, the light distribution of the
cluster MS~1224 is not circular, and it cannot be excluded that this
cluster is not in virial equilibrium.

\cite{SQ96.1} compared the mass profiles derived from weak lensing
data and the X-ray emission of the cluster A~2218. Under the
assumption that the hot X-ray-emitting intra-cluster gas is in
hydrostatic equilibrium between gravity and thermal pressure support,
the mass profile of the cluster can be constrained. The reconstructed mass map
qualitatively agrees with the optical and
X-ray light distributions. Using the aperture mass estimate, a mass-to-light
ratio of $M/L=(440\pm80)\,h$ in solar units is found. The radial
mass profile appears to be flatter than isothermal. Within the
error bars, it agrees with the mass profile obtained from the X-ray
analysis, with a slight indication that at large radii the lensing
mass is larger than the mass inferred from X-rays.

Abell~2218 also contains a large number of arcs and multiply-imaged
galaxies which have been used by \cite{KN96.1} to construct a detailed
mass model of the cluster's central region. In addition to the main
mass concentration, there is a secondary clump of cluster galaxies
whose effects on the arcs is clearly visible. The separation of these
two mass centres is $67''$. Whereas the resolution of the weak lensing
mass map as obtained by \citename{SQ96.1} is not sufficient to reveal
a distinct secondary peak, the elongation of the central density
contours extend towards the secondary galaxy clump.

General agreement between the reconstructed mass map and the
distribution of cluster galaxies and X-ray emission has also been
found for the two clusters Cl~1455$+$22 ($z=0.26$) and Cl~0016$+$16
($z=0.55$) by \cite{SM95.1}. Both are highly X-ray luminous clusters
in the {\em Einstein\/} Extended Medium Sensitivity Survey (EMSS;
\cite*{smg91}). The orientation and ellipticity of the central mass
peak is in striking agreement with those of the galaxy distribution
and the X-ray map. However, the authors find some indication that the
mass is more centrally condensed than the other two distributions. In
addition, given the finite angular resolution of the mass map, the
core size derived from weak lensing is most likely only an upper bound
to the true value, and in both clusters the derived core size is
significantly larger than found in clusters with giant luminous arcs
(see, e.g., \cite*{FO94.1}).

The mass-to-light ratios for the two clusters are $\sim1000\,h$ and
$\sim740\,h$, respectively. However, at least for Cl~0016$+$16, the
mass scale is fairly uncertain, owing to the high cluster redshift and
the unknown redshift distribution of the faint galaxies. The mean
value of $D_\mathrm{ds}/D_\mathrm{s}$ must be estimated from an
assumed distribution $p(z)$.

\begin{figure}[ht]
  \includegraphics[width=0.49\hsize]{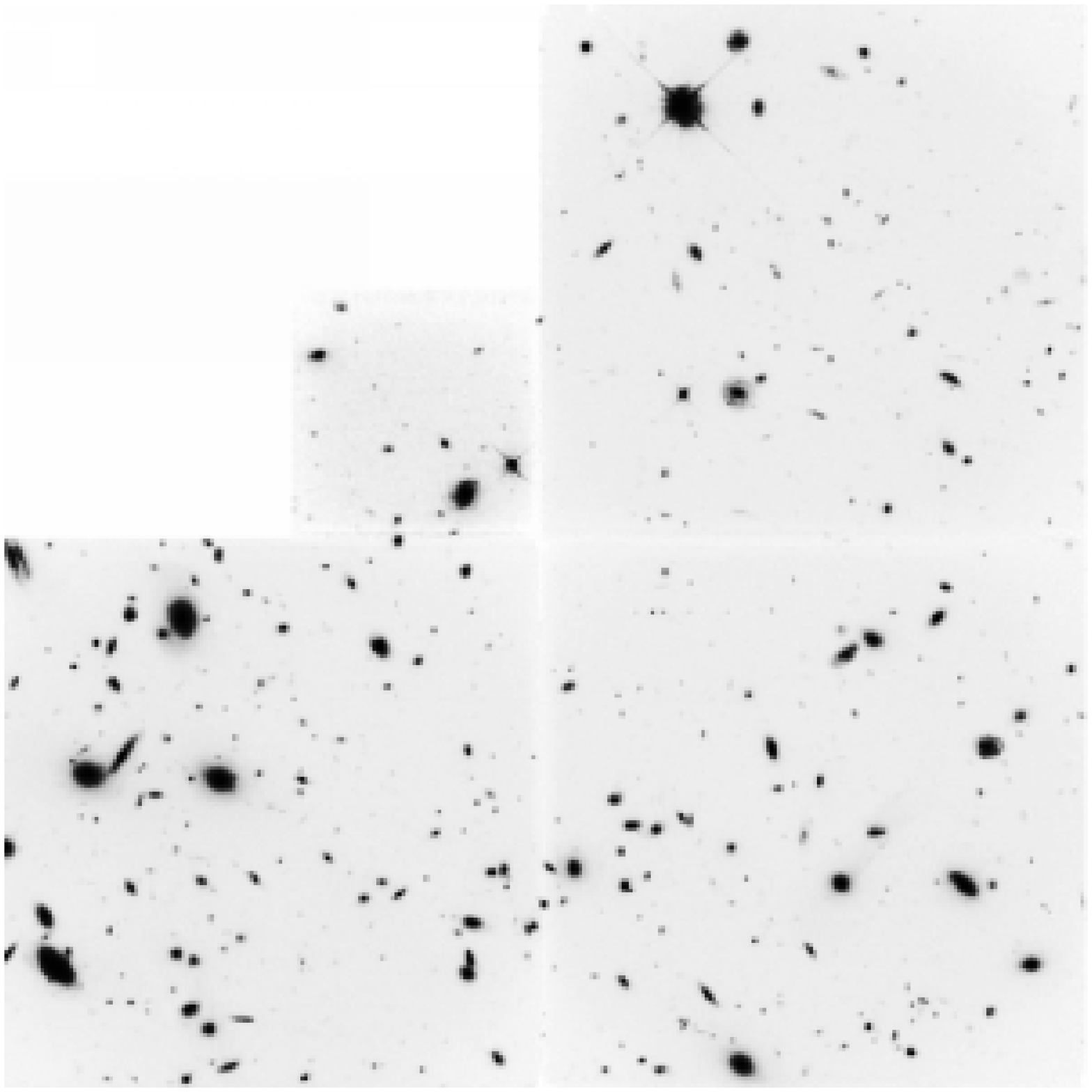}\hfill
  \includegraphics[width=0.49\hsize]{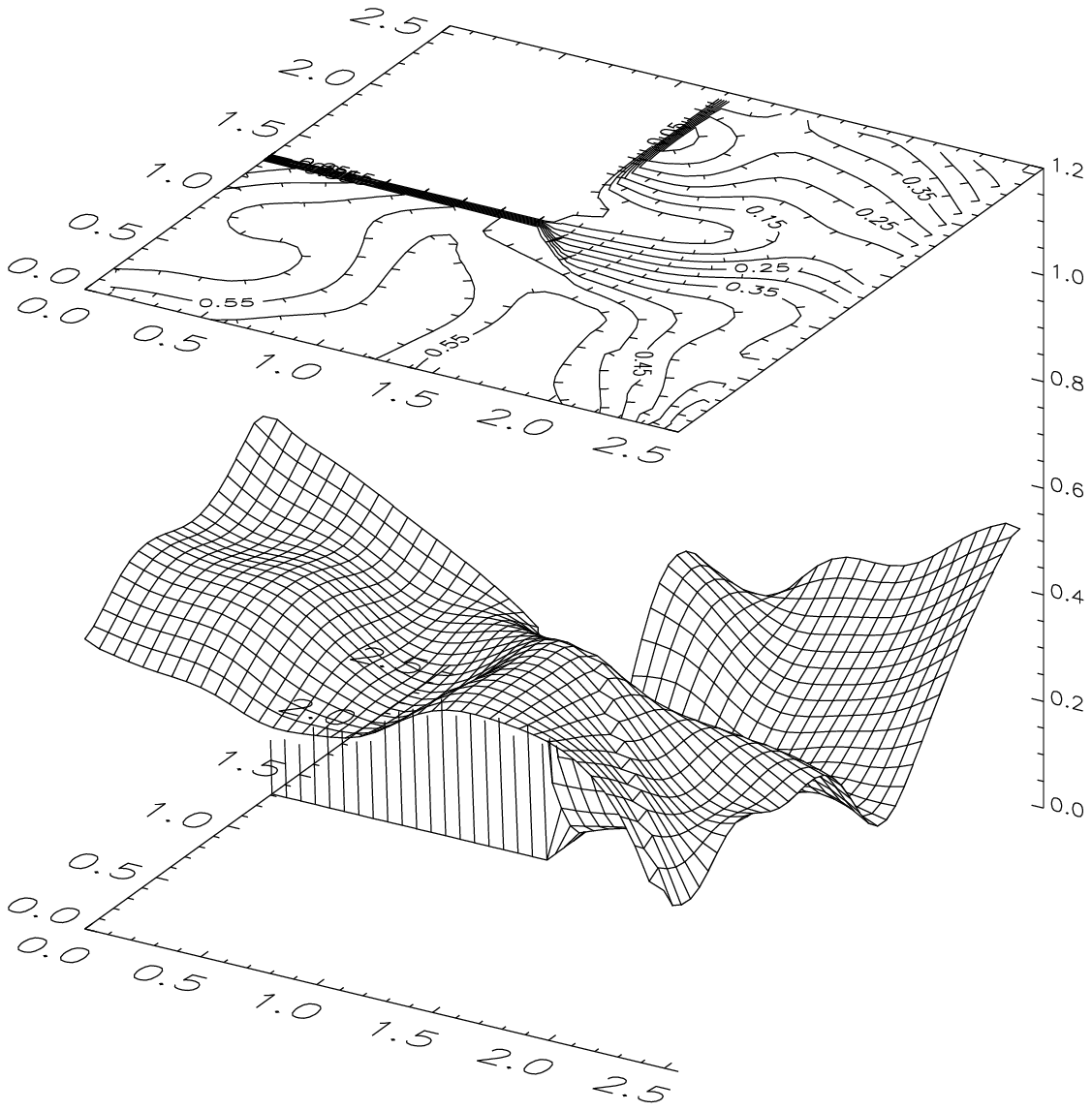}
\caption{{\em Left panel:\/} WFPC2 image of the cluster Cl0939$+$4713
(A~851); North is at the bottom, East to the right. The cluster centre
is located at about the upper left corner of the left CCD, a secondary
maximum of the bright (cluster) galaxies is seen close to the
interface of the two lower CCDs, and a minimum in the cluster light is
at the interface between the two right CCDs. In the lensing analysis,
the data from the small CCD (the Planetary Camera) were not used. {\em
Right panel:\/} The reconstructed mass distribution of A~851, assuming
a mean redshift of the $N=295$ galaxies with $24\le R\le25.5$ of
$\langle z\rangle=1$.}
\label{fig:5.1}
\end{figure}

The unprecedented imaging quality of the refurbished {\em Hubble Space
Telescope\/} (HST) can be used profitably for weak lensing
analyses. Images taken with the {\em Wide Field Planetary Camera 2\/}
(WFPC2) have an angular resolution of order $0\arcsecf1$, limited by
the pixel size. Because of this superb resolution and the lower sky
background, the number density of galaxy images for which a shape can
reliably be measured is considerably larger than from the ground, so
that higher-resolution mass maps can be determined. The drawback is
the small field covered by the WFPC2, which consists of 3 CCD chips
with $80''$ side-length each. Using the first publicly available deep
image of a cluster obtained with the WFPC2, \cite{SE96.2} have
constructed a mass map of the cluster Cl~0939$+$47 ($z=0.41$).
Figure~\ref{fig:5.1} clearly shows a mass peak near the left boundary
of the frame shown. This maximum coincides with the cluster centre as
determined from the cluster galaxies (\cite*{drg92}). Furthermore, a
secondary maximum is clearly visible in the mass map, as well as a
pronounced minimum. When compared to the optical image, a clear
correlation with the bright (cluster) galaxies is obvious. In
particular, the secondary maximum and the minimum correspond to the
same features in the bright galaxy distribution. A formal correlation
test confirms this similarity. Applying the maximum-likelihood mass
reconstruction technique (\cite*{ssb97}; see Sect.~\ref{sc:5.4}) to
the same HST image, \cite{ges99} constructed a higher-resolution map
of this cluster. The angular resolution achieved is much higher in the
cluster centre, predicting a region in which strong lensing effects
may occur. Indeed, \cite{TR97.1} reported on a highly elongated arc
and a triple image, with both source galaxies having a redshift
$z\approx3.97$.

The X-ray map of this cluster (\cite*{SC97.5}) shows that the two mass
peaks are also close to two X-ray components. The determination of the
total mass inside the WFPC2 frame is difficult, for two reasons:
First, the high redshift of the cluster implies that the mean value of
$D_\mathrm{ds}/D_\mathrm{s}$ depends quite sensitively on the assumed
redshift distribution of the background galaxies. Second, the small
field of the WFPC2 precludes the measurement of the surface mass
density at large distance where $\kappa$ tends to zero, and thus the
mass-sheet degeneracy implies a considerable uncertainty in the mass
scale. Attempting to lift the mass sheet degeneracy with the
number-density effect -- see \ref{sc:4.4.1} --, a mass-to-light ratio
of $\sim250\,h$ was derived within the WFPC2 aperture. This value is
also affected by the unknown fraction of cluster members in the
catalog of faint galaxies. \cite{SE96.2} assumed that the spatial
distribution of faint cluster galaxies follows that of brighter
cluster galaxies. The striking difference between the $M/L$ ratios for
this and the other clusters described above may be related to the fact
that Cl~0939$+$47 is the highest-redshift cluster in the Abell catalog
(A~851). Hence, it was selected by its high optical luminosity,
whereas the previously mentioned clusters are all X-ray selected. The
X-ray luminosity of Cl~0939$+$47 is fairly small for such a rich
cluster (\cite*{SC96.2}). Since X-ray luminosity and cluster mass are
generally well correlated, the small $M/L$-ratio found from the weak
lensing analysis is in agreement with the expectations based on the
high optical flux and the low X-ray flux. Note that the large spread
of mass-to-light ratios as found by the existing cluster mass
reconstructions is unexpected in the frame of hierarchical models of
structure formation and thus poses an interesting astrophysical
problem.

\cite{hfk98} reconstructed the mass distribution in the cluster
MS~1358$+$62 from a mosaic of HST images, so that their data field in
substantially larger than for a single HST pointing (about
$8'\times8'$). This work uses the correction method presented in
Sect.~\ref{sc:4.6.2}, thus accounting for the relatively strong PSF
anisotropy at the edges of each WFPC2 chip. A weak-lensing signal out
to $1.5\,\mathrm{Mpc}$ is found. The X-ray mass is found to be
slightly lower than the dynamical mass estimate, but seems to agree
well with the lensing mass determination.

\cite{LU97.1} found a surprisingly strong weak-lensing signal in the
field of the high-redshift cluster MS~1054$-$03 ($z=0.83$). This
implies that the sheared galaxies must have an appreciably higher
redshift than the cluster, thus strongly constraining their redshift
distribution. In fact, unless the characteristic redshift of these
faint background galaxies is $\gtrsim1.5$, this cluster would have an
unrealistically large mass. It was also found that the lensing signal
from the bluer galaxies is stronger than from the redder ones,
indicating that the characteristic redshift of the bluer sample is
higher. In fact, the mass estimated assuming $\langle
z_\mathrm{s}\rangle=1.5$ agrees well with results from analyses of the
X--ray emission (\cite*{DO98.1}) and galaxy kinematics
(\cite*{tkd99}). \cite{CL98.1} derived weak lensing maps for two
additional clusters at $z\sim0.8$, namely MS~1137$+$66 at $z=0.783$
and RXJ~1716$+$67 at $z=0.813$.

The mass distribution in the supercluster MS~0302$+$17 at $z=0.42$ was
reconstructed by \cite{kwl98} in a wide-field image of size $\sim30'$.
The supercluster consists of three clusters which are very close
together on the sky and in redshift. The image contains about 30,000
galaxies from which a shear can be measured. This shear was found to
correlate strongly with the distribution of the early-type
(foreground) galaxies in the field, provided that the overall
mass-to-light ratio is about $250\,h$. Each of the three clusters,
which are also seen in X-rays, is recovered in the mass map. The
ratios between mass and light or X-ray emission differ slightly across
the three clusters, but the differences are not highly significant.

A magnification effect was detected from the depletion of the number
counts (see Sect.~\ref{sc:4.4.1}) in two clusters. \cite{FO97.1}
discovered that the number density of very faint galaxies drops
dramatically near the critical curve of the cluster Cl~0024$+$16, and
remains considerably lower than the mean number density out to about
twice the Einstein radius. This is seen in photometric data with two
filters. \cite{FO97.1} interpret this broad depletion curve in terms
of a broad redshift distribution of the background galaxies, so that
the location of the critical curve of the cluster varies over a large
angular scale. A spatially-dependent number depletion was detected in
the cluster A~1689 by \cite{TA98.1}.

These examples should suffice to illustrate the current status of weak
lensing cluster mass reconstructions. For additional results, see
\cite{SQ96.2}, \cite{SQ97.1}, \cite{FI97.1}, \cite{FI97.2}. Many of
the difficulties have been overcome; e.g., the method presented in
Sect.~\ref{sc:4.6.2} appears to provide an accurate correction method
for PSF effects. The quantitative results, for example for the
$M/L$-ratios, are somewhat uncertain due to the lack of sufficient
knowledge on the source redshift distribution, which applies in
particular to the high-redshift clusters.

Further large-format HST mosaic images either are already or will soon
become available, e.g.~for the clusters A~2218, A~1689, and
MS~1054$-$03. Their analysis will substantially increase the accuracy
of cluster mass determinations from weak lensing compared to
ground-based imaging.

\subsection{\label{sc:5.5}Outlook}

We have seen in the preceding subsection that first results on the
mass distribution in clusters were derived with the methods described
earlier. Because weak lensing is now widely regarded as the most
reliable method to determine the mass distribution of clusters, since
it does not rely on assumptions on the physical state and symmetries
of the matter distribution, further attempts at improving the method
are in progress, and some of them will briefly be outlined below.

In particular, we describe a method which simultaneously accounts for
shear and magnification information, and which can incorporate
constraints from strong-lensing features (such as arcs and multiple
images of background sources). A method for the determination of the
local shear is described next which does not rely on the detection and
the quadrupole measurement of individual galaxies, and instead makes
use of the light from very faint galaxies which need not be
individually detected. We will finally consider the potential of weak
lensing for determining the redshift distribution of galaxies which
are too faint to be investigated spectroscopically, and report on
first results.

\subsubsection{\label{sc:5.5.1}Maximum-Likelihood Cluster
  Reconstructions}

The mass reconstruction method described above is a direct method: The
locally averaged observed image ellipticities $\langle\epsilon\rangle$
are inserted into an inversion equation such as (\ref{eq:5.8}) to find
the mass map $\kappa(\vec\theta)$. The beauty of this method is its
simplicity and computational speed. Mass reconstructions from the
observed image ellipticities are performed in a few CPU seconds.

The drawback of this method is its lack of flexibility. No additional
information can be incorporated into the inversion process. For
example, if strong-lensing features like giant arcs or multiple galaxy
images are observed, they should be included in the mass
reconstruction. Since such strong-lensing features typically occur in
the innermost parts of the clusters (at $\lesssim30''$ from cluster
centres), they strongly constrain the mass distribution in cluster
cores which can hardly be probed by weak lensing alone due to its
finite angular resolution. A further example is the incorporation of
magnification information, as described in Sect.~\ref{sc:4.4}, which
can in principle not only be used to lift the mass-sheet degeneracy,
but also provides local information on the shape of the mass
distribution.

An additional problem of direct inversion techniques is the choice of
the smoothing scale which enters the weight factors $u_i$ in
(\ref{eq:5.16}). We have not given a guideline on how this scale
should be chosen. Ideally, it should be adapted to the data.  In
regions of strong shear, the signal-to-noise ratio of a shear
measurement for a fixed number of galaxy images is larger than in
regions of weak shear, and so the smoothing scale can be smaller
there.

Recently, these problems have been attacked with inverse methods.
Suppose the mass distribution of a cluster is parameterised by a set
of model parameters $p_k$. These model parameters could then be varied
until the best-fitting model for the observables is found. Considering
for example the observed image ellipticities $\epsilon_i$ and assuming
a non-critical cluster, the expectation value of $\epsilon_i$ is the
reduced shear $g$ at the image position, and the dispersion is
determined (mainly) by the intrinsic dispersion of galaxy
ellipticities $\sigma_\epsilon$. Hence, one can define a
$\chi^2$-function
\begin{equation}
 \chi^2 = \sum_{i=1}^{N_\mathrm{g}}
 \frac{|\epsilon_i-g(\vec\theta_i)|^2}{\sigma_\epsilon^2} 
 \label{eq:5.35}
\end{equation}
and minimise it with respect to the $p_k$. A satisfactory model is
obtained if $\chi^2$ is of order $N_\mathrm{g}$ at its minimum, as
long as the number of parameters is much smaller than
$N_\mathrm{g}$. If the chosen parameterisation does not achieve this
minimum value, another one must be tried. However, the resulting mass
model will depend on the parameterisation which is a serious drawback
relative to the parameter-free inversion methods discussed before.

This problem can be avoided with `generic' mass models. For instance,
the deflection potential $\psi(\vec\theta)$ can be composed of a
finite sum of Fourier modes (\cite*{SQ96.3}), whose amplitudes are the
parameters $p_k$.\footnote{\label{fn:5.5}It is important to note that
the deflection potential $\psi$ rather than the surface mass density
$\kappa$ (as in \cite{SQ96.3}) should be parameterised, because shear
and surface mass density depend on the local behaviour of $\psi$,
while the shear {\em cannot\/} be obtained from the local $\kappa$,
and not even from $\kappa$ on a finite field. In addition, the local
dependence of $\kappa$ and $\gamma$ on $\psi$ is computationally much
more efficient than calculating $\gamma$ by integrating over $\kappa$
as in \cite{bhl98}.} The number of Fourier modes can be chosen such
that the resulting $\chi^2$ per degree of freedom is approximately
unity. Additional modes would then start to fit the noise in the data.

Alternatively, the values of the deflection potential $\psi$ on a
(regular) grid can be used as the $p_k$. \cite{BA96.2} employed the
locally averaged image ellipticities and the size ratios
$\langle\omega\rangle/\langle\omega\rangle_0$ -- see (\ref{eq:4.47})
-- on a grid. The corresponding expectation values of these
quantities, the reduced shear $g$ and the magnification $\mu$, were
calculated by finite differencing of the discretised deflection
potential $\psi$. Since both $\gamma$ and $\kappa$, and thus $\mu$,
are unchanged under the transformation $\psi(\vec\theta) \to
\psi(\vec\theta)+\psi_0+\vec a\cdot\vec\theta$, the deflection
potential has to be kept fixed at three grid points. If no
magnification information is used, the mass-sheet degeneracy allows a
further transformation of $\psi$ which leaves the expected image
ellipticities invariant, and the potential has to be kept fixed at
four grid points.

A $\chi^2$-function was defined using the local dispersion of the
image ellipticities and image sizes relative to unlensed sizes of
galaxies with the same surface brightness, and it was minimised with
respect to the values of $\psi$ on the grid points. The grid spacing
was chosen such that the resulting minimum $\chi^2$ has approximately
the correct value. Tests with synthetic data sets, using a numerically
generated cluster mass distribution, showed that this method
reconstructs very satisfactory mass maps, and the total mass of the
cluster was accurately reproduced.

If a finer grid is used, the model for the deflection potential will
reproduce noise features in the data. On the other hand, the choice of
a relatively coarse grid which yields a satisfactory $\chi^2$ implies
that the resolution of the mass map is constant over the data
field. Given that the signal increases towards the centre of the
cluster, one would like to use a finer grid there. To avoid
over-fitting of noise, the maximum-likelihood method can be
complemented by a regularisation term (see \cite*{PR86.1},
Chap.~18). As shown by \cite{ssb97}, a maximum-entropy regularisation
(\cite*{nan86}) is well suited for the problem at hand. As in
maximum-entropy image restoration (e.g., \cite*{luc94}), a prior is
used in the entropy term which is a smoothed version of the current
density field, and thus is being adapted during the minimisation. The
relative weight of the entropy term is adjusted such that the
resulting minimum $\chi^2$ is of order unity per degree of freedom.

In this scheme, the expectation values and dispersions of the
individual image ellipticities and sizes are found by bi-linear
interpolation of $\kappa$ and $\gamma$ on the grid which themselves
are obtained by finite differencing of the potential. When tested on
synthetic data sets, this refined maximum-likelihood method produces
mass maps with considerably higher resolution near the cluster centre
without over-fitting the noise at larger cluster-centric distances. The
practical implementation of this method is somewhat complicated. In
particular, if critical clusters are studied, some modifications have
to be included to allow the minimisation algorithm to move critical
curves across galaxy images in the lens plane. However, the quality of
the reconstruction justifies the additional effort, especially if
high-quality data from HST images are available. A first application
of this method is presented by \cite{ges99}.

Inverse methods such as the ones described here are likely to become
the standard tool for cluster mass profile reconstruction, owing to
their flexibility. As mentioned before, additional constraints from
strong lensing signatures such as arcs and multiply-imaged sources,
can straightforwardly be incorporated into these methods. The
additional numerical effort is negligible compared to the efforts
needed to gain the observational data. Direct inversion methods will
certainly retain an important role in this field, to obtain quick mass
maps during the galaxy image-selection process (e.g., cuts in colour
and brightness can be applied). Also, a mass map obtained by a direct
method as a starting model in the inverse methods reduces the
computational effort.

\subsubsection{\label{sc:5.5.2}The Auto-Correlation Function of the
  Extragalactic Background Light}

So far, we described how shear can be determined from ellipticities of
individual galaxy images on a CCD. In that context, a galaxy image is
a statistically significant flux enhancement on the CCD covering
several contiguous pixels and being more extended than the PSF as
determined from stars. Reducing the threshold for the signal-to-noise
per object, the number density of detected galaxies increases, but so
does the fraction of misidentifications.  Furthermore, the measured
ellipticity of faint galaxies has larger errors than that of brighter
and larger images. The detection threshold therefore is a compromise
between high number density of images and significance per individual
object.

Even the faintest galaxy images whose ellipticity cannot be measured
reliably still contain information on the lens distortion. It is
therefore plausible to use this information, by `adding up' the
faintest galaxies statistically. For instance, one could co-add their
brightness profiles and measure the shear of the combined
profiled. This procedure, however, is affected by the uncertainties in
defining the centres of the faint galaxies. Any error in the position
of the centre, as defined in (\ref{eq:5.1}), will affect the resulting
ellipticity.

To avoid this difficulty, and also the problem of faint object
definition at all, \cite{VA97.1} have suggested considering the
auto-correlation function (ACF) of the `background' light. Most of the
sky brightness is due to atmospheric scattering, but this contribution
is uniform. Fluctuations of the brightness on the scale of arc seconds
is supposedly mainly due to very faint galaxies. Therefore, these
fluctuations should intrinsically be isotropic. If the light from the
faint galaxies propagates through a tidal gravitational field, the
isotropy will be perturbed, and this provides a possibility to measure
this tidal field.

Specifically, if $I(\vec\theta)$ denotes the brightness distribution
as measured on a CCD, and $\bar I$ is the brightness averaged over the
CCD (or a part of it, see below), the auto-correlation function
$\xi(\vec\theta)$ of the brightness is defined as
\begin{equation}
  \xi(\vec\theta)= \left\langle
    \left(I(\vec\vartheta)-\bar I\right)
    \left(I(\vec\vartheta+\vec\theta)-\bar I\right)
  \right\rangle_{\vec\vartheta} \;,
\label{eq:5.36}
\end{equation}
where the average is performed over all pairs of pixels with
separation $\vec\theta$. From the invariance of surface brightness
(\ref{eq:3.10}, page~\pageref{eq:3.10}) and the locally linearised
lens mapping, $I(\vec\theta)= I^{(\mathrm{s})}(\A\vec\theta)$, one
finds that the observed ACF is related to the intrinsic ACF
$\xi^{(\mathrm{s})}$, defined in complete analogy to (\ref{eq:5.36}),
by
\begin{equation}
  \xi(\vec\theta) = \xi^{(\mathrm{s})}(\A\vec\theta)\;.
\label{eq:5.37}
\end{equation}
Thus the transformation from intrinsic to observed ACF has the
same functional form as the transformation of surface brightness. In
analogy to the definition of the quadrupole tensor $Q$ for galaxy
images -- see (\ref{eq:5.2}) -- the tensor of second
moments of the ACF is defined as
\begin{equation}
  \mathcal{M}_{ij} =
  \frac{\int\d^2\theta\,\xi(\vec\theta)\,\theta_i\theta_j}
  {\int\d^2\theta\;\xi(\vec\theta)}\;.
\label{eq:5.38}
\end{equation}
The transformation between the observed quadrupole tensor
$\mathcal{M}$ and the intrinsic one, $\mathcal{M}^{(\mathrm{s})}$, is
the same as for the moment tensor of image ellipticities,
(\ref{eq:5.5}), $\mathcal{M}^\s = \A\mathcal{M}\A$. As shown by
\cite{VA97.1}, the tensor $\mathcal{M}$ directly determines the
distortion $\delta$,
\begin{equation}
  \delta = \frac{\mathcal{M}_{11}- \mathcal{M}_{22}
  +2\mathrm{i}\mathcal{M}_{12}}{\mathcal{M}_{11}+\mathcal{M}_{22}}\;.
\label{eq:5.39}
\end{equation}
Hence, $\delta$ is related to $\mathcal{M}$ in the same way as the
complex ellipticity $\chi$ is related to $Q$. In some sense, the ACF
plays the role of a single `equivalent' image from which the
distortion can be determined, instead of an ensemble average over
individual galaxy ellipticities.

Working with the ACF has several advantages. First, centres of galaxy
images do not need to be determined, which avoids a potential source
of error. Second, the ACF can be used with substantial flexibility.
For instance, one can use all galaxy images which are detected with
high significance, determine their ellipticity, and obtain an estimate
of $\delta$ from them. Sufficiently large circles containing these
galaxies can be cut out of the data frame, so that the remaining frame
is reminiscent of a Swiss cheese. The ACF on this frame provides
another estimate of $\delta$, which is independent information and can
statistically be combined with the estimate from galaxy
ellipticities. Or one can use the ACF only on galaxy images detected
within a certain magnitude range, still avoiding the need to determine
centres.

Third, on sufficiently deep images with the brighter objects cut out
as just described, one might assume that the intrinsic ACF is due to a
very large number of faint galaxies, so that the intrinsic ACF becomes
a universal function. This function can in principle be determined
from deep HST images. In that case, one also knows the width of the
intrinsic ACF, as measured by the trace or determinant of
$\mathcal{M}$, and can determine the magnification from the width of
the observed ACF, very similar to the method discussed in
Sect.~\ref{sc:4.4.2}, but with the advantage of dealing with a single
`universal source'.

If this universal intrinsic ACF does exist, corrections of the
measured $\mathcal{M}$ for a PSF considerably simplify compared to the
case of individual image ellipticities, as shown by
\cite{VA97.1}. They performed several tests on synthetic data to
demonstrate the potential of the ACF method for the recovery of the
shear applied to the simulated images. \citename{VA97.1} determined
shear fields of two clusters, with several magnitude thresholds for
the images which were punched out. A comparison of these shear fields
with those obtained from the standard method using galaxy
ellipticities clearly shows that the ACF method is at least
competitive, but since it provides additional information from those
parts of the CCD which are unused by the standard method, it should in
be employed any case. The optimal combination of standard method and
ACF still needs to be investigated, but detailed numerical experiments
indicate that the ACF may be the best method for measuring very weak
shear amplitudes (L.~van Waerbeke \& Y.~Mellier, private
communication).

\subsubsection{\label{sc:5.5.3}The Redshift Distribution of Very
  Faint Galaxies}

Galaxy redshifts are usually determined spectroscopically. A
successful redshift measurement depends on the magnitude of the
galaxy, the exposure time, and the spectral type of the galaxy. If it
shows strong emission or absorption lines, as star-forming galaxies
do, a redshift can much easier be determined than in absence of strong
spectral features. The recently completed Canadian-French Redshift
Survey (CFRS) selected 730 galaxies in the magnitude interval $17.5\le
I\le22.5$ (see \cite*{llc95} and references therein). For 591 of them
(81\%), redshifts were secured with multi-slit spectroscopy on a 3.6m
telescope (CFHT) with a typical exposure time of
$\sim8$~hours. Whereas the upcoming 10m-class telescopes will be able
to perform redshift surveys to somewhat fainter magnitude limits, it
will be difficult to secure fairly complete redshift information of a
flux-limited galaxy sample fainter than $I\sim24$. In addition, it can
be expected that many galaxies in a flux-limited sample with fainter
threshold will have redshifts between $\sim1.2$ and $\sim2.2$, where
the cleanest spectral features, the OII emission line at
$\lambda=372.7$~nm and the $\lambda=400$~nm break are shifted beyond
the region where spectroscopy can easily be done from the ground.

As we have seen, the calibration of cluster mass distributions depends
on the assumed redshift distribution of the background galaxies. Most
of the galaxies used for the reconstruction are considerably fainter
than those magnitude limits for which complete redshift samples are
available, so that this mass calibration requires an extrapolation of
the redshift distribution from brighter galaxy samples. The fact that
lensing is sensitive to the redshift distribution is not only a source
of uncertainty, but also offers the opportunity to investigate the
redshift distribution of galaxies too faint to be investigated
spectroscopically. Several approaches towards a redshift estimate of
faint galaxies by lensing have been suggested, and some of them have
already shown spectacular success, as will be discussed next.

First of all, a strongly lensed galaxy (e.g.~a giant luminous arc) is
highly magnified, and so the gravitational lens effect allows to
obtain spectra of objects which would be too faint for a spectroscopic
investigation without lensing. It was possible in this way to measure
the redshifts of several arcs, e.g., the giant arc in A~370 at
$z=0.724$ (\cite*{SO88.1}), the arclet A~5 in A~370 at $z=1.305$
(\cite*{ME91.1}), the giant arc in Cl~2244$-$02 at $z=2.237$
(\cite*{ME91.1}), and the `straight arc' in A~2390 at $z=0.913$. In
the latter case, even the rotation curve of the source galaxy was
determined (\cite*{PE91.1}). For a more complete list of arc
redshifts, see \cite{FO94.1}. If the cluster contains several
strong-lensing features, the mass model can be sufficiently well
constrained to determine the arc magnifications (if they are resolved
in width, which has become possible only from imaging with the
refurbished HST), and thus to determine the unlensed magnitude of the
source galaxies, some of which are fainter that $B\sim25$.

Some clusters, such as A~370 and A~2218, were observed in great detail
both from the ground and with HST, and show a large number of strongly
lensed images. They can be used to construct very detailed mass models
of the cluster centre (e.g., \cite*{KN93.1}; \cite*{KN96.1}). An
example is A~2218, in which at least five multiply imaged systems were
detected (\cite*{KN96.1}), and several giant arcs were clearly
seen. Refining the mass model for A~2218 constructed from ground-based
data (\cite*{KN95.2}) with the newly discovered or confirmed strong
lensing features on the WFPC2 image, a strongly constrained mass model
for the cluster can be computed and calibrated by two arc redshifts (a
five-image system at $z=0.702$, and at $z=1.034$).

Visual inspection of the WFPC2 image immediately shows a large number
of arclets in A~2218, which surround the cluster centre in a nearly
perfect circular pattern. These arclets have very small axis ratios,
and most of them are therefore highly distorted. The strength of the
distortion depends on the redshift of the corresponding
galaxy. Assuming that the sources have a considerably smaller
ellipticity than the observed images, one can then estimate a redshift
range of the galaxy.

To be more specific, let $p^\s(\epsilon^\s)$ be the probability
density of the intrinsic source ellipticity, assumed for simplicity to
be independent of redshift. The corresponding probability distribution
for the image ellipticity is then
\begin{equation}
  p(\epsilon) = p^\s\left(\epsilon^\s(\epsilon)\right)\,
  \det\left(\frac{\partial \epsilon^\s}{\partial \epsilon}\right)\;,
\label{eq:5.40}
\end{equation}
where the transformation $\epsilon^\s(\epsilon)$ is given by
eq.~(\ref{eq:4.12}, page~\pageref{eq:4.12}), and the final term is the
Jacobian of this transformation. For each arclet near the cluster
centre where the mass profile is well constrained, the value of the
reduced shear $g$ is determined up to the unknown redshift of the
source -- see eq.~(\ref{eq:5.37}).

One can now try to maximise $p(\epsilon)$ with respect to the source
redshift, and in that way find the most likely redshift for the
arc.\footnote{\label{fn:5.6} This simplified treatment neglects the
magnification bias, i.e.~the fact that at locations of high
magnification the redshift probability distribution is changed -- see
Sect.~\ref{sc:4.3.2}.} Depending on the ellipticity of the arclet and
the local values of shear and surface mass density, three cases have
to be distinguished: (1) the arclet has the `wrong' orientation
relative to the local shear, i.e., if the source lies behind the
cluster, it must be even more elliptical than the observed arclet. For
the arclets in A~2218, this case is very rare. (2) The most probable
redshift is `at infinity', i.e., even if the source is placed at very
high redshift, the maximum of $p(\epsilon)$ is not reached. (3)
$p(\epsilon)$ attains a maximum at a finite redshift. This is by far
the most common case in A~2218.

This method, first applied to A~370 (\cite*{KN94.1}), was used to
estimate the redshifts of $\sim80$ arclets in A~2218 brighter than
$R\sim 25$. Their typical redshifts are estimated to be of order
unity, with the fainter sub-sample $24\le R\le25$ extending to
somewhat higher redshifts. For one of them, a redshift range
$2.6\lesssim z\lesssim3.3$ was estimated, and a spectroscopic redshift
of $z=2.515$ was later measured (\cite*{EB96.1}), providing
spectacular support for this method. Additional spectroscopic
observations of arclets in A~2218 were conducted and further confirmed
the reliability of the method for the redshift estimates of individual
arclets (\cite*{EB98.1}).

Another success of this arclet redshift estimate was recently achieved
in the cluster A~2390, which can also be modelled in great detail from
HST data. There, two arclets with very strong elongation did not fit
into the cluster mass model unless they are at very high redshift.
Spectroscopic redshifts of $z\sim 4.05$ were recently measured for
these two arclets (\cite*{FR98.1}, \cite*{pkb99}).

However, several issues should be kept in mind. First, the arclets for
which a reliable estimate of the redshift can be obtained are clearly
magnified, and thus the sample is magnification biased. Since it is
well known that the galaxy number counts are considerable steeper in
the blue than in the red (see, e.g., \cite*{SM95.1}), blue galaxies
are preferentially selected as arclets -- see
eq.~(\ref{eq:4.42}). This might also provide the explanation why most
of the giant arcs are blue (\cite*{bro95}). Therefore, the arclets
represent probably a biased sample of faint galaxies. Second, the
redshift dependence of $p(\epsilon)$ enters through the ratio
$D_\mathrm{ds}/D_\mathrm{s}$. For a cluster at relatively low
redshift, such as A~2218 ($z_\mathrm{d}=0.175$), this ratio does not
vary strongly with redshift once the source redshift is larger than
$\sim1$. Hence, to gain more accurate redshift estimates for
high-redshift galaxies, a moderately-high redshift cluster should be
used.

The method just described is not a real `weak lensing' application,
but lies on the borderline between strong and weak lensing. With weak
lensing, the redshifts of {\em individual\/} galaxy images cannot be
determined, but some statistical redshift estimates can be
obtained. Suppose the mass profile of a cluster has been reconstructed
using the methods described in Sect.~\ref{sc:5.2} or
Sect.~\ref{sc:5.5.1}, for which galaxy images in a certain magnitude
range were used. If the cluster contains strong-lensing features with
spectroscopic information (such as a giant luminous arc with measured
redshift), then the overall mass calibration can be determined, i.e.,
the factor $\langle w\rangle$ -- see Sect.~\ref{sc:4.3.2} -- can be
estimated, which provides a first integral constraint on the redshift
distribution.

Repeating this analysis with several such clusters at different
redshifts, further estimates of $\langle w\rangle$ with different
$D_\mathrm{d}$ are obtained, and thus additional constraints on the
redshift distribution. In addition, one can group the faint galaxy
images into sub-samples, e.g., according to their apparent
magnitude. Ignoring for simplicity the magnification bias (which can
safely be done in the outer parts of clusters), one can determine
$\langle w\rangle$ for each magnitude bin. Restricting our treatment
to the regions of weak lensing only, such that $|\gamma|\ll1$,
$\kappa\ll1$, the expectation value of the ellipticity $\epsilon_i$ of
a galaxy at position $\vec\theta_i$ is $\langle w\rangle
\gamma(\vec\theta_i)$, and so an estimate of $\langle w\rangle$ for
the galaxy sub-sample under consideration is
\begin{equation}
  \langle w\rangle =
  \frac{\sum \Re\left(\gamma(\vec\theta_i)\epsilon_i^*\right)}
  {\sum|\gamma(\vec\theta_i)|^2}\;.
\label{eq:5.41}
\end{equation}
In complete analogy, \cite{BA95.5} suggested the `lens parallax
method', an algorithm for determining mean redshifts for galaxy
sub-samples at fixed surface brightness, using the magnification
effect as described in Sect.~\ref{sc:4.4.2}. Since the surface
brightness $I$ is most likely much more strongly correlated with
galaxy redshift than the apparent magnitude (due to the $(1+z)^{-4}$
decrease of bolometric surface brightness with redshift), a narrow bin
in $I$ will probably correspond to a fairly narrow distribution in
redshift, allowing to relate $\langle w\rangle$ of a surface
brightness bin fairly directly to a mean redshift in that bin, while
$\langle w\rangle$ in magnitude bins can only be translated into
redshift information with a parameterised model of the redshift
distribution. On the other hand, apparent magnitudes are easier to
measure than surface brightness and are much less affected by seeing.

Even if a cluster without strong lensing features is considered, the
two methods just described can be applied. The mass reconstruction
then gives the mass distribution up to an overall multiplicative
constant. We assume here that the mass-sheet degeneracy can be lifted,
either using the magnification effect as described in
Sect.~\ref{sc:5.4}, or by extending the observations so sufficiently
large distances so that $\kappa\approx 0$ near the boundary of the
data field. The mass scale can then be fixed by considering the
brightest sub-sample of galaxy images for which a shear signal is
detected if they are sufficiently bright for their redshift
probability distribution to be known from spectroscopic redshift
surveys (\cite*{BA95.5}).

Whereas these methods have not yet rigourously been applied, there is
one observational result which indicates that the faint galaxy
population has a relatively high median redshift. In a sequence of
clusters with increasing redshift, more and more of the faint galaxies
will lie in the foreground or very close behind the cluster and
therefore be unlensed. The dependence of the observed lensing strength
of clusters on their redshift can thus be used as a rough indication
of the median redshift of the faint galaxies. This idea was put
forward by \cite{SM94.1}, who observed three clusters with redshifts
$z=0.26$, $z=0.55$ and $z=0.89$. In the two lower-redshift clusters, a
significant weak lensing signal was detected, but no significant
signal in the high-redshift cluster. From the detection, models for
the redshift distribution of faint $I\le25$ can be ruled out which
predict a large fraction to be dwarf galaxies at low redshift. The
non-detection in the high-redshift cluster cannot easily be
interpreted since little information (e.g., from X-ray maps) is
available for this cluster, and thus the absence of a lensing signal
may be due to the cluster being not massive enough.

However, the detection of a strong shear signal in the cluster
MS~1054$-$03 at $z=0.83$ (\cite*{LU97.1}) implies that a large
fraction of galaxies with $I\le25.5$ must lie at redshifts larger than
$z\sim1.5$. They split their galaxy sample into red and blue
sub-samples, as well as into brighter and fainter sub-samples, and
found that the shear signal is mainly due to the fainter and the blue
galaxies. If all the faint blue galaxies have a redshift
$z_\mathrm{s}=1.5$, the mass-to-light ratio of this cluster is
estimated to be $M/L\sim580\,h$, and if they all lie at redshift
$z_\mathrm{s}=1$, $M/L$ exceeds $\sim1000\,h$. This observational
result, which is complemented by several additional shear detections
in high-redshift clusters, one of them at $z=0.82$ (G.~Luppino,
private communication), provides the strongest evidence for the
high-redshift population of faint galaxies. In addition, it strongly
constrains cosmological models; an $\Omega_0=1$ cosmological model
predicts the formation of massive clusters only at relatively low
redshifts (e.g., \cite*{rlt92}; \cite*{bes93}) and has difficulties to
explain the presence of strong lensing clusters at redshift
$z\sim0.8$.

Recently, \cite{lob99} and \cite{gfm98} suggested that weak lensing by
galaxy clusters can be used to constrain the cosmological parameters
$\Omega_0$ and $\Omega_\Lambda$. Both of these two different methods
assume that the redshift of background galaxies can be estimated,
e.g.~with sufficiently precise photometric-redshift techniques. Owing
to the dependence of the lensing strength on the angular-diameter
distance ratio $D_\mathrm{ds}/D_\mathrm{s}$, sufficiently detailed
knowledge of the mass distribution in the lens and of the source
redshifts can be employed to constrain these cosmological
parameters. Such a determination through purely geometrical methods
would be very valuable, although the observational requirements for
applying these methods appear fairly demanding at present.

  % -*- LaTeX -*-  
  
\section{\label{sc:6}Weak Cosmological Lensing}

In this section, we review how weak density perturbations in otherwise
homogeneous and isotropic Friedmann-Lema{\^\i}tre model universes
affect the propagation of light. We first describe how light
propagates in the homogeneous and isotropic background models, and
then discuss how local density inhomogeneities can be taken into
account. The result is a propagation equation for the transverse
separation between the light rays of a thin light bundle.

The solution of this equation leads to the deflection angle
$\vec\alpha$ of weakly deflected light rays. In close analogy to the
thin-lens situation, half the divergence of the deflection angle can
be identified with an effective surface-mass density
$\kappa_\mathrm{eff}$. The power spectrum of $\kappa_\mathrm{eff}$ is
closely related to the power spectrum of the matter fluctuations, and
it forms the central physical object of the further discussion. Any
two-point statistics of cosmic magnification and cosmic shear can then
be expressed in a fairly simple manner in terms of the
effective-convergence power spectrum.

We discuss several applications, among which are the uncertainty in
brightness determinations of cosmologically distant objects due to
cosmic magnification, and several measures for cosmic shear, one of
which is particularly suited for determining the effective-convergence
power spectrum. At the end of this chapter, we turn to higher-order
statistical measures of cosmic lensing effects, which reflect the
non-Gaussian nature of the non-linearly evolved density
perturbations.

When we give numerical examples, we generally employ four different
model universes. All have the CDM power spectrum for density
fluctuations, but different values for the cosmological
parameters. They are summarised in Tab.~\ref{tab:6.1}. We choose two
Einstein-de Sitter models, SCDM and $\sigma$CDM, normalised either to
the local abundance of rich clusters or to $\sigma_8=1$, respectively,
and two low-density models, OCDM and $\Lambda$CDM, which are cluster
normalised and either open or spatially flat, respectively.

\begin{table}[ht]
\caption{Cosmological models and their parameters used for numerical
  examples}
\label{tab:6.1}
\begin{center}
\medskip
\begin{tabular}{|l|rrrlr|}
\hline
Model & $\Omega_0$ & $\Omega_\Lambda$ & $h$ & Normalisation &
$\sigma_8$ \\
\hline
SCDM         & $1.0$ & $0.0$ & $0.5$ & cluster    & $0.5$  \\
$\sigma$CDM  & $1.0$ & $0.0$ & $0.5$ & $\sigma_8$ & $1.0$  \\
OCDM         & $0.3$ & $0.0$ & $0.7$ & cluster    & $0.85$ \\
$\Lambda$CDM & $0.3$ & $0.7$ & $0.7$ & cluster    & $0.9$  \\
\hline
\end{tabular}
\end{center}
\end{table}

Light propagation in inhomogeneous model universes has been the
subject of numerous studies. Among them are \cite{ZE64.1},
\cite{DA65.1}, \cite{KR66.1}, \cite{GU67.2}, \cite{JA90.1},
\cite{BA91.1}, \cite{BA91.2}, \cite{BL91.1}, \cite{MI91.3}, and
\cite{KA92.1}. Non-linear effects were included analytically by
\cite{JA97.2}, who also considered statistical effects of higher than
second order, as did \cite{BE97.5}. A particularly suitable measure
for cosmic shear was introduced by \cite{svj98}.

\subsection{\label{sc:6.1}Light Propagation; Choice of Coordinates}

As outlined in Sect.~\ref{sc:3.2.1} (page~\pageref{sc:3.2.1}), the
governing equation for the propagation of thin light bundles through
arbitrary space times is the equation of geodesic deviation
(e.g.~\cite*{MI73.1}, \S~11; \cite*{sef92}, \S~3.5), or Jacobi
equation (\ref{eq:3.23}, page~\pageref{eq:3.23}). This equation
implies that the transverse physical separation $\vec\xi$ between
neighbouring rays in a thin light bundle is described by the
second-order differential equation
\begin{equation}
  \frac{\d^2\vec\xi}{\d\lambda^2} = \mathcal{T}\,\vec\xi\;,
\label{eq:6.1}
\end{equation}
where $\mathcal{T}$ is the {\em optical tidal matrix\/}
(\ref{eq:3.25}, page~\pageref{eq:3.25}) which describes the influence
of space-time curvature on the propagation of light. The affine
parameter $\lambda$ has to be chosen such that it locally reproduces
the proper distance and increases with decreasing time, hence
$\d\lambda=-ca\,\d t$. The elements of the matrix $\mathcal{T}$ then
have the dimension [length]$^{-2}$.

We already discussed in Sect.~\ref{sc:3.2.1} that the optical tidal
matrix is proportional to the unit matrix in a Friedmann-Lema{\^\i}tre
universe,
\begin{equation}
  \mathcal{T} = \mathcal{R}\,\mathcal{I}\;,
\label{eq:6.2}
\end{equation}
where the factor $\mathcal{R}$ is determined by the Ricci tensor as in
eq.~(\ref{eq:3.26}, page~\pageref{eq:3.26}). For a model universe
filled with a perfect pressure-less fluid, $\mathcal{R}$ can be
written in the form (\ref{eq:3.28}, page~\pageref{eq:3.28}).

It will prove convenient for the following discussion to replace the
affine parameter $\lambda$ in eq.~(\ref{eq:6.1}) by the comoving
distance $w$, which was defined in eq.~(\ref{eq:2.3},
page~\pageref{eq:2.3}) before. This can be achieved using
eqs.~(\ref{eq:3.31}) and (\ref{eq:3.32}) together with the definition
of Hubble's parameter, $H(a)=\dot a\,a^{-1}$. Additionally, we
introduce the {\em comoving\/} separation vector $\vec
x=a^{-1}\vec\xi$. These substitutions leave the propagation equation
(\ref{eq:6.1}) in the exceptionally simple form
\begin{equation}
  \frac{\d^2\vec x}{\d w^2} + K\,\vec x = 0\;,
\label{eq:6.8}
\end{equation}
where $K$ is the spatial curvature given in eq.~(\ref{eq:2.29},
page~\pageref{eq:2.29}). Equation (\ref{eq:6.8}) has the form of an
oscillator equation, hence its solutions are trigonometric or
hyperbolic functions, depending on whether $K$ is positive or
negative. In the special case of spatial flatness, $K=0$, the comoving
separation between light rays is a linear function of distance.

\subsection{\label{sc:6.2}Light Deflection}

We now proceed by introducing density perturbations into the
propagation equation (\ref{eq:6.8}). We assume throughout that the
Newtonian potential $\Phi$ of these inhomogeneities is small,
$|\Phi|\ll c^2$, that they move with velocities much smaller than the
speed of light, and that they are localised, i.e.~that the typical
scales over which $\Phi$ changes appreciably are much smaller than the
curvature scale of the background Friedmann-Lema{\^\i}tre model. Then,
there exists a local neighbourhood around each density perturbation
which is large enough to contain the perturbation completely and still
small enough to be considered flat. Under these circumstances, the
metric is well approximated by the first post-Newtonian order of the
Minkowski metric (\ref{eq:3.36}, page~\pageref{eq:3.36}). It then
follows from eq.~(\ref{eq:3.36}) that the effective local index of
refraction in the neighbourhood of the perturbation is
\begin{equation}
  \frac{\d l}{\d t} = n = 1-\frac{2\Phi}{c^2}\;.
\label{eq:6.10}
\end{equation}
Fermat's principle (e.g.~\cite*{BL86.1}; \cite*{SC85.3}) demands that
the light travel time along actual light paths is stationary, hence
the variation of $\int n\d l$ must vanish. This condition implies that
light rays are deflected locally according to
\begin{equation}
  \frac{\d^2\vec x}{\d w^2} = -\frac{2}{c^2}\,\nabla_\perp\Phi\;.
\label{eq:6.11}
\end{equation}
In weakly perturbed Minkowski space, this equation describes how an
{\em actual\/} light ray is curved away from a straight line in
unperturbed Minkowski space. It is therefore appropriate for
describing light propagation through e.g.~the Solar system and other
well-localised mass inhomogeneities.

This interpretation needs to be generalised for large-scale mass
inhomogeneities embedded in an expanding cosmological background,
since the meaning of a ``straight'' fiducial ray is then no longer
obvious. In general, any physical fiducial ray will also be deflected
by potential gradients along its way. We can, however, interpret $\vec
x$ as the comoving separation vector between an arbitrarily chosen
fiducial light ray and a closely neighbouring light ray. The
right-hand side of eq.~(\ref{eq:6.11}) must then contain the {\em
difference\/} $\Delta(\nabla_\perp\Phi)$ of the perpendicular
potential gradients between the two rays to account for the {\em
relative\/} deflection of the two rays.

Let us therefore imagine a fiducial ray starting at the observer
($w=0$) into direction $\vec\theta=0$, and a neighbouring ray starting
at the same point but into direction $\vec\theta\ne\vec0$. Let further
$\vec x(\vec\theta,w)$ describe the comoving separation between these
two light rays at comoving distance $w$. Combining the cosmological
contribution given in eq.~(\ref{eq:6.8}) with the modified local
contribution (\ref{eq:6.11}) leads to the propagation equation
\begin{equation}
  \frac{\d^2\vec x}{\d w^2} + K\,\vec x = 
  -\frac{2}{c^2}\Delta\left\{
    \nabla_\perp\Phi[\vec x(\vec\theta,w),w]\right\}\;.
\label{eq:6.12}
\end{equation}
The notation on the right-hand side indicates that the difference of
the perpendicular potential gradients has to be evaluated between the
two light rays which have comoving separation $\vec x(\vec\theta,w)$
at comoving distance $w$ from the observer.

Linearising the right-hand side of eq.~(\ref{eq:6.12}) in $\vec x$
immediately returns the geodesic deviation equation (\ref{eq:6.1})
with the full optical tidal matrix, which combines the homogeneous
cosmological contribution (\ref{eq:3.28}, page~\pageref{eq:3.28}) with
the contributions of local perturbations (\ref{eq:3.37},
page~\pageref{eq:3.37}).

Strictly speaking, the comoving distance $w$, or the affine parameter
$\lambda$, are changed in the presence of density perturbations. Here,
we assume that the global properties of the weakly perturbed
Friedmann-Lema{\^\i}tre models remain the same as in the homogeneous
and isotropic case, and under this assumption the comoving distance
$w$ remains the same as in the unperturbed model.

To solve eq.~(\ref{eq:6.12}), we first construct a Green's function
$G(w,w')$, which has to be a suitable linear combination of either
trigonometric or hyperbolic functions since the homogeneous equation
(\ref{eq:6.12}) is an oscillator equation. We further have to specify
two boundary conditions. According to the situation we have in mind,
these boundary conditions read
\begin{equation}
  \vec x = 0\;,\quad\frac{\d\vec x}{\d w} = \vec\theta
\label{eq:6.13}
\end{equation}
at $w=0$. The first condition states that the two light rays start
from the same point, so that their initial separation is zero, and the
second condition indicates that they set out into directions which
differ by $\vec\theta$.

The Green's function is then uniquely determined by
\begin{equation}
  G(w,w') = \left\{
  \begin{array}{ll}
    f_K(w-w') & \quad\hbox{for}\quad w>w' \\
    0         & \quad\hbox{otherwise} \\
  \end{array}\right.\;,
\label{eq:6.14}
\end{equation}
with $f_K(w)$ given in eq.~(\ref{eq:2.4}, page~\pageref{eq:2.4}). As a
function of distance $w$, the comoving separation between the two
light rays is thus
\begin{equation}
  \vec x(\vec\theta,w) = f_K(w)\vec\theta - 
  \frac{2}{c^2}\,\int_0^w\,dw'\,f_K(w-w')\,
  \Delta\left\{
    \nabla_\perp\Phi[\vec x(\vec\theta,w'),w']\right\}\;.
\label{eq:6.15}
\end{equation}
The perpendicular gradients of the Newtonian potential are to be
evaluated along the true paths of the two light rays. In its exact
form, eq.~(\ref{eq:6.15}) is therefore quite involved.

Assuming that the change of the comoving separation vector $\vec x$
between the two {\em actual\/} rays due to light deflection is small
compared to the comoving separation of {\em unperturbed\/} rays,
\begin{equation}
  \frac{|\vec x(\vec\theta,w')-f_K(w')\vec\theta|}
  {|f_K(w')\vec\theta|}\ll1\;,
\label{eq:6.15a}
\end{equation}
we can replace $\vec x(\vec\theta,w')$ by $f_K(w')\vec\theta$ in the
integrand to arrive at a much simpler expression which corresponds to
the Born approximation of small-angle scattering. The Born
approximation allows us to replace the difference of the perpendicular
potential gradients with the perpendicular gradient of the potential
difference. Taking the potential difference then amounts to adding a
term to the potential which depends on the comoving distance $w'$ from
the observer only. For notational simplicity, we can therefore rename
the potential difference $\Delta\Phi$ between the two rays to $\Phi$.

It is an important consequence of the Born approximation that the
Jacobian matrix of the lens mapping (\ref{eq:3.11},
page~\pageref{eq:3.11}; \ref{eq:6.31} below) remains symmetric even in
the case of cosmological weak lensing. In a general multiple
lens-plane situation, this is not the case (\cite*{sef92}, chapter~9).

If the two light rays propagated through unperturbed space-time, their
comoving separation at distance $w$ would simply be $\vec
x'(\vec\theta,w)=f_K(w)\vec\theta$, which is the first term on the
right-hand side of eq.~(\ref{eq:6.15}). The net deflection angle at
distance $w$ between the two rays is the difference between $\vec x'$
and $\vec x$, divided by the angular diameter distance to $w$, hence
\begin{equation}
  \vec\alpha(\vec\theta,w) =
  \frac{f_K(w)\vec\theta-\vec x(\vec\theta,w)}{f_K(w)}
  \frac{2}{c^2}\,\int_0^w\,\d w'\,\frac{f_K(w-w')}{f_K(w)}
  \nabla_\perp\Phi[f_K(w')\vec\theta,w']\;.
\label{eq:6.16}
\end{equation}
Again, this is the deflection angle of a light ray that starts out at
the observer into direction $\vec\theta$ relative to a nearby fiducial
ray. Absolute deflection angles cannot be measured. All measurable
effects of light deflection therefore only depend on {\em
derivatives\/} of the deflection angle (\ref{eq:6.16}), so that the
choice of the fiducial ray is irrelevant for practical purposes. For
simplicity, we call $\vec\alpha(\vec\theta,w)$ the deflection angle at
distance $w$ of a light ray starting into direction $\vec\theta$ on
the observer's sky, bearing in mind that it is the deflection angle
relative to an arbitrarily chosen fiducial ray, so that
$\vec\alpha(\vec\theta,w)$ is far from unique.

In an Einstein-de Sitter universe, $f_K(w)=w$. Defining $y=w'/w$,
eq.~(\ref{eq:6.16}) simplifies to
\begin{equation}
  \vec\alpha(\vec\theta,w) = \frac{2w}{c^2}\,\int_0^1\,\d y\,(1-y)\,
  \nabla_\perp\Phi(wy\vec\theta,wy)\;.
\label{eq:6.17}
\end{equation}
Clearly, the deflection angle $\vec\alpha$ depends on the direction
$\vec\theta$ on the sky into which the light rays start to propagate,
and on the comoving distance $w$ to the sources.

Recall the various approximations adopted in the derivation of
eq.~(\ref{eq:6.16}): (i) The density perturbations are well localised
in an otherwise homogeneous and isotropic background, i.e.~each
perturbation can be surrounded by a spatially flat neighbourhood which
can be chosen small compared to the curvature radius of the background
model, and yet large enough to encompass the entire perturbation. In
other words, the largest scale on which the density fluctuation
spectrum $P_\delta(k)$ has appreciable power must be much smaller than
the Hubble radius $c/H_0$. (ii) The Newtonian potential of the
perturbations is small, $\Phi\ll c^2$, and typical velocities are much
smaller than the speed of light. (iii) Relative deflection angles
between neighbouring light rays are small enough so that the
difference of the transverse potential gradient can be evaluated at
the unperturbed path separation $f_K(w)\vec\theta$ rather than the
actual one. Reassuringly, these approximations are very comfortably
satisfied even under fairly extreme conditions. The curvature radius
of the Universe is of order $cH_0^{-1}=3000\,h^{-1}\,\mathrm{Mpc}$ and
therefore much larger than perturbations of even several tens of Mpc's
in size. Typical velocities in galaxy clusters are of order
$10^3\,\mathrm{km\,s}^{-1}$, much smaller than the speed of light, and
typical Newtonian potentials are of order $\Phi\lesssim10^{-5}\,c^2$.

\subsection{\label{sc:6.3}Effective Convergence}

\subsubsection{\label{sc:6.3.1}Definition and Derivation}

In the thin-lens approximation, convergence $\kappa$ and deflection
angle $\vec\alpha$ are related by
\begin{equation}
  \kappa(\vec\theta) =
  \frac{1}{2}\,\nabla_\theta\cdot\vec\alpha(\vec\theta) =
  \frac{1}{2}\,\frac{\partial\alpha_i(\vec\theta)}{\partial\theta_i}\;,
\label{eq:6.18}
\end{equation}
where summation over $i$ is implied. In exact analogy, an effective
convergence $\kappa_\mathrm{eff}(w)$ can be defined for cosmological
weak lensing,
\begin{eqnarray}
  \kappa_\mathrm{eff}(\vec\theta,w) &=&
  \frac{1}{2}\,\nabla_\theta\cdot\vec\alpha(\vec\theta,w)
  \nonumber\\ &=&
  \frac{1}{c^2}\,\int_0^w\,\d w'\,\frac{f_K(w-w')f_K(w')}{f_K(w)}\,
  \frac{\partial^2}{\partial x_i\partial x_i}\,
  \Phi[f_K(w')\vec\theta,w']\;.
\label{eq:6.19}
\end{eqnarray}

Had we not replaced $\vec x(\vec\theta,w')$ by $f_K(w')\vec\theta$
following eq.~(\ref{eq:6.15}), eq.~(\ref{eq:6.19}) would have
contained second and higher-order terms in the potential
derivatives. Since eq.~(\ref{eq:6.15}) is a Volterra integral equation
of the second kind, its solution (and derivatives thereof) can be
expanded in a series, of which the foregoing expression for
$\kappa_\mathrm{eff}$ is the first term. Equation~(\ref{eq:6.21})
below shows that this term is of the order of the line-of-sight
average of the density contrast $\delta$. The next higher-order term,
explicitly written down in the Appendix of \cite{svj98}, is determined
by the product $\delta(w')\,\delta(w'')$, averaged along the
line-of-sight over $w'<w''$. Analogous estimates apply to higher-order
terms. Whereas the density contrast may be large for individual
density perturbations passed by a light ray, the average of $\delta$
is small compared to unity for most rays, hence
$\kappa_\mathrm{eff}\ll1$, and higher-order terms are accordingly
negligible.

The effective convergence $\kappa_\mathrm{eff}$ in eq.~(\ref{eq:6.19})
involves the two-dimensional Laplacian of the potential. We can
augment it by $(\partial^2\Phi/\partial x_3^2)$ which involves only
derivatives along the light path, because these average to zero in the
limit to which we are working. The three-dimensional Laplacian of the
potential can then be replaced by the density contrast via Poisson's
equation (\ref{eq:2.64}, page~\pageref{eq:2.64}),
\begin{equation}
  \Delta\Phi = \frac{3H_0^2\Omega_0}{2a}\,\delta\;.
\label{eq:6.20}
\end{equation}
Hence, we find for the effective convergence,
\begin{equation}
  \kappa_\mathrm{eff}(\vec\theta,w) = \frac{3H_0^2\Omega_0}{2c^2}\,
  \int_0^w\,\d w'\,\frac{f_K(w')f_K(w-w')}{f_K(w)}\,
  \frac{\delta[f_K(w')\vec\theta,w']}{a(w')}\;.
\label{eq:6.21}
\end{equation}
The effective convergence along a light ray is therefore an integral
over the density contrast along the (unperturbed) light path, weighted
by a combination of comoving angular-diameter distance factors, and
the scale factor $a$. The amplitude of $\kappa_\mathrm{eff}$ is
proportional to the cosmic density parameter $\Omega_0$.

Expression~(\ref{eq:6.21}) gives the effective convergence for a fixed
source redshift corresponding to the comoving source distance
$w$. When the sources are distributed in comoving distance,
$\kappa_\mathrm{eff}(\vec\theta,w)$ needs to be averaged over the
(normalised) source-distance distribution $G(w)$,
\begin{equation}
  \bar\kappa_\mathrm{eff}(\vec\theta) = 
  \int_0^{w_\mathrm{H}}\,
  \d w\,G(w)\,\kappa_\mathrm{eff}(\vec\theta,w)\;,
\label{eq:6.22}
\end{equation}
where $G(w)\d w=p_z(z)\d z$. Suitably re-arranging the integration
limits, we can then write the source-distance weighted effective
convergence as
\begin{equation}
  \bar\kappa_\mathrm{eff}(\vec\theta) =
  \frac{3H_0^2\Omega_0}{2c^2}\,
  \int_0^{w_\mathrm{H}}\,\d w\,\bar{W}(w)\,f_K(w)\,
  \frac{\delta[f_K(w)\vec\theta,w]}{a(w)}\;,
\label{eq:6.23}
\end{equation}
where the weighting function $\bar{W}(w)$ is now
\begin{equation}
  \bar{W}(w) \equiv \int_w^{w_\mathrm{H}}\d w'\,G(w')\,
  \frac{f_K(w'-w)}{f_K(w')}\;.
\label{eq:6.24}
\end{equation}
The upper integration boundary $w_\mathrm{H}$ is the horizon distance,
defined as the comoving distance obtained for infinite redshift. In
fact, it is easily shown that the effective convergence can be written
as
\begin{equation}
  \kappa_\mathrm{eff}=\int\d z\,\frac{4\pi G}{c^2}\,
  \frac{D_\mathrm{d}D_\mathrm{ds}}{D_\mathrm{s}}\,
  \frac{\d D_\mathrm{prop}}{\d z}\,(\rho-\bar\rho)\;,
\label{eq:6.24a}
\end{equation}
and the weighting function $\bar{W}$ is the distance ratio $\langle
D_\mathrm{ds}/D_\mathrm{s}\rangle$, averaged over the source distances
at fixed lens distance. Naively generalising the definition of the
dimension-less surface-mass density (\ref{eq:3.7},
page~\pageref{eq:3.7}) to a three-dimensional matter distribution
would therefore directly have led to the cosmologically correct
expression for the effective convergence.

\subsection{\label{sc:6.4}Effective-Convergence Power Spectrum}

\subsubsection{\label{sc:6.4.1}The Power Spectrum from Limber's
  Equation}

Here, we are interested in the statistical properties of the effective
convergence $\kappa_\mathrm{eff}$, especially its power spectrum
$P_\kappa(l)$. We refer the reader to Sect.~\ref{sc:2.4}
(page~\pageref{sc:2.4}) for the definition of the power spectrum. We
also note that the expression for
$\bar\kappa_\mathrm{eff}(\vec\theta)$ is of the form (\ref{eq:2.76},
page~\pageref{eq:2.76}), and so the power spectrum $P_\kappa(l)$ is
given in terms of $P_\delta(k)$ by eq.~(\ref{eq:2.83},
page~\pageref{eq:2.83}), if one sets
\begin{equation}
  q_1(w)=q_2(w)=\frac{3}{2}\,\frac{H_0^2}{c^2}\,\Omega_0\,
  \bar W(w)\,\frac{f_K(w)}{a(w)}\;.
\label{eq:6.24b}
\end{equation}
We therefore obtain
\begin{equation}
  P_\kappa(l) = \frac{9H_0^4\Omega_0^2}{4c^4}\,
  \int_0^{w_\mathrm{H}}\,\d w\,\frac{\bar{W}^2(w)}{a^2(w)}\,
  P_\delta\left(\frac{l}{f_K(w)},w\right)\;,
\label{eq:6.25}
\end{equation}
with the weighting function $\bar{W}$ given in
eq.~(\ref{eq:6.24}). This power spectrum is the central quantity for
the discussion in the remainder of this chapter.

\begin{figure}[ht]
  \includegraphics[width=\hsize]{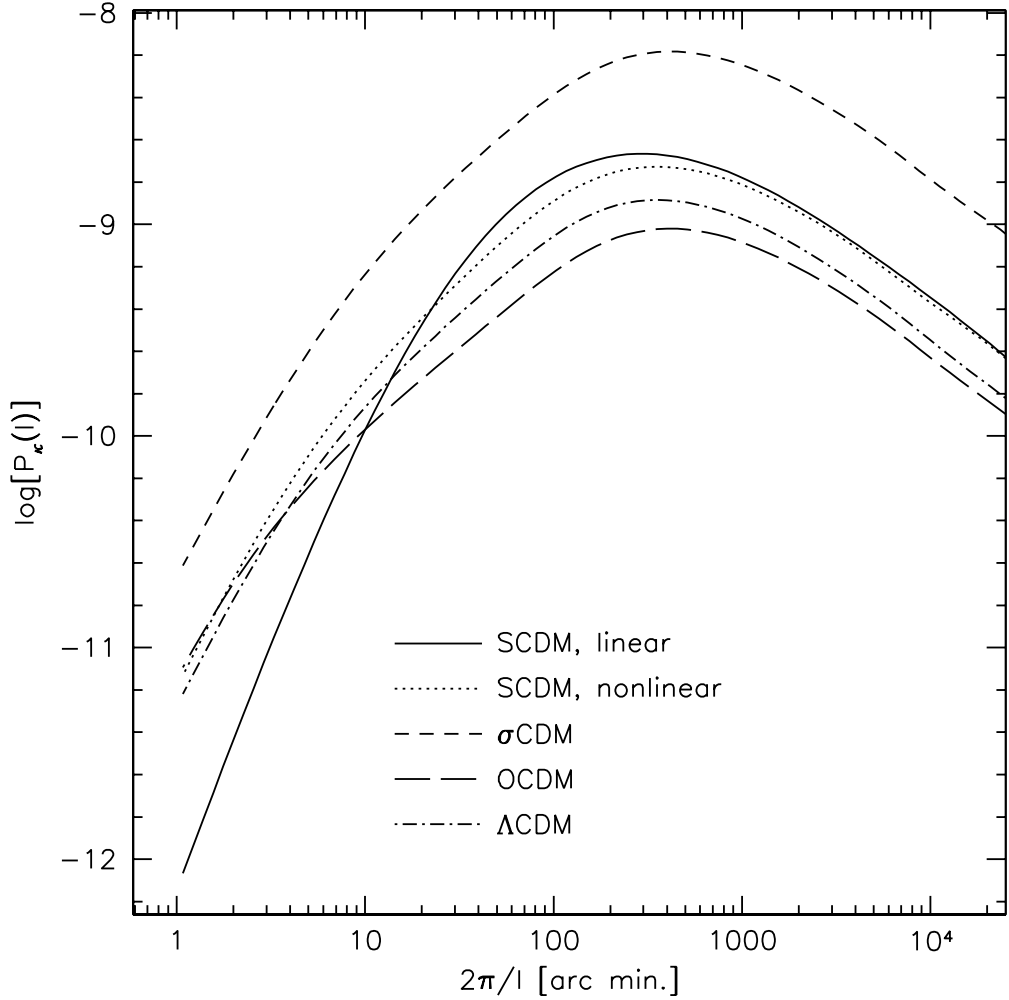}
\caption{Five effective-convergence power spectra $P_\kappa(l)$ are
shown as functions of the angular scale $2\pi l^{-1}$, expressed in
arc minutes. All sources were assumed to lie at $z_\mathrm{s}=1$. The
five curves represent the four realisations of the CDM cosmogony
listed in Tab.~\ref{tab:6.1}, all with non-linearly evolving
density-perturbation power spectra $P_\delta$, plus the SCDM model
with linearly evolving $P_\delta$. Solid curve (1): Linearly evolving
SCDM model; dotted curve (2): non-linearly evolving SCDM; short-dashed
curve (3): non-linearly evolving $\sigma$CDM; dashed-dotted and
long-dashed curves (4 and 5): non-linearly evolving OCDM and
$\Lambda$CDM, respectively.}
\label{fig:6.1}
\end{figure}

Figure~\ref{fig:6.1} shows $P_\kappa(l)$ for five different
realisations of the CDM cosmogony. These are the four models whose
parameters are detailed in Tab.~\ref{tab:6.1}, all with non-linearly
evolving density power spectrum $P_\delta$, using the prescription of
\cite{PE96.3}, plus the SCDM model with linearly evolving
$P_\delta$. Sources are assumed to be at redshift
$z_\mathrm{s}=1$. Curves 1 and 2 (solid and dotted; SCDM with linear
and non-linear evolution, respectively) illustrate the impact of
non-linear density evolution in an Einstein-de Sitter universe with
cluster-normalised density fluctuations. Non-linear effects set in on
angular scales below a few times $10'$, and increase the amplitude of
$P_\kappa(l)$ by more than an order of magnitude on scales of
$\approx1'$. Curve 3 (short-dashed; $\sigma$CDM), obtained for CDM
normalised to $\sigma_8=1$ rather than the cluster abundance,
demonstrates the potential influence of different choices for the
power-spectrum normalisation. Curves 4 and 5 (dashed-dotted and
long-dashed; OCDM and $\Lambda$CDM, respectively) show $P_\kappa(l)$
for cluster-normalised CDM in an open universe ($\Omega_0=0.3$,
$\Omega_\Lambda=0$) and in a spatially flat, low-density universe
($\Omega_0=0.3$, $\Omega_\Lambda=0.7$). It is a consequence of the
normalisation to the local cluster abundance that the various
$P_\kappa(l)$ are very similar for the different cosmologies on
angular scales of a few arc minutes. For the low-density universes,
the difference between the cluster- and the $\sigma_8$ normalisation
is substantially smaller than for the Einstein-de Sitter model.

\begin{figure}[ht]
  \includegraphics[width=\hsize]{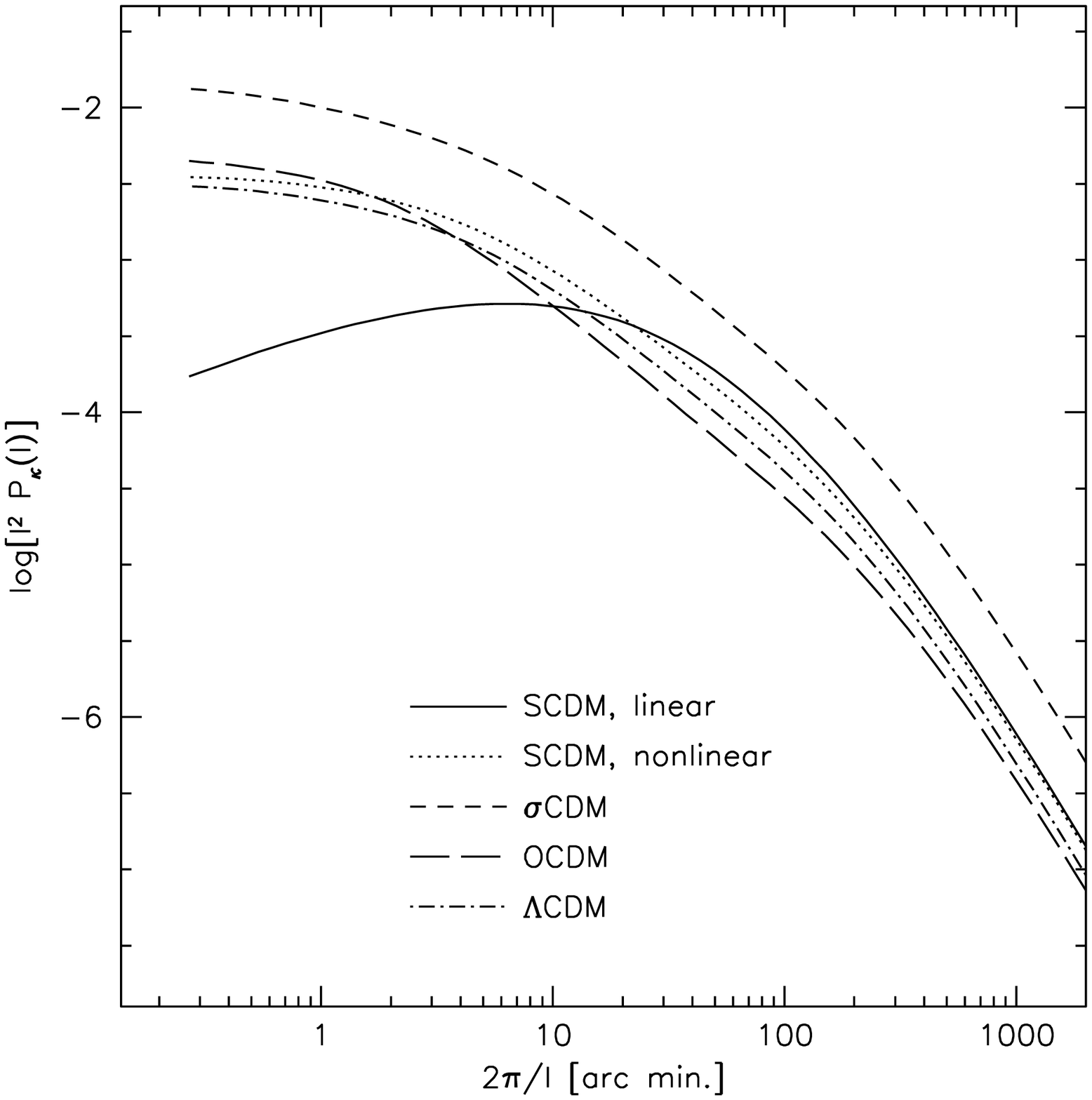}
\caption{Different representation of the curves in
Fig.~\ref{fig:6.1}. We plot here $l^2\,P_\kappa(l)$, representing the
total power in the effective convergence per logarithmic $l$
interval. See the caption of Fig.~\ref{fig:6.1} for the meaning of the
different line types. The figure demonstrates that the total power
increases monotonically towards small angular scales when non-linear
evolution is taken into account (i.e.~with the exception of the solid
curve). On angular scales still smaller than $\approx1'$, the curves
level off and decrease very slowly. This shows that weak lensing by
cosmological mass distributions is mostly sensitive to structures
smaller than $\approx10'$.}
\label{fig:6.2}
\end{figure}

Figure~\ref{fig:6.2} gives another representation of the curves in
Fig.~\ref{fig:6.1}. There, we plot $l^2\,P_\kappa(l)$, i.e.~the total
power in the effective convergence per logarithmic $l$ interval. This
representation demonstrates that density fluctuations on angular
scales smaller than $\approx10'$ contribute most strongly to weak
gravitational lensing by large-scale structures. On angular scales
smaller than $\approx1'$, the curves level off and then decrease very
gradually. The solid curve in Fig~\ref{fig:6.2} shows that, when
linear density evolution is assumed, most power is contributed by
structures on scales above $10'$, emphasising that it is crucial to
take non-linear evolution into account to avoid misleading
conclusions.

\subsubsection{\label{sc:6.4.3}Special Cases}

In the approximation of linear density evolution, applicable on large
angular scales $\gtrsim30'$, the density contrast grows in proportion
with $a\,g(a)$, as described following eq.~(\ref{eq:2.51}) on
page~\pageref{eq:2.51}. The power spectrum of the density contrast
then evolves $\propto a^2\,g^2(a)$. Inserting this into
eq.~(\ref{eq:6.25}), the squared scale factor $a^2(w)$ cancels, and we
find
\begin{equation}
  P_\kappa(l) = \frac{9H_0^4\Omega_0^2}{4c^4}\,
  \int_0^{w_\mathrm{H}}\,\d w\,g^2[a(w)]\,\bar{W}^2(w)\,
  P_\delta^{0}\left(\frac{l}{f_K(w)}\right)\;.
\label{eq:6.26}
\end{equation}
Here, $P_\delta^0(k)$ is the density-contrast power spectrum linearly
extrapolated to the present epoch.

In an Einstein-de Sitter universe, the growth function $g(a)$ is unity
since $P_\delta$ grows like the squared scale factor. In that special
case, the expression for the power spectrum of
$\bar\kappa_\mathrm{eff}$ further reduces to
\begin{equation}
  P_\kappa(l) = \frac{9H_0^4}{4c^4}\,
  \int_0^{w_\mathrm{H}}\,\d w\,\bar{W}^2(w)\,
  P_\delta^{0}\left(\frac{l}{w}\right)\;,
\label{eq:6.27}
\end{equation}
and the weight function $\bar{W}$ simplifies to
\begin{equation}
  \bar{W}(w) = \int_w^{w_\mathrm{H}}\,\d w'\,G(w')\,
  \left(1-\frac{w}{w'}\right)\;.
\label{eq:6.28}
\end{equation}
In some situations, the distance distribution of the sources can be
approximated by a delta peak at some distance $w_\mathrm{s}$,
$G(w)=\delta_\mathrm{D}(w-w_\mathrm{s})$. A typical example is weak
lensing of the Cosmic Microwave Background, where the source is the
surface of last scattering at redshift
$z_\mathrm{s}\approx1000$. Under such circumstances,
\begin{equation}
  \bar{W}(w) = \left(1-\frac{w}{w_\mathrm{s}}\right)\,
  \mathrm{H}(w_\mathrm{s}-w)\;,
\label{eq:6.29}
\end{equation}
where the Heaviside step function $\mathrm{H}(x)$ expresses the fact
that sources at $w_\mathrm{s}$ are only lensed by mass distributions
at smaller distance $w$. For this specific case, the
effective-convergence power spectrum reads
\begin{equation}
  P_\kappa(l) = \frac{9H_0^4}{4c^4}\,w_\mathrm{s}\,
  \int_0^1\,\d y\,(1-y)^2\,
  P_\delta^0\left(\frac{l}{w_\mathrm{s}y}\right)\;,
\label{eq:6.30}
\end{equation}
where $y=w/w_\mathrm{s}$ is the distance ratio between lenses and
sources. This equation illustrates that all density-perturbation modes
whose wave numbers are larger than $k_\mathrm{min}=w_\mathrm{s}^{-1}l$
contribute to $P_\kappa(l)$, or whose wavelengths are smaller than
$\lambda_\mathrm{max}=w_\mathrm{s}\theta$. For example, the power
spectrum of weak lensing on angular scales of $\theta\approx10'$ on
sources at redshifts $z_\mathrm{s}\approx2$ originates from all
density perturbations smaller than
$\approx7\,h^{-1}\,\mathrm{Mpc}$. This result immediately illustrates
the limitations of the foregoing approximations. Density perturbations
on scales smaller than a few Mpc become non-linear even at moderate
redshifts, and the assumption of linear evolution breaks down.

\subsection{\label{sc:6.5}Magnification and Shear}

In analogy to the Jacobian matrix $\mathcal{A}$ of the conventional
lens equation (\ref{eq:3.11}, page~\pageref{eq:3.11}), we now form the
matrix
\begin{equation}
  \mathcal{A}(\vec\theta,w) = \mathcal{I} - 
  \frac{\partial\vec\alpha(\vec\theta,w)}{\partial\vec\theta} =
  \frac{1}{f_K(w)}\,
  \frac{\partial\vec x(\vec\theta,w)}{\partial\vec\theta}\;.
\label{eq:6.31}
\end{equation}
The magnification is the inverse of the determinant of $\mathcal{A}$
(see eq.~\ref{eq:3.14}, page~\pageref{eq:3.14}). To first order in the
perturbations, we obtain for the magnification of a source at distance
$w$ seen in direction $\vec\theta$
\begin{eqnarray}
  \mu(\vec\theta,w) &=& \frac{1}{\det\mathcal{A}(\vec\theta,w)}
  \approx 1 + \nabla_\theta\cdot\vec\alpha(\vec\theta,w) =
  1 + 2\kappa_\mathrm{eff}(\vec\theta,w)
  \nonumber\\ &\equiv&
  1 + \delta\mu(\vec\theta,w)\;.
\label{eq:6.32}
\end{eqnarray}
In the weak-lensing approximation, the magnification fluctuation
$\delta\mu$ is simply twice the effective convergence
$\kappa_\mathrm{eff}$, just as in the thin-lens approximation.

We emphasise again that the approximations made imply that the matrix
$\mathcal{A}$ is symmetric. In general, when higher-order terms in the
Newtonian potential are considered, $\mathcal{A}$ attains an
asymmetric contribution. \cite{jsw99} used ray-tracing simulations
through the density distribution of the Universe computed in very high
resolution $N$-body simulations to show that the symmetry of
$\mathcal{A}$ is satisfied to very high accuracy. Only for those light
rays which happen to propagate close to more than one strong deflector
can the deviation from symmetry be appreciable. Further estimates of
the validity of the various approximations have been carried out
analytically by \cite{BE97.5} and \cite{svj98}.

Therefore, as in the single lens-plane situation, the anisotropic
deformation, or shear, of a light bundle is determined by the
trace-free part of the matrix $\mathcal{A}$ (cf.~eq.~\ref{eq:3.11},
page~\pageref{eq:3.11}). As explained there, the shear makes
elliptical images from circular sources. Let $a$ and $b$ be the major
and minor axes of the image ellipse of a circular source,
respectively, then the ellipticity is
\begin{equation}
  |\chi|= \frac{a^2-b^2}{a^2+b^2} \approx 2|\gamma|\;,
\label{eq:6.34}
\end{equation}
where the latter approximation is valid for weak lensing,
$|\gamma|\ll1$; cf.~eq.~(\ref{eq:4.18}). The quantity $2\gamma$ was
sometimes called {\em polarisation\/} in the literature
(\cite*{BL91.1}, \cite*{MI91.3}, \cite*{KA92.1}).

In the limit of weak lensing which is relevant here, the two-point
statistical properties of $\delta\mu$ and of $2\gamma$ are identical
(e.g.~\cite*{BL91.1}). To see this, we first note that the first
derivatives of the deflection angle occurring in eqs.~(\ref{eq:6.32})
can be written as second derivatives of an effective deflection
potential $\psi$ which is defined in terms of the effective surface
mass density $\kappa_\mathrm{eff}$ in the same way as in the single
lens-plane case; see (\ref{eq:3.9}, page~\pageref{eq:3.9}). We then
imagine that $\delta\mu$ and $\gamma$ are Fourier transformed,
whereupon the derivatives with respect to $\theta_i$ are replaced by
multiplications with components of the wave vector $\vec l$ conjugate
to $\vec\theta$. In Fourier space, the expressions for the averaged
quantities $\langle\delta\mu^2\rangle$ and
$4\,\langle|\gamma|^2\rangle$ differ only by the combinations of $l_1$
and $l_2$ which appear under the average. We have
\begin{equation}
  \begin{array}{ll}
    \left(l_1^2+l_2^2\right)^2 = |\vec l|^4 & 
    \quad\hbox{for}\quad\langle\delta\mu^2\rangle \\
    \left(l_1^2-l_2^2\right)^2 + 4\,l_1^2l_2^2= |\vec l|^4 & 
    \quad\hbox{for}\quad4\,\langle|\gamma|^2\rangle =
    4\,\langle\gamma_1^2+\gamma_2^2\rangle
  \end{array}\;,
\label{eq:6.35}
\end{equation}
and hence the two-point statistical properties of $\delta\mu$ and
$2\,\gamma$ agree identically. Therefore, the power spectra of
effective convergence and shear agree,
\begin{equation}
  \langle\hat{\kappa}_\mathrm{eff}(\vec l)
         \hat{\kappa}^*_\mathrm{eff}(\vec l')\rangle=
  \langle\hat{\gamma}(\vec l)
         \hat{\gamma}^*(\vec l')\rangle\quad\Rightarrow\quad
  P_\kappa(l)=P_\gamma(l)\;.
\label{eq:6.35a}
\end{equation}
Thus we can concentrate on the statistics of either the magnification
fluctuations or the shear only. Since
$\delta\mu=2\kappa_\mathrm{eff}$, the magnification power spectrum
$P_\mu$ is $4P_\kappa$, and we can immediately employ the convergence
power spectrum $P_\kappa$.

\subsection{\label{sc:6.6}Second-Order Statistical Measures}

We aim at the statistical properties of the magnification fluctuation
and the shear. In particular, we are interested in the amplitude of
these quantities and their angular coherence. Both can be described by
their angular auto-correlation functions, or other second-order
statistical measures that will turn out to be more practical later. As
long as the density fluctuation field $\delta$ remains Gaussian, the
probability distributions of $\delta\mu$ and $\gamma$ are also
Gaussians with mean zero, and two-point statistical measures are
sufficient for their complete statistical description. When non-linear
evolution of the density contrast sets in, non-Gaussianity develops,
and higher-order statistical measures become important.

\subsubsection{\label{sc:6.6.1}Angular Auto-Correlation Function}

The angular autocorrelation function $\xi_q(\phi)$ of some isotropic
quantity $q(\vec\theta)$ is the Fourier transform of the power
spectrum $P_q(l)$ of $q(\vec\theta)$. In particular, the
auto-correlation function of the magnification fluctuation,
$\xi_\mu(\phi)$, is related to the effective-convergence power
spectrum $P_\kappa(l)$ through
\begin{eqnarray}
  \xi_\mu(\phi) &=& \langle
  \delta\mu(\vec\theta)\delta\mu(\vec\theta+\vec\phi)
  \rangle =
      4\,\langle\kappa_\mathrm{eff}(\vec\theta)
                \kappa_\mathrm{eff}(\vec\theta+\vec\phi)
         \rangle =
      4\,\langle\gamma(\vec\theta)
                \gamma^*(\vec\theta+\vec\phi)
         \rangle\nonumber\\
  &=& 4\,\int\frac{\d^2l}{(2\pi)^2}\,P_\kappa(l)\,
  \exp(-\mathrm{i}\,\vec l\cdot\vec\phi) =
  4\,\int_0^\infty\frac{l\d l}{2\pi}\,P_\kappa(l)\,\J(l\phi)\;,
\label{eq:6.36}
\end{eqnarray}
where $\vec\phi$ is a vector with norm $\phi$. The factor four in
front of the integral accounts for the fact that
$\delta\mu=2\kappa_\mathrm{eff}$ in the weak-lensing
approximation. For the last equality in (\ref{eq:6.36}), we integrated
over the angle enclosed by $\vec l$ and $\vec\phi$, leading to the
zeroth-order Bessel function of the first kind,
$\J(x)$. Equation~(\ref{eq:6.36}) shows that the magnification (or
shear) auto-correlation function is an integral over the power
spectrum of the effective convergence $\kappa_\mathrm{eff}$, filtered
by the Bessel function $\J(x)$. Since the latter is a broad-band
filter, the magnification auto-correlation function is not well suited
for extracting information on $P_\kappa$. It would be desirable to
replace $\xi_\mu(\phi)$ by another measurable quantity which involves
a narrow-band filter.

Nonetheless, inserting eq.~(\ref{eq:6.25}) into eq.~(\ref{eq:6.36}),
we obtain the expression for the magnification auto-correlation
function,
\begin{eqnarray}
  \xi_\mu(\phi) = &\frac{9H_0^4\Omega_0^2}{c^4}&\,
  \int_0^{w_\mathrm{H}}\,\d w\,f_K^2(w)\,\bar{W}^2(w,w)\,a^{-2}(w)
  \nonumber\\ &\times&
  \int_0^\infty\,\frac{k\d k}{2\pi}\,P_\delta(k,w)\,
  \J[f_K(w)k\phi]\;.
\label{eq:6.37}
\end{eqnarray}
The magnification autocorrelation function therefore turns out to be
an integral over the density-fluctuation power spectrum weighted by a
$k$--space window function which selects the contributing density
perturbation modes.

\subsubsection{\label{sc:6.6.2}Special Cases and Qualitative
  Expectations}

In order to gain some insight into the expected behaviour of the
magnification auto-correlation function $\xi_\mu(\phi)$, we now make a
number of simplifying assumptions. Let us first specialise to linear
density evolution in an Einstein-de Sitter universe, and assume
sources are at a single distance
$w_\mathrm{s}$. Equation~(\ref{eq:6.37}) then immediately simplifies
to
\begin{equation}
  \xi_\mu(\phi) = \frac{9H_0^4}{c^4}\,w_\mathrm{s}^3\,
  \int_0^1\,\d y\,y^2(1-y)^2\,
  \int_0^\infty\,\frac{k\d k}{2\pi}\,P_\delta^0(k)\J(wyk\phi)\;,
\label{eq:6.38}
\end{equation}
with $y\equiv w_\mathrm{s}^{-1}w$.

We now introduce two model spectra $P_\delta^0(k)$, one of which has
an exponential cut-off above some wave number $k_0$, while the other
falls off like $k^{-3}$ for $k>k_0$. For small $k$, both spectra
increase like $k$. They approximately describe two extreme cases of
popular cosmogonies, the HDM and the CDM model. We choose the
functional forms
\begin{equation}
  P_{\delta,\mathrm{HDM}}^0 = A\,k\,
  \exp\left(-\frac{k}{k_0}\right)\;,\quad
  P_{\delta,\mathrm{CDM}}^0 = A\,k\,
  \frac{9k_0^4}{(k^2+3k_0^2)^2}\;,
\label{eq:6.39}
\end{equation}
where $A$ is the normalising amplitude of the power spectra. The
numerical coefficients in the CDM model spectrum are chosen such that
both spectra peak at the same wave number $k=k_0$. Inserting these
model spectra into eq.~(\ref{eq:6.38}), performing the $k$
integration, and expanding the result in a power series in $\phi$, we
obtain (\cite*{BA95.4})
\begin{eqnarray}
  \xi_{\mu,\mathrm{HDM}}(\phi) &=&
  \frac{3\,A'}{10\pi}\,(w_\mathrm{s}k_0)^3 - 
  \frac{9\,A'}{35\pi}\,(w_\mathrm{s}k_0)^5\,\phi^2 + 
  \mathcal{O}(\phi^4)\;,\nonumber\\
  \xi_{\mu,\mathrm{CDM}}(\phi) &=& 
  \frac{9\sqrt{3}\,A'}{80}\,(w_\mathrm{s}k_0)^3 - 
  \frac{27\,A'}{40\pi}\,(w_\mathrm{s}k_0)^4\,\phi + 
  \mathcal{O}(\phi^2)\;,
\label{eq:6.40}
\end{eqnarray}
where $A'=(H_0c^{-1})^4\,A$. We see from eq.~(\ref{eq:6.40}) that the
magnification correlation function for the HDM spectrum is flat to
first order in $\phi$, while it decreases linearly with $\phi$ for the
CDM spectrum. This demonstrates that the shape of the magnification
autocorrelation function $\xi_\mu(\phi)$ reflects the shape of the
dark-matter power spectrum. Motivated by the result of a large number
of cosmological studies showing that HDM models have the severe
problem of structure on small scales forming at times much later than
observed (see e.g.~\cite*{pea99}), we now neglect the HDM model and
focus on the CDM power spectrum only.

We can then expect $\xi_\mu(\phi)$ to increase linearly with $\phi$ as
$\phi$ goes to zero. Although we assumed linear evolution of the power
spectrum to achieve this result, this qualitative behaviour remains
valid when non-linear evolution is assumed, because for large wave
numbers $k$, the non-linear CDM power spectra also asymptotically fall
off $\propto k^{-3}$ for large $k$.

Although the model spectra (\ref{eq:6.39}) are of limited validity, we
can extract some useful information from the small-angle
approximations given in eq.~(\ref{eq:6.40}).  First, the correlation
amplitude $\xi_\mu(0)$ scales with the comoving distance to the
sources $w_\mathrm{s}$ as $w_\mathrm{s}^3$. In the Einstein-de Sitter
case, for which eq.~(\ref{eq:6.40}) was derived,
$w_\mathrm{s}=(2c/H_0)\,[1-(1+z_\mathrm{s})^{-1/2}]$. For low source
redshifts, $z_\mathrm{s}\ll1$,
$w_\mathrm{s}\approx(c/H_0)\,z_\mathrm{s}$, so that $\xi_\mu(0)\propto
z_\mathrm{s}^3$. For $z_\mathrm{s}\gg1$, $w_\mathrm{s}\to(2c/H_0)$,
and $\xi_\mu(0)$ becomes independent of source redshift. For
intermediate source redshifts, progress can be made by defining
$\zeta_\mathrm{s}\equiv\ln(z_\mathrm{s})$ and expanding $\ln
w[\exp(\zeta_\mathrm{s})]$ in a power series in
$\zeta_\mathrm{s}$. The result is an approximate power-law expression,
$w(z_\mathrm{s})\propto z_\mathrm{s}^\epsilon$, valid in the vicinity
of the zero point of the expansion. The exponent $\epsilon$ changes
from $\approx0.6$ at $z_\mathrm{s}\approx1$ to $\approx0.38$ at
$z_\mathrm{s}\approx3$.

Second, typical source distances are of order $2\,\mathrm{Gpc}$. Since
$k_0$ is the wave number corresponding to the horizon size when
relativistic and non-relativistic matter had equal densities,
$k_0^{-1}=d_\mathrm{H}(a_\mathrm{eq}) \approx
12\,(\Omega_0\,h^2)^{-1}\,\mathrm{Mpc}$. Therefore,
$w_\mathrm{s}k_0\approx150$. Typically, the spectral amplitude $A'$
ranges between $10^{-8}$--$10^{-9}$. A rough estimate for the
correlation amplitude $\xi_\mu(0)$ thus ranges between
$10^{-2}$--$10^{-3}$ for `typical' source redshifts
$z_\mathrm{s}\gtrsim1$.

Third, an estimate for the angular scale $\phi_0$ of the magnification
correlation is obtained by determining the angle where $\xi_\mu(\phi)$
has dropped to half its maximum. From the small-angle approximation
(\ref{eq:6.40}), we find
$\phi_0=\pi\sqrt{3}(12\,w_\mathrm{s}k_0)^{-1}$. Inserting as before
$w_\mathrm{s}k_0\approx150$, we obtain $\phi_0\approx10'$, {\em
decreasing\/} with increasing source redshift.

Summarising, we expect $\xi_\mu(\phi)$ in a CDM universe to
\begin{enumerate}
\item start at $10^{-2}$--$10^{-3}$ at $\phi=0$ for source redshifts
$z_\mathrm{s}\sim1$;
\item decrease linearly for small $\phi$ on an angular scale of
$\phi_0\approx10'$; and
\item increase with source redshift roughly as $\propto
z_\mathrm{s}^{0.6}$ around $z_\mathrm{s}=1$.
\end{enumerate}

\subsubsection{\label{sc:6.6.3}Realistic Cases}

\begin{figure}[ht]
  \includegraphics[width=\hsize]{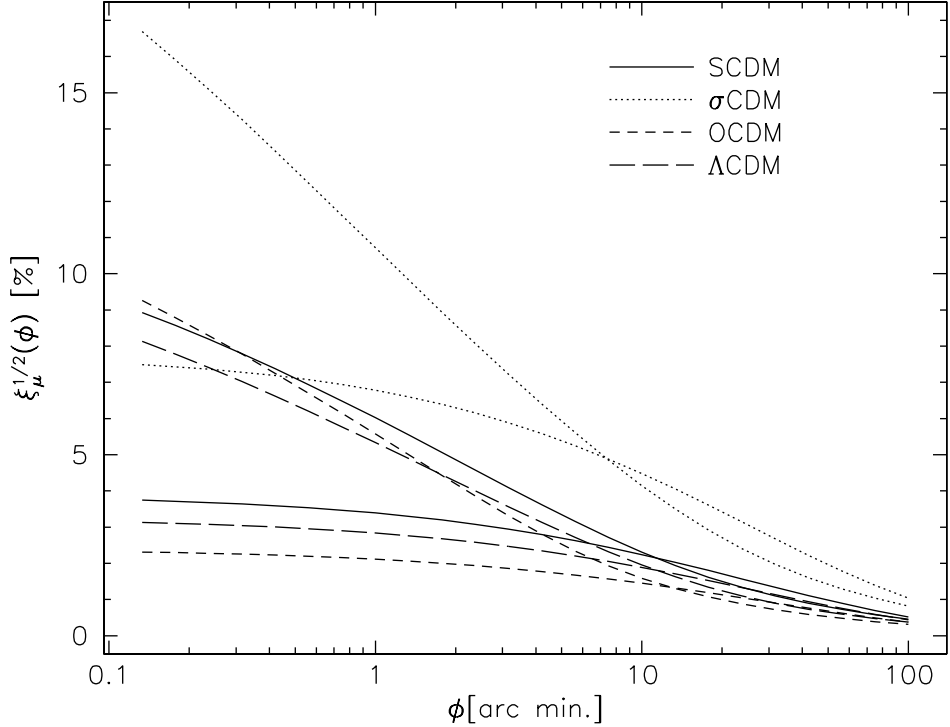}
\caption{Four pairs of magnification auto-correlation functions are
shown for the cosmological model parameter sets listed in
Tab.~\ref{tab:6.1}, and for an assumed source redshift
$z_\mathrm{s}=1$. For each pair, plotted with the same line type, the
curve with lower amplitude at small angular scale was calculated
assuming linear, and the other one non-linear density evolution. Solid
curves: SCDM; dotted curves: $\sigma$CDM; short-dashed curve: OCDM;
and long-dashed curve: $\Lambda$CDM. Non-linear evolution increases
the amplitude of $\xi_\mu^{1/2}(\phi)$ on small angular scales by
factors of three to four. The results for the cluster-normalised
models differ fairly little. At $\phi\approx1'$,
$\xi_\mu^{1/2}(\phi)\approx6\%$ for non-linear density evolution. For
the Einstein-de Sitter models, the difference between cluster- and
$\sigma_8=1$ normalisation amounts to about a factor of two in
$\xi_\mu^{1/2}(\phi)$.}
\label{fig:6.3}
\end{figure}

After this digression, we now return to realistic CDM power spectra
normalised to fit observational constraints. Some representative
results are shown in Fig.~\ref{fig:6.3} for the model parameter sets
listed in Tab.~\ref{tab:6.1}.

The figure shows that typical values for $\xi_\mu^{1/2}(\phi)$ in
cluster-normalised CDM models with non-linear density evolution are
$\approx6\%$ at $\phi\approx1'$, quite independent of the cosmological
model. The effects of non-linear evolution are considerable.
Non-linear evolution increases the $\xi_\mu^{1/2}$ by factors of three
to four. The uncertainty in the normalisation is illustrated by the
two curves for the Einstein-de Sitter model, one of which was
calculated with the cluster-, the other one with the $\sigma_8=1$
normalisation, which yields about a factor of two larger results for
$\xi_\mu^{1/2}$. For the other cosmological models (OCDM and
$\Lambda$CDM), the effects of different normalisations (cluster
vs.~COBE) are substantially smaller.

\subsubsection{\label{sc:6.6.4}Application: Magnification
  Fluctuations}

\begin{figure}[ht]
  \includegraphics[width=\hsize]{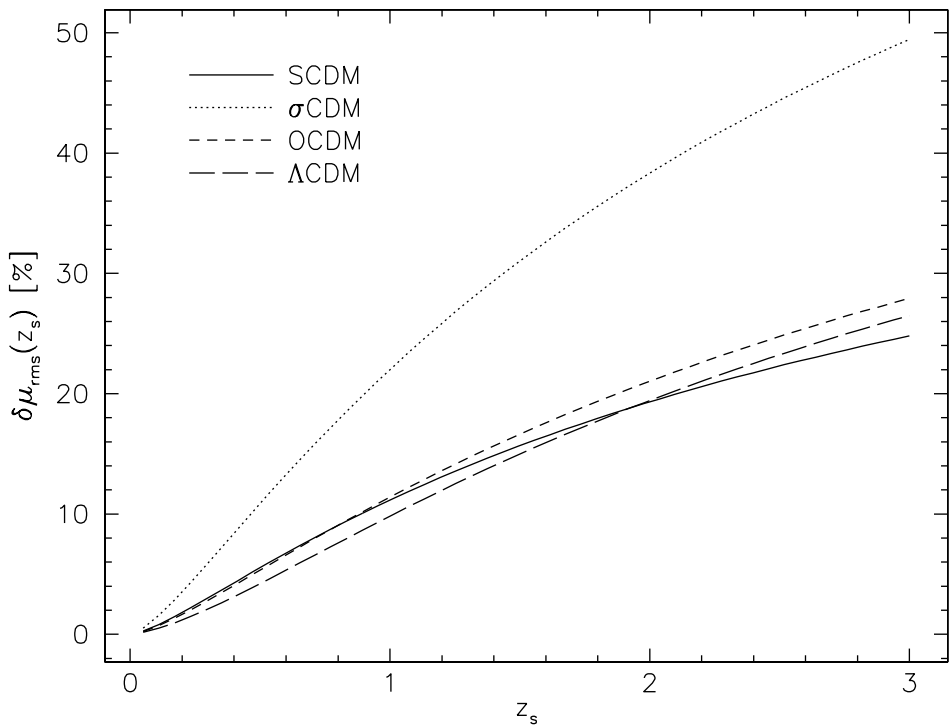}
\caption{The {\em rms\/} magnification fluctuation
$\delta\mu_\mathrm{rms}$ is shown as a function of source redshift
$z_\mathrm{s}$ for non-linearly evolving density fluctuations in the
four different realisations of the CDM cosmogony detailed in
Tab.~\ref{tab:6.1}. Solid curve: SCDM; dotted curve: $\sigma$CDM;
short-dashed curve: OCDM; and long-dashed curve: $\Lambda$CDM. Except
for the $\sigma$CDM model, typical {\em rms\/} magnification
fluctuations are of order $20\%$ at $z_\mathrm{s}=2$, and $25\%$ for
$z_\mathrm{s}=3$.}
\label{fig:6.4}
\end{figure}

At zero lag, the magnification autocorrelation function reads
\begin{equation}
  \xi_\mu(0) = \left\langle
    \left[\mu(\vec\theta) - 1\right]^2
  \right\rangle \equiv \left\langle\delta\mu^2\right\rangle\;,
\label{eq:6.42}
\end{equation}
which is the variance of the magnification fluctuation
$\delta\mu$. Consequently, the {\em rms\/} magnification fluctuation
is
\begin{equation}
  \delta\mu_\mathrm{rms} = 
  \left\langle\delta\mu^2\right\rangle^{1/2} = \xi_\mu^{1/2}(0)\;.
\label{eq:6.43}
\end{equation}
Figure~\ref{fig:6.4} shows $\delta\mu_\mathrm{rms}$ as a function of
source redshift for four different realisations of the CDM cosmogony.
For cluster-normalised CDM models, the {\em rms\/} magnification
fluctuation is of order $\delta\mu_\mathrm{rms}\approx 20\%$ for
sources at $z_\mathrm{s}\approx2$, and increases to
$\delta\mu_\mathrm{rms}\approx 25\%$ for $z_\mathrm{s}\approx3$. The
strongest effect occurs for open CDM (OCDM) because there non-linear
evolution sets in at the highest redshifts.

The results shown in Fig.~\ref{fig:6.4} indicate that for any
cosmological source, gravitational lensing causes a statistical
uncertainty of its brightness. In magnitudes, a typical effect at
$z_\mathrm{s}\approx2$ is $\delta
m\approx2.5\times\log(1.2)\approx0.2$. This can be important for
e.g.~high-redshift supernovae of type Ia, which are used as
cosmological standard candles. Their intrinsic magnitude scatter is of
order $\delta m\approx0.1-0.2$ magnitudes (e.g.~\cite*{phi93};
\cite*{rpk95}, \citeyear{rpk96}; \cite*{hps96}). Therefore, the
lensing-induced brightness fluctuation is comparable to the intrinsic
uncertainty at redshifts $z_\mathrm{s}\gtrsim2$ (\cite*{fri96};
\cite*{WA97.2}; \cite*{hol98}; \cite*{mes99}).

Since the magnification probability can be highly skewed, the {\em
most probable\/} observed flux of a high-redshift supernova can
deviate from the {\em mean\/} flux at given redshift, even if the
intrinsic luminosity distribution is symmetric. This means that
particular care needs to be taken in the analysis of future large SN
surveys. However, if SNe~Ia are quasi standard candles also at high
redshifts, with an intrinsic scatter of $\Delta L=4\pi
D_\mathrm{lum}^2(z) \Delta S(z)$ around the mean luminosity $L_0=4\pi
D_\mathrm{lum}^2(z) S_0(z)$, then it is possible to obtain {\em
volume-limited samples\/} (in contrast to flux-limited samples) of
them.

If, for a given redshift, the sensitivity limit is chosen to be
$S_\mathrm{min}\lesssim\mu_\mathrm{min}\,(S_0-3\Delta S)$, one can be
sure to find all SNe~Ia at the redshift considered. Here,
$\mu_\mathrm{min}$ is the minimum magnification of a source at the
considered redshift. Since no source can be more de-magnified than one
that is placed behind a hypothetical empty cone (see \cite*{DY73.1}
and the discussion in Sect.~4.5 of \cite*{sef92}), $\mu_\mathrm{min}$
is not much smaller than unity. Flux conservation
(e.g.~\cite*{WE76.1}) implies that the mean magnification of all
sources at given redshift is unity, $\langle\mu(z)\rangle=1$, and so
the expectation value of the observed flux at given redshift is the
unlensed flux, $\langle S(z)\rangle=S_0(z)$. It should be pointed out
here that a similar relation for the magnitudes does {\em not\/} hold,
since magnitude is a logarithmic measure of the flux, and so $\langle
m(z)\rangle\ne m_0(z)$. This led to some confusing conclusions in the
literature claiming that lensing introduces a bias in cosmological
parameter estimates from lensing, but this is not true: One just has
to work in terms of fluxes rather than magnitudes.

However, a broad magnification probability distribution increases the
confidence contours for $\Omega_0$ and $\Omega_\Lambda$
(e.g.~\cite*{hol98}). If the probability distribution was known, more
sensitive estimators of the cosmological model than the mean flux at
given redshift could be constructed. Furthermore, if the intrinsic
luminosity distribution of the SNe was known, the normalisation of the
power spectrum as a function of $\Omega_0$ and $\Omega_\Lambda$ could
be inferred from the broadened observed flux distribution
(\cite*{met99}). If the dark matter is in the form of compact objects
with mass $\gtrsim10^{-2}M_\odot$, these objects can individually
magnify a SN (\cite*{SC87.6}), additionally broadening the
magnification probability distribution and thus enabling the nature of
dark matter to be tested through SN observations (\cite*{mes99},
\cite*{seh99}).

\subsubsection{\label{sc:6.6.5}Shear in Apertures}

\begin{figure}[ht]
  \includegraphics[width=\hsize]{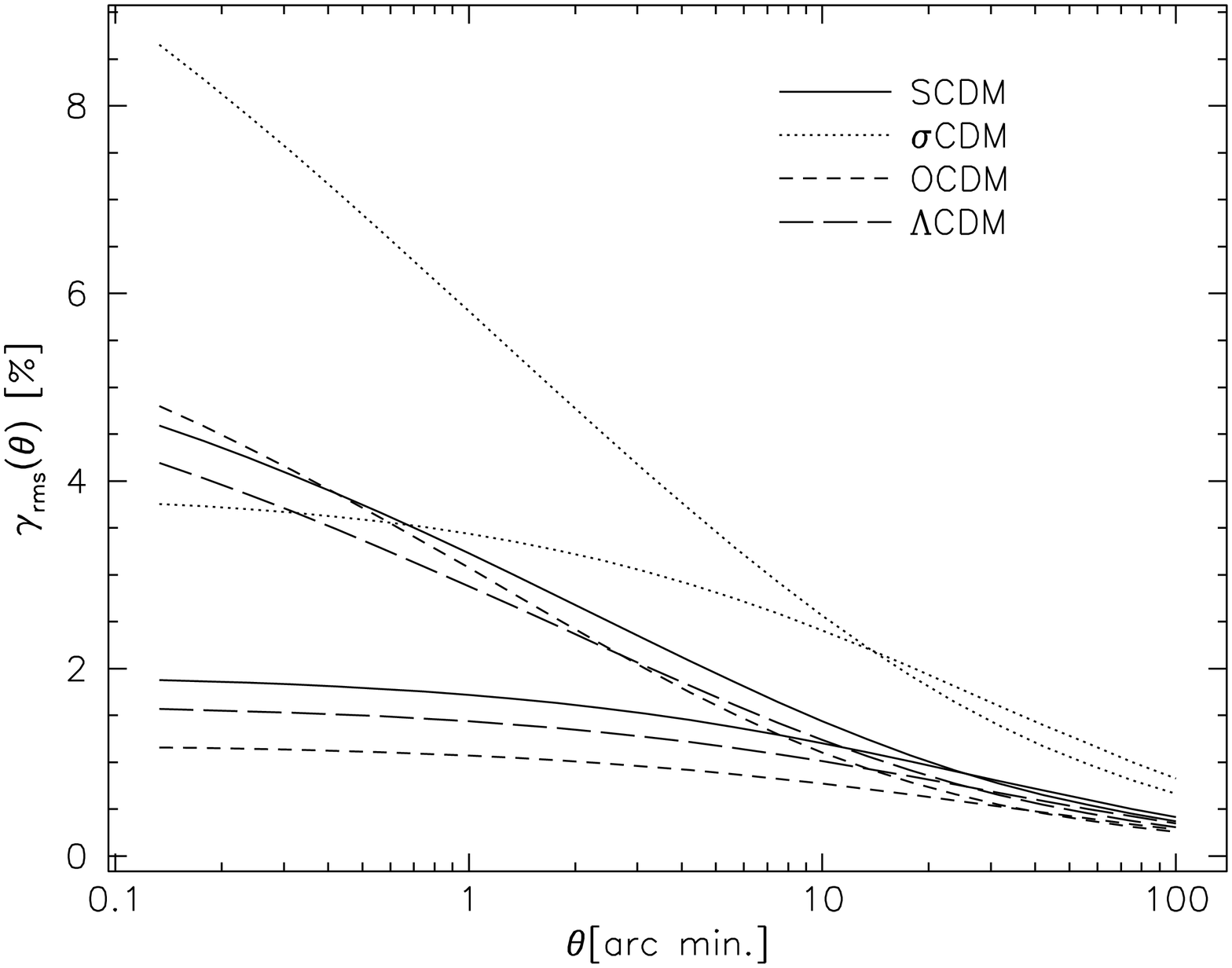}
\caption{The {\em rms\/} shear $\gamma_\mathrm{rms}(\theta)$ in
circular apertures of radius $\theta$ is plotted as a function of
$\theta$ for the four different realisations of the CDM cosmogony
detailed in Tab.~\ref{tab:6.1}, where all sources are assumed to be at
redshift $z_\mathrm{s}=1$. A pair of curves is plotted for each
realisation, where for each pair the curve with lower amplitude at
small $\theta$ is for linearly, the other one for non-linearly
evolving density fluctuations. Solid curves: SCDM; dotted curves:
$\sigma$CDM; short-dashed curves: OCDM; and long-dashed curves:
$\Lambda$CDM. For the cluster-normalised models, typical {\em rms\/}
shear values are $\approx3\%$ for $\theta\approx1'$. Non-linear
evolution increases the amplitude by about a factor of two at
$\theta\approx1'$ over linear evolution.}
\label{fig:6.5}
\end{figure}

We mentioned below eq.~(\ref{eq:6.36}) that measures of cosmic
magnification or shear other than the angular auto-correlation
function which filter the effective-convergence power spectrum
$P_\kappa$ with a function narrower than the Bessel function $\J(x)$
would be desirable. In practice, a convenient measure would be the
variance of the effective convergence within a circular aperture of
radius $\theta$. Within such an aperture, the averaged effective
convergence and shear are
\begin{equation}
  \kappa_\mathrm{av}(\theta) = \int_0^\theta
  \frac{\d^2\phi}{\pi\theta^2}\,\bar\kappa_\mathrm{eff}(\vec\phi)
  \;,\quad
  \gamma_\mathrm{av}(\theta) = \int_0^\theta
  \frac{\d^2\phi}{\pi\theta^2}\,\gamma(\vec\phi)\;,
\label{eq:6.44}
\end{equation}
and their variance is
\begin{equation}
  \langle {\kappa}_\mathrm{av}^2\rangle(\theta) = 
    \int_0^\theta\frac{\d^2\phi}{\pi\theta^2}
    \int_0^\theta\frac{\d^2\phi'}{\pi\theta^2}\,
    \left\langle
      \bar{\kappa}_\mathrm{eff}(\vec\phi)
      \bar{\kappa}_\mathrm{eff}(\vec\phi')
    \right\rangle=
  \langle|\gamma_\mathrm{av}|^2\rangle(\theta)\;.
\label{eq:6.45}
\end{equation}
The remaining average is the effective-convergence auto-correlation
function $\xi_\kappa(|\vec\phi-\vec\phi'|)$, which can be expressed in
terms of the power spectrum $P_\kappa$. The final equality follows
from $\xi_\kappa=\xi_\gamma$. Inserting (\ref{eq:6.45}) and performing
the angular integrals yields
\begin{equation}
  \langle {\kappa}_\mathrm{av}^2\rangle(\theta) =
  2\pi\int_0^\infty l\d l\,P_\kappa(l)\,\left[
    \frac{\mathrm{J}_1(l\theta)}{\pi l\theta}
  \right]^2 = \langle|\gamma_\mathrm{av}|^2\rangle(\theta)\;,
\label{eq:6.46}
\end{equation}
where $\mathrm{J}_1(x)$ is the first-order Bessel function of the
first kind. Results for the {\em rms\/} shear in apertures of varying
size are shown in Fig.~\ref{fig:6.5} (cf.~\cite*{BL91.1}).

\subsubsection{\label{sc:6.6.6}Aperture Mass}

Another measure for the effects of weak lensing, the {\em aperture
mass\/} $M_\mathrm{ap}(\theta)$ (cf.~Sect.~\ref{sc:5.3.1}), was
introduced for cosmic shear by \cite{svj98} as
\begin{equation}
  M_\mathrm{ap}(\theta) = \int_0^\theta\d^2\phi\,U(\phi)\,
  \bar\kappa_\mathrm{eff}(\vec\phi)\;,
\label{eq:6.47}
\end{equation}
where the weight function $U(\phi)$ satisfies the criterion
\begin{equation}
  \int_0^\theta\phi\d\phi\,U(\phi) = 0\;.
\label{eq:6.48}
\end{equation}
In other words, $U(\phi)$ is taken to be a {\em compensated\/} radial
weight function across the aperture. For such weight functions, the
aperture mass can be expressed in terms of the tangential component of
the observable shear relative to the aperture centre,
\begin{equation}
  M_\mathrm{ap}(\theta) = \int_0^\theta\d^2\phi\,Q(\phi)\,
  \gamma_\mathrm{t}(\vec\phi)\;,
\label{eq:6.48a}
\end{equation}
where $Q(\phi)$ is related to $U(\phi)$ by
(\ref{eq:5.25}). $M_\mathrm{ap}$ is a scalar quantity directly
measurable in terms of the shear. The variance of $M_\mathrm{ap}$
reads
\begin{equation}
  \langle M_\mathrm{ap}^2\rangle(\theta) =
  2\pi\,\int_0^\infty l\d l\,P_\kappa(l)\,\left[
    \int_0^\theta\phi\d\phi\,U(\phi)\,\J(l\phi)
  \right]^2\;.
\label{eq:6.49}
\end{equation}
Equations~(\ref{eq:6.46}) and (\ref{eq:6.49}) provide alternative
observable quantities which are related to the effective-convergence
power spectrum $P_\kappa$ through narrower filters than the
auto-correlation function $\xi_\kappa$. The $M_\mathrm{ap}$ statistic
in particular permits one to tune the filter function through
different choices of $U(\phi)$ within the constraint
(\ref{eq:6.48}). It is important that $M_\mathrm{ap}$ can also be
expressed in terms of the shear [see eq.~(\ref{eq:5.28},
page~\pageref{eq:5.28})], so that $M_\mathrm{ap}$ can directly be
obtained from the observed galaxy ellipticities.

\begin{figure}[ht]
  \includegraphics[width=\hsize]{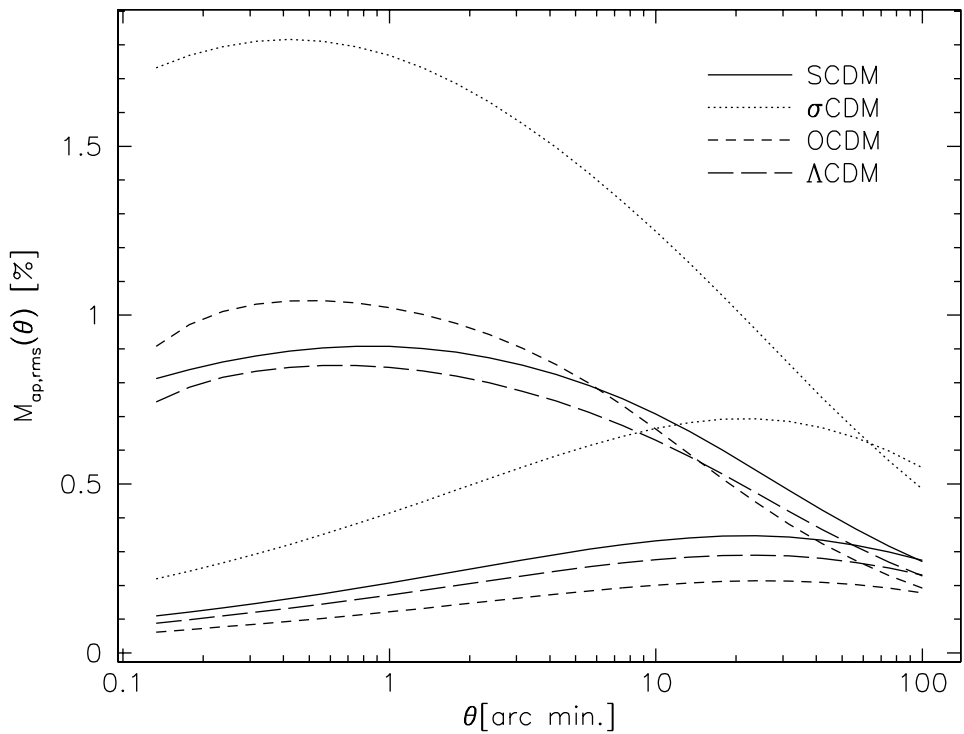}
\caption{The {\em rms\/} aperture mass, $M_\mathrm{ap,rms}(\theta)$,
is shown in dependence of aperture radius $\theta$ for the four
different realisations of the CDM cosmogony detailed in
Tab.~\ref{tab:6.1} where all sources are assumed to be at redshift
$z_\mathrm{s}=1$. For each realisation, a pair of curves is plotted;
one curve with lower amplitude for linear, and the second curve for
non-linear density evolution. Solid curves: SCDM; dotted curves:
$\sigma$CDM; short-dashed curves: OCDM; and long-dashed curves:
$\Lambda$CDM. Non-linear evolution has a pronounced effect: The
amplitude is approximately doubled, and the peak shifts from degree-
to arc-minute scales.}
\label{fig:6.6}
\end{figure}

\cite{svj98} suggested a family of radial filter functions $U(\phi)$,
the simplest of which is
\begin{equation}
  U(\phi) = \frac{9}{\pi\theta^2}\,(1-x^2)\,
  \left(\frac{1}{3}-x^2\right)\;,\quad
  Q(\phi) = \frac{6}{\pi\theta^2}\,x^2\,(1-x^2)\;,
\label{eq:6.50}
\end{equation}
where $x\theta=\phi$. With this choice, the variance $\langle
M_\mathrm{ap}^2\rangle(\theta)$ becomes
\begin{equation}
  \langle M_\mathrm{ap}^2\rangle(\theta) =
  2\pi\,\int_0^\infty l\d l\,P_\kappa(l)\,J^2(l\theta)\;,
\label{eq:6.51}
\end{equation}
with the filter function
\begin{equation}
  J(\eta) = \frac{12}{\pi\eta^2}\,\mathrm{J}_4(\eta)\;,
\label{eq:6.52}
\end{equation}
where $\mathrm{J}_4(\eta)$ is the fourth-order Bessel function of the
first kind. Examples for the {\em rms\/} aperture mass,
$M_\mathrm{ap,rms}(\theta)=\langle
M_\mathrm{ap}^2\rangle^{1/2}(\theta)$, are shown in
Fig.~\ref{fig:6.6}.

The curves look substantially different from those shown in
Figs.~\ref{fig:6.3} and \ref{fig:6.5}. Unlike there, the aperture mass
does not increase monotonically as $\theta\to0$, but reaches a maximum
at finite $\theta$ and drops for smaller angles. When non-linear
evolution of the density fluctuations is assumed, the maximum occurs
at much smaller $\theta$ than for linear evolution: Linear evolution
predicts the peak at angles of order one degree, non-linear evolution
around $1'$~! The amplitude of $M_\mathrm{ap,rms}(\theta)$ reaches
$\approx1\%$ for cluster-normalised models, quite independent of the
cosmological parameters.

\begin{figure}[ht]
  \includegraphics[width=\hsize]{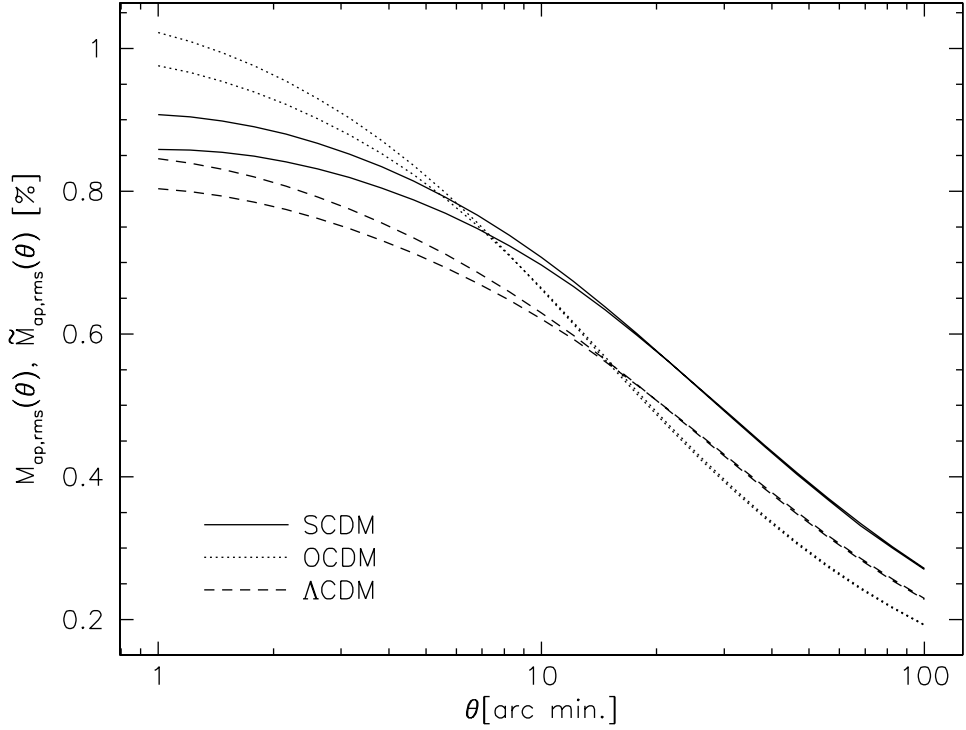}
\caption{The {\em rms\/} aperture mass $M_\mathrm{ap,rms}(\theta)$ is
shown together with the approximation
$\tilde{M}_\mathrm{ap,rms}(\theta)$ of eq.~(\ref{eq:6.55}). The three
curves correspond to the three cluster-normalised cosmological models
(SCDM, OCDM and $\Lambda$CDM) introduced in Tab.~\ref{tab:6.1} for
non-linearly evolving matter perturbations. All sources were assumed
to be at redshift $z_\mathrm{s}=1$. Clearly, the {\em rms\/} aperture
mass is very accurately approximated by $\tilde{M}_\mathrm{ap,rms}$ on
angular scales $\theta\gtrsim10'$, and even for smaller aperture sizes
of order $\sim1'$ the deviation between the curves is smaller than
$\approx5\%$. The observable {\em rms\/} aperture mass therefore
provides a very direct measure for the effective-convergence power
spectrum $P_\kappa(l)$.}
\label{fig:6.7}
\end{figure}

Some insight into the expected amplitude and shape of $\langle
M_\mathrm{ap}^2\rangle(\theta)$ can be gained by noting that
$J^2(\eta)$ is well approximated by a Gaussian,
\begin{equation}
  J^2(\eta) \approx
  A\,\exp\left[-\frac{(\eta-\eta_0)^2}{2\,\sigma^2}\right]\;,
\label{eq:6.53}
\end{equation}
with mean $\eta_0\approx4.11$, amplitude $A\approx4.52\times10^{-3}$,
and width $\sigma\approx1.24$. At aperture radii of $\theta\approx1'$,
the peak $\eta_0\approx4.11$ corresponds to angular scales of $2\pi
l^{-1}\approx1.6'$, where the total power $l^2P_\kappa(l)$ in the
effective convergence is close to its broad maximum
(cf.~Fig.~\ref{fig:6.2}). The filter function $J^2(\eta)$ is therefore
fairly narrow. Its relative width corresponds to an $l$ range of
$\delta l/l \approx\sigma/\eta_0\sim 0.3$. Thus, the contributing
range of modes $l$ in the integral (\ref{eq:6.51}) is very
small. Crudely approximating the Gaussian by a delta distribution,
\begin{equation}
  J^2(\eta) \approx
  A\,\sqrt{2\pi}\sigma\,\delta_\mathrm{D}(\eta-\eta_0)\;,
\label{eq:6.54}
\end{equation}
we are led to
\begin{equation}
  \langle M_\mathrm{ap}^2\rangle \approx
  \langle\tilde{M}_\mathrm{ap}^2\rangle \equiv
  \frac{(2\pi)^{3/2}\,A\,\sigma}{\eta_0}\,
  \left(\frac{\eta_0}{\theta}\right)^2\,
  P_\kappa\left(\frac{\eta_0}{\theta}\right) \approx
  2.15\times10^{-2}\,l_0^2P_\kappa(l_0)\;,
\label{eq:6.55}
\end{equation}
with $l_0\equiv \eta_0\theta^{-1}$. Hence, the mean-square aperture
mass is expected to directly yield the total power in the
effective-convergence power spectrum, scaled down by a factor of
$\approx2.15\times10^{-2}$. We saw in Fig.~\ref{fig:6.2} that
$l^2P_\kappa(l)\approx3\times10^{-3}$ for $2\pi l^{-1}\approx1'$ in
cluster-normalised CDM models, so that
\begin{equation}
  \langle M_\mathrm{ap}^2\rangle^{1/2} \approx 0.8\%
  \quad\hbox{at}\quad \theta\approx1'
\label{eq:6.56}
\end{equation}
for sources at redshift unity. We compare $M_\mathrm{ap,rms}(\theta)$
and the approximation $\tilde{M}_\mathrm{ap,rms}(\theta)$ in
Fig.~\ref{fig:6.7}. Obviously, the approximation is excellent for
$\theta\gtrsim10'$, but even for smaller aperture radii of $\sim1'$
the relative deviation is less than $\approx5\%$. At this point, the
prime virtue of the narrow filter function $J(\eta)$ shows up most
prominently. Up to relatively small errors of a few per cent, the {\em
rms\/} aperture mass very accurately reflects the
effective-convergence power spectrum $P_\kappa(l)$. Observations of
$M_\mathrm{ap,rms}(\theta)$ are therefore most suitable to obtain
information on the matter power spectrum (cf.~\cite*{bas99}).

\subsubsection{\label{sc:6.6.7}Power Spectrum and Filter Functions}

The three statistical measures discussed above, the magnification (or,
equivalently, the shear) auto-correlation function $\xi_\mu$, the
mean-square shear in apertures $\langle\gamma^2\rangle$, and the
mean-square aperture mass $\langle M_\mathrm{ap}^2\rangle$, are
related to the effective-convergence power spectrum $P_\kappa$ in very
similar ways. According to eqs.~(\ref{eq:6.36}), (\ref{eq:6.46}), and
(\ref{eq:6.51}), they can all be written in the form
\begin{equation}
  Q(\theta) = 2\pi\int_0^\infty\,l\d l\,P_\kappa(l)\,F(l\theta)\;,
\label{eq:6.57}
\end{equation}
where the filter functions $F(\eta)$ are given by
\begin{equation}
  F(\eta) = \left\{\begin{array}{l@{\quad}l@{\quad}l}
    \displaystyle
    \frac{\mathrm{J}_0(\eta)}{\pi^2} & \hbox{for} & Q=\xi_\mu \\
    \displaystyle
    \left[\frac{\mathrm{J}_1(\eta)}{\pi\eta}\right]^2 & \hbox{for} &
      Q=\langle\gamma_\mathrm{av}^2\rangle \\
    \displaystyle
    \left[\frac{12\,\mathrm{J}_4(\eta)}{\pi\eta^2}\right]^2 &
      \hbox{for} & Q=\langle M_\mathrm{ap}^2\rangle \\
  \end{array}\right.\;.
\label{eq:6.58}
\end{equation}

\begin{figure}[ht]
  \includegraphics[width=\hsize]{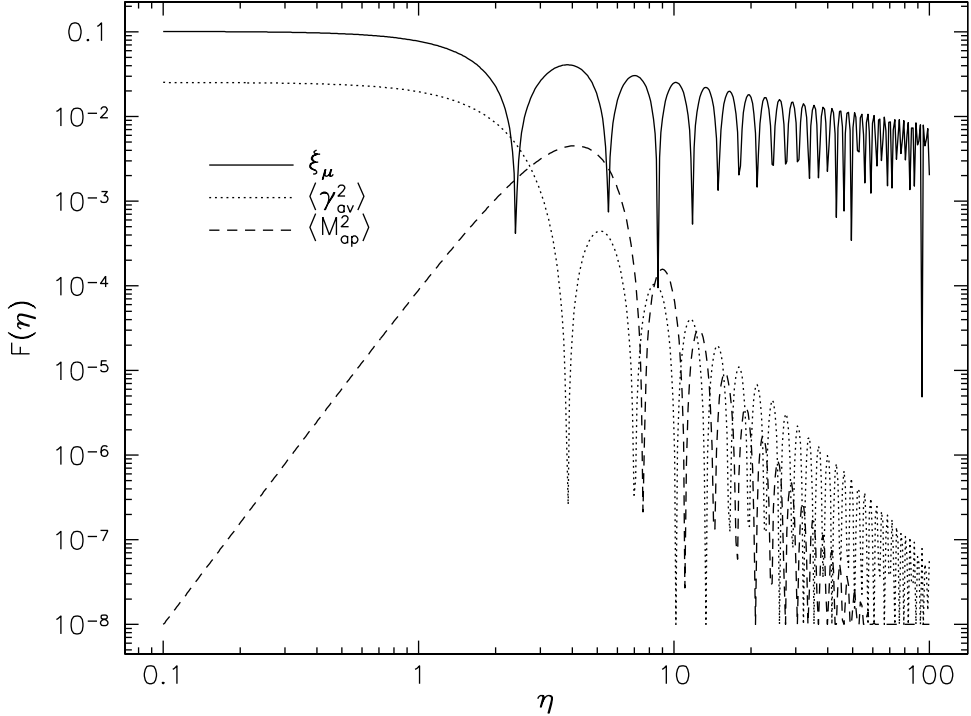}
\caption{The three filter functions $F(\eta)$ defined in
eq.~(\ref{eq:6.58}) are shown as functions of $\eta=l\theta$. They
occur in the expressions for the magnification auto-correlation
function, $\xi_\mu$ (solid curve), the mean-square shear in apertures,
$\langle\gamma^2\rangle$ (dotted curve), and the mean-square aperture
mass, $\langle M_\mathrm{ap}^2\rangle$ (dashed curve).}
\label{fig:6.8}
\end{figure}

Figure~\ref{fig:6.8} shows these three filter functions as functions
of $\eta=l\theta$.  Firstly, the curves illustrate that the amplitude
of $\xi_\mu$ is largest (owing to the factor of four relative to the
definition of $\xi_\gamma$), and that of $\langle
M_\mathrm{ap}^2\rangle$ is smallest because the amplitudes of the
filter functions themselves decrease. Secondly, it becomes evident
that, for given $\theta$, the range of $l$ modes of the
effective-convergence power spectrum $P_\kappa(l)$ convolved into the
weak-lensing estimator is largest for $\xi_\mu$ and smallest for
$\langle M_\mathrm{ap}^2\rangle$. Thirdly, the envelope of the filter
functions for large $\eta$ decreases most slowly for $\xi_\mu$ and
most rapidly for $\langle M_\mathrm{ap}^2\rangle$. Although the
aperture mass has the smallest signal amplitude, it is a much better
probe for the effective-convergence power spectrum $P_\kappa(l)$ than
the other measures because it picks up the smallest range of $l$ modes
and most strongly suppresses the $l$ modes smaller or larger than its
peak location.

We can therefore conclude that, while the strongest weak-lensing
signal is picked up by the magnification auto-correlation function
$\xi_\mu$, the aperture mass is the weak-lensing estimator most
suitable for extracting information on the effective-convergence power
spectrum.

\subsubsection{\label{sc:6.6.8}Signal-to-Noise Estimate of
  Aperture-Mass Measurements}

The question then arises whether the aperture mass can be measured
with sufficient significance in upcoming wide-field imaging
surveys. In practice, $M_\mathrm{ap}$ is derived from observations of
image distortions of faint background galaxies, using
eq.~(\ref{eq:5.28}, page~\pageref{eq:5.28}) and replacing the integral
by a sum over galaxy ellipticities. If we consider $N_\mathrm{ap}$
independent apertures with $N_i$ galaxies in the $i$-th aperture, an
unbiased estimator of $\langle M_\mathrm{ap}^2 \rangle$ is
\begin{equation}
  \mathcal{M} = \frac{(\pi\theta^2)^2}{N_\mathrm{ap}}\,
  \sum_{i=1}^{N_\mathrm{ap}}
  \frac{1}{N_i (N_i-1)}\sum_{j\ne k}^{N_i}Q_{ij}Q_{ik}
  \epsilon_{\mathrm{t},ij}\,\epsilon_{\mathrm{t},ik}\;,
\label{eq:6.N1}
\end{equation}
where $Q_{ij}$ is the value of the weight function at the position of
the $j$-th galaxy in the $i$-th aperture, and
$\epsilon_{\mathrm{t},ij}$ is defined accordingly.

The noise properties of this estimator were investigated in
\cite{svj98}. One source of noise comes from the fact that galaxies
are not intrinsically circular, but rather have an intrinsic
ellipticity distribution. A second contribution to the noise is due to
the random galaxy positions, and a third one to cosmic (or sampling)
variance. Under the assumptions that the number of galaxies $N_i$ in
the apertures is large, $N_i\gg1$, it turns out that the second of
these contributions can be neglected compared to the other two. For
this case, and assuming for simplicity that all $N_i$ are equal,
$N_i\equiv N$, the signal-to-noise of the estimator $\mathcal{M}$
becomes
\begin{equation}
  \frac{\mathrm{S}}{\mathrm{N}} \equiv
  \frac{\langle M_\mathrm{ap}^2\rangle}{\sigma(\mathcal{M})}
  = N_\mathrm{ap}^{1/2}\left[\mu_4+\left(\sqrt{2}+
  \frac{6\sigma_\epsilon^2}{5\sqrt{2}N\langle M_\mathrm{ap}^2\rangle}
  \right)^2\right]^{-1}\;,
\label{eq:6.59}
\end{equation}
where $\sigma_\epsilon\approx0.2$ (e.g.~\cite*{hgd98}) is the
dispersion of the intrinsic galaxy ellipticities, and $\mu_4=\langle
M_\mathrm{ap}^4\rangle/ \langle M_\mathrm{ap}^2\rangle^2-3$ is the
curtosis of $M_\mathrm{ap}$, which vanishes for a Gaussian
distribution. The two terms of (\ref{eq:6.59}) in parentheses
represent the noise contributions from Gaussian sampling variance and
the intrinsic ellipticity distribution, respectively, and $\mu_4$
accounts for sampling variance in excess of that for a Gaussian
distribution. On angular scales of a few arc minutes and smaller, the
intrinsic ellipticities dominate the noise, while the cosmic variance
dominates on larger scales.

\begin{figure}
  \includegraphics[width=\hsize]{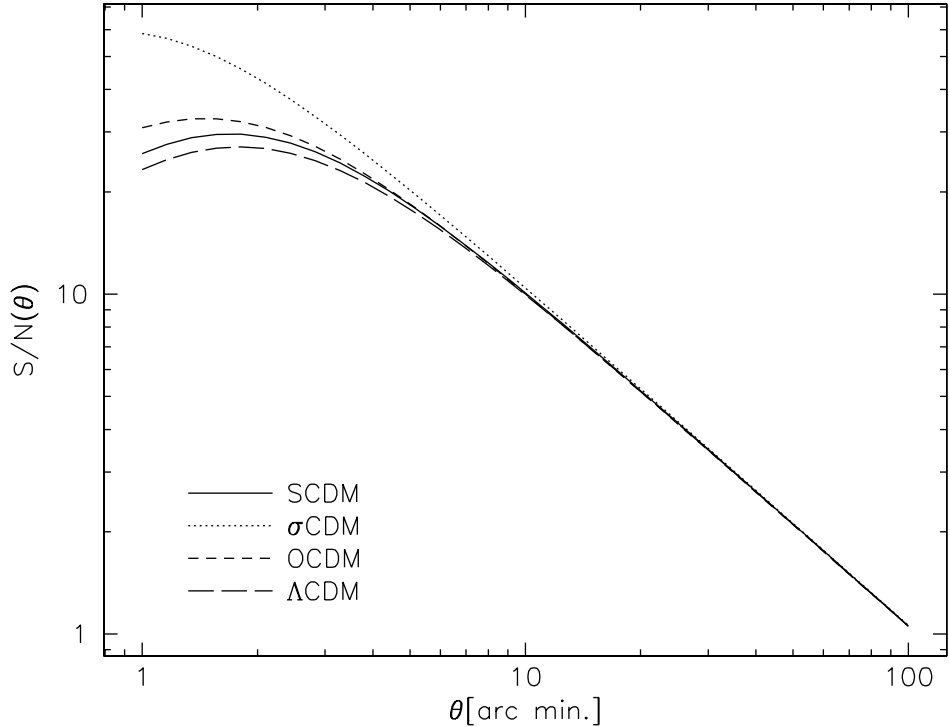}
\caption{The signal-to-noise ratio $\mathrm{S}/\mathrm{N}(\theta)$ of
measurements of mean-square aperture masses $\langle
M_\mathrm{ap}^2\rangle$ is plotted as a function of aperture radius
$\theta$ for an experimental setup as described in the text. The
curtosis was set to zero here. The four curves are for the four
different realisations of the CDM cosmogony listed in
Tab.~\ref{tab:6.1}. Solid curve: SCDM; dotted curve: $\sigma$CDM;
short-dashed curve: OCDM; and long-dashed curve: $\Lambda$CDM. Quite
independently of the cosmological parameters, the signal-to-noise
ratio $\mathrm{S}/\mathrm{N}$ reaches values of $>10$ on scales of
$\approx1'-2'$.}
\label{fig:6.9}
\end{figure}

Another convenient and useful property of the aperture mass
$M_\mathrm{ap}$ follows from its filter function being narrow, namely
that $M_\mathrm{ap}$ is a well localised measure of cosmic weak
lensing. This implies that $M_\mathrm{ap}$ measurements in
neighbouring apertures are almost uncorrelated even if the aperture
centres are very close (\cite*{svj98}). It is therefore possible to
gain a large number of (almost) independent $M_\mathrm{ap}$
measurements from a single large data field by covering the field
densely with apertures. This is a significant advantage over the other
two measures for weak lensing discussed above, whose broad filter
functions introduce considerable correlation between neighbouring
measurements, implying that for their measurement imaging data on
widely separated fields are needed to ensure statistical independence.
Therefore, a meaningful strategy to measure cosmic shear consists in
taking a large data field, covering it densely with apertures of
varying radius $\theta$, and determining $\langle
M_\mathrm{ap}^2\rangle$ in them via the ellipticities of galaxy
images. Figure~\ref{fig:6.9} shows an example for the signal-to-noise
ratio of such a measurement that can be expected as a function of
aperture radius $\theta$.

Computing the curves in Fig.~\ref{fig:6.9}, we assumed that a data
field of size $5^\circ\times5^\circ$ is available which is densely
covered by apertures of radius $\theta$, hence the number of (almost)
independent apertures is $N_\mathrm{ap}=(300'/2\theta)^2$. The number
density of galaxies was taken as $30\,\mathrm{arcmin}^{-2}$, and the
intrinsic ellipticity dispersion was assumed to be
$\sigma_\epsilon=0.2$. Evidently, high signal-to-noise ratios of $>10$
are reached on angular scales of $\approx1'$ in cluster-normalised
universes quite independent of the cosmological parameters. The
decline of $\mathrm{S}/\mathrm{N}$ for large $\theta$ is due to the
decreasing number of independent apertures on the data field, whereas
the decline for small $\theta$ is due to the decrease of the signal
$\langle M_\mathrm{ap}^2\rangle$, as seen in Fig.~\ref{fig:6.6}. We
also note that for calculating the curves in Fig.~\ref{fig:6.9}, we
have put $\mu_4=0$. This is likely to be an overly optimistic
assumption for small angular scales where the density field is highly
non-linear. Unfortunately, $\mu_4$ cannot easily be estimated
analytically. It was numerically derived from ray-tracing through
$N$-body simulations of large-scale matter distributions by
\cite{rkj99}. The curtosis exceeds unity even on scales as large as
$10'$, demonstrating the highly non-Gaussian nature of the
non-linearly developed density perturbations.

Although the aperture mass is a very convenient measure of cosmic
shear and provides a localised estimate of the projected power
spectrum $P_\kappa(l)$ [see (\ref{eq:6.55})], it is by no means clear
that it is an optimal measure for the projected power
spectrum. \cite{KA98.1} considered the case of a square-shaped data
field and employed the Fourier-transformed Kaiser \& Squires inversion
formula, eq.~(\ref{eq:5.3}, page~\pageref{eq:5.3}). The Fourier
transform of the shear is then replaced by a sum over galaxy
ellipticities $\epsilon_i$, so that $\hat\kappa_\mathrm{eff}(\vec l)$
is expressed directly in terms of the $\epsilon_i$. The square
$|\hat\kappa_\mathrm{eff}(\vec l)|^2$ yields an estimate for the power
spectrum which allows a simple determination of the noise coming from
the intrinsic ellipticity distribution.  As \cite{KA98.1} pointed out
that, while this noise is very small for angular scales much smaller
than the size of the data field, the sampling variance is much larger,
so that different sampling strategies should be explored. For example,
he suggests to use a sparse sampling strategy. \cite{sel98} developed
an estimator for the power spectrum which achieves minimum variance in
the case of a Gaussian field. Since the power spectrum $P_\kappa(l)$
deviates significantly from its linear prediction on angular scales
below one degree, one expects that the field attains significant
non-Gaussian features on smaller angular scales, so that this
estimator does no longer need to have minimum variance.

\subsection{\label{sc:6.7}Higher-Order Statistical Measures}

\subsubsection{\label{sc:6.7.1}The Skewness}

As the density perturbation field $\delta$ grows with time, it
develops non-Gaussian features. In particular, $\delta$ is bounded by
$-1$ from below and unbounded from above, and therefore the
distribution of $\delta$ is progressively skewed while evolution
proceeds. The same then applies to quantities like the effective
convergence $\kappa_\mathrm{eff}$ derived from $\delta$
(cf.~\cite*{JA97.2}; \cite*{BE97.5}; \cite*{svj98}). Skewness of the
effective convergence can be quantified by means of the three-point
correlator of $\kappa_\mathrm{eff}$. In order to compute that, we use
expression (\ref{eq:6.23}), Fourier transform it, and also express the
density contrast $\delta$ in terms of its Fourier
transform. Additionally, we employ the same approximation used in
deriving Limber's equation in Fourier space, namely that correlations
of the density contrast {\em along\/} the line-of-sight are negligibly
small. After carrying out this lengthy but straightforward procedure,
the three-point correlator of the Fourier transform of
$\kappa_\mathrm{eff}$ reads (suppressing the subscript `eff' for
brevity)
\begin{eqnarray}
  \langle
    \hat\kappa(\vec l_1)\hat\kappa(\vec l_2)\hat\kappa(\vec l_3)
  \rangle &=& \frac{27H_0^6\Omega_0^3}{8c^6}\,
  \int_0^{w_\mathrm{H}}\,\d w\,\frac{\bar{W}^3(w)}{a^3(w)\,f_K^3(w)}\,
  \int_{-\infty}^\infty\frac{\d k_3}{2\pi}\,\exp({\rm i}k_3w)\,
  \nonumber\\
  &\times&
  \left\langle
    \hat\delta\left(\frac{\vec l_1}{f_K(w)},k_3\right)
    \hat\delta\left(\frac{\vec l_2}{f_K(w)},0\right)
    \hat\delta\left(\frac{\vec l_3}{f_K(w)},0\right)
  \right\rangle\;.
\label{eq:6.61}
\end{eqnarray}
Hats on symbols denote Fourier transforms. Note the fairly close
analogy between (\ref{eq:6.61}) and (\ref{eq:6.25}): The three-point
correlator of $\hat\kappa$ is a distance-weighted integral over the
three-point correlator of the Fourier-transformed density contrast
$\hat\delta$. The fact that the three-component $k_3$ of the wave
vector $\vec k$ appears only in the first factor $\hat\delta$ reflects
the approximation mentioned above, i.e.~that correlations of $\delta$
along the line-of-sight are negligible.

\begin{figure}[ht]
  \includegraphics[width=\hsize]{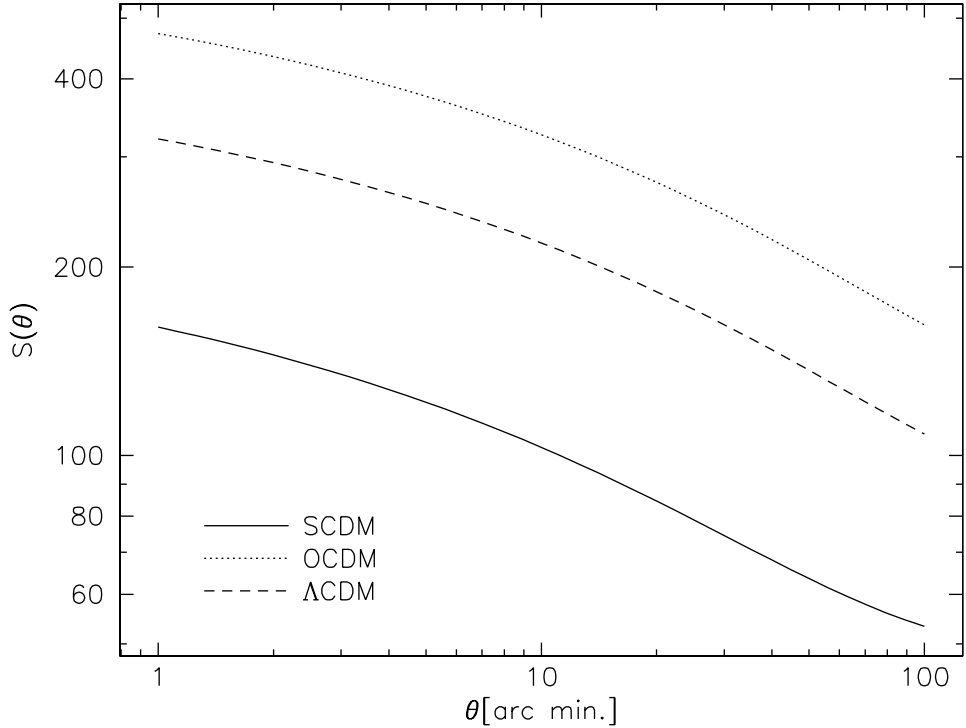}
\caption{The skewness $\mathcal{S}(\theta)$ of the aperture mass
$M_\mathrm{ap}(\theta)$ is shown as a function of aperture radius
$\theta$ for three of the realisations of the cluster-normalised CDM
cosmogony listed in Tab.~\ref{tab:6.1}: SCDM (solid curve); OCDM
(dotted curve); and $\Lambda$CDM (dashed curve). The source redshift
was assumed to be $z_\mathrm{s}=1$.}
\label{fig:6.10}
\end{figure}

Suppose now that the density contrast $\delta$ is expanded in a
perturbation series, $\delta=\sum\delta^{(i)}$ such that
$\delta^{(i)}=\mathcal{O}([\delta^{(1)}]^i)$, and truncated after the
second order. The three-point correlator of $\hat\delta^{(1)}$
vanishes because $\delta$ remains Gaussian to first perturbation
order. The lowest-order, non-vanishing three-point correlator of
$\delta$ can therefore symbolically be written
$\langle\hat\delta^{(1)}\hat\delta^{(1)}\hat\delta^{(2)}\rangle$, plus
two permutations of that expression. The second-order density
perturbation is related to the first order through (\cite*{fry84};
\cite*{ggr86}; \cite*{bjc92})
\begin{equation}
  \hat\delta^{(2)}(\vec k,w) = D_+^2(w)\,\int\frac{\d^3k'}{(2\pi)^3}\,
  \hat\delta_0^{(1)}(\vec k')\hat\delta_0^{(1)}(\vec k-\vec k')\,
  F(\vec k',\vec k-\vec k')\;,
\label{eq:6.62}
\end{equation}
where $\delta_0^{(1)}$ is the first-order density perturbation
linearly extrapolated to the present epoch, and $D_+(w)$ is the linear
growth factor, $D_+(w)=a(w)\,g[a(w)]$ with $g(a)$ defined in
eq.~(\ref{eq:2.51}) on page~\pageref{eq:2.51}. The function $F(\vec
x,\vec y)$ is given by
\begin{equation}
  F(\vec x,\vec y) = \frac{5}{7} + \frac{1}{2}\,\left(
    \frac{1}{|\vec x|^2}+\frac{1}{|\vec y|^2}
  \right)\vec x\cdot\vec y +
  \frac{2}{7}\frac{(\vec x\cdot\vec y)^2}{|\vec x|^2|\vec y|^2}\;.
\label{eq:6.63}
\end{equation}
Relation (\ref{eq:6.62}) implies that the lowest-order three-point
correlator
$\langle\hat\delta^{(1)}\hat\delta^{(1)}\hat\delta^{(2)}\rangle$
involves four-point correlators of $\hat\delta^{(1)}$. For Gaussian
fields like $\delta^{(1)}$, four-point correlators can be decomposed
into sums of products of two-point correlators, which can be expressed
in terms of the linearly extrapolated density power spectrum
$P_\delta^{(0)}$. This leads to
\begin{eqnarray}
  \langle
    \hat\delta^{(1)}(\vec k_1)
    \hat\delta^{(1)}(\vec k_2)
    \hat\delta^{(2)}(\vec k_3)
  \rangle &=& 2\,(2\pi)^3\,D_+^4(w)\,
  P_\delta^{(0)}(k_1)P_\delta^{(0)}(k_2)\,\nonumber\\
  &\times&
  \delta_\mathrm{D}(\vec k_1+\vec k_2+\vec k_3)\,
  F(\vec k_1,\vec k_2)\;.
\label{eq:6.64}
\end{eqnarray}
The complete lowest-order three-point correlator of $\hat\delta$ is a
sum of three terms, namely the left-hand side of (\ref{eq:6.64}) and
two permutations thereof. Each permutation yields the same result, so
that the complete correlator is three times the right-hand side of
(\ref{eq:6.64}). We can now work our way back, inserting the
three-point density correlator into eq.~(\ref{eq:6.61}) and
Fourier-transforming the result with respect to $\vec l_{1,2,3}$. The
three-point correlator of the effective convergence so obtained can
then in a final step be used to compute the third moment of the
aperture mass. The result is (\cite*{svj98})
\begin{eqnarray}
  \langle M_\mathrm{ap}^3(\theta)\rangle &=&
  \frac{81H_0^6\Omega_0^3}{8\pi c^6}\,\int_0^{w_\mathrm{H}}\d w\,
  \frac{\bar{W}^3(w)D_+^4(w)}{a^3(w)f_K(w)}\nonumber\\
  &\times&
  \int\d^2l_1\,
    P_\delta^{(0)}\left(\frac{l_1}{f_K(w)}\right)\,J^2(l_1\theta)
  \nonumber\\
  &\times&
  \int\d^2l_2\,
    P_\delta^{(0)}\left(\frac{l_2}{f_K(w)}\right)\,J^2(l_2\theta)\,
  J^2(|\vec l_1+\vec l_2|\theta)\,F(\vec l_1,\vec l_2)\;,
\label{eq:6.65}
\end{eqnarray}
with the filter function $J(\eta)$ defined in eq.~(\ref{eq:6.52}).
Commonly, third-order moments are expressed in terms of the skewness,
\begin{equation}
  \mathcal{S}(\theta) \equiv 
  \frac{\langle M_\mathrm{ap}^3(\theta)\rangle}
  {\langle M_\mathrm{ap}^2(\theta)\rangle^2}\;,
\label{eq:6.66}
\end{equation}
where $\langle M_\mathrm{ap}^2(\theta)\rangle$ is calculated with the
linearly evolved power spectrum.  As seen earlier in
eq.~(\ref{eq:6.51}), $\langle M_\mathrm{ap}^2\rangle$ scales with the
amplitude of the power spectrum, while $\langle
M_\mathrm{ap}^3\rangle$ scales with the square of it. In this
approximation, the skewness $\mathcal{S}(\theta)$ is therefore
independent of the normalisation of the power spectrum, removing that
major uncertainty and leaving cosmological parameters as primary
degrees of freedom. For instance, the skewness $\mathcal{S}(\theta)$
is expected to scale approximately with
$\Omega_0^{-1}$. Figure~\ref{fig:6.10} shows three examples.

As expected, lower values of $\Omega_0$ yield larger skewness, and the
skewness is reduced when $\Omega_\Lambda$ is increased keeping
$\Omega_0$ fixed. Despite the sensitivity of $\mathcal{S}(\theta)$ to
the cosmological parameters, it should be noted that the source
redshift distribution [entering through $\bar{W}(w)$] needs to be
known sufficiently well before attempts can be made at constraining
cosmological parameters through measurements of the aperture-mass
skewness. However, photometric redshift estimates are expected to
produce sufficiently well-constrained redshift distributions in the
near future (\cite*{ccs95}; \cite*{gwh96}; \cite*{hcb98}).

We have confined the discussion of the skewness to the aperture mass
since $M_\mathrm{ap}$ is a scalar measure of the cosmic shear which
can directly be expressed in terms of the observed image
ellipticities. One can of course also consider the skewness directly
in terms of $\kappa$, since $\kappa$ can be obtained from the observed
image ellipticities through a mass reconstruction algorithm as
described in Sect.~\ref{sc:5}. Analytical and numerical results for
this skewness have been presented in, e.g., \cite{BE97.5},
\cite{vbm99}, \cite{jsw99} and \cite{rkj99}. We shall discuss some of
their results in Sect.~\ref{sec:6.9.1}.

\subsubsection{\label{sc:6.7.2}Number density of (dark) haloes}

In Sect.~\ref{sc:5.3.1}, we discussed the possibility to detect mass
concentrations by their weak lensing effects on background galaxies by
means of the aperture mass. The number density of mass concentrations
that can be detected at a given threshold of $M_\mathrm{ap}$ depends
on the cosmological model. Fixing the normalisation of the power
spectrum so that the the local abundance of massive clusters is
reproduced, the evolution of the density field proceeds differently in
different cosmologies, and so the abundances will differ at redshifts
$z\sim0.3$ where the aperture-mass method is most sensitive.

The number density of haloes above a given threshold of
$M_\mathrm{ap}(\theta)$ can be estimated analytically, using two
ingredients. First, the spatial number density of haloes at redshift
$z$ with mass $M$ can be described by the Press-Schechter theory
(\cite*{prs74}), which numerical simulations (\cite*{LA93.3},
\cite*{LA94.2}) have shown to be a fairly accurate approximation.
Second, in a series of very large $N$-body simulations,
\citename{nfw96} (\citeyear{nfw96}, \citeyear{nfw97}) found that dark
matter haloes have a universal density profile which can be described
by two parameters, the halo mass and a characteristic scale length,
which depends on the cosmological model and the redshift. Combining
these two results from cosmology, \cite{krs99} calculated the number
density of haloes exceeding $M_\mathrm{ap}$. Using the signal-to-noise
estimate eq.~(\ref{eq:6.59}), a threshold value of $M_\mathrm{ap}$ can
be directly translated into a signal-to-noise threshold
$S_\mathrm{c}$. For an assumed number density of
$n=30\,\mathrm{arcmin}^{-2}$ and an ellipticity dispersion
$\sigma_\epsilon=0.2$, one finds $S_\mathrm{c}\approx 0.016
(\theta/1\mathrm{arcmin})M_\mathrm{ap}(\theta)$.

For the redshift distribution (\ref{eq:2.68}, page~\pageref{eq:2.68})
with $\beta=3/2$ and $z_0=1$, the number density of haloes with
$S_\mathrm{c}\ge 5$ exceeds 10 per square degree for
cluster-normalised cosmologies, across angular scales $1'\lesssim
\theta\lesssim 10'$, and these haloes have a broad redshift
distribution which peaks at $z_\mathrm{d}\sim 0.3$.  This implies that
a wide-field imaging survey should be able to detect a statistically
interesting sample of medium redshift haloes, thus allowing the
definition of a {\em mass-selected\/} sample of haloes. Such a sample
will be of utmost interest for cosmology, since the halo abundance is
considered to be one of the most sensitive cosmological probes (e.g.,
\cite*{EK96.1}, \cite*{bah98}). Current attempts to apply this tool
are hampered by the fact that haloes are selected either by the X-ray
properties or by their galaxy content. These properties are much more
difficult to predict than the dark matter distribution of haloes which
can directly be determined from cosmological $N$-body
simulations. Thus, these mass-selected haloes will provide a much
closer link to cosmological predictions than currently
possible. \cite{krs99} estimated that an imaging survey of several
square degrees will allow one to distinguish between the cosmological
models given in Table~\ref{tab:6.1}, owing to the different number
density of haloes that they predict. Using the aperture-mass
statistics, \cite{ewm99} recently detected a highly significant matter
concentration on two independent wide-field images centred on the
galaxy cluster A~1942. This matter concentration $7'$ South of A~1942
is not associated with an overdensity of bright foreground galaxies,
which sets strong lower limits on the mass-to-light ratio of this
putative cluster.

\subsection{\label{sc:6.8}Cosmic Shear and Biasing}

Up to now, we have only considered the mass properties of the
large-scale structure and tried to measure them with weak lensing
techniques. An interesting question arises when the luminous
constituents of the Universe are taken into account. Most importantly,
the galaxies are supposed to be strongly tied to the distribution of
dark matter. In fact, this assumption underlies all attempts to
determine the power spectrum of cosmic density fluctuations from the
observed distribution of galaxies.  The relation between the galaxy
and dark-matter distributions is parameterised by the so-called
biasing factor $b$ (\cite*{kai84}), which is defined such that the
relative fluctuations in the spatial number density of galaxies are
$b$ times the relative density fluctuations $\delta$,
\begin{equation}
  \frac{n(\vec x)-\langle n\rangle}{\langle n\rangle} =
  b\,\delta(\vec x)\;,
\label{eq:6.67}
\end{equation}
where $\langle n\rangle$ denotes the mean spatial number density of
galaxies at the given redshift. The bias factor $b$ is not really a
single number, but generally depends on redshift, on the spatial
scale, and on the galaxy type (see, e.g., \cite*{EF96.1},
\cite*{pea97}, \cite*{kns97}, \cite*{clm98}). Typical values for the
bias factor are assumed to be $b\sim1-2$ at the current epoch, but can
increase towards higher redshifts. The clustering properties of UV
dropout galaxies (\cite*{sad98}) indicate that $b$ can be as large as
5 at redshifts $z\sim3$, depending on the cosmology.

The projected surface mass density $\kappa_\mathrm{eff}(\vec\theta)$
should therefore be correlated with the number density of (foreground)
galaxies in that direction. Let $G_\mathrm{G}(w)$ be the distribution
function of a suitably chosen population of galaxies in comoving
distance (which can be readily converted to a redshift probability
distribution). Then, assuming that $b$ is independent of scale and
redshift, the number density of the galaxies is
\begin{equation}
  n_\mathrm{G}(\vec\theta)=\langle n_\mathrm{G} \rangle
  \left[ 1+b\int \d w\;G_\mathrm{G}(w)\,
  \delta(f_K(w)\vec\theta,w)\right]\;,
\label{eq:6.68}
\end{equation}
where $\langle n_\mathrm{G}\rangle$ is the mean number density of the
galaxy population. The distribution function $G_\mathrm{G}(w)$ depends
on the selection of galaxies. For example, for a flux-limited sample
it may be of the form (\ref{eq:2.68}). Narrower distribution functions
can be achieved by selecting galaxies in multi-colour space using
photometric redshift techniques. The correlation function between
$n_\mathrm{G}(\vec\theta)$ and $\kappa_\mathrm{eff}(\vec\theta)$ can
directly be obtained from eq.~(\ref{eq:2.82}) by identifying $q_1(w)=
3 H_0^2\Omega_0 \bar W(w) f_K(w)/[2 c^2 a(w)]$ [see
eq.~(\ref{eq:6.23})], and $q_2(w)=\langle n_\mathrm{G} \rangle b
G_\mathrm{G}(w)$. It reads
\begin{eqnarray}
  \xi_{\mathrm{G}\kappa}(\theta) &\equiv &
  \langle n_\mathrm{G}\kappa_\mathrm{eff} \rangle (\theta)
  =\frac{3 H_0^2\Omega_0}{2c^2}\,b\langle n_\mathrm{G} \rangle 
  \int\d w\,\frac{\bar W(w) f_K(w)}{a(w)} G_\mathrm{G}(w)\nonumber \\
  &\times &\int\frac{\d k\;k}{2\pi}\,P_\delta(k,w)\,
  \mathrm{J}_0(f_K(w)\theta  k)\;.
\label{eq:6.69}
\end{eqnarray}
Similar equations were derived by, e.g., \cite{KA92.1}, \cite{BA95.4},
\cite{DO97.1}, \cite{SA97.3}.

One way to study the correlation between foreground galaxies and the
projected density field consists in correlating the aperture mass
$M_\mathrm{ap}(\theta)$ with a similarly filtered galaxy number
density, defined as
\begin{equation}
  \mathcal{N}(\theta)=\int\d^2\vartheta\;U(|\vec\vartheta|)\,
  n_\mathrm{G}(\vec\vartheta)\;,
\label{eq:6.70}
\end{equation}
with the same filter function $U$ as in $M_\mathrm{ap}$. The
correlation between $M_\mathrm{ap}(\theta)$ and $\mathcal{N}(\theta)$
then becomes
\begin{eqnarray}
  \xi(\theta) &\equiv&
  \langle M_\mathrm{ap}(\theta)\mathcal{N}(\theta)\rangle =
  \int\d^2\vartheta\;U(|\vec\vartheta|)
  \int\d^2\vartheta'\;U(|\vec\vartheta'|)
  \xi_{\mathrm{G}\kappa}(|\vec\vartheta-\vec\vartheta'|)
  \label{eq:6.71}\\ &=&
  3\pi\left(\frac{H_0}{c}\right)^2\Omega_0 b
  \langle n_\mathrm{G}\rangle
  \int\d w\,\frac{\bar W(w) G_\mathrm{G}(w)}{a(w) f_K(w)}
  \int\d l\,l\,P_\delta\left(\frac{l}{f_K(w)},w\right)\,
  J^2(l\theta)\;,\nonumber
\end{eqnarray}
where we used eq.~(\ref{eq:2.82}) for the correlation function
$\xi_{\mathrm{G}\kappa}$ in the final step. The filter function $J$ is
defined in eq.~(\ref{eq:6.52}). Note that this correlation function
filters out the power spectrum $P_\delta$ at redshifts where the
foreground galaxies are situated. Thus, by selecting galaxy
populations with narrow redshift distribution, one can study the
cosmological evolution of the power spectrum or, more accurately, the
product of the power spectrum and the bias factor.

The convenient property of this correlation function is that one can
define an unbiased estimator for $\xi$ in terms of observables. If
$N_\mathrm{b}$ galaxies are found in an aperture of radius $\theta$ at
positions $\vec\vartheta_i$ with tangential ellipticity
$\epsilon_{\mathrm{t}i}$, and $N_\mathrm{f}$ foreground galaxies at
positions $\vec\varphi_i$, then
\begin{equation}
  \tilde\xi(\theta)=\frac{\pi\theta^2}{N_\mathrm{b}}
  \sum_{i=1}^{N_\mathrm{b}}Q(|\vec\vartheta_i|)\,
  \epsilon_{\mathrm{t}i}
  \sum_{k=1}^{N_\mathrm{f}}U(|\varphi_k|)
\label{eq:6.72}
\end{equation}
is an unbiased estimator for $\xi(\theta)$. \cite{SC98.2} calculated
the noise properties of this estimator, concentrating on an
Einstein-de Sitter model and a linearly evolving power spectrum which
can locally be approximated by a power law in $k$. A more general and
thorough treatment is given in \cite{VA98.1}, where various
cosmological models and the non-linear power spectrum are considered.
\cite{VA98.1} assumed a broad redshift distribution for the background
galaxies, but a relatively narrow redshift distribution for the
foreground galaxies, with $\delta z_\mathrm{d}/z_\mathrm{d}\sim
0.3$. For an open model with $\Omega_0=0.3$, $\xi(\theta)$ declines
much faster with $\theta$ than for flat models, implying that open
models have relatively more power on small scales at intermediate
redshift. This is a consequence of the behaviour of the growth factor
$D_+(w)$; see Fig.~\ref{fig:2.3} on page~\pageref{fig:2.3}. For
foreground redshifts $z_\mathrm{d}\gtrsim0.2$, the signal-to-noise
ratio of the estimator (\ref{eq:6.72}) for a single aperture is
roughly constant for $\theta\gtrsim5'$, and relatively independent of
the exact value of $z_\mathrm{d}$ over a broad redshift interval, with
a characteristic value of $\sim0.4$.

\cite{VA98.1} also considered the ratio
\begin{equation}
  R\equiv\frac{\xi(\theta)}{\langle\mathcal{N}^2(\theta)\rangle}
\label{eq:6.73}
\end{equation}
and found that it is nearly independent of $\theta$. This result was
shown in \cite{SC98.2} to hold for linearly evolving power spectra
with power-law shape, but surprisingly it also holds for the fully
non-linear power spectrum. Indeed, varying $\theta$ between $1'$ and
$100'$, $R$ varies by less than 2\% for the models considered in
\cite{VA98.1}. This is an extremely important result, in that any
observed variation of $R$ with angular scale indicates a corresponding
scale dependence of the bias factor $b$. A direct observation of this
variation would provide valuable constraints on the models for the
formation and evolution of galaxies.

\subsection{\label{sec:6.9}Numerical Approach to Cosmic Shear, 
  Cosmological Parameter Estimates, and Observations}

\subsubsection{\label{sec:6.9.1}Cosmic Shear Predictions from
  Cosmological Simulations}

So far, we have treated the lensing effect of the large-scale
structure with analytic means. This was possible because of two
assumptions. First, we considered only the lowest-order lensing
effect, by employing the Born approximation and neglecting lens-lens
coupling in going from eq.~(\ref{eq:6.15}) to
eq.~(\ref{eq:6.16}). Second, we used the prescription for the
non-linear power spectrum as given by \cite{PE96.3}, assuming that it
is a sufficiently accurate approximation. Both of these approximations
may become less accurate on small angular scales. Providing a
two-point quantity, the analytic approximation of $P_\kappa$ is
applicable only for two-point statistical measures of cosmic shear.
In addition, the error introduced with these approximations cannot be
controlled, i.e., we cannot attach `error bars' to the analytic
results.

A practical way to avoid these approximations is to study the
propagation of light in a model universe which is generated by
cosmological structure-formation simulations. They typically provide
the three-dimensional mass distribution at different redshifts in a
cube whose side-length is much smaller than the Hubble radius. The
mass distribution along a line-of-sight can be generated by combining
adjacent cubes from a sequence of redshifts. The cubes at different
redshifts should either be taken from different realisations of the
initial conditions, or, if this requires too much computing time, they
should be translated and rotated such as to avoid periodicity along
the line-of-sight. The mass distribution in each cube can then be
projected along the line-of-sight, yielding a surface mass density
distribution at that redshift. Finally, by employing the multiple
lens-plane equations, which are a discretisation of the propagation
equation (\ref{eq:6.15}; \cite*{SE94.5}), shear and magnification can
be calculated along light rays within a cone whose size is determined
by the side length of the numerical cube. This approach was followed
by many authors (e.g., \cite*{JA90.1}, \cite*{JA91.1}, \cite*{BA91.2},
\cite*{BL91.1}, \cite*{WA95.3}), but the rapid development of $N$-body
simulations of the cosmological dark matter distribution render the
more recent studies particularly useful (\cite*{WA98.1},
\cite*{vbm99}, \cite*{jsw99}).

As mentioned below eq.~(\ref{eq:6.34}), the Jacobian matrix
$\mathcal{A}$ is generally asymmetric when the propagation equation is
not simplified to (\ref{eq:6.16}). Therefore, the degree of asymmetry
of $\mathcal{A}$ provides one test for the accuracy of this
approximation. \cite{jsw99} found that the power spectrum of the
asymmetric component is at least three orders of magnitude smaller
than that of $\kappa_\mathrm{eff}$. For a second test, we have seen
that the power spectrum of $\kappa_\mathrm{eff}$ should equal that of
the shear in the frame of our approximations. This analytic prediction
is very accurately satisfied in the numerical simulations.

\cite{jsw99} and \cite{rkj99} found that analytic predictions of the
dispersions of $\kappa$ and $M_\mathrm{ap}$ respectively, are very
accurate when compared to numerical results. For both cosmic shear
measures, however, the analytic predictions of the skewness are not
satisfactory on angular scales below $\sim 10'$. This discrepancy
reflects the limited accuracy of the second-order Eulerian
perturbation theory employed in deriving the analytic
results. \cite{hui99} showed that the accuracy of the analytic
predictions can be much increased by using a prescription for the
highly-nonlinear three-point correlation function of the cosmic
density contrast, as developed by \cite{scf99}.

The signal-to-noise ratio of the dispersion of the cosmic shear, given
explicitly for $M_\mathrm{ap}$ in eq.~(\ref{eq:6.59}), is determined
by the intrinsic ellipticity dispersion of galaxies and the sampling
variance, expressed in terms of the curtosis. As shown in \cite{vbm99}
and \cite{rkj99}, this curtosis is remarkably large. For instance, the
curtosis of the aperture mass exceeds unity even on scales larger than
$10'$, revealing non-Gaussianity on such large scales. Unfortunately,
this large sampling variance implies not only that the area over which
cosmic shear needs to be measured to achieve a given accuracy for its
dispersion must be considerably larger than estimated for a Gaussian
density field, but also that numerical estimates of cosmic shear
quantities need to cover large solid angles for an accurate numerical
determination of the relevant quantities.

From such numerical simulations, one can not only determine moments of
the shear distribution, but also consider its full probability
distribution. For example, the predictions for the number density of
dark matter haloes that can be detected through highly significant
peaks of $M_\mathrm{ap}$ -- see Sect.~\ref{sc:6.7.2} -- have been
found by \cite{rkj99} to be fairly accurate, perhaps surprisingly so,
given the assumptions entering the analytic results. Similarly, the
extreme tail (say more than 5 standard deviations from the mean) of
the probability distribution for $M_\mathrm{ap}$, calculated
analytically in \cite{krs00}, does agree with the numerical results;
it decreases exponentially.

\subsubsection{\label{sec:6.9.2}Cosmological Parameter Estimates}

Since the cosmic shear described in this section directly probes the
total matter content of the universe, i.e., without any reference to
the relation between mass and luminosity, it provides an ideal tool to
investigate the large-scale structure of the cosmological density
field. Assuming the dominance of cold dark matter, the statistical
properties of the cosmic mass distribution are determined by a few
parameters, the most important of which are $\Omega_0$,
$\Omega_\Lambda$, the shape parameter of the power spectrum, $\Gamma$,
and the normalisation of the power spectrum expressed in terms of
$\sigma_8$. For each set of these parameters, the corresponding cosmic
shear signals can be predicted, and a comparison with observations
then constrains the cosmological parameters.

Several approaches to this parameter estimation have been discussed in
the literature. For example, \cite{vbm99} used numerical simulations
to generate synthetic cosmic shear data, fixing the normalisation of
the density fluctuations to $\sigma_8\,\Omega_0=0.6$, which is
essentially the normalisation by cluster abundance. A moderately wide
and deep weak-lensing survey, covering 25~square degrees and reaching
a number density of 30~galaxies per arcmin$^2$ with characteristic
redshift $z_\mathrm{s}\sim1$, will enable the distinction between an
Einstein-de Sitter model and an open universe with $\Omega_0=0.3$ at
the 6-$\sigma$ level, though each of these models is degenerate in the
$\Omega_0$ vs.~$\Omega_\Lambda$ plane. For this conclusion, only the
skewness of the reconstructed effective surface mass density or the
aperture mass was used. \cite{krs00} instead considered the highly
non-Gaussian tail of the aperture mass statistics to constrain
cosmological parameters, whereas \cite{krs99} considered the abundance
of highly significant peaks of $M_\mathrm{ap}$ as a probe of the
cosmological models. The peak statistics of reconstructed surface
density maps (\cite*{jvw99}) also provides a valuable means to
distinguish between various cosmological models.

Future work will also involve additional information on the redshifts
of the background galaxies. \cite{huu99} pointed out that splitting up
the galaxy sample into several redshift bins substantially increases
the ability to constrain cosmological parameters. He considered the
power spectrum of the projected density and found that the accuracy of
the corresponding cosmological parameters improves by a factor of
$\sim7$ for $\Omega_\Lambda$, and by a factor of $\sim3$ for
$\Omega_0$, estimated for a median redshift of unity.

All of the quoted work concentrated mainly on one particular measure
of cosmic shear. One goal of future theoretical investigations will
certainly be the construction of a method which combines the various
measures into a `global' statistics, designed to minimise the volume
of parameter space allowed by the data of future observational weak
lensing surveys. Future, larger-scale numerical simulations will guide
the search for such a statistics and allow one to make accurate
predictions.

In addition to a pure cosmic shear investigation, cosmic shear
constraints can be used in conjunction with other measures of
cosmological parameters. One impressive example has been given by
\cite{hut98}, who showed that even a relatively small weak lensing
survey could dramatically improve the accuracy of cosmological
parameters measured by future Cosmic Microwave Background missions.

\subsubsection{\label{sec:6.9.3}Observations}

We are not aware of any convincing and cosmologically useful
measurement of cosmic shear yet obtained. One of the first attempts
was reported in \cite{MO94.1}, where the mean shear was investigated
across a field of $9\arcminf6\times9\arcminf6$, observed with the Hale
5-meter Telescope. The image is very deep and has good quality (i.e., a
seeing of $0\arcsecf87$ FWHM). It is the same data as used by
\cite{BR96.1} for the first detection of galaxy-galaxy lensing (see
Sect.~\ref{sc:8}). The mean ellipticity of the 4363 galaxies within a
circle of $4\arcminf8$ radius with magnitudes $23\le r\le26$ was found
to be $(0.5\pm0.5)\%$. A later, less conservative reanalysis of these
data by Villumsen (unpublished), where an attempt was made to account
for the seeing effects, yielded a 3-$\sigma$ detection of a
non-vanishing mean ellipticity.

Following the suggestion that the observed large-angle QSO-galaxy
associations are due to weak lensing by the large-scale structure in
which the foreground galaxies are embedded (see Sect.~\ref{sc:7}),
\cite{FO96.1} searched for shear around five luminous radio
quasars. In one of the fields, the number density of stars was so high
that no reasonable shear measurement on faint background galaxies
could be performed.\footnote{This field was subsequently used to
demonstrate the superb image quality of the SUSI instrument on the ESO
NTT.} In the remaining four QSO fields, they found a shear signal on a
scale of $\sim1'$ for three of the QSOs (those which were observed
with SUSI, which has a field-of-view of $\sim2\arcminf2$), and on a
somewhat larger angular scale for the fourth QSO. Taken at face value,
these observations support the suggestion of magnification bias caused
by the large-scale structure. A reanalysis of the three SUSI fields by
\cite{SC98.3}, considering the {\em rms\/} shear over the fields,
produced a positive value for $\langle|\gamma|^2\rangle$ at the 99\%
significance level, as determined by numerous simulations randomising
the orientation angles of the galaxy ellipticities. The amplitude of
the {\em rms\/} shear, when corrected for the dilution by seeing, is
of the same magnitude as expected from cluster-normalised
models. However, if the magnification bias hypothesis is true, these
three lines-of-sight are not randomly selected, and therefore this
measurement is of no cosmological use.

Of course, one or a few narrow-angle fields cannot be useful for a
measurement of cosmic shear, owing to cosmic variance. Therefore, a
meaningful measurement of cosmic shear must either include many small
fields, or must be obtained from a wide-field survey. Using the first
strategy, several projects are under way: The Hubble Space Telescope
has been carrying out so-called parallel surveys, where one or more of
the instruments not used for primary observations are switched on to
obtain data of a field located a few arc minutes away from the primary
pointing. Over the past few years, a considerable database of such
parallel data sets has accumulated. Two teams are currently analysing
parallel data sets taken with WFPC2 and STIS, respectively (see
\cite*{scp98}, \cite*{rrg99}). In addition, a cosmic-shear survey is
currently under way, in which randomly selected areas of the sky are
mapped with the FORS instrument ($\sim6\arcminf7\times6\arcminf7$) on
the VLT. Some of these areas include the fields from the STIS parallel
survey.

The alternative approach is to map big areas and measure the cosmic
shear on a wide range of scales. The wide-field cameras currently
being developed and installed are ideally suited for this purpose, and
several groups are actively engaged in this work (see the proceedings
of the Boston lens conference, July 1999). At present, no conclusive
results are available, which is perhaps not too surprising given the
smallness of the expected effect, the infancy of the research area,
and the relatively small amount of high-quality data collected and
analysed so far. Nevertheless, upper limits on the cosmic shear have
been derived by several groups which apparently exclude a
COBE-normalised SCDM model.

There is nothing special about weak lensing being carried out
predominantly in the optical wavelength regime, except that the
optical sky is full of faint extended sources, whereas the radio sky
is relatively empty. The FIRST radio survey covers at present about
4200~square degrees and contains $4\times10^5$ sources, i.e., the
number density is smaller by about a factor $\sim1000$ than in deep
optical images. However, this radio survey covers a much larger solid
angle than current or foreseeable {\em deep\/} optical surveys. As
discussed in \cite{rbk98}, this survey may yield a significant
measurement of the two-point correlation function of image
ellipticities on angular scales $\gtrsim10'$. On smaller angular
scales, sources with intrinsic double-lobe structure cannot be
separated from individual independent sources. The Square Kilometer
Array (\cite*{vhh99}) currently being discussed will yield such a
tremendous increase in sensitivity for cm-wavelength radio astronomy
that the radio sky will then be as crowded as the current optical
sky. Finally, the recently commissioned Sloan telescope will map a
quarter of the sky in five colours. Although the imaging survey will
be much shallower than current weak-lensing imaging, the huge area
surveyed can compensate for the reduced galaxy number density and
their smaller mean redshift \cite{smf96}. Indeed, first weak-lensing
results were already reported at the Boston lensing conference (July
1999) from commissioning data of the telescope (see also
\cite*{sdss99}).

  % -*- LaTeX -*-

\section{\label{sc:7}QSO Magnification Bias and Large-Scale Structure}

\subsection{\label{sc:7.1}Introduction}

Magnification by gravitational lenses is a purely geometrical
phenomenon. The solid angle spanned by the source is enlarged, or
equivalently, gravitational focusing directs a larger fraction of the
energy radiated by the source to the observer. Sources that would have
been too faint without magnification can therefore be seen in a
flux-limited sample. However, these sources are now distributed over a
larger patch of the sky because the solid angle is stretched by the
lens, so that the number density of the sources on the sky is
reduced. The net effect on the number density depends on how many
sources are added to the sample because they appear brighter. If the
number density of sources increases steeply with decreasing flux, many
more sources appear due to a given magnification, and the simultaneous
dilution can be compensated or outweighed.

This magnification bias was described in Sect.~\ref{sc:4.4.1}
(page~\pageref{sc:4.4.1}) and quantified in eq.~(\ref{eq:4.38}). As
introduced there, let $\mu(\vec\theta)$ denote the magnification into
direction $\vec\theta$ on the sky, and $n_0(>S)$ the intrinsic counts
of sources with observed flux exceeding $S$. In the limit of weak
lensing, $\mu(\vec\theta)\gtrsim1$, and the flux will not change by a
large factor, so that it is sufficient to know the behaviour of
$n_0(>S)$ in a small neighbourhood of $S$. Without loss of generality,
we can assume the number-count function to be a power law in that
neighbourhood, $n_0(>S)\propto S^{-\alpha}$. We can safely ignore any
redshift dependence of the intrinsic source counts here because we aim
at lensing effects of moderate-redshift mass distributions on
high-redshift sources. Equation~(\ref{eq:4.43},
page~\pageref{eq:4.43}) then applies, which relates the cumulative
source counts $n(>S,\vec\theta)$ observed in direction $\vec\theta$ to
the intrinsic source counts,
\begin{equation}
  n(>S,\vec\theta) = \mu^{\alpha-1}(\vec\theta)\,n_0(>S)\;.
\label{eq:7.1}
\end{equation}
Hence, if $\alpha>1$, the observed number density of objects is
increased by lensing, and reduced if $\alpha<1$. This effect is called
{\em magnification bias\/} or {\em magnification anti-bias\/}
(e.g.~\cite*{sef92}).

The intrinsic number-count function of QSOs is well fit by a broken
power law with a slope of $\alpha\sim0.64$ for QSOs fainter than
$\sim19$th blue magnitude, and a steeper slope of $\alpha\sim2.52$ for
brighter QSOs (\cite*{bsp88}; \cite*{has90}; \cite*{pei95}). Faint
QSOs are therefore anti-biased by lensing, and bright QSOs are
biased. In the neighbourhood of gravitational lenses, the number
density of bright QSOs is thus expected to be higher than average, in
other words, more bright QSOs should be observed close to foreground
lenses than expected without lensing. According to eq.~(\ref{eq:7.1}),
the overdensity factor is
\begin{equation}
  q(\vec\theta) = \frac{n(>S,\vec\theta)}{n_0(>S)} =
  \mu^{\alpha-1}(\vec\theta)\;.
\label{eq:7.2}
\end{equation}
If the lenses are individual galaxies, the magnification
$\mu(\vec\theta)$ drops rapidly with increasing distance from the
lens. The natural scale for the angular separation is the Einstein
radius, which is of order an arc second for galaxies. Therefore,
individual galaxies are expected to increase the number density of
bright QSOs only in a region of radius a few arc seconds around them.

\cite{FU90.1} reported an observation which apparently contradicts
this expectation. He correlated bright, radio-loud QSOs at moderate
and high redshifts with galaxies from the Lick catalogue
(\cite*{ssg77}) and found that there is a significant overdensity of
galaxies around the QSOs of some of his sub-samples. This is
intriguing because the Lick catalogue contains the counts of galaxies
brighter than $\sim19$th magnitude in square-shaped cells with $10'$
side length. Galaxies of $\lesssim19$th magnitude are typically at
much lower redshifts than the QSOs, $z\lesssim0.1-0.2$, so that the
QSOs with redshifts $z\gtrsim0.5-1$ are in the distant background of
the galaxies, with the two samples separated by hundreds of
megaparsecs. Physical correlations between the QSOs and the galaxies
are clearly ruled out. Can the observed overdensity be expected from
gravitational lensing? By construction, the angular resolution of the
Lick catalogue is of order $10'$, exceeding the Einstein radii of
individual galaxies by more than two orders of magnitude. The result
that Lick galaxies are correlated with bright QSOs can thus neither be
explained by physical correlations nor by gravitational lensing due to
individual galaxies.

On the other hand, the angular scale of $\sim10'$ is on the right
order of magnitude for lensing by large-scale structures. The question
therefore arises whether the magnification due to lensing by
large-scale structures is sufficient to cause a magnification bias in
flux-limited QSO samples which is large enough to explain the observed
QSO-galaxy correlation. The idea is that QSOs are then expected to
appear more abundantly behind matter overdensities. More galaxies are
expected where the matter density is higher than on average, and so
the galaxies would act as tracers for the dark material responsible
for the lensing magnification. This could then cause foreground
galaxies to be overdense around background QSOs. This exciting
possibility clearly deserves detailed investigation.

Even earlier than \citename{FU90.1}, \cite{TY86.3} had inferred that
galaxies apparently underwent strong luminosity evolution from a
detection of significant galaxy overdensities on scales of $30''$
around 42~QSOs with redshifts $1\le z\le1.5$, assuming that the excess
galaxies were at the QSO redshifts. In the light of later observations
and theoretical studies, he probably was the first to detect
weak-lensing induced associations of distant sources with foreground
galaxies.

\subsection{\label{sc:7.2}Expected Magnification Bias from
  Cosmological Density Perturbations}

To estimate the magnitude of the effect, we now calculate the angular
cross-correlation function $\xi_\mathrm{QG}(\phi)$ between background
QSOs and foreground galaxies expected from weak lensing due to
large-scale structures (\cite*{BA95.4}; \cite*{DO97.1};
\cite*{SA97.3}). We employ a simple picture for the relation between
the number density of galaxies and the density contrast of dark
matter, the linear biasing scheme (e.g.~\cite*{kai84}; \cite*{bbk86};
\cite*{wde87}). Within this picture, and assuming weak lensing, we
shall immediately see that the desired correlation function
$\xi_\mathrm{QG}$ is proportional to the cross-correlation function
$\xi_{\mu\delta}$ between magnification $\mu$ and density contrast
$\delta$. The latter correlation can straightforwardly be computed
with the techniques developed previously.

\subsubsection{\label{sc:7.2.1}QSO-Galaxy Correlation Function}

The angular cross-correlation function $\xi_\mathrm{QG}(\phi)$ between
galaxies and QSOs is defined by
\begin{equation}
  \xi_\mathrm{GQ}(\phi) = \frac{1}
    {\langle n_\mathrm{Q}\rangle\langle n_\mathrm{G}\rangle}\,
  \left\langle\left[
    n_\mathrm{Q}(\vec\theta) - \langle n_\mathrm{Q}\rangle
  \right]
  \left[
    n_\mathrm{G}(\vec\theta+\vec\phi) -
    \langle n_\mathrm{G}\rangle
  \right]\right\rangle\;,
\label{eq:7.3}
\end{equation}
where $\langle n_\mathrm{Q,G}\rangle$ are the mean number densities of
QSOs and galaxies averaged over the whole sky. Assuming isotropy,
$\xi_\mathrm{QG}(\phi)$ does not depend on the direction of the lag
angle $\vec\phi$. All number densities depend on flux (or galaxy
magnitude), but we leave out the corresponding arguments for brevity.

We saw in eq.~(\ref{eq:7.1}) in the introduction that
$n_\mathrm{Q}(\vec\theta)=\mu^{\alpha-1}(\vec\theta)\,\langle
n_\mathrm{Q}\rangle$. Since the magnification expected from
large-scale structures is small, $\mu=1+\delta\mu$ with
$|\delta\mu|\ll1$, we can expand $\mu^{\alpha-1} \approx
1+(\alpha-1)\delta\mu$. Hence, we can approximate
\begin{equation}
  \frac{n_\mathrm{Q}(\vec\theta)-\langle n_\mathrm{Q}\rangle}
  {\langle n_\mathrm{Q}\rangle} \approx
  (\alpha-1)\,\delta\mu(\vec\theta)\;,
\label{eq:7.4}
\end{equation}
so that the relative fluctuation of the QSO number density is
proportional to the magnification fluctuation, and the factor of
proportionality quantifies the magnification bias. Again, for
$\alpha=1$, lensing has no effect on the number density.

The linear biasing model for the fluctuations in the galaxy density
asserts that the relative fluctuations in the galaxy number counts are
proportional to the density contrast $\delta$,
\begin{equation}
  \frac{n_\mathrm{G}(\vec\theta)-\langle n_\mathrm{G}\rangle}
  {\langle n_\mathrm{G}\rangle} = b\,\bar{\delta}(\vec\theta)\;,
\label{eq:7.5}
\end{equation}
where $\bar{\delta}(\vec\theta)$ is the line-of-sight integrated
density contrast, weighted by the galaxy redshift distribution,
i.e.~the $w$-integral in eq.~(\ref{eq:6.68}),
page~\pageref{eq:6.68}. The proportionality factor $b$ is the
effective biasing factor appropriately averaged over the
line-of-sight. Typical values for the biasing factor are assumed to be
$b\gtrsim1-2$. Both the relative fluctuations in the galaxy number
density and the density contrast are bounded by $-1$ from below, so
that the right-hand side should be replaced by
$\mathrm{max}[b\bar{\delta}(\vec\theta),-1]$ in places where
$\bar{\delta}(\vec\theta)<-b^{-1}$. For simplicity we use
(\ref{eq:7.5}), keeping this limitation in mind.

Using eqs.~(\ref{eq:7.4}) and (\ref{eq:7.5}), the QSO-galaxy
cross-correlation function (\ref{eq:7.3}) becomes 
\begin{equation}
  \xi_\mathrm{QG}(\phi) = (\alpha-1)\,b\,
  \langle\delta\mu(\vec\theta)
  \bar{\delta}(\vec\theta+\vec\phi)\rangle\;.
\label{eq:7.6}
\end{equation}
Hence, it is proportional to the cross-correlation function
$\xi_{\mu\delta}$ between magnification and density contrast, and the
proportionality factor is given by the steepness of the intrinsic QSO
number counts and the bias factor (\cite*{BA95.4}). As expected from
the discussion of the magnification bias, the magnification bias is
ineffective for $\alpha=1$, and QSOs and galaxies are anti-correlated
for $\alpha<1$. Furthermore, if the number density of galaxies does
not reflect the dark-matter fluctuations, $b$ would vanish, and the
correlation would disappear. In order to find the QSO-galaxy
cross-correlation function, we therefore have to evaluate the angular
cross-correlation function between magnification and density contrast.

\subsubsection{\label{sc:7.2.2}Magnification-Density Correlation
  Function}

We have seen in Sect.~\ref{sc:6} that the magnification fluctuation is
twice the effective convergence
$\delta\mu(\vec\theta)=2\kappa_\mathrm{eff}(\vec\theta)$ in the limit
of weak lensing, see eq.~(\ref{eq:6.32}, page~\pageref{eq:6.32}). The
latter is given by eq.~(\ref{eq:6.24}, page~\pageref{eq:6.24}), in
which the average over the source-distance distribution has already
been performed. Therefore, we can immediately write down the
source-distance averaged magnification fluctuation as
\begin{equation}
  \delta\bar\mu(\vec\theta) = \frac{3H_0^2\Omega_0}{c^2}\,
  \int_0^{w_\mathrm{H}}\,\d w\,\bar{W}_\mathrm{Q}(w)\,f_K(w)\,
  \frac{\delta[f_K(w)\vec\theta,w]}{a(w)}\;.
\label{eq:7.7}
\end{equation}
Here, $\bar{W}_\mathrm{Q}(w)$ is the modified QSO weight function
\begin{equation}
  \bar{W}_\mathrm{Q}(w) \equiv \int_w^{w_\mathrm{H}}\,\d w'\,
  G_\mathrm{Q}(w')\,\frac{f_K(w'-w)}{f_K(w')}\;,
\label{eq:7.8}
\end{equation}
and $G_\mathrm{Q}(w)$ is the normalised QSO distance distribution.

Both the average density contrast $\bar\delta$ and the average
magnification fluctuation $\delta\bar\mu$ are weighted projections of
the density fluctuations along the line-of-sight, which is assumed to
be a homogeneous and isotropic random field. As in the derivation of
the effective-convergence power spectrum in Sect.~\ref{sc:6}, we can
once more employ Limber's equation in Fourier space to find the cross
power spectrum $P_{\mu\delta}(l)$ for projected magnification and
density contrast,
\begin{equation}
  P_{\mu\delta}(l) = \frac{3H_0^2\Omega_0}{c^2}\,
  \int_0^{w_\mathrm{H}}\,\d w\,
  \frac{\bar{W}_\mathrm{Q}(w)\,G_\mathrm{G}(w)}{a(w)\,f_K(w)}\,
  P_\delta\left(\frac{l}{f_K(w)}\right)\;.
\label{eq:7.10}
\end{equation}
The cross-correlation function between magnification and density
contrast is obtained from eq.~(\ref{eq:7.10}) via Fourier
transformation, which can be carried out and simplified to yield
\begin{eqnarray}
  \xi_{\mu\delta}(\phi) = &\frac{3H_0^2\Omega_0}{c^2}&\,
  \int_0^{w_\mathrm{H}}\,\d w'\,f_K(w')\,
  \bar{W}_\mathrm{Q}(w')\,G_\mathrm{G}(w')\,a^{-1}(w')\nonumber\\
  &\times&\int_0^\infty\,\frac{k\d k}{2\pi}\,
  P_\delta(k,w')\J[f_K(w')k\phi]\;.
\label{eq:7.11}
\end{eqnarray}
Quite obviously, there is a strong similarity between this equation
and that for the magnification autocorrelation function,
eq.~(\ref{eq:6.37}, page~\pageref{eq:6.37}). We note that
eq.~(\ref{eq:7.11}) automatically accounts for galaxy autocorrelations
through the matter power spectrum $P_\delta(k)$.

\subsubsection{\label{sc:7.2.3}Distance Distributions and Weight
  Functions}

The QSO and galaxy weight functions $G_\mathrm{Q,G}(w)$ are normalised
representations of their respective redshift distributions, where the
redshift needs to be transformed to comoving distance $w$.

\begin{figure}[ht]
  \includegraphics[width=\hsize]{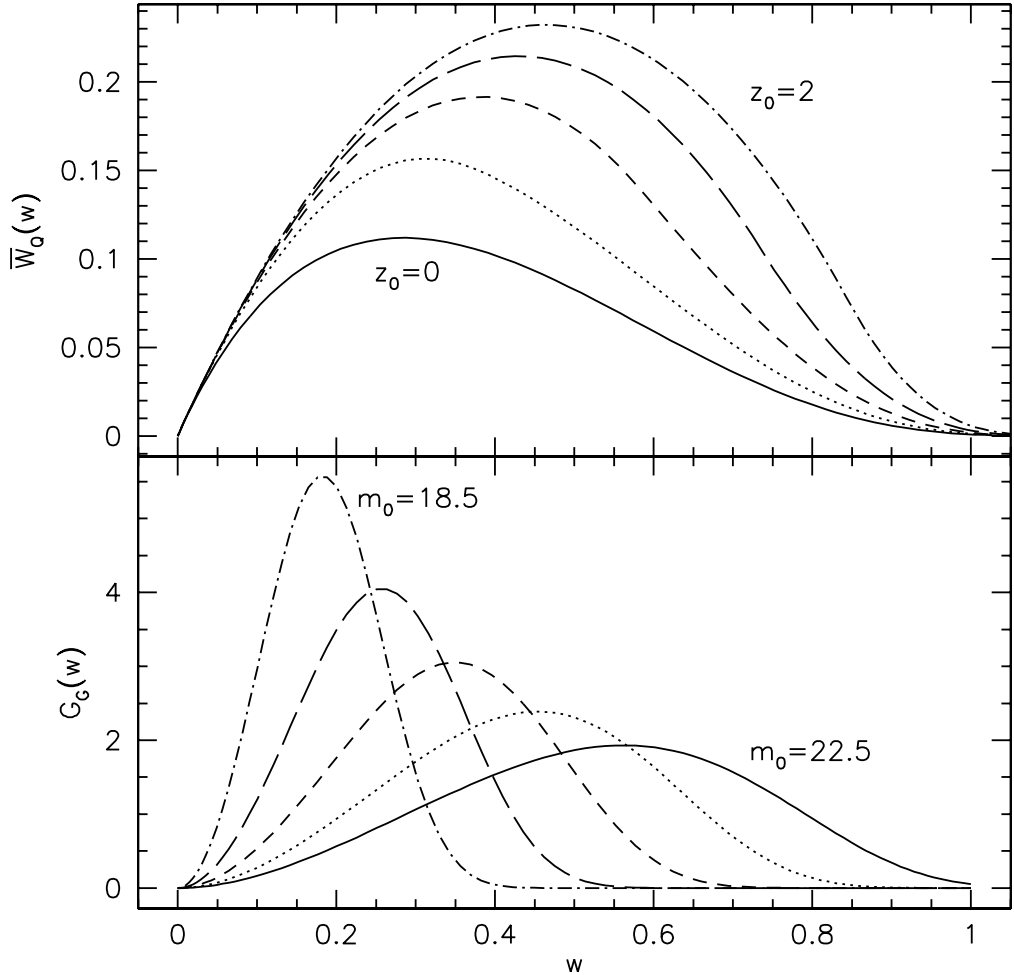}
\caption{QSO and galaxy weight functions, $\bar{W}_\mathrm{Q}(w)$ and
$G_\mathrm{G}(w)$, respectively. Top panel: $\bar{W}_\mathrm{Q}(w)$
for five different choices of the lower cut-off redshift $z_0$ imposed
on the QSO sample; $z_0$ increases from $0.0$ (solid curve) to $2.0$
in steps of $0.5$. The peak in $\bar{W}_\mathrm{Q}(w)$ shifts to
larger distances for increasing $z_0$. Bottom panel: $G_\mathrm{G}(w)$
for five different galaxy magnitude limits $m_0$, increasing from
$18.5$ to $22.5$ (solid curve) in steps of one magnitude. The peak in
the galaxy distance distribution shifts towards larger distances with
increasing $m_0$, i.e.~with decreasing brightness of the galaxy
sample.}
\label{fig:7.1}
\end{figure}

The redshift distribution of QSOs has frequently been measured and
parameterised. Using the functional form and the parameters determined
by \cite{pei95}, the modified QSO weight function
$\bar{W}_\mathrm{Q}(w)$ has the shape illustrated in the top panel of
Fig.~\ref{fig:7.1}. It is necessary for our present purposes to be
able to impose a lower redshift limit on the QSO sample. Since we want
to study lensing-induced correlations between background QSOs and
foreground galaxies, there must be a way to exclude QSOs physically
associated with galaxy overdensities. This is observationally achieved
by choosing a lower QSO redshift cut-off high enough to suppress any
redshift overlap between the QSO and galaxy samples. This procedure
must be reproduced in theoretical calculations of the QSO-galaxy
cross-correlation function. This can be achieved by cutting off the
observed redshift distribution $G_\mathrm{Q}$ below some redshift
$z_0$, re-normalising it, and putting the result into
eq.~(\ref{eq:7.8}) to find $\bar{W}_\mathrm{Q}$. The five curves shown
in the top panel of Fig.~\ref{fig:7.1} are for cut-off redshifts $z_0$
increasing from $0.0$ (solid curve) to $2.0$ in steps of
$0.5$. Obviously, the peak in $\bar{W}_\mathrm{Q}$ shifts to larger
$w$ for increasing $z_0$.

Galaxy redshift distributions $G_\mathrm{G}$ can be obtained by
extrapolating local galaxy samples to higher redshifts, adopting a
constant comoving number density and a Schechter-type luminosity
function. For the present purposes, this is a safe procedure because
the galaxies to be correlated with the QSOs {\em should\/} be at
sufficiently lower redshifts than the QSOs to avoid overlap between
the samples. Thus the extrapolation from the local galaxy population
is well justified. In order to convert galaxy luminosities to observed
magnitudes, $k$-corrections need to be taken into
account. Conveniently, the resulting weight functions should be
parameterised by the brightness cut-off of the galaxy sample, in
practice by the maximum galaxy magnitude $m_0$ (i.e.~the minimum
luminosity) required for a galaxy to enter the sample. The five
representative curves for $G_\mathrm{G}(w)$ in the lower panel of
Fig.~\ref{fig:7.1} are for $m_0$ increasing from $18.5$ to $22.5$
(solid curve) in steps of one magnitude. $R$-band magnitudes are
assumed. For increasing cut-off magnitude $m_0$, i.e.~for fainter
galaxy samples, the distributions broaden, as expected. The
correlation amplitude as a function of $m_0$ peaks if $m_0$ is chosen
such that the median distance to the galaxies is roughly half the
distance to the bulk of the QSO population considered.

\subsubsection{\label{sc:7.2.4}Simplifications}

It turns out in practice that the exact shapes of the QSO and galaxy
weight functions $\bar{W}_\mathrm{Q}(w)$ and $G_\mathrm{G}(w)$ are of
minor importance for the results. Allowing inaccuracies of order 10\%,
we can replace the functions $G_\mathrm{Q,G}(w)$ by delta
distributions centred on typical QSO and galaxy distances
$w_\mathrm{Q}$ and $w_\mathrm{G}<w_\mathrm{Q}$. Then, from
eq.~(\ref{eq:7.8}),
\begin{equation}
  \bar{W}_\mathrm{Q}(w) =
  \frac{f_K(w_\mathrm{Q}-w)}{f_K(w_\mathrm{Q})}\,
  \mathrm{H}(w_\mathrm{Q}-w)\;,
\label{eq:7.12}
\end{equation}
where $\mathrm{H}(x)$ is the Heaviside step function, and the
line-of-sight integration in eq.~(\ref{eq:7.7}) becomes trivial. It is
obvious that matter fluctuations at redshifts higher than the QSO
redshift do not contribute to the cross-correlation function
$\xi_{\mu\delta}(\phi)$: Inserting (\ref{eq:7.12}) together with
$G_\mathrm{G}=\delta(w-w_\mathrm{G})$ into eq.~(\ref{eq:7.11}), we
find $\xi_{\mu\delta}(\phi)=0$ if $w_\mathrm{G}>w_\mathrm{Q}$, as it
should be.

The expression for the magnification-density cross-correlation
function further simplifies if we specialise to a model universe with
zero spatial curvature, $K=0$, such that $f_K(w)=w$. Then,
\begin{equation}
  \bar{W}_\mathrm{Q}(w) = \left(1-\frac{w}{w_\mathrm{Q}}\right)\,
  \mathrm{H}(w_\mathrm{Q}-w)\;,
\label{eq:7.13}
\end{equation}
and the cross-correlation function $\xi_{\mu\delta}(\phi)$ reduces to
\begin{equation}
  \xi_{\mu\delta}(\phi) = \frac{3H_0^2\Omega_0}{c^2}\,
  \frac{w_\mathrm{G}}{a(w_\mathrm{G})}\,
  \left(1-\frac{w_\mathrm{G}}{w_\mathrm{Q}}\right)\,
  \int_0^\infty\,\frac{k\d k}{2\pi}\,P_\delta(k,w_\mathrm{G})\,
  \J(w_\mathrm{G}k\phi)
\label{eq:7.14}
\end{equation}
for $w_\mathrm{Q}>w_\mathrm{G}$, and $\xi_{\mu\delta}(\phi)=0$
otherwise.

\subsection{\label{sc:7.3}Theoretical Expectations}

\subsubsection{\label{sc:7.3.1}Qualitative Behaviour}

\begin{figure}[ht]
  \includegraphics[width=\hsize]{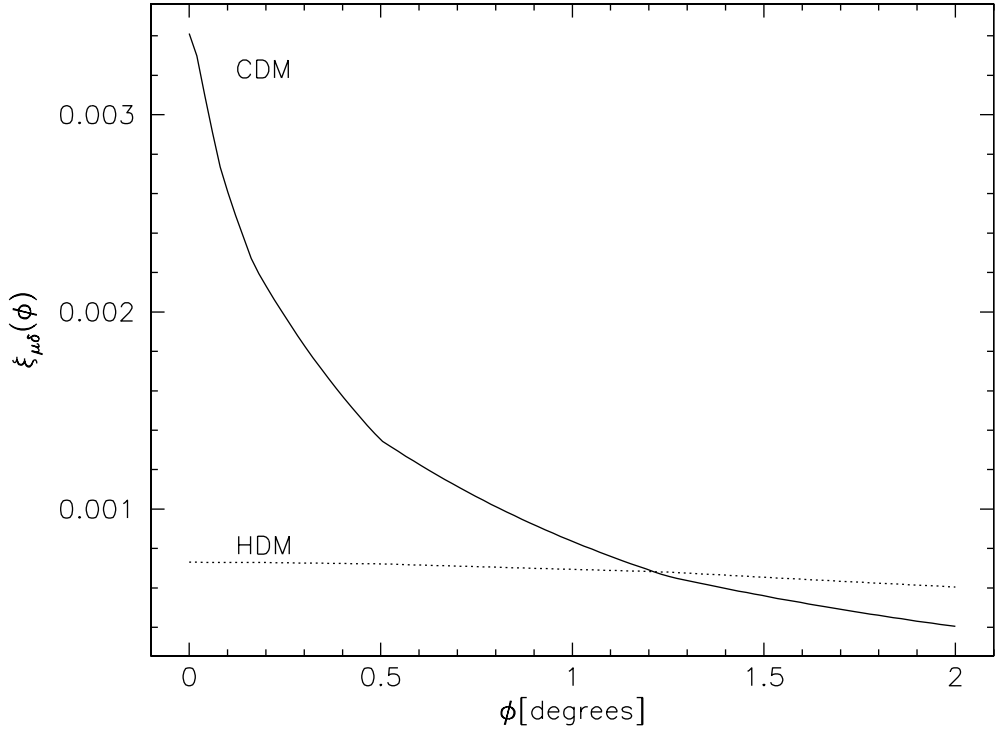}
\caption{Cross-correlation functions between magnification and density
contrast, $\xi_{\mu\delta}(\phi)$, are shown for an Einstein-de Sitter
universe with $h=0.5$, adopting CDM (solid curve) and HDM (dotted
curve) density fluctuation spectra. Both spectra are normalised to the
local cluster abundance, and linear density evolution is assumed. The
lower cut-off redshift of the QSOs is $z_0=0.3$, the galaxy magnitude
limit is $m_0=20.5$. In agreement with the expectation derived from
the CDM and HDM model spectra (\ref{eq:6.39}, page~\pageref{eq:6.39}),
the CDM cross-correlation function decreases linearly with increasing
$\phi$ for small $\phi$, while it is flat to first order in $\phi$ for
HDM. The small-scale matter fluctuations in CDM compared to HDM cause
$\xi_{\mu\delta}(\phi)$ to increase more steeply as $\phi\to0$.}
\label{fig:7.2}
\end{figure}

Before we evaluate the magnification-density cross-correlation
function fully numerically, we can gain some insight into its expected
behaviour by inserting the CDM and HDM model spectra defined in
eq.~(\ref{eq:6.39}, page~\pageref{eq:6.39}) into eq.~(\ref{eq:7.11})
and expanding the result into a power series in $\phi$
(\cite*{BA95.4}). As in the case of the magnification auto-correlation
function before, the two model spectra produce qualitatively different
results. To first order in $\phi$, $\xi_{\mu\delta}(\phi)$ decreases
linearly with increasing $\phi$ for CDM, while it is flat for HDM. The
reason for this different appearance is the lack of small-scale power
in HDM, and the abundance thereof in CDM. The two curves shown in
Fig.~\ref{fig:7.2} illustrate this for an Einstein-de Sitter universe
with Hubble constant $h=0.5$. The underlying density-perturbation
power spectra were normalised by the local abundance of rich clusters,
and linear density evolution was assumed.

The {\em linear\/} correlation amplitude, $\xi_{\mu\delta}(0)$, for
CDM is of order $3\times10^{-3}$, and about a factor of five smaller
for HDM. The magnification-density cross-correlation function for CDM
drops to half its peak value within a few times 10 arc minutes. This,
and the monotonic increase of $\xi_{\mu\delta}$ towards small $\phi$,
indicate that density perturbations on angular scales below $10'$
contribute predominantly to $\xi_{\mu\delta}$. At typical lens
redshifts, such angular scales correspond to physical scales up to a
few Mpc. Evidently therefore, the non-linear evolution of the density
perturbations needs to be taken into account, and its effect is
expected to be substantial.

\subsubsection{\label{sc:7.3.3}Results}

\begin{figure}[ht]
  \includegraphics[width=\hsize]{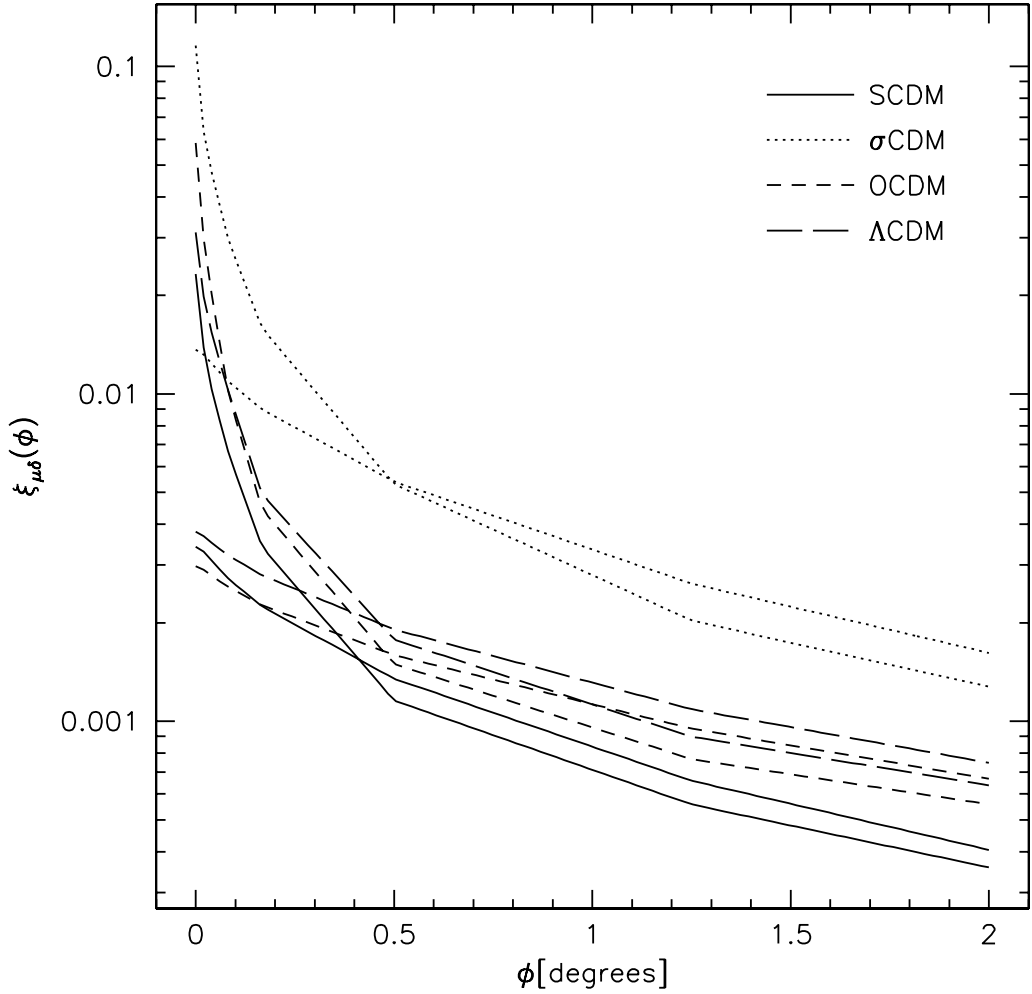}
\caption{Angular magnification-density cross-correlation functions
$\xi_{\mu\delta}(\phi)$ are shown for the four cosmological models
specified in Table~\ref{tab:6.1} on page~\pageref{tab:6.1}. Two curves
are shown for each cosmological model; those with the higher (lower)
amplitude at $\phi=0$ were calculated with the non-linearly (linearly)
evolving density-perturbation power spectra, respectively. The models
are: SCDM (solid curves), $\sigma$CDM (dotted curves), OCDM
(short-dashed curves), and $\Lambda$CDM (long-dashed
curves). Obviously, non-linear evolution has a substantial effect. It
increases the correlation amplitude by about an order of
magnitude. The Einstein-de Sitter model normalised to $\sigma_8=1$ has
a significantly larger cross-correlation amplitude than the
cluster-normalised Einstein-de Sitter model. For the low-density
models, the difference is much smaller. The curves for the
cluster-normalised models are very similar, quite independent of
cosmological parameters.}
\label{fig:7.3}
\end{figure}

Figure~\ref{fig:7.3} confirms this expectation; it shows
magnification-density cross-correlation functions for the four
cosmological models detailed in Tab.~\ref{tab:6.1} on
page~\pageref{tab:6.1}. Two curves are shown for each model, one for
linear and the other for non-linear density evolution. The two curves
of each pair are easily distinguished because non-linear evolution
increases the cross-correlation amplitude at small $\phi$ by about an
order of magnitude above linear evolution, quite independent of the
cosmological model. At the same time, the angular cross-correlation
scale is reduced to a few arc minutes. At angular scales
$\lesssim30'$, the non-linear cross-correlation functions are above
the linear results, falling below at larger scales. The correlation
functions for the three cluster-normalised models (SCDM, OCDM and
$\Lambda$CDM; see Tab.~\ref{tab:6.1} on page~\pageref{tab:6.1}) are
very similar in shape and amplitude. The curve for the $\sigma$CDM
model lies above the other curves by a factor of about five, but for
low-density universes, the influence of different power-spectrum
normalisations are much less prominent.

The main results to be extracted from Fig.~\ref{fig:7.3} are that the
amplitude of the magnification-density cross-correlation function,
$\xi_{\mu\delta}(0)$, reaches approximately $5\times10^{-2}$, and that
$\xi_{\mu\delta}$ drops by an order of magnitude within about
$20'$. This behaviour is quite independent of the cosmological
parameters if the density-fluctuation power spectrum is normalised by
the local abundance of rich galaxy clusters. More detailed results can
be found in \cite{DO97.1} and \cite{SA97.3}.

\subsubsection{\label{sc:7.3.4}Signal-to-Noise Estimate}

The QSO-galaxy correlation function $\xi_\mathrm{QG}(\phi)$ is larger
than $\xi_{\mu\delta}(\phi)$ by the factor $(\alpha-1)b$. The value of
the bias factor $b$ is yet unclear, but it appears reasonable to
assume that it is between 1 and 2. For optically selected QSOs,
$\alpha\approx2.5$, so that $(\alpha-1)b\approx2-3$. Combining this
with the correlation amplitude for CDM read off from
Fig.~\ref{fig:7.3}, we can expect $\xi_\mathrm{QG}(0)\lesssim0.1$.

Given the meaning of $\xi_\mathrm{QG}(\phi)$, the probability to find
a foreground galaxy close to a background QSO is increased by a factor
of $[1+\xi_\mathrm{QG}(\phi)]\lesssim1.1$ above random. In a small
solid angle $\d^2\omega$ around a randomly selected background QSO, we
thus expect to find
\begin{equation}
  N_\mathrm{G} \approx [1+\xi_\mathrm{QG}(0)]\,
  \langle n_\mathrm{G}\rangle\,\d^2\omega \equiv
  [1+\xi_\mathrm{QG}(0)]\,\langle N_\mathrm{G}\rangle
\label{eq:7.15}
\end{equation}
galaxies, where $\langle N_\mathrm{G}\rangle$ is the average number of
galaxies within a solid angle of $\d^2\omega$. In a sample of
$N_\mathrm{Q}$ fields around randomly selected QSOs, the
signal-to-noise ratio for the detection of a galaxy overdensity is
then
\begin{equation}
  \frac{\mathrm{S}}{\mathrm{N}} \approx
  \frac{N_\mathrm{Q}(N_\mathrm{G}-\langle N_\mathrm{G}\rangle)}
  {(N_\mathrm{Q}\langle N_\mathrm{G}\rangle)^{1/2}} =
  (N_\mathrm{Q}\langle N_\mathrm{G}\rangle)^{1/2}\,
  \xi_\mathrm{QG}(0)\;.
\label{eq:7.16}
\end{equation}
Typical surface number densities of reasonably bright galaxies are of
order $n_\mathrm{G}\sim10$ per square arc minute. Therefore, there
should be of order $\langle N_\mathrm{G}\rangle\sim30$ galaxies within
a randomly selected disk of one arc minute radius, in which the
QSO-galaxy cross correlation is sufficiently strong. If we require a
certain minimum signal-to-noise ratio such that
$\mathrm{S}/\mathrm{N}\ge(\mathrm{S}/\mathrm{N})_0$, the number of QSO
fields to be observed in order to meet this criterion is
\begin{eqnarray}
  N_\mathrm{Q}
  &\ge& \left(\frac{\mathrm{S}}{\mathrm{N}}\right)_0^2\,
    \xi_\mathrm{QG}^{-2}(0)\,\langle N_\mathrm{G}\rangle^{-1}
    \nonumber\\
  &=& \left(\frac{\mathrm{S}}{\mathrm{N}}\right)_0^2\,
    [(\alpha-1)\,b]^{-2}\,\xi_{\mu\delta}^{-2}(0)\,
    \langle N_\mathrm{G}\rangle^{-1}
    \nonumber\\
  &=& 20\,\left[\frac{(\mathrm{S}/\mathrm{N})_0}{5}\right]^2\,
    \left[\frac{(\alpha-1)\,b}{4}\right]^{-2}\,
    \left(\frac{\xi_{\mu\delta}(0)}{0.05}\right)^{-2}\,
    \left(\frac{30}{\langle N_\mathrm{G}\rangle}\right)\;,
\label{eq:7.17}
\end{eqnarray}
where we have inserted typical numbers in the last step. This estimate
demonstrates that gravitational lensing by non-linearly evolving
large-scale structures in cluster-normalised CDM can produce
correlations between background QSOs and foreground galaxies at the
$5\,\sigma$ level on arc minute scales in samples of $\gtrsim20$
QSOs. The angular scale of the correlations is expected to be of order
1 to 10 arc minutes. Equation~(\ref{eq:7.17}) makes it explicit that
more QSO fields need to be observed in order to establish the
significance of the QSO-galaxy correlations if (i) the QSO number
count function is shallow ($\alpha$ close to unity), and (ii) the
galaxy bias factor $b$ is small. In particular, no correlations are
expected if $\alpha=1$, because then the dilution of the sources and
the increase in QSO number exactly cancel. Numerical simulations
(\cite*{BA95.4}) confirm the estimate (\ref{eq:7.17}).

\citename{FU90.1}'s (\citeyear{FU90.1}) observation was also tested in
a numerical model universe based on the adhesion approximation to
structure formation (\cite*{BA92.4}). This model universe was
populated with QSOs and galaxies, and QSO-galaxy correlations on
angular scales on the order of $\sim10'$ were investigated using
Spearman's rank-order correlation test (\cite*{BA93.1}). Light
propagation in the model universe was described with the multiple
lens-plane approximation of gravitational lensing. In agreement with
the analytical estimate presented above, it was found that lensing by
large-scale structures can indeed account for the observed
correlations between high-redshift QSOs and low-redshift galaxies,
provided the QSO number-count function is steep. Lensing by individual
galaxies was confirmed to be entirely negligible.

\subsubsection{\label{sc:7.3.5}Multiple-Waveband Magnification Bias}

The magnification bias quantified by the number-count slope $\alpha$
can be substantially increased if QSOs are selected in two or more
mutually uncorrelated wave bands rather than one (\cite*{BO91.1}). To
see why, suppose that optically bright {\em and\/} radio-loud QSOs
were selected, and that their fluxes in the two wave bands are
uncorrelated. Let $S_{1,2}$ be the flux thresholds in the optical and
in the radio regimes, respectively, and $n_{1,2}$ the corresponding
number densities of either optically bright or radio-loud QSOs on the
sky. As in the introduction, we assume that $n_{1,2}$ can be written
as power laws in $S_{1,2}$, with exponents $\alpha_{1,2}$.

In a small solid angle $\d^2\omega$, the probability to find an
optically bright {\em or\/} radio-loud QSO is then
$p_i(S_i)=n_i(S_i)\,\d^2\omega$, and the joint probability to find an
optically bright {\em and\/} radio-loud QSO is the product of the
individual probabilities, or
\begin{equation}
  p(S_1,S_2) = p_1(S_1)\,p_2(S_2) =
  [n_1(S_1)\,n_2(S_2)]\,\d^2\omega \propto
  S_1^{-\alpha_1}S_2^{-\alpha_2}\,\d^2\omega\;,
\label{eq:7.18}
\end{equation}
provided there is no correlation between the fluxes $S_{1,2}$ so that
the two probabilities are independent. Suppose now that lensing
produces a magnification factor $\mu$ across $\d^2\omega$. The joint
probability is then changed to
\begin{equation}
  p'(S_1,S_2) \propto
  \left(\frac{\mu}{S_1}\right)^{\alpha_1}\,
  \left(\frac{\mu}{S_2}\right)^{\alpha_2}\,
  \frac{\d^2\omega}{\mu} =
  \mu^{\alpha_1+\alpha_2-1}\,p(S_1,S_2)\;.
\label{eq:7.19}
\end{equation}
Therefore, the magnification bias in the optically bright {\em and\/}
radio-loud QSO sample is as efficient as if the number-count function
had a slope of $\alpha=\alpha_1+\alpha_2$. 

More generally, the effective number-count slope for the magnification
bias in a QSO sample that is flux limited in $m$ mutually uncorrelated
wave bands is
\begin{equation}
  \alpha = \sum_{i=1}^m\,\alpha_i\;,
\label{eq:7.20}
\end{equation}
where $\alpha_i$ are the number-count slopes in the individual wave
bands. Then, the QSO-galaxy cross-correlation function is
\begin{equation}
  \xi_\mathrm{QG}(\phi) = \left(\sum_{i=1}^m\,\alpha_i-1\right)\,b\,
  \xi_{\mu\delta}(\phi)\;,
\label{eq:7.21}
\end{equation}
and can therefore be noticeably larger than for a QSO sample which is
flux limited in one wave band only.

\subsection{\label{sc:7.4}Observational Results}

After this theoretical investigation, we turn to observations of
QSO-galaxy cross-correlations on large angular scales. The existence
of QSO-galaxy correlations was tested and verified in several studies
using some very different QSO- and galaxy samples.

\cite{BA93.2} repeated Fugmann's analysis with a well-defined sample
of background QSOs, namely the optically identified QSOs from the
1-Jansky catalogue (\cite*{kwp81}; \cite*{skf93};
\cite*{stk93}). Optically identified QSOs with measured redshifts need
to be bright enough for detection and spectroscopy, hence the chosen
sample is implicitly also constrained by an optical flux
limit. Optical and radio QSO fluxes are generally not strongly
correlated, so that the sample is affected by a double-waveband
magnification bias, which can further be strengthened by explicitly
imposing an optical flux (or magnitude) limit.

Although detailed results differ from Fugmann's, the presence of the
correlation is confirmed at the $98\%$ confidence level for QSOs with
redshifts $\ge0.75$ and brighter than $18$th magnitude. The number of
QSOs matching these criteria is $56$. The correlation significance
decreases both for lower- and higher-redshift QSO samples, and also
for optically fainter ones. This is in accordance with an explanation
in terms of a (double-waveband) magnification bias due to
gravitational lensing. For low-redshift QSOs, lensing is not efficient
enough to produce the correlations. For high-redshift QSOs, the most
efficient lenses are at higher redshifts than the galaxies, so that
the {\em observed\/} galaxies are uncorrelated with the structures
which magnify the QSOs. Hence, the correlation is expected to
disappear for increasing QSO redshifts. For an optically unconstrained
QSO sample, the effective slope of the number-count function is
smaller, reducing the strength of the magnification bias and therefore
also the significance of the correlation.

With a similar correlation technique, correlations between the
1-Jansky QSO sample and IRAS galaxies (\cite*{BA94.1}) and diffuse
X--ray emission (\cite*{BA94.5}) were investigated, leading to
qualitatively similar results. IRAS galaxies are correlated with
optically bright, high-redshift $z\ge1.5$ 1-Jansky sources at the
$99.8\%$ confidence level. The higher QSO redshift for which the
correlation becomes significant can be understood if the IRAS galaxy
sample is deeper than the Lick galaxy sample, so that the structures
responsible for the lensing can be traced to higher redshift. 

\cite{BA97.2} re-analysed the correlation between IRAS galaxies and
1-Jansky QSOs using a more advanced statistical technique which can be
optimised to the correlation function expected from lensing by
large-scale structures. In agreement with \cite{BA94.1}, they found
significant correlations between the QSOs and the IRAS galaxies on
angular scales of $\sim5'$, but the correlation amplitude is higher
than expected from large-scale structure lensing, assuming linear
evolution of the density-perturbation power spectrum. Including
non-linear evolution, however, the results by \cite{BA97.2} can well
be reproduced (\cite*{DO97.1}).

X--ray photons from the ROSAT {\em All-Sky Survey\/}
(e.g.~\cite*{vog92}) are correlated with optically bright 1-Jansky
sources both at low ($0.5\le z\le1.0$) and at high redshifts ($1.5\le
z\le2.0$), but there is no significant correlation with QSOs in the
intermediate redshift regime. A plausible explanation for this is that
the correlation of X--ray photons with low-redshift 1-Jansky QSOs is
due to hot gas which is physically associated with the QSOs,
e.g.~which resides in the host clusters of these QSOs. Increasing the
source redshift, the flux from these clusters falls below the
detection threshold of the {\em All-Sky Survey\/}, hence the
correlation disappears. Upon further increasing the QSO redshift,
lensing by large-scale structures becomes efficient, and the X--ray
photons trace hot gas in the lenses.

\cite{RO94.1} found a highly significant correlation between
optically-selected, high-redshift QSOs and Zwicky clusters. Their
cluster sample was fairly bright, which indicates that the clusters
are in the foreground of the QSOs. This rules out that the clusters
are physically associated with the QSOs and thus exert environmental
effects on them which might lead to the observed
association. \citename{RO94.1} discussed lensing as the most probable
reason for the correlations, although simple mass models for the
clusters yield lower magnifications than required to explain the
significance of the effect. \cite{SE95.2} repeated their analysis with
the 1-Jansky sample of QSOs. They found agreement with
\citename{RO94.1}'s result for intermediate-redshift ($z\sim1$) QSOs,
but failed to detect significant correlations for higher-redshift
sources. In addition, a significant under-density of low-redshift QSOs
close to Zwicky clusters was found, for which environmental effects
like dust absorption are the most likely explanation. A
variability-selected QSO sample was correlated with Zwicky clusters by
\cite{roh95}. They detected a significant correlation between QSOs
with $0.4\le z\le2.2$ with foreground Zwicky clusters (with $\langle
z\rangle\sim0.15$) and interpreted it in terms of gravitational
lensing. Again, the implied average QSO magnification is substantially
larger than that inferred from simple lens models for clusters with
velocity dispersions of $\sim10^3\;\mathrm{km\,s}^{-1}$. \cite{WU95.1}
searched for associations between distant 1-Jansky and 2-Jansky QSOs
and foreground Abell clusters. They found no correlations with the
1-Jansky sources, and a marginally significant correlation with
2-Jansky sources. They argue that lensing by individual clusters is
insufficient if cluster velocity dispersions are of order
$10^3\;\mathrm{km\,s}^{-1}$, and that lensing by large-scale
structures provides a viable explanation.

\begin{figure}[ht]
  \includegraphics[width=\hsize]{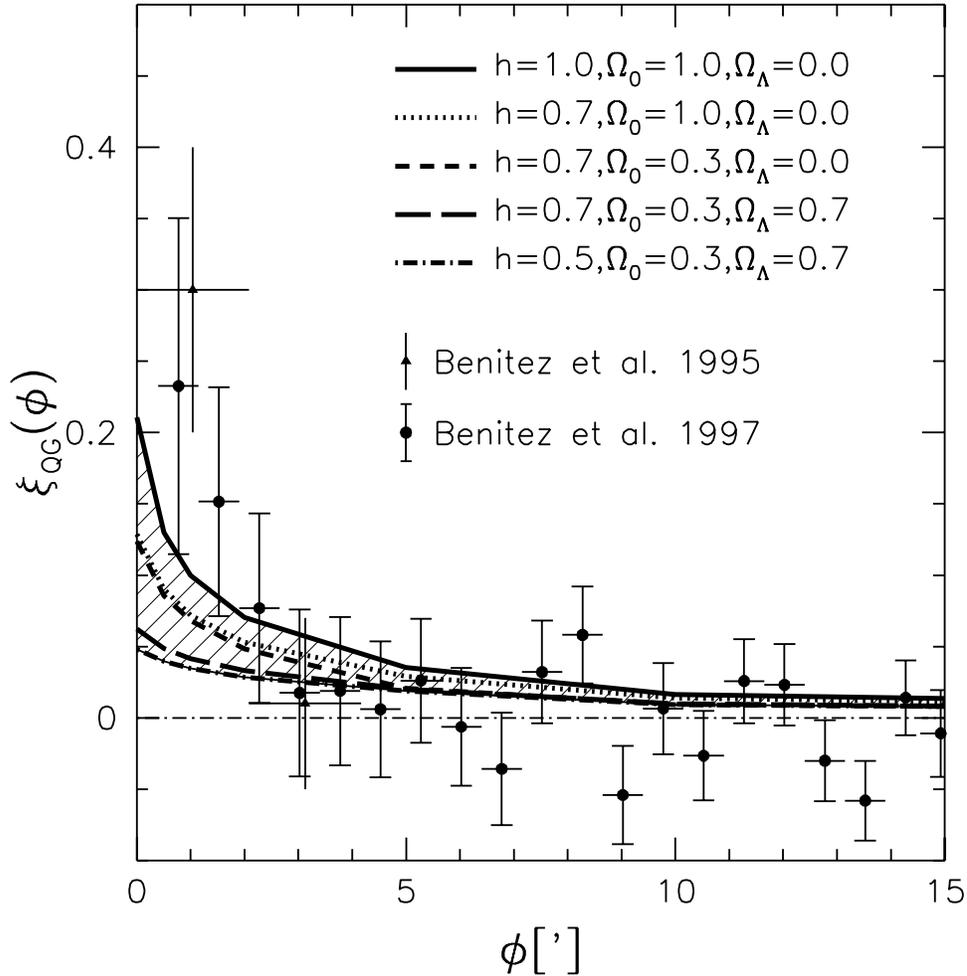}
\caption{QSO-galaxy cross-correlation measurements are plotted
together with theoretical cross-correlation functions
$\xi_\mathrm{QG}(\phi)$ for various cosmological models as indicated
by line type. The CDM density-perturbation power spectrum was
cluster-normalised, and non-linear evolution was taken into
account. The figure shows that the measurements fall above the
theoretical predictions at small angular scales,
$\phi\lesssim2'$. This excess can be attributed to gravitational
lensing by individual galaxy clusters (see the text for more detail).
The theoretical curves depend on the Hubble constant $h$ through the
shape parameter $\Gamma=\Omega_0h$, which determines the peak location
of the power spectrum.}
\label{fig:7.4}
\end{figure}

\cite{BE95.2} found an excess of red galaxies from the APM catalog
with moderate-redshift ($z\sim1$) 1-Jansky QSOs on angular scales
$<5'$ at the $99.1\%$ significance level. Their colour selection
ensures that the galaxies are most likely at redshifts $0.2\le
z\le0.4$, well in the foreground of the QSOs. The amplitude and
angular scale of the excess is compatible with its originating from
lensing by large-scale structures. The measurements by \cite{BE95.2}
are plotted together with various theoretical QSO-galaxy
cross-correlation functions in Fig.~\ref{fig:7.4}, which clearly shows
that the QSO-galaxy cross-correlation measurements agree quite well
with the cross-correlation functions $\xi_\mathrm{QG}(\phi)$, but they
fall above the range of theoretical predictions at small angular
scales, $\phi\lesssim2'$. This can be attributed to the magnification
bias due to gravitational lensing by individual clusters. Being based
on the weak-lensing approximation, our approach breaks down when the
magnification becomes comparable to unity, $\mu\gtrsim1.5$, say. This
amount of magnification occurs for QSOs closer than $\sim3$~Einstein
radii to cluster cores. Depending on cosmological parameters, QSO and
galaxy redshifts, $\sim3$~Einstein radii correspond to
$\sim1'-2'$. Hence, we {\em expect\/} the theoretical expectations
from lensing by large-scale structures alone to fall below the
observations on angular scales $\phi\lesssim1'-2'$.

\cite{noi99} took wide-field R-band images centred on a subsample of
1-Jansky QSOs with redshifts between 1 and 2. They searched for an
excess of galaxies in the magnitude range $19.5<R<21$ on angular
scales of $\gtrsim10'$ around these QSOs and found a correlation at
the $99\%$ significance level. The redshift distribution of the
galaxies is likely to peak around $z\sim0.2$. The angular
cross-correlation function between the QSOs and the galaxies agrees
well with the theoretical expectations, although the error bars are
fairly large.

All these results indicate that there are correlations between
background QSOs and foreground `light', with light either in the
optical, the infrared, or the (soft) X--ray wave bands. The angular
scale of the correlations is compatible with that expected from
lensing by large-scale structures, and the amplitude is either
consistent with that explanation or somewhat larger. \cite{WU96.4}
discussed whether the autocorrelation of clusters modelled as singular
isothermal spheres can produce sufficient magnification to explain
this result. They found that this is not the case, and argued that
large-scale structures must contribute substantially.

If lensing is indeed responsible for the correlations detected, other
signatures of lensing should be found in the vicinity of distant
QSOs. Indeed, \cite{FO96.1} searched for the shear induced by weak
lensing in the fields of five luminous QSOs with $z\approx1$ and found
coherent shear signals in four of them (see also \cite*{SC98.3}). In
addition, they detected galaxy groups in three of their
fields. Earlier, \cite{BO93.1} had found evidence for coherent weak
shear in the field of the potentially multiply-imaged QSO~2345$+$007,
which was later identified with a distant cluster (\cite*{ME94.1};
\cite*{FI94.2}).

\cite{BO97.1} searched for weak-lensing signals in fields around eight
luminous radio sources at redshifts $\sim1$. They confirmed the
coherent shear detected earlier by \cite{FO96.1} around one of the
sources (3C336 at $z=0.927$), but failed to find signatures of weak
lensing in the combined remaining seven fields.

A cautionary note was recently added to this discussion by
\cite{wii98} and \cite{now99}. Cross-correlating LBQS and 1-Jansky
quasars with APM galaxies, they claimed significant galaxy
overdensities around QSOs on angular scales of order one degree. As
discussed above, lensing by currently favoured models of large-scale
structures is not able to explain such large correlation scales. Thus,
if these results hold up, they would provide evidence that there is a
fundamental difficulty with the current models of large-scale
structure formation.

\subsection{\label{sc:7.5}Outlook}

Cross correlations between distant QSOs and foreground galaxies on
angular scales of about ten arc minutes have been observed, and they
can be attributed to the magnification bias due to gravitational
lensing by large-scale structures. Coherent shear patterns have been
detected around QSOs which are significantly correlated with
galaxies. The observations so far are in reasonable agreement with
theoretical expectations, except for the higher observed signal in the
innermost few arc minutes, and the claimed correlation signal on
degree scales. While the excess cross-correlation on small scales can
be understood by the lensing effects of individual galaxy clusters,
correlations on degree scales pose a severe problem for the lensing
explanation if they persist, because the lensing-induced
cross-correlation quickly dies off beyond scales of approximately
$10'$.

QSO-galaxy cross-correlations have the substantial advantage over
other diagnostics of weak lensing by large-scale structures that they
do not pose any severe observational problems. In particular, it is
not necessary to measure either shapes or sizes of faint background
galaxies accurately, because it is sufficient to detect and count
comparatively bright foreground galaxies near QSOs. However, such
counting requires homogeneous photometry, which is difficult to
achieve in particular on photographic plates, and requires careful
calibration.

Since the QSO-galaxy cross-correlation function involves filtering the
density-perturbation power spectrum with a fairly broad function, the
zeroth-order Bessel function $\mathrm{J}_0(x)$
[cf.~eq.~(\ref{eq:7.11})], these correlations are not well suited for
constraining the power spectrum. If the cluster normalisation is close
to the correct one, the QSO-galaxy cross-correlation function is also
fairly insensitive to cosmological parameters.

Rather, QSO-galaxy cross correlations are primarily important for
measuring the bias parameter $b$. The rationale of future observations
of QSO-galaxy correlations should therefore be to accurately measure
the correlation amplitude on scales between a few and 10 arc
minutes. On smaller scales, the influence of individual galaxy
clusters sets in, and on larger scales, the correlation signal is
expected to be weak. Once it becomes possible to reliably constrain
the density-fluctuation power spectrum, such observations can then be
used to quantify the bias parameter, and thereby provide most valuable
information for theories of galaxy formation. A possible dependence of
the bias parameter on scale and redshift can also be extracted.

Sufficiently large data fields for this purpose will soon become
available, in particular through wide-field surveys like the 2dF
Survey (\cite*{col98}) and the Sloan Digital Sky Survey
(\cite*{guk93}, \cite*{lop98}). It therefore appears feasible that
within a few years weak lensing by large-scale structures will be able
to quantify the relation between the distributions of galaxies and the
dark matter.

  % -*- LaTeX -*-

\section{\label{sc:8}Galaxy-Galaxy Lensing}

\subsection{\label{sc:8.1}Introduction}

Whereas the weak lensing techniques described in Sect.~\ref{sc:5} are
adequate to map the projected matter distribution of galaxy clusters,
individual galaxies are not sufficiently massive to show up in the
distortion of the images of background galaxies. From the
signal-to-noise ratio (\ref{eq:4.55}, page~\pageref{eq:4.55}) we see
that individual isothermal haloes with a velocity dispersion in excess
of $\sim600\,\mathrm{km\,s}^{-1}$ can be detected at a high
significance level with the currently achievable number densities of
faint galaxy images. Galaxies have haloes of much lower velocity
dispersion: The velocity dispersion of an $L_*$ elliptical galaxy is
$\sim220\,\mathrm{km\,s}^{-1}$, that of an $L_*$ spiral
$\sim145\,\mathrm{km\,s}^{-1}$.

However, if one is not interested in the mass properties of individual
galaxies, but instead in the statistical properties of massive haloes
of a population of galaxies, the weak lensing effects of several such
galaxies can statistically be superposed. For example, if one
considers $N_\mathrm{f}$ identical foreground galaxies, the
signal-to-noise ratio of the combined weak lensing effect increases as
$N_\mathrm{f}^{1/2}$, so that for a typical velocity dispersion for
spiral galaxies of $\sigma_v\sim160\,\mathrm{km\,s}^{-1}$, a few
hundred foreground galaxies are sufficient to detect the distortion
they induce on the background galaxy images.

Of course, detection alone does not yield new insight into the mass
properties of galaxy haloes. A quantitative analysis of the lensing
signal must account for the fact that `identical' foreground galaxies
cannot be observed.  Therefore, the mass properties of galaxies have
to be parameterised in order to allow the joint analysis of the
foreground galaxy population. In particular, one is interested in the
velocity dispersion of a typical ($L_*$, say) galaxy. Furthermore, the
rotation curves of (spiral) galaxies which have been observed out to
$\sim30 h^{-1}$~kpc show no hint of a truncation of the dark halo out
to this distance. Owing to the lack of dynamical tracers, with the
exception of satellite galaxies (\cite*{zaw94}), a direct observation
of the extent of the dark halo towards large radii is not feasible
with conventional methods. The method described in this section uses
the light bundles of background galaxies as dynamical tracers, which
are available at all distances from the galaxies' centres, and are
therefore able, at least in principle, to probe the size (or the
truncation radius) of the haloes. Methods for a quantitative analysis
of galaxy haloes will be described in Sect.~\ref{sc:8.2}.

The first attempt at detecting this galaxy-galaxy lensing effect was
reported by \cite{TY84.1}, but the use of photographic plates and the
relatively poor seeing prevented them from observing a galaxy-galaxy
lensing signal. The first detection was reported by \cite{BR96.1}, and
as will be described in Sect.~\ref{sc:8.3}, several further
observational results have been derived.

Gravitational light deflection can also be used to study the dark
matter haloes of galaxies in clusters. The potential influence of the
environment on the halo properties of galaxies can provide a strong
hint on the formation and lifetimes of clusters. One might expect that
galaxy haloes are tidally stripped in clusters and therefore
physically smaller than those of field galaxies. In
Sect.~\ref{sc:8.4}, we consider galaxy-galaxy lensing in clusters, and
report on some first results.

\subsection{\label{sc:8.2}The Theory of Galaxy-Galaxy Lensing}

A light bundle from a distant galaxy is affected by the tidal field of
many foreground galaxies. Therefore, in order to describe the image
distortion, the whole population of foreground galaxies has to be
taken into account. But first we shall consider the simple case that
the image shape is affected (mainly) by a single foreground
galaxy. Throughout this section we assume that the shear is weak, so
that we can replace (\ref{eq:4.12}, page~\pageref{eq:4.12}) by
\begin{equation}
  \epsilon^\s=\epsilon-\gamma\;.
\label{eq:8.1}
\end{equation}
Consider an axi-symmetric mass distribution for the foreground galaxy,
and background images at separation $\theta$ from its centre. The
expectation value of the image ellipticity then is the shear at
$\theta$, which is oriented tangentially. If $p(\epsilon)$ and
$p^\s(\epsilon^\s)$ denote the probability distributions of the image
and source ellipticities, then according to (\ref{eq:8.1}),
\begin{equation}
  p(\epsilon) = p^\s(\epsilon-\gamma) =
  p^\s(\epsilon)-\gamma_\alpha
  \frac{\partial}{\partial\epsilon_\alpha}p^\s(\epsilon)\;,
\label{eq:8.2}
\end{equation}
where the second equality applies for $|\gamma|\ll1$. If $\varphi$ is
the angle between the major axis of the image ellipse and the line
connecting source and lens centre, one finds the probability
distribution of $\varphi$ by integrating (\ref{eq:8.2}) over the
modulus of $\epsilon$,
\begin{equation}
  p(\varphi) = \int\d|\epsilon|\,|\epsilon|\,p(\epsilon) =
  \frac{1}{2\pi}-\gamma_{\rm t}\,\cos(2\varphi)\frac{1}{2\pi}\,
  \int\d|\epsilon|\;p^\s(\epsilon)\;,
\label{eq:8.3}
\end{equation}
where $\varphi$ ranges within $[0,2\pi]$. Owing to the symmetry of the
problem, we can restrict $\varphi$ to within 0 and $\pi/2$, so that
the probability distribution becomes
\begin{equation}
  p(\varphi) = \frac{2}{\pi}\left[
    1-\gamma_\mathrm{t}
    \left\langle\frac{1}{\epsilon^\s}\right\rangle\cos(2\varphi)
  \right]\;,
\label{eq:8.4}
\end{equation}
i.e., the probability distribution is skewed towards values larger
than $\pi/4$, showing preferentially a tangential alignment.

Lensing by additional foreground galaxies close to the line-of-sight
to the background galaxy does not substantially change the probability
distribution (\ref{eq:8.4}). First of all, since we assume weak
lensing throughout, the effective shear acting on a light bundle can
well be approximated by the sum of the shear contributions from the
individual foreground galaxies. This follows either from the linearity
of the propagation equation in the mass distribution, or from the
lowest-order approximation of multiple-deflection gravitational
lensing (e.g., \cite*{BL86.1}; \cite*{SE92.1}). Second, the additional
lensing galaxies are placed at random angles around the line-of-sight,
so that the expectation value of their combined shear averages to
zero. Whereas they slightly increase the dispersion of the observed
image ellipticities, this increase is negligible since the dispersion
of the intrinsic ellipticity distribution is by far the dominant
effect. However, if the lens galaxy under consideration is part of a
galaxy concentration, such as a cluster, the surrounding galaxies are
not isotropically distributed, and the foregoing argument is
invalid. We shall consider galaxy-galaxy lensing in clusters in
Sect.~\ref{sc:8.4}, and assume here that the galaxies are generally
isolated.

For an ensemble of foreground-background pairs of galaxies, the
probability distribution for the angle $\varphi$ simply reads
\begin{equation}
  p(\varphi) = \frac{2}{\pi}\left[
    1-\langle\gamma_\mathrm{t}\rangle
    \left\langle\frac{1}{\epsilon^\s}\right\rangle
    \cos(2\varphi)\right]\;,
\label{eq:8.5}
\end{equation}
where $\langle\gamma_{\rm t}\rangle$ is the mean tangential shear of
all pairs considered. The function $p(\varphi)$ is an observable. A
significant deviation from a uniform distribution signals the presence
of galaxy-galaxy lensing. To obtain quantitative information on the
galaxy haloes from the amplitude of the cosine term, one needs to know
$\langle1/\epsilon^\s\rangle$. It can directly be derived from
observations because the weak shear assumed here does not
significantly change this average between source and image
ellipticities, from a parameterised relation between observable galaxy
properties, and from the mean shear
$\langle\gamma_\mathrm{t}\rangle$. Although in principle fine binning
in galaxy properties (like colour, redshift, luminosity, morphology)
and angular separation of foreground-background pairs is possible in
order to probe the shear as a function of angular distance from a
well-defined set of foreground galaxies and thus to obtain its radial
mass profile without any parameterisation, this approach is currently
unfeasible owing to the relatively small fields across which
observations of sufficient image quality are available.

A convenient parameterisation of the mass profile is the truncated
isothermal sphere with surface mass density
\begin{equation}
  \Sigma(\xi) = \frac{\sigma_v^2}{2G\xi}\left(
    1-\frac{\xi}{\sqrt{s^2+\xi^2}}\right)\;, 
\label{eq:8.6}
\end{equation}
where $s$ is the truncation radius. This is a special case of the mass
distribution (\ref{eq:3.20}, page~\pageref{eq:3.20}).  \cite{BR96.1}
showed that this mass profile corresponds to a physically realisable
dark-matter particle distribution.\footnote{It is physically
realisable in the sense that there exists an isotropic, non-negative
particle distribution function which gives rise to a spherical density
distribution corresponding to (\ref{eq:8.6}).} The velocity dispersion
is assumed to scale with luminosity according to (\ref{eq:2.67},
page~\pageref{eq:2.67}), which is supported by observations. A similar
scaling of $s$ with luminosity $L$ or velocity dispersion $\sigma_v$
is also assumed,
\begin{equation}
  s = s_*\left(\frac{\sigma_v}{\sigma_{v,*}}\right)^2 = 
  s_*\left(\frac{L}{L_*}\right)^{2/\alpha}\;,
\label{eq:8.7}
\end{equation}
where the choice of the exponent is largely arbitrary. The scaling in
(\ref{eq:8.7}) is such that the ratio of truncation radius and
Einstein radius at fixed redshift is independent of $L$. If, in
addition, $\alpha=4$, the total mass-to-light ratio is identical for
all galaxies. The fiducial luminosity $L_*$ may depend on
redshift. For instance, if the galaxies evolve passively, their mass
properties are unaffected, but aging of the stellar population cause
them to become fainter with decreasing redshift. This effect may be
important for very deep observations, such as the Hubble Deep Field
(\cite*{hgd98}), in which the distribution of lens galaxies extends to
high redshifts.

The luminosity $L$ of a lens galaxy can be inferred from the observed
flux and an assumed redshift. Since the scaling relation
(\ref{eq:2.67}) applies to the luminosity measured in a particular
waveband, the calculation of the luminosity from the apparent
magnitude in a specified filter needs to account for the
k-correction. If data are available in a single waveband only, an
approximate average k-correction relation has to be chosen. For
multi-colour data, the k-correction can be estimated for individual
galaxies more reliably. In any case, one assumes a relation between
luminosity, apparent magnitude, and redshift,
\begin{equation}
  L = L(m,z)\;.
\label{eq:8.8}
\end{equation}

The final aspect to be discussed here is the redshift of the
galaxies. Given that a galaxy-galaxy analysis involves at least
several hundred foreground galaxies, and even more background
galaxies, one cannot expect that all of them have spectroscopically
determined redshifts. In a more favourable situation, multi-colour
data are given, from which a redshift estimate can be obtained, using
the photometric redshift method (e.g., \cite*{ccs95}; \cite*{gwh96};
\cite*{hcb98}). These redshift estimates are characteristically
accurate to $\Delta z\sim0.1$, depending on the photometric accuracy
and the number of filter bands in which photometric data are
measured. For a single waveband only, one can still obtain a redshift
estimate, but a quite unprecise one.  One then has to use the redshift
distribution of galaxies at that particular magnitude, obtained from
spectroscopic or multi-colour redshift surveys in other fields. Hence,
one assumes that the redshift probability distribution $p_z(z;m)$ as a
function of magnitudes is known sufficiently accurately.

Suppose for a moment that all galaxy redshifts were known. Then, one
can predict the effective shear for each galaxy, caused by all the
other galaxies around it,
\begin{equation}
  \gamma_i = \sum_j\gamma_{ij}
  (\vec\theta_i-\vec\theta_j,z_i,z_j,m_j)\;,
\label{eq:8.9}
\end{equation}
where $\gamma_{ij}$ is the shear produced by the $j$-th galaxy on the
$i$-th galaxy image, which depends on the angular separation and the
mass properties of the $j$-th galaxy. From its magnitude and redshift,
the luminosity can be inferred from (\ref{eq:8.8}), which fixes
$\sigma_v$ and the halo size $s$ through the scaling relations
(\ref{eq:2.67}) and (\ref{eq:8.7}). Of course, for $z_i\le z_j$,
$\gamma_{ij}=0$. Although the sum in (\ref{eq:8.9}) should in
principle extend over the whole sky, the lensing effect of all
foreground galaxies with angular separation larger than some
$\theta_\mathrm{max}$ will average to zero. Therefore, the sum can be
restricted to separations $\le\theta_\mathrm{max}$. We shall discuss
the value of $\theta_\mathrm{max}$ further below.

In the realistic case of unknown redshifts, but known probability
distribution $p_z(z;m)$, the shear $\gamma_i$ cannot be
determined. However, by averaging (\ref{eq:8.9}) over $p_z(z;m)$, the
mean and dispersion, $\langle\gamma_i\rangle$ and $\sigma_{\gamma,i}$,
of the shear for the $i$-th galaxy can be calculated. Instead of
performing the high-dimensional integration explicitly, this averaging
can conveniently be done by a Monte-Carlo integration. One can
generate multiple realisations of the redshift distribution by
randomly drawing redshifts from the probability density
$p_z(z;m)$. For each realisation, the $\gamma_i$ can be calculated
from (\ref{eq:8.9}). By averaging over the realisations, the mean
$\langle\gamma_i\rangle$ and dispersion $\sigma_{\gamma,i}$ of
$\gamma_i$ can be estimated.

\subsection{\label{sc:8.3}Results}

The first attempt at detecting galaxy-galaxy lensing was made by
\cite{TY84.1}. They analysed a deep photographic survey consisting of
35 prime-focus plates with the 4-meter Mayall Telescope at Kitt
Peak. An area of 36~(arc min.)$^2$ on each plate was digitised. After
object detection, $\sim12,000$ `foreground' and $\sim47,000$
`background' galaxies were selected by their magnitudes, such that the
faintest object in the `foreground' class was one magnitude brighter
than the brightest `background' galaxy. This approach assumes that the
apparent magnitude of an object provides a good indication for its
redshift, which seems to be valid, although the redshift distributions
of `foreground' and `background' galaxies will substantially
overlap. There were $\sim28,000$ foreground-background pairs with
$\Delta\theta\le63''$ in their sample, but no significant tangential
alignment could be measured. By comparing their observational results
with Monte-Carlo simulations, \citename{TY84.1} concluded that the
characteristic velocity dispersion of a foreground galaxy in their
sample must be smaller than about $120\,\mathrm{km\,s}^{-1}$. This
limit was later revised upwards to $\sim230\,\mathrm{km\,s}^{-1}$ by
\cite{KO87.6} who noted that the assumption made in
\citename{TY84.1}'s analysis that all background galaxies are at
infinite distance (i.e., $D_\mathrm{ds}/D_\mathrm{s}=1$) was
critical. This upper limit is fully compatible with our knowledge of
galaxy masses.

This null-detection of galaxy-galaxy lensing in a very large sample of
objects apparently discouraged other attempts for about a
decade. After the first weak-lensing results on clusters became
available, it was obvious that this method requires deep data with
superb image quality. In particular, the non-linearity of photographic
plates and mediocre seeing conditions are probably fatal to the
detection of this effect, owing to its smallness. The shear at $5''$
from an $L_*$ galaxy with $\sigma_v=160\,\mathrm{km\,s}^{-1}$ is less
than 5\%, and pairs with smaller separations are very difficult to
investigate as the bright galaxy will affect the ellipticity
measurement of its close neighbour on ground-based images.

\begin{figure}[ht]
  \includegraphics[width=\hsize]{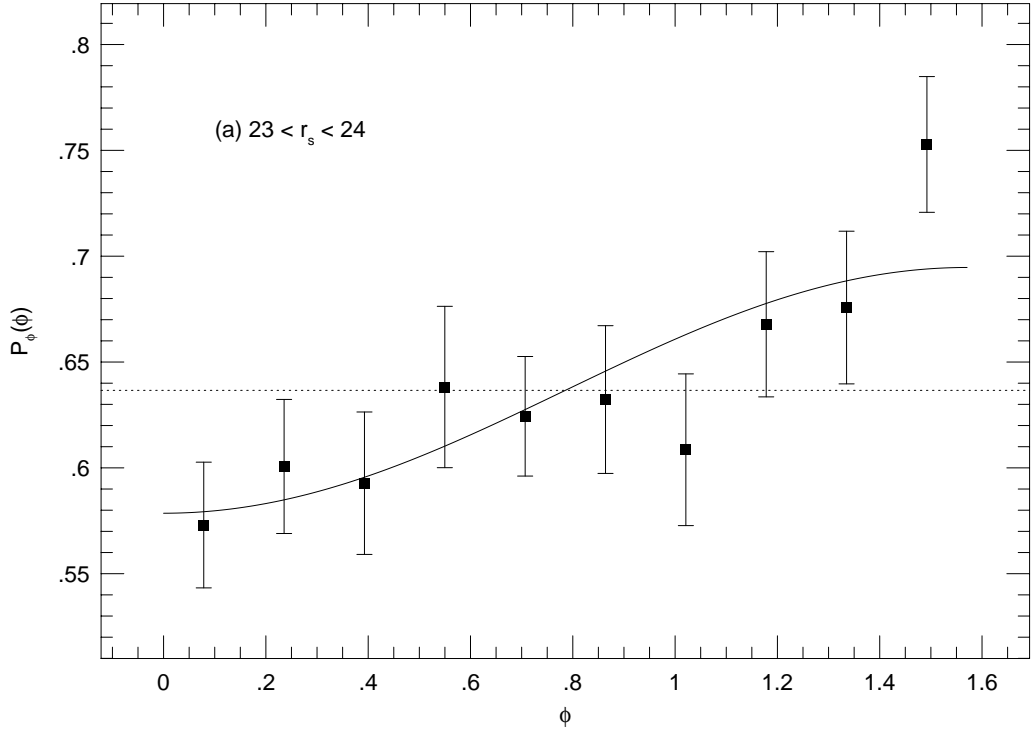}
\caption{The probability distribution $p(\varphi)$ for the 3202
foreground-background pairs ($20\le r\le 23$ and $23\le r\le 24$,
respectively) with $5''\le\Delta\theta\le34''$ in the sample used by
\protect\cite{BR96.1}, together with the best fit according to
(\ref{eq:8.5}). The observed distribution is incompatible with a flat
distribution (dotted line) at a high confidence level of 99.9\%
(Fig.~2a of \protect\citename{BR96.1}).}
\label{fig:8.1}
\end{figure}

Using a single $9\arcminf6\times9\arcminf6$ blank field, with a total
exposure time of nearly seven hours on the 5-meter Hale Telescope on
Mount Palomar, \cite{BR96.1} reported the first detection of
galaxy-galaxy lensing. Their co-added image had a seeing of
$0\arcsecf87$ at FWHM, and the 97\% completeness limit was
$r=26$. They considered `foreground' galaxies in the magnitude range
$20\le r\le23$, and several fainter bins for defining the `background'
population, and investigated the distribution function $p(\varphi)$
for pairs with separation $5''\le\Delta\theta\le34''$. The most
significant deviation of $p(\varphi)$ from a flat distribution occurs
for `background' galaxies in the range $23\le r\le24$. For fainter
(and thus smaller) galaxies, the accuracy of the shape determination
deteriorates, as \citename{BR96.1} explicitly show. The number of
`foreground' galaxies, `background' galaxies, and pairs, is
$N_\mathrm{f}=439$, $N_\mathrm{b}=506$, and $N_\mathrm{pairs}=3202$.
The binned distribution for this `background' sample is shown in
Fig.~\ref{fig:8.1}, together with a fit according to (\ref{eq:8.5}). A
Kolmogorov-Smirnov test rejects a uniform distribution of $p(\varphi)$
at the 99.9\% level, thus providing the first detection of
galaxy-galaxy lensing.

\citename{BR96.1} performed a large number of tests to check for
possible systematic errors, including null tests (e.g., replacing the
positions of `foreground' galaxies by random points, or stars),
splitting the whole sample into various subsamples (e.g., inner part
vs.~outer part of the image, upper half vs.~lower half etc.), and
these tests were passed satisfactorily. Also a slight PSF anisotropy
in the data, or contamination of the ellipticity measurement of faint
galaxies by brighter neighbouring galaxies, cannot explain the
observed relative alignment, as tested with extensive simulations, so
that the detection must be considered real.

\begin{figure}
  \centerline{\includegraphics[width=0.6\hsize]{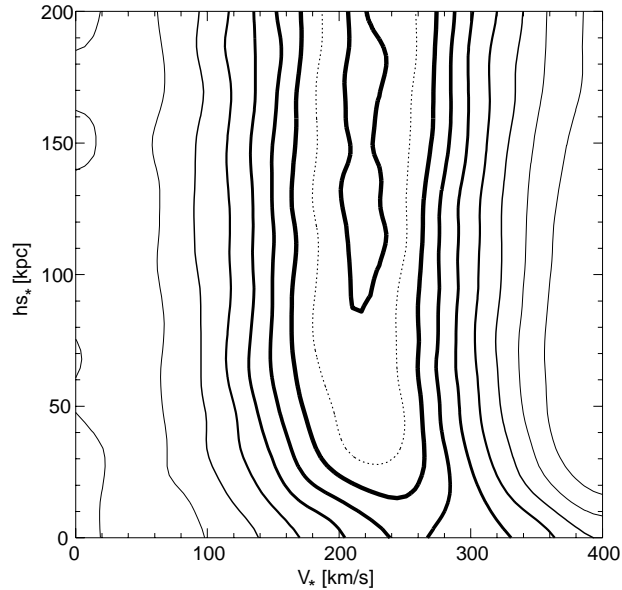}}
\caption{Contours of constant $\chi^2$ in the $V_*$--$h\,s_*$
parameter plane, where $V_*=\sqrt{2}\sigma_{v,*}$, obtained from a
comparison of the observed tangential alignment
$\langle\gamma_\mathrm{t}\rangle$ with the distribution found in
Monte-Carlo simulations. The solid contours range from 0.8 (innermost)
to 8 per degree of freedom; the dotted curve displays $\chi^2=1$ per
degree of freedom. (Fig.~7 of \protect\citename{BR96.1}).}
\label{fig:8.2}
\end{figure}

\citename{BR96.1} then quantitatively analysed their observed
alignment, using the model outlined in Sect.~\ref{sc:8.2}, with
$\alpha=4$. The predictions of the model were inferred from
Monte-Carlo simulations, in which galaxies were randomly distributed
with the observed number density, and redshifts were assigned
according to a probability distribution $p_z(z;m)$, for which they
used a slight extrapolation from existing redshift surveys, together
with a simple prescription for the k-correction in (\ref{eq:8.8}) to
assign luminosities to the galaxies. The ellipticity for each
background galaxy image was then obtained by randomly drawing an
intrinsic ellipticity, adding shear according to (\ref{eq:8.9}). The
simulated probability distribution $p(\varphi)$ was discretised into
several bins in angular separation $\Delta\theta$, and compared to the
observed orientation distribution, using $\chi^2$-minimisation with
respect to the model parameters $\sigma_{v,*}$ and $s_*$. The result
of this analysis is shown in Fig.~\ref{fig:8.2}. The shape of the
$\chi^2$-contours is characteristic in that they form a valley which
is relatively narrow in the $\sigma_{v,*}$-direction, but extends very
far out into the $s_*$-direction. Thus, the velocity dispersion
$\sigma_{v,*}$ can significantly be constrained with these
observations, while only a lower limit on $s_*$ can be derived. Formal
90\% confidence limits on $\sigma_{v,*}$ are
$\sim100\,\mathrm{km\,s}^{-1}$ and $\sim210\,\mathrm{km\,s}^{-1}$,
with a best-fitting value of about $160\,\mathrm{km\,s}^{-1}$, whereas
the 1- and 2-$\sigma$ lower limits on $s_*$ are $25\,h^{-1}$~kpc and
$\sim10\,h^{-1}$~kpc, respectively.

Finally, \citename{BR96.1} studied the dependence of the lensing
signal $\langle\gamma_\mathrm{t}\rangle$ on the colour of their
`background' sample, by splitting it into a red and a blue half. The
lensing signal of the former is compatible with zero on all scales,
while the blue sample reveals a strong signal which decreases with
angular separation as expected. This result is in accordance with that
discussed in Sect.~\ref{sc:5.5.3}, where the blue galaxies showed a
stronger lensing signal as well, indicating that their redshift
distribution extends to larger distances.

We have discussed the work of \cite{BR96.1} in some detail since it
provided the first detection of galaxy-galaxy lensing, and since it is
so far remains the only one obtained from the ground. Also, their
careful analysis exemplifies the difficulties in deriving a convincing
result.

\cite{GR96.2} analysed the images from the {\em Hubble Space
Telescope\/} Medium Deep Survey (MDS) in terms of galaxy-galaxy
lensing. The MDS is an imaging survey, using parallel data obtained
with the WFPC2 camera on-board HST. They identified 1600 `foreground'
($15<I<22$) and 14000 `background' ($22<I<26$) galaxies. Owing to the
spatial resolution of the HST, a morphological classification of the
foreground galaxies could be performed, and spiral and elliptical
galaxies could separately be analysed. They considered the mean
orientation angle $\langle\varphi\rangle = \pi/4+\pi^{-1}
\langle\gamma_\mathrm{t}\rangle\langle1/|\epsilon^\s|\rangle$ as a
statistical variable, and scaled the truncation radius in their mass
models in proportion to the half-light radius. They found that
$\sigma_{v,*}=220\,\mathrm{km\,s}^{-1}$ and
$\sigma_{v,*}=160\,\mathrm{km\,s}^{-1}$ are compatible with their
shear data for elliptical and spiral galaxies, respectively. For their
sample of elliptical foreground galaxies, they claim that the
truncation radius must be more than ten times the half-light radius to
fit their data, and that a de Vaucouleurs mass profile is
excluded. Unfortunately, no significance levels are quoted.

A variant of the method for a quantitative analysis of galaxy-galaxy
lensing was developed by \cite{SC97.2}. Instead of a $\chi^2$-analysis
of $\langle\gamma_\mathrm{t}\rangle$ in angular separation bins, they
suggested a maximum-likelihood analysis, using the individual galaxy
images. In their Monte-Carlo approach, the galaxy positions (and
magnitudes) are kept fixed, and only the redshifts of the galaxies are
drawn from their respective probability distribution $p_z(z;m)$, as
described at the end of Sect.~\ref{sc:8.2}. The resulting
log-likelihood function
\begin{equation}
  \ell = -\sum_i\frac{|\epsilon_i-\langle\gamma_i\rangle|^2}
  {\rho^2+\sigma_{\gamma,i}^2} -\sum_i\ln\left[
    \pi(\rho^2+\sigma_{\gamma,i}^2)\right]\;,
\label{eq:8.10}
\end{equation}
where $\rho$ is the dispersion of intrinsic ellipticity distribution,
here assumed to be a Gaussian, can then be maximised with respect to
the model parameters, e.g., $\sigma_{v,*}$ and $s_*$. Extensive
simulations demonstrated that this approach, which utilises all of the
information provided by observations, yields an unbiased estimate of
these model parameters. Later, \cite{erb97} showed that this remains
valid even if the lens galaxies have elliptical projected mass
profiles.

This method was applied to the deep multi-colour imaging data of the
Hubble Deep Field (HDF; \cite*{wbd96}) by \cite{hgd98}, after
\cite{DE96.1} detected a galaxy-galaxy lensing signal in the HDF on an
angular scale of $\lesssim5''$. The availability of data in four
wavebands allows an estimate of photometric redshifts, a method
demonstrated to be quite reliable by spectroscopy of HDF galaxies
(e.g., \cite*{hcb98}). The accurate redshift estimates, and the depth
of the HDF, compensates for the small field-of-view of
$\sim5\,\mathrm{arcmin}^2$. A similar study of the HDF data was
carried out by the Caltech group (see \cite*{bck98}).

In order to avoid k-corrections, using the multi-colour photometric
data to relate all magnitudes to the rest-frame B-band,
\citename{hgd98} considered lens galaxies with redshift $z\lesssim
0.85$ only, leaving 208 galaxies. Only such source-lens pairs for
which the estimated redshifts differ by at least 0.5 were included in
the analysis, giving about $10^4$ foreground-background pairs. They
adopted the same parameterisation for the lens population as described
in Sect.~\ref{sc:8.2}, except that the depth of the HDF suggests that
the fiducial luminosity $L_*$ should be allowed to depend on redshift,
$L_*\propto(1+z)^\zeta$. Assuming no evolution, $\zeta=0$, and a
Tully-Fisher index of $1/\alpha=0.35$, they found
$\sigma_{v,*}=(160\pm 30)\,\mathrm{km\,s}^{-1}$. Various control tests
were performed to demonstrate the robustness of this result, and
potential systematic effects were shown to be negligible.

As in the previous studies, halo sizes could not be significantly
constrained. The lensing signal is dominated by spiral galaxies at a
redshift of $z\sim0.6$. Comparing the Tully-Fisher relation at this
redshift to the local relation, the lensing results indicate that
intermediate-redshift galaxies are fainter than local spirals by
$1\pm0.6$ magnitudes in the B-band, at fixed circular velocity.

Hence, all results reported so far yield compatible values of
$\sigma_{v,*}$, but do not allow upper bounds on the halo size to be
set. The flatness of the likelihood surface in the $s_*$-direction
shows that a measurement of $s_*$ requires much larger samples than
used before. We can understand the insensitivity to $s_*$ in the
published analyses at least qualitatively. The shear caused by a
galaxy at a distance of, say, $100\,\mathrm{kpc}$ is very small, of
order 1\%. This implies that the difference in shear caused by
galaxies with truncation radius of $20\,\mathrm{kpc}$ and
$s=100\,\mathrm{kpc}$ is very small indeed. In addition, there are
typically other galaxies closer to the line-of-sight to background
galaxies which produce a larger shear, making it more difficult to
probe the shear of widely separated foreground galaxies. Hence, to
probe the halo size, many more foreground-background pairs must be
considered. In addition, the angular scale $\theta_\mathrm{max}$
within which pairs are considered needs to be larger than the angular
scale of the truncation radius at typical redshifts of the galaxies,
and on the other hand, $\theta_\mathrm{max}$ should be much smaller
than the size of the data field available. Hence, to probe large
scales of the halo, wide-field imaging data are needed.

There is a related problem which needs to be understood in greater
detail. Since galaxies are clustered, and probably (biased) tracers of
an underlying dark matter distribution (e.g., most galaxies may live
in groups), it is not evident whether the shear caused by a galaxy at
a spatial separation of, say, $100\,\mathrm{kpc}$ is caused mainly by
the dark matter halo of the galaxy itself, or rather by the
dark-matter halo associated with the group. Here, numerical
simulations of the dark matter may indicate to which degree these two
effects can be separated, and observational strategies for this need
to be developed.

\subsection{\label{sc:8.4}Galaxy-Galaxy Lensing in Galaxy Clusters}

An interesting extension of the work described above aims at the
investigation of the dark-matter halo properties of galaxies within
galaxy clusters. In the hierarchical model for structure formation,
clusters grow by mergers of less massive haloes, which by themselves
formed by merging of even smaller substructures. Tidal forces in
clusters, possible ram-pressure stripping by the intra-cluster medium,
and close encounters during the formation process, may affect the
haloes of galaxies, most of which presumably formed at an early
epoch. Therefore, it is unclear at present whether the halo properties
of galaxies in clusters are similar to those of field galaxies.

Galaxy-galaxy lensing offers an exciting opportunity to probe the dark
galaxy haloes in clusters. There are several differences between the
investigation of field and of cluster galaxies. First, the number of
massive galaxies in a cluster is fairly small, so the statistics for a
single cluster will be limited. This can be compensated by
investigating several clusters simultaneously. Second, the image
distortion is determined by the reduced shear,
$g=\gamma/(1-\kappa)$. For field galaxies, where the shear and the
surface mass density is small, one can set $g\approx\gamma$, but this
approximation no longer holds for galaxies in clusters, where the
cluster provides $\kappa$ substantially above zero. This implies that
one needs to know the mass distribution of the cluster before the
statistical properties of the massive galaxy haloes can be
investigated. On the other hand, it magnifies the lensing signal from
the galaxies, so that fewer cluster galaxies are needed to derive
significant lensing results compared to field galaxies of similar
mass. Third, most cluster galaxies are of early type, and thus their
$\sigma_{v,*}$ -- and consequently, their lensing effect -- is
expected to be larger than for typical field galaxies.

In fact, the lensing effect of individual cluster galaxies can even be
seen from strong lensing. Modelling clusters with many strong-lensing
constraints (e.g., several arcs, multiple images of background
galaxies), the incorporation of individual cluster galaxies turns out
to be necessary (e.g., \cite*{KA92.6}; \cite*{WA95.1};
\cite*{KN96.1}). However, the resulting constraints are relevant only
for a few cluster galaxies which happen to be close to the
strong-lensing features, and mainly concern the mass of these galaxies
within $\sim10\,h^{-1}$~kpc.

The theory of galaxy-galaxy lensing in clusters was developed in
\cite{NA97.3} and \cite{GE98.1}, using several different
approaches. The simplest possibility is related to the aperture mass
method discussed in Sect.~\ref{sc:5.3.1}. Measuring the tangential
shear within an annulus around each cluster galaxy, perhaps including
a weight function, permits a measurement of the aperture mass, and
thus to constrain the parameters of a mass model for the
galaxies. Provided the scale of the aperture is sufficiently small,
the tidal field of the cluster averages out to first order, and the
local influence of the cluster occurs through the local surface mass
density $\kappa$. In particular, the scale of the aperture should be
small enough in order to exclude neighbouring cluster galaxies.

A more sophisticated analysis starts from a mass model of the cluster,
as obtained by one of the reconstruction techniques discussed in
Sect.~\ref{sc:5}, or by a parameterised mass model constructed from
strong-lensing constraints. Then, parameterised galaxy models are
added, again with a prescription similar to that of
Sect.~\ref{sc:8.2}, and simultaneously the mass model of the cluster
is multiplied by the relative mass fraction in the smoothly
distributed cluster mass (compared to the total mass). In other words,
the mass added by inserting galaxies into the cluster is subtracted
from the smooth density profile. From the observed galaxy
ellipticities, a likelihood function can be defined and maximised with
respect to the parameters ($\sigma_{v,*}$, $s_*$) of the galaxy model.

\cite{NA98.1} applied this method to WFPC2 images of the cluster
AC~114 ($z_\mathrm{d}=0.31$). They concluded that most of the mass of
a fiducial $L_*$ cluster galaxy is contained in a radius of
$\sim15\,\mathrm{kpc}$, indicating that the halo size of galaxies in
this cluster is smaller than that of field galaxies.

Once the mass contained in the cluster galaxies is a significant
fraction of the total mass of the cluster, this method was found to
break down, or give strongly biased results. \cite{ges99} modified
this approach by performing a maximum-likelihood cluster mass
reconstruction for each parameter set of the cluster galaxies,
allowing the determination of the best representation of the global
underlying cluster component that is consistent with the presence of
the cluster galaxies and the observed image ellipticities of
background galaxies.

\begin{figure}
  \centerline{\includegraphics[width=0.8\hsize]{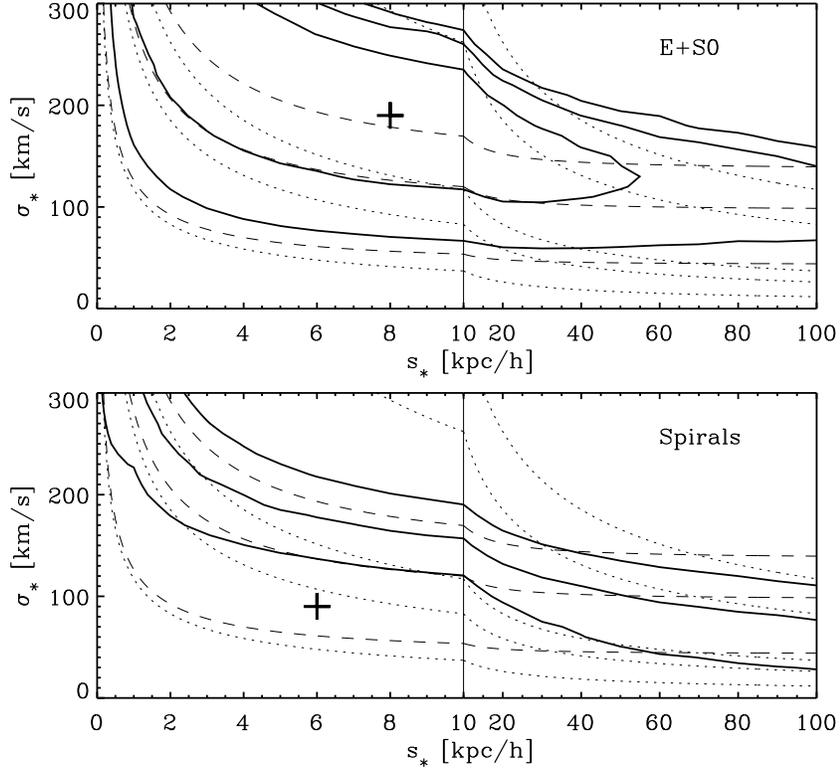}}
\caption{Results of applying the entropy-regularised
maximum-likelihood method for galaxy-galaxy lensing to the WFPC2 image
of the cluster Cl0939$+$4713. The upper and lower panels correspond to
early-type and spiral galaxies, respectively. The solid lines are
confidence contours at 68.3\%, 95.4\% and 99.7\%, and the cross marks
the maximum of the likelihood function. Dashed lines correspond to
galaxy models with equal aperture mass $M_*(<8
h^{-1}\,\mathrm{kpc})=(0.1, 0.5, 1.0)\times
10^{11}\,h^{-1}\,M_\odot$. Similarly, the dotted lines connect models
of constant total mass for an $L_*$-galaxy, of $M_*=(0.1, 0.5, 1.0,
5.0, 10)\times 10^{11}\,h^{-1}\,M_\odot$, which corresponds to a mass
fraction contained in galaxies of $(0.15, 0.75, 1.5, 7.5, 15)$\%,
respectively (from \protect\cite*{ges99}).}
\label{fig:8.3}
\end{figure}

This method was then applied to the WFPC-2 image of the cluster
Cl0939$+$4713, already described in Sect.~\ref{sc:5.4}. The
entropy-regularised maximum-likelihood mass reconstruction of the
cluster is very similar to the one shown in Fig.~\ref{fig:5.1}
(page~\pageref{fig:5.1}), except that the cluster centre is much
better resolved, with a peak very close to the observed strong lensing
features (\cite*{TR97.1}). Cluster galaxies were selected according to
their magnitudes, and divided by morphology into two subsamples,
viz.~early-type galaxies and spirals. In Fig.~\ref{fig:8.3} we show
the likelihood contours in the $s_*$--$\sigma_{v,*}$ plane, for both
subsets of cluster galaxies. Whereas there is no statistically
significant detection of lensing by spiral galaxies, the lensing
effect of early-type galaxies is clearly detected. Although no firm
upper limit of the halo size $s_*$ can be derived from this analysis
owing to the small angular field of the image (the maximum of the
likelihood function occurs at $8\,h^{-1}\,\mathrm{kpc}$, and a
1-$\sigma$ upper limit would be $\sim50\,h^{-1}\,\mathrm{kpc}$), the
contours `close' at smaller values of $s_*$ compared to the results
obtained from field galaxies. By statistically combining several
cluster images, a significant upper limit on the halo size can be
expected.

It should be noted that the results presented above still contain some
uncertainties, most notably the unknown redshift distribution of the
background galaxies and the mass-sheet degeneracy, which becomes
particularly severe owing to the small field-of-view of
WFPC2. Changing the assumed redshift distribution and the scaling
parameter $\lambda$ in (\ref{eq:5.8}, page~\pageref{eq:5.8}) shifts
the likelihood contours in Fig.~\ref{fig:8.3} up or down, i.e., the
determination of $\sigma_{v,*}$ is affected. As for galaxy-galaxy
lensing of field galaxies, the accuracy can be increased by using
photometric redshift estimates. Similarly, the allowed range of the
mass-sheet transformation can be constrained by combining these
small-scale images with larger scale ground-based images, or, if
possible, by using magnification information to break the
degeneracy. Certainly, these improvements of the method will be a
field of active research in the immediate future.

  % -*- LaTeX -*-

\section{\label{sc:9}The Impact of Weak Gravitational Light Deflection
  on the Microwave Background Radiation}

\subsection{\label{sc:9.1}Introduction}

The Cosmic Microwave Background originated in the hot phase after the
Big Bang, when photons were created in thermal equilibrium with
electromagnetically interacting particles. While the Universe expanded
and cooled, the photons remained in thermal equilibrium until the
temperature was sufficiently low for electrons to combine with the
newly formed nuclei of mainly hydrogen and helium. While the formation
of atoms proceeded, the photons decoupled from the matter due to the
rapidly decreasing abundance of charged matter. Approximately
$300,000$ years after the Big Bang, corresponding to a redshift of
$z\approx1,000$, the universe became transparent for the radiation,
which retained the Planck spectrum it had acquired while it was in
thermal equilibrium, and the temperature decreased in proportion with
the scale factor as the Universe expanded. This relic radiation,
cooled to $T=2.73\,$K, forms the Cosmic Microwave Background
(hereafter CMB). \cite{pew65} detected it as an ``excess antenna
temperature'', and \cite{fcg96} used the COBE-FIRAS instrument to
prove its perfect black-body spectrum.

Had the Universe been ideally homogeneous and isotropic, the CMB would
have the intensity of black-body radiation at $2.73\,$K in all
directions on the sky, and would thus be featureless. Density
perturbations in the early Universe, however, imprinted their
signature on the CMB through various mechanisms, which are thoroughly
summarised and discussed in \cite{huw95}. Photons in potential wells
at the time of decoupling had to climb out, thus losing energy and
becoming slightly cooler than the average CMB. This effect, now called
the {\em Sachs-Wolfe effect\/} was originally studied by
\cite{SA67.1}, who found that the temperature anisotropies in the CMB
trace the potential fluctuations on the `surface' of decoupling. CMB
fluctuations were first detected by the COBE-DMR experiment
(\cite*{sbk92}) and subsequently confirmed by numerous ground-based
and balloon-borne experiments (see \cite*{smo97} for a review).

The interplay between gravity and radiation pressure in perturbations
of the cosmic `fluid' before recombination gave rise to another
important effect. Radiation pressure is only effective in
perturbations smaller than the horizon. Upon entering the horizon,
radiation pressure provides a restoring force against gravity, leading
to acoustic oscillations in the tightly coupled fluid of photons and
charged particles, which cease only when radiation pressure drops
while radiation decouples. Therefore, for each physical perturbation
scale, the acoustic oscillations set in at the same time, i.e.~when
the horizon size becomes equal the perturbation size, and they end at
the same time, i.e.~when radiation decouples. At fixed physical scale,
these oscillations are therefore coherent, and they show up as
distinct peaks (the so-called {\em Doppler peaks\/}) and troughs in
the power spectrum of the CMB fluctuations. Perturbations large enough
to enter the horizon after decoupling never experience these
oscillations. Going through the CMB power spectrum from large to small
scales, there should therefore be a `first' Doppler peak at a location
determined by the horizon scale at the time of decoupling.

A third important effect sets in on the smallest scales. If a density
perturbation is small enough, radiation pressure can blow it apart
because its self-gravity is too weak. This effect is comparable to the
Jeans' criterion for the minimal mass required for a pressurised
perturbation to collapse. It amounts to a suppression of small-scale
fluctuations and is called {\em Silk damping\/}, leading to an
exponential decline at the small-scale end of the CMB fluctuation
power spectrum.

Other effects arise between the `surface' of decoupling and the
observer. \cite{res68} pointed out that large non-linear density
perturbations between the last-scattering surface and us can lead to a
distinct effect if those fluctuations change while the photons
traverse them. Falling into the potential wells, they experience a
stronger blue-shift than climbing out of them because expansion makes
the wells shallower in the meantime, thus giving rise to a net
blue-shift of photons. Later, this effect was re-examined in the
framework of the `Swiss-Cheese' (\cite*{DY76.1}) and `vacuole'
(\cite*{not84}) models of density perturbations in an expanding
background space-time. The masses of such perturbations have to be
very large for this effect to become larger than the
Sunyaev-Zel'dovich effect\footnote{The (thermal) Sunyaev-Zel'dovich
effect is due to Compton-upscattering of CMB photons by thermal
electrons in the hot plasma in galaxy clusters. Since the temperature
of the electrons is much higher than that of the photons, CMB photons
are effectively re-distributed towards higher energies. At frequencies
lower than $\approx272\,$GHz, the CMB intensity is thus decreased
towards galaxy clusters; in effect, they cast shadows on the surface
of the CMB.} due to the hot gas contained in them; \cite{DY76.1}
estimated that masses beyond $10^{19}\,M_\odot$ would be necessary, a
value four to five orders of magnitude larger than that of typical
galaxy clusters.

The gravitational lens effect of galaxy clusters moving transverse to
the line-of-sight was investigated by \cite{big83} who found that a
cluster with $\sim10^{15}\,M_\odot$ and a transverse velocity of
$\sim6000\,\mathrm{km\,s}^{-1}$ should change the CMB temperature by
$\sim10^{-4}\,$K. Later, \cite{GU86.1} re-investigated this effect and
found it to be about an order of magnitude smaller.

Cosmic strings as another class of rapidly moving gravitational lenses
were studied by \cite{KA84.1} who discussed that they would give rise
to step-like features in the CMB temperature pattern.

\subsection{\label{sc:9.2}Weak Lensing of the CMB}

The introduction shows that the CMB is expected to display distinct
features in a hierarchical model of structure formation. The CMB power
spectrum should be featureless on large scales, then exhibit
pronounced Doppler peaks at scales smaller than the horizon at the
time of decoupling, and an exponential decrease due to Silk damping at
the small-scale end. We now turn to investigate whether and how
gravitational lensing by large-scale structures can alter these
features.

The literature on the subject is rich (see \cite*{bls87},
\cite*{CA93.1}, \cite*{CA93.3}, \cite*{CO89.1}, \cite*{FU92.1},
\cite*{KA88.5}, \cite*{LI88.3}, \cite*{LI90.1}, \cite*{LI90.2},
\cite*{mss90}, \cite*{SA89.1}, \cite*{tom89}, \cite*{WA91.1}), but
different authors have sometimes arrived at contradicting
conclusions. Perhaps the most elegant way of studying weak lensing of
the CMB is the power-spectrum approach, which was most recently
advocated by \citename{SE94.2} (\citeyear{SE94.2}, \citeyear{SE96.1}).

We should like to start our discussion by clearly stating two facts
concerning the effect of lensing on fluctuations in the Cosmic
Microwave Background which clarify and resolve several apparently
contradictory discussions and results in the literature.

\begin{enumerate}

\item {\em If the CMB was completely isotropic, gravitational lensing
would have no effect whatsoever because it conserves surface
brightness.\/} In this case, lensing would only magnify certain
patches in the sky and de-magnify others, but since it would not alter
the surface brightness in the magnified or de-magnified patches, the
temperature remained unaffected. An analogy would be observers facing
an infinitely extended homogeneously coloured wall, seeing some parts
of it enlarged and others shrunk. Regardless of the magnification,
they would see the same colour everywhere, and so they would notice
nothing despite the magnification.

\item {\em It is not the absolute value of the light deflection due to
lensing which matters, but the relative deflection of neighbouring
light rays.\/} Imagine a model universe in which all light rays are
isotropically deflected by the same arbitrary amount. The pattern of
CMB anisotropies seen by an observer would then be coherently shifted
relative to the intrinsic pattern, but remain unchanged otherwise. It
is thus merely the {\em dispersion\/} of deflection angles what is
relevant for the impact of lensing on the observed CMB fluctuation
pattern.

\end{enumerate}

\subsection{\label{sc:9.3}CMB Temperature Fluctuations}

In the absence of any lensing effects, we observe at the sky position
$\vec\theta$ the intrinsic CMB temperature $T(\vec\theta)$. There are
fluctuations $\Delta T(\vec\theta)$ in the CMB temperature about its
average value $\langle T\rangle=2.73\,\mathrm{K}$. We abbreviate the
relative temperature fluctuations by
\begin{equation}
  \frac{\Delta T(\vec\theta)}{\langle T\rangle} \equiv
  \tau(\vec\theta)
\label{eq:9.1}
\end{equation}
in the following. They can statistically be described by their angular
auto-correlation function
\begin{equation}
  \xi_\mathrm{T}(\phi) = \left\langle
    \tau(\vec\theta)\,\tau(\vec\theta + \vec\phi)
  \right\rangle\;,
\label{eq:9.2}
\end{equation}
with the average extending over all positions $\vec\theta$. Due to
statistical isotropy, $\xi_\mathrm{T}(\phi)$ depends neither on the
position $\theta$ nor on the direction of $\vec\phi$, but only on the
absolute separation $\phi$ of the correlated points.

Commonly, CMB temperature fluctuations are also described in terms of
the coefficients $a_{lm}$ of an expansion into spherical harmonics,
\begin{equation}
  \tau(\theta,\phi) = \sum_{l=0}^\infty\sum_{m=-l}^l\,
  a_{lm}\,Y_l^m(\theta,\phi)\;,
\label{eq:9.3}
\end{equation}
and the averaged expansion coefficients constitute the angular power
spectrum $C_l$ of the CMB fluctuations,
\begin{equation}
  C_l = \left\langle|a_{lm}|^2\right\rangle\;.
\label{eq:9.4}
\end{equation}
It can then be shown that the correlation function
$\xi_\mathrm{T}(\phi)$ is related to the power-spectrum coefficients
$C_l$ through
\begin{equation}
  C_l = \int_0^\pi\d\phi\,\sin(\phi)\,P_l(\cos\phi)\,
  \xi_\mathrm{T}(\phi)\;,
\label{eq:9.5}
\end{equation}
with the Legendre functions $P_l(\cos\phi)$.

\subsection{\label{sc:9.4}Auto-Correlation Function of the
  Gravitationally Lensed CMB}

\subsubsection{\label{sc:9.4.1}Definitions}

If there are any density inhomogeneities along the line-of-sight
towards the last-scattering surface at $z\approx1,000$ (the `source
plane' of the CMB), a light ray starting into direction $\vec\theta$
at the observer will intercept the last-scattering surface at the
deflected position
\begin{equation}
  \vec\beta = \vec\theta - \vec\alpha(\vec\theta)\;,
\label{eq:9.6}
\end{equation}
where $\vec\alpha(\vec\theta)$ is the (position-dependent) deflection
angle experienced by the light ray. We will therefore observe, at
position $\vec\theta$, the temperature of the CMB at position
$\vec\beta$, or
\begin{equation}
  T(\vec\beta) \equiv T'(\vec\theta) = 
  T[\vec\theta-\vec\alpha(\vec\theta)]\;.
\label{eq:9.7}
\end{equation}

The intrinsic temperature autocorrelation function is thus changed by
lensing to
\begin{equation}
  \xi_\mathrm{T}'(\phi) = \left\langle
    \tau[\vec\theta-\alpha(\vec\theta)]\,
    \tau[(\vec\theta+\vec\phi)-\vec\alpha(\vec\theta+\vec\phi)]
  \right\rangle\;.
\label{eq:9.8}
\end{equation}
For simplicity of notation, we further abbreviate
$\vec\alpha(\vec\theta)\equiv\vec\alpha$ and
$\vec\alpha(\vec\theta+\vec\phi)\equiv\vec\alpha'$ in the following.

\subsubsection{\label{sc:9.4.2}Evaluation}

In this section we evaluate the modified correlation function
(\ref{eq:9.8}) and quantify the lensing effects. For this purpose, it
is convenient to decompose the relative temperature fluctuation
$\tau(\vec\theta)$ into Fourier modes,
\begin{equation}
  \tau(\vec\theta) = \int\frac{\d^2\vec l}{(2\pi)^2}\,
  \hat\tau(\vec l)\,\exp(\mathrm{i}\,\vec l\,\vec\theta)\;.
\label{eq:9.9}
\end{equation}
The expansion of $\tau(\vec\theta)$ into Fourier modes rather than
into spherical harmonics is permissible because we do not expect any
weak-lensing effects on large angular scales, so that we can consider
$T(\vec\theta)$ on a plane locally tangential to the sky rather than
on a sphere.

We insert the Fourier decomposition (\ref{eq:9.9}) into the expression
for the correlation function (\ref{eq:9.8}) and perform the
average. We need to average over ensembles and over the random angle
between the wave vector $\vec l$ of the temperature modes and the
angular separation $\vec\phi$ of the correlated points. The ensemble
average corresponds to averaging over realisations of the CMB
temperature fluctuations in a sample of universes or, since we focus
on small scales, over a large number of disconnected regions on the
sky. This average introduces the CMB fluctuation spectrum
$P_\mathrm{T}(l)$, which is defined by
\begin{equation}
  \left\langle
    \hat\tau(\vec l)\,\hat\tau^*(\vec l')
  \right\rangle \equiv
  (2\pi)^2\,\delta^{(2)}(\vec l-\vec l')\,P_\mathrm{T}(l)\;.
\label{eq:9.10}
\end{equation}
Averaging over the angle between $\vec l$ and the position angle
$\vec\phi$ gives rise to the zeroth-order Bessel function of the first
kind, $\J(x)$. These manipulations leave eq.~(\ref{eq:9.8}) in the
form
\begin{equation}
  \xi_\mathrm{T}'(\phi) = \int_0^\infty\,\frac{l\d l}{2\pi}\,
  P_\mathrm{T}(l)\,\left\langle\exp\left[
    \mathrm{i}\,\vec l\left(\vec\alpha-\vec\alpha'\right)
  \right]\right\rangle\,\J(l\phi)\;.
\label{eq:9.11}
\end{equation}

The average over the exponential in eq.~(\ref{eq:9.11}) remains to be
performed. To do so, we first expand the exponential into a power
series,
\begin{equation}
  \left\langle
    \exp(\mathrm{i}\,\vec l\,\delta\vec\alpha)
  \right\rangle =
  \sum_{j=0}^\infty\,\frac{\langle
    (\mathrm{i}\,\vec l\,\delta\vec\alpha)^j
  \rangle}{j!}\;,
\label{eq:9.12}
\end{equation}
where $\delta\vec\alpha\equiv\vec\alpha-\vec\alpha'$ is the
deflection-angle difference between neighbouring light rays with
initial angular separation $\vec\phi$. We now assume that the
deflection angles are Gaussian random fields. This is reasonable
because (i) deflection angles are due to Gaussian random fluctuations
in the density-contrast field as long as the fluctuations evolve
linearly, and (ii) the assumption of linear evolution holds well for
redshifts where most of the deflection towards the last-scattering
surface occurs. Of course, this makes use of the commonly held view
that the initial density fluctuations are of Gaussian nature. Under
this condition, the odd moments in eq.~(\ref{eq:9.12}) all vanish. It
can then be shown that
\begin{equation}
  \left\langle
    \exp(\mathrm{i}\,\vec l\,\delta\vec\alpha)
  \right\rangle =
  \exp\left(-\frac{1}{2}l^2\sigma^2(\phi)\right)
\label{eq:9.13}
\end{equation}
holds exactly, where $\sigma^2(\phi)$ is the deflection-angle
dispersion,
\begin{equation}
  \sigma^2(\phi) \equiv
  \left\langle(\vec\alpha-\vec\alpha')^2\right\rangle\;.
\label{eq:9.14}
\end{equation}
Even if the assumption that $\delta\vec\alpha$ is a Gaussian random
field fails, eq.~(\ref{eq:9.13}) still holds approximately. To see
this, we note that the CMB power spectrum falls sharply on scales
$l\gtrsim l_\mathrm{c}\approx(10'\,\Omega_0^{1/2})^{-1}$. The scale
$l_\mathrm{c}$ is set by the width of the last-scattering surface at
redshift $z\sim1,000$. Smaller-scale fluctuations are efficiently
damped by acoustic oscillations of the coupled photon-baryon
fluid. Typical angular scales $l^{-1}$ in the CMB fluctuations are
therefore considerably larger than the difference between
gravitational deflection angles of neighbouring rays,
$\delta\vec\alpha$, so that $\vec l(\vec\alpha-\vec\alpha')$ is a
small number. Hence, ignoring fourth-order terms in $\vec
l\delta\vec\alpha$, the remaining exponential in (\ref{eq:9.11}) can
be {\em approximated\/} by
\begin{equation}
  \left\langle
    \exp(\mathrm{i}\,\vec l\,\delta\vec\alpha)
  \right\rangle \approx 1-\frac{1}{2}l^2\sigma^2(\phi) \approx
  \exp\left[-\frac{1}{2}l^2\sigma^2(\phi)\right]\;.
\label{eq:9.15}
\end{equation}
Therefore, the temperature auto-correlation function modified by
gravitational lensing can safely be written,
\begin{equation}
  \xi_\mathrm{T}'(\phi) = 
  \int_0^\infty\,\frac{l\d l}{2\pi}\,P_\mathrm{T}(l)\,
  \exp\left[-\frac{1}{2}l^2\sigma^2(\phi)\right]\,
  \J(l\phi)\;.
\label{eq:9.16}
\end{equation}
This equation shows that the intrinsic temperature-fluctuation power
spectrum is convolved with a Gaussian function in wave number $l$ with
dispersion $\sigma^{-1}(\phi)$. The effect of lensing on the CMB
temperature fluctuations is thus to smooth fluctuations on angular
scales of order or smaller than $\sigma(\phi)$.

\subsubsection{\label{sc:9.4.3}Alternative Representations}

Equation~(\ref{eq:9.16}) relates the unlensed CMB power spectrum to
the lensed temperature auto-correlation function. Noting that
$P_\mathrm{T}(l)$ is the Fourier transform of $\xi_\mathrm{T}(\phi)$,
\begin{equation}
  P_\mathrm{T}(l) = \int\d^2\phi\,\xi_\mathrm{T}(\phi)\,
  \exp(-\mathrm{i}\,\vec l\vec\phi) =
  2\pi\,\int\phi\d\phi\,\xi_\mathrm{T}(\phi)\,\J(l\phi)\;,
\label{eq:9.17}
\end{equation}
we can substitute one for the other. Isotropy permitted us to perform
the integration over the (random) angle between $\vec l$ and
$\vec\phi$ in the last step of (\ref{eq:9.17}). Inserting
(\ref{eq:9.17}) into (\ref{eq:9.16}) leads to
\begin{equation}
  \xi_\mathrm{T}'(\phi) = \int\phi'\d\phi'\,\xi_\mathrm{T}(\phi')\,
  K(\phi,\phi')\;.
\label{eq:9.18}
\end{equation}
The kernel $K(\phi,\phi')$ is given by
\begin{eqnarray}
  K(\phi,\phi') &\equiv& \int_0^\infty l\d l\,
  \J(l\phi)\,\J(l\phi')\,
  \exp\left[-\frac{1}{2}l^2\,\sigma^2(\phi)\right]
  \nonumber\\
  &=& \frac{1}{\sigma^2(\phi)}\,
  \exp\left[-\frac{\phi^2+\phi^{\prime\,2}}{2\sigma^2(\phi)}\right]\,
  \I\left[\frac{\phi\phi'}{\sigma^2(\phi)}\right]\;,
\label{eq:9.19}
\end{eqnarray}
where $\I(x)$ is the modified zeroth-order Bessel function.
Equation~(6.663.2) of \cite{grr94} was used in the last step. As will
be shown below, $\sigma(\phi)\ll1$, so that the argument of $\I$ is
generally a very large number. Noting that $\I(x)\approx(2\pi
x)^{-1/2}\exp(x)$ for $x\to\infty$, we can write eq.~(\ref{eq:9.16})
in the form
\begin{equation}
  \xi_\mathrm{T}'(\phi) \approx
  \frac{1}{(2\pi\phi)^{1/2}\,\sigma(\phi)}\,
  \int\d\phi'\,\phi^{\prime\,1/2}\,\xi_\mathrm{T}(\phi')\,
  \exp\left[-\frac{(\phi-\phi')^2}{2\sigma^2(\phi)}\right]\;.
\label{eq:9.20}
\end{equation}
Like eq.~(\ref{eq:9.16}), this expression shows that lensing smoothes
the intrinsic temperature auto-correlation function
$\xi_\mathrm{T}(\phi)$ on angular scales of $\phi\approx\sigma(\phi)$
and smaller. Note in particular that, if $\sigma(\phi)\to0$, the
exponential in (\ref{eq:9.20}) tends towards a Dirac delta
distribution,
\begin{equation}
  \lim_{\sigma(\phi)\to0}
  \frac{1}{\sqrt{2\pi}\,\sigma(\phi)}\,
  \exp\left[-\frac{(\phi-\phi')^2}{2\sigma^2(\phi)}\right]
  = \delta(\phi-\phi')\;,
\label{eq:9.21}
\end{equation}
so that the lensed and unlensed temperature auto-correlation functions
agree, $\xi_\mathrm{T}(\phi)=\xi_\mathrm{T}'(\phi)$.

Likewise, one can Fourier back-transform eq.~(\ref{eq:9.16}) to obtain
a relation between the lensed and the un-lensed CMB power spectra. To
evaluate the resulting integral, it is convenient to assume
$\sigma(\phi)=\epsilon\phi$, with $\epsilon$ being either a constant
or a slowly varying function of $\phi$. This assumption will be
justified below. One then finds
\begin{equation}
  P'_\mathrm{T}(l') = \int_0^\infty
  \frac{\d l}{\epsilon^2l}\,
  P_\mathrm{T}(l)\,\exp\left(
    -\frac{l^2+l^{\prime\,2}}{2\epsilon^2l^2}
  \right)\,\I\left(\frac{l'}{\epsilon^2l}\right)\;.
\label{eq:9.22}
\end{equation}
For $\epsilon\ll1$, this expression can be simplified to
\begin{equation}
  P'_\mathrm{T}(l') = \int_0^\infty
  \frac{\d l}{\sqrt{2\pi}\epsilon l}\,
  P_\mathrm{T}(l)\,\exp\left[
    -\frac{(l-l')^2}{2\epsilon^2l^2}
  \right]\;.
\label{eq:9.23}
\end{equation}

\subsection{\label{sc:9.5}Deflection-Angle Variance}

\subsubsection{\label{sc:9.5.1}Auto-Correlation Function of
  Deflection Angles}

We proceed by evaluating the dispersion $\sigma^2(\phi)$ of the
deflection angles. This is conveniently derived from the
deflection-angle auto-correlation function,
\begin{equation}
  \xi_{\vec\alpha}(\phi) \equiv \left\langle
  \vec\alpha\,\vec\alpha'\right\rangle\;.
\label{eq:9.24}
\end{equation}
Note that the correlation function of $\vec\alpha$ is the sum of the
correlation functions of the components of $\vec\alpha$,
\begin{equation}
  \xi_{\vec\alpha} = 
  \langle\vec\alpha\,\vec\alpha'\rangle =
  \langle\alpha_1\alpha_1'\rangle + \langle\alpha_2\alpha_2'\rangle =
  \xi_{\alpha_1} + \xi_{\alpha_2}\;.
\label{eq:9.25}
\end{equation}

In terms of the autocorrelation function, the dispersion
$\sigma^2(\phi)$ can be written
\begin{equation}
  \sigma^2(\phi) = \left\langle\left[
  \vec\alpha-\vec\alpha'\right]^2\right\rangle = 
  2\left[\xi_{\vec\alpha}(0)-\xi_{\vec\alpha}(\phi)\right]\;.
\label{eq:9.26}
\end{equation}
The deflection angle is given by eq.~(\ref{eq:6.16}) on
page~\pageref{eq:6.16} in terms of the Newtonian potential $\Phi$ of
the density fluctuations $\delta$ along the line-of-sight. For lensing
of the CMB, the line-of-sight integration extends along the
(unperturbed) light ray from the observer at $w=0$ to the
last-scattering surface at $w(z\approx1000)$; see the derivation in
Sect.~\ref{sc:6.2} leading to eq.~(\ref{eq:6.16},
page~\pageref{eq:6.16}).

We introduced the effective convergence in (\ref{eq:6.19},
page~\pageref{eq:6.19}) as half the divergence of the deflection
angle. In Fourier space, this equation can be inverted to yield the
Fourier transform of the deflection angle,
\begin{equation}
  \hat{\vec\alpha}(\vec l)=
  -\frac{2\mathrm{i}\,\hat\kappa_\mathrm{eff}(\vec l)}
        {|\vec l|^2}\,\vec l\;.
\label{eq:9.30}
\end{equation}
The deflection-angle power spectrum can therefore be written as
\begin{equation}
  P_{\vec\alpha}(\phi) = \frac{4}{l^2}\,P_\kappa(l)\;.
\label{eq:9.31}
\end{equation}

\begin{figure}[ht]
  \includegraphics[width=\hsize]{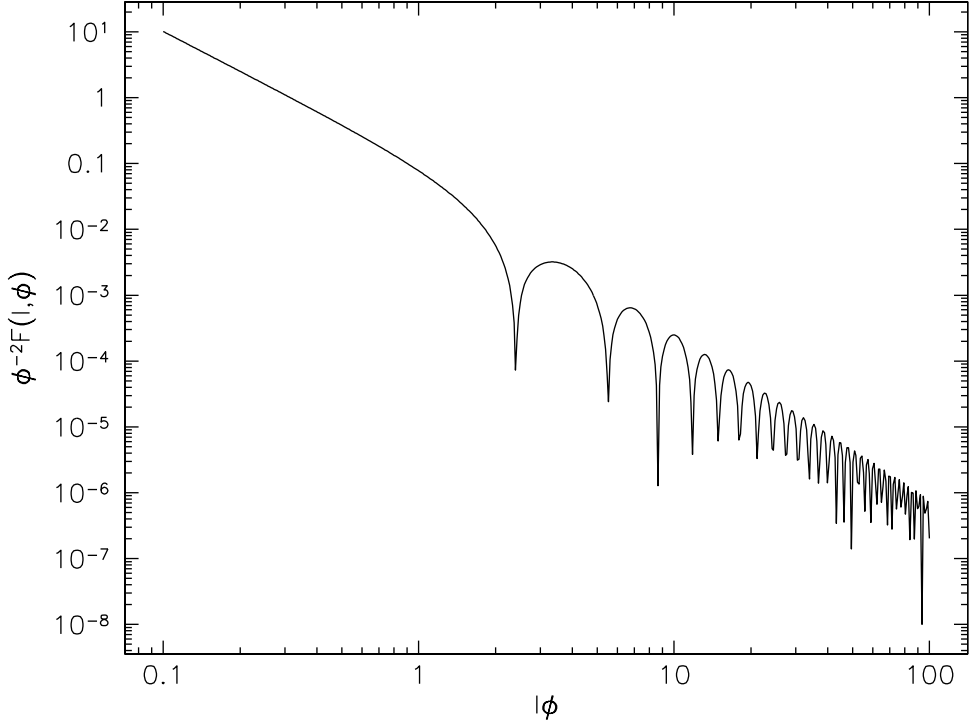}
\caption{The filter function $F(l,\phi)$ as defined in
eq.~(\ref{eq:9.33}), divided by $\phi^2$, is shown as a function of
$l\phi$. Compare Fig.~\ref{fig:6.8} on page~\pageref{fig:6.8}. For
fixed $\phi$, the filter function emphasises large-scale projected
density perturbations (i.e.~structures with small $l$).}
\label{fig:9.1}
\end{figure}

The deflection-angle autocorrelation function is obtained from
eq.~(\ref{eq:9.31}) via Fourier transformation. The result is
\begin{equation}
  \xi_{\vec\alpha}(\phi) = \int\frac{\d^2\vec l}{(2\pi)^2}\,
  P_{\vec\alpha}(l)\,\exp(-\mathrm{i}\,\vec l\vec\phi) =
  2\pi\,\int_0^\infty l\d l\,
  P_\kappa(l)\,\frac{\J(l\phi)}{(\pi l)^2}\;,
\label{eq:9.32}
\end{equation}
similar to the form (\ref{eq:6.61}, page~\pageref{eq:6.61}), but here
the filter function is no longer a function of the {\em product\/}
$l\phi$ only, but of $l$ and $\phi$ separately,
\begin{equation}
  F(l,\phi) = \frac{\J(l\phi)}{(\pi l)^2} =
  \phi^2\,\frac{\J(l\phi)}{(\pi l\phi)^2}\;.
\label{eq:9.33}
\end{equation}
We plot $\phi^{-2}\,F(l,\phi)$ in Fig.~\ref{fig:9.1}. For fixed
$\phi$, the filter function suppresses small-scale fluctuations, and
it tends towards $F(l,\phi)\to(\pi l)^{-2}$ for $l\to0$.

Inserting $P_\kappa(l)$ into (\ref{eq:9.32}), we find the explicit
expression for the deflection-angle auto-correlation function,
\begin{eqnarray}
  \xi_{\vec\alpha}(\phi) &=& \frac{9H_0^4\Omega_0^2}{c^4}\,
  \int_0^w\d w'\,W^2(w,w')\,a^{-2}(w')\,\nonumber\\
  &\times&
  \int_0^\infty\frac{\d k}{2\pi k}\,P_\delta(k,w')\,
  \J[f_K(w')k\phi]\;.
\label{eq:9.34}
\end{eqnarray}
Despite the obvious similarity between this result and the
magnification auto-correlation function (\ref{eq:6.37} on
page~\pageref{eq:6.37}), it is worth noting two important
differences. First, the weighting of the integrand along the
line-of-sight differs by a factor of $f_K^2(w')$ because we integrate
deflection-angle components rather than the convergence, i.e.~first
rather than second-order derivatives of the potential
$\Phi$. Consequently, structures near the observer are weighted more
strongly than for magnification or shear effects. Secondly, the
wave-number integral is weighted by $k^{-1}$ rather than $k$, giving
most weight to the largest-scale structures. Since their evolution
remains linear up to the present, it is expected that non-linear
density evolution is much less important for lensing of the CMB than
it is for cosmic magnification or shear.

\subsubsection{\label{sc:9.5.2}Typical Angular Scale}

A typical angular scale $\phi_\mathrm{g}$ for the coherence of
gravitational light deflection can be obtained as
\begin{equation}
  \phi_\mathrm{g} \equiv \left[\frac{1}{\xi_{\vec\alpha}(0)}\,
  \left(\left|
    \frac{\partial^2\xi_{\vec\alpha}(\phi)}{\partial\phi^2}
  \right|_{\phi=0}\right)\right]^{-1/2}\;.
\label{eq:9.35}
\end{equation}
As eq.~(\ref{eq:9.34}) shows, the deflection-angle auto-correlation
function depends on $\phi$ only through the argument of the Bessel
function $\J(x)$. For small arguments $x$, the second-order derivative
of the $\J(x)$ is approximately
$\J''(x)\approx-\J(x)/2$. Differentiating $\xi_{\vec\alpha}(\phi)$
twice with respect to $\phi$, and comparing the result to the
expression for the magnification auto-correlation function
$\xi_\mu(\phi)$ in eq.~(\ref{eq:6.37}, page~\pageref{eq:6.37}), we
find
\begin{equation}
  \frac{\partial^2\xi_{\vec\alpha}(\phi)}{\partial\phi^2} \approx
  -\frac{1}{2}\,\xi_\mu(\phi)\;,
\label{eq:9.36}
\end{equation}
and thus
\begin{equation}
  \phi_\mathrm{g}^2 \approx
  2\,\frac{\xi_{\vec\alpha}(0)}{\xi_\mu(0)}\;.
\label{eq:9.37}
\end{equation}
We shall estimate $\phi_\mathrm{g}$ later after giving a simple
expression for $\xi_{\vec\alpha}(\phi)$. The angle $\phi_\mathrm{g}$
gives an estimate of the scale over which gravitational light
deflection is coherent.

\subsubsection{\label{sc:9.5.3}Special Cases and Qualitative
  Expectations}

We mentioned before that it is less critical here to assume linear
density evolution because large-scale density perturbations dominate
in the expression for $\xi_{\vec\alpha}(\phi)$. Specialising further
to an Einstein-de Sitter universe so that $w\approx2c/H_0$,
eq.~(\ref{eq:9.34}) simplifies to
\begin{equation}
  \xi_{\vec\alpha}(\phi) = \frac{9H_0^4}{c^4}\,
  w\,\int_0^1\,\d y\,(1-y)^2\,
  \int_0^\infty\,\frac{\d k}{2\pi
  k}\,P_\delta^{(0)}(k)\,\J(wyk\phi)\;,
\label{eq:9.38}
\end{equation}
with $wy\equiv w'$.

Adopting the model spectra for HDM and CDM specified in
eq.~(\ref{eq:6.39}, page~\pageref{eq:6.39}) and expanding
$\xi_{\vec\alpha}(\phi)$ in a power series in $\phi$, we find, to
second order in $\phi$,
\begin{equation}
  \xi_{\vec\alpha}(\phi) = A'\,wk_0\,\left\{
  \begin{array}{ll}
    \displaystyle\frac{3}{2\pi}\,
    \left[1-\frac{\phi^2}{20}\,(wk_0)^2\right] &
    \quad\hbox{for HDM}\\
    \displaystyle\frac{3\sqrt{3}}{8}\,
    \left[1-\frac{3\,\phi^2}{40}(wk_0)^2\right] &
    \quad\hbox{for CDM}\\
  \end{array}\right.\;.
\label{eq:9.39}
\end{equation}
Combining these expressions with eqs.~(\ref{eq:9.37}) and
(\ref{eq:6.40}, page~\pageref{eq:6.40}), we find for the
deflection-angle coherence scale $\phi_\mathrm{g}$
\begin{equation}
  \phi_\mathrm{g} \approx 3\,(wk_0)^{-1}\;.
\label{eq:9.40}
\end{equation}
It is intuitively clear that $\phi_\mathrm{g}$ should be determined by
$(wk_0)^{-1}$. Since $k_0^{-1}$ is the typical length scale of
light-deflecting density perturbations, it subtends an angle
$(wk_0)^{-1}$ at distance $w$. Thus the coherence angle of light
deflection is given by the angle under which the deflecting density
perturbation typically appears. The source distance $w$ in the case of
the CMB is the comoving distance to $z=1,000$. In the Einstein-de
Sitter case, $w=2$ in units of the Hubble length. Hence, with
$k_0^{-1}\approx12\,(\Omega_0h^2)\,$Mpc [cf.~eq.~(\ref{eq:2.48}),
page~\pageref{eq:2.48}], we have $wk_0\approx500$. Therefore, the
angular scale of the deflection-angle auto-correlation is of order
\begin{equation}
  \phi_\mathrm{g}\approx6\times10^{-3}\approx20'\;.
\label{eq:9.41}
\end{equation}
To lowest order in $\phi$, the deflection-angle dispersion
(\ref{eq:9.26}) reads
\begin{equation}
  \sigma^2(\phi) \propto (wk_0)^3\,\phi^2\;.
\label{eq:9.42}
\end{equation}
The dispersion $\sigma(\phi)$ is plotted in Fig.~\ref{fig:9.2} for
the four cosmological models specified in Tab.~\ref{tab:6.1} on
page~\pageref{tab:6.1} for linear and non-linear evolution of the
density fluctuations.

The behaviour of $\sigma(\phi)$ expressed in eq.~(\ref{eq:9.42}) can
qualitatively be understood describing the change in the transverse
separation between light paths as a random walk. Consider two light
paths separated by an angle $\phi$ such that their comoving transverse
separation at distance $w$ is $w\phi$. Let $k^{-1}$ be the typical
scale of a potential fluctuation $\Phi$. We can then distinguish two
different cases depending on whether $w\phi$ is larger or smaller than
$k^{-1}$. If $w\phi>k^{-1}$, the transverse separation between the
light paths is much larger than the typical potential fluctuations,
and their deflection will be incoherent. It will be coherent in the
opposite case, i.e.~if $w\phi<k^{-1}$.

When the light paths are coherently scattered passing a potential
fluctuation, their angular separation changes by $\delta\phi_1\approx
w\phi\,\nabla_\perp(2k^{-1}\nabla_\perp\Phi/c^2)$, which is the change
in the deflection angle across $w\phi$. If we replace the gradients by
the inverse of the typical scale, $k$, we have
$\delta\phi_1\approx2\,w\phi\,k\Phi/c^2$. Along a distance $w$, there
are $N\approx kw$ such potential fluctuations, so that the total
change in angular separation is expected to be $\delta\phi\approx
N^{1/2}\delta\phi_1$.

In case of incoherent scattering, the total deflection of each light
path is expected to be $\delta\phi\approx
N^{1/2}\,(2k^{-1}\nabla_\perp\Phi/c^2)\approx N^{1/2}\,2\Phi/c^2$,
independent of $\phi$. Therefore,
\begin{equation}
  \sigma^2(\phi) \approx \left\{
  \begin{array}{ll}
    N\delta\phi_1^2 \approx (2\Phi/c^2)^2\,(wk)^3\,\phi^2 &
    \quad\hbox{for}\quad \phi<(wk)^{-1} \\
    N\,(2\Phi/c^2)^2 \approx (2\Phi/c^2)^2\,(wk) &
    \quad\hbox{for}\quad \phi>(wk)^{-1} \\
  \end{array}\right.\;.
\label{eq:9.43}
\end{equation}
This illustrates that the dependence of $\sigma^2(\phi)$ on
$(wk)^3\,\phi^2$ for small $\phi$ is merely a consequence of the
random coherent scattering of neighbouring light rays at potential
fluctuations. For large $\phi$, $\sigma(\phi)$ becomes constant, and
so $\sigma(\phi)\phi^{-1}\to0$. As Fig.~\ref{fig:9.2} shows, the
dispersion $\sigma(\phi)$ increases linearly with $\phi$ for small
$\phi$ and flattens gradually for
$\phi>\phi_\mathrm{g}\approx(10-20)'$ as expected, because
$\phi_\mathrm{g}$ divides coherent from incoherent scattering.

\subsubsection{\label{sc:9.5.4}Numerical Results}

\begin{figure}[ht]
  \includegraphics[width=\hsize]{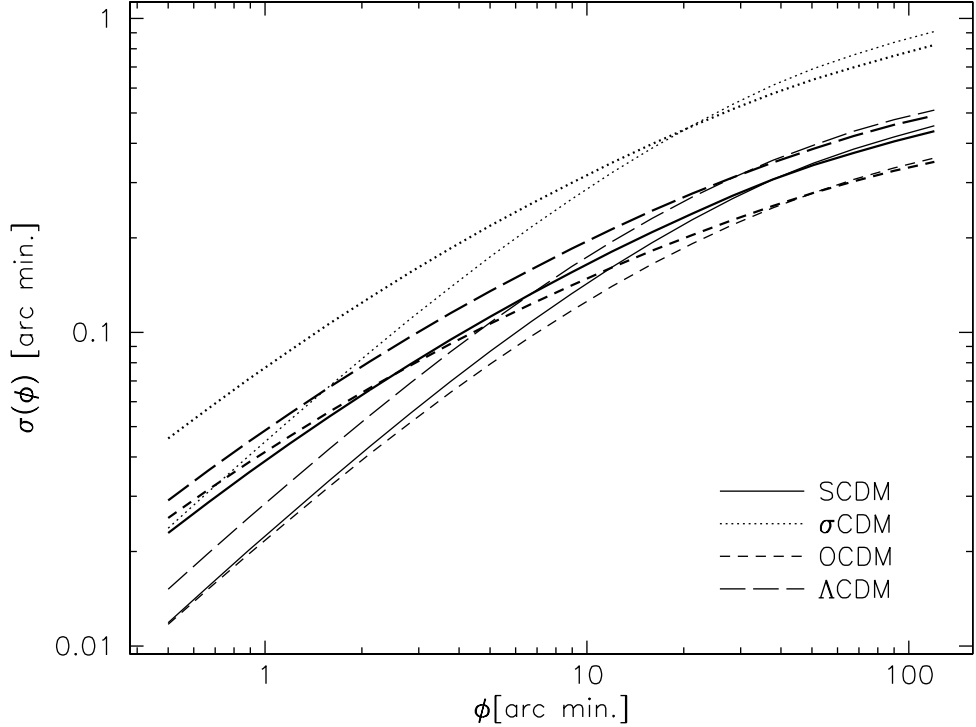}
\caption{The deflection-angle variance $\sigma(\phi)$ is shown for the
four cosmological models specified in Tab.~\ref{tab:6.1} on
page~\pageref{tab:6.1}. Two curves are shown for each model, one for
linear and one for non-linear evolution of the density
fluctuations. Solid curves: SCDM; dotted curve: $\sigma$CDM;
short-dashed curves: OCDM; and long-dashed curves: $\Lambda$CDM. The
somewhat steeper curves are for linear density evolution. Generally,
the deflection-angle variance increases linearly with $\phi$ for small
$\phi$, and flattens gradually for $\phi\gtrsim20'$. At
$\phi\approx10'$, $\sigma(\phi)$ reaches $\approx0.1'$, or
$\approx0.01\phi$, for the cluster-normalised model universes (all
except $\sigma$CDM; dotted curves). As expected, the effect of
non-linear density evolution is fairly moderate, and most pronounced
on small angular scales, $\phi\lesssim10'$.}
\label{fig:9.2}
\end{figure}

The previous results were obtained by specialising to linear evolution
of the density contrast in an Einstein-de Sitter universe. For
arbitrary cosmological parameters, the deflection-angle dispersion has
to be computed numerically. We show in Fig.~\ref{fig:9.2} examples for
$\sigma(\phi)$ numerically calculated for the four cosmological models
detailed in Tab.~\ref{tab:6.1} on page~\pageref{tab:6.1}. Two curves
are plotted for each model. The somewhat steeper curves were obtained
for linear, the others for non-linear density evolution.

Figure~\ref{fig:9.2} shows that typical values for the
deflection-angle variance in cluster-normalised model universes are of
order $\sigma(\phi)\approx(0.03-0.1)'$ on angular scales between
$\phi\approx(1-10)'$. While the results for different cosmological
parameters are fairly close for cluster-normalised CDM, $\sigma(\phi)$
is larger by about a factor of two for CDM in an Einstein-de Sitter
model normalised to $\sigma_8=1$. For the other cosmological models,
the differences between different choices for the normalisation are
less pronounced. The curves shown in Fig.~\ref{fig:9.2} confirm the
qualitative behaviour estimated in the previous section: The variance
$\sigma(\phi)$ increases approximately linearly with $\phi$ as long as
$\phi$ is small, and it gradually flattens off at angular scales
$\phi\gtrsim\phi_\mathrm{g}\approx20'$.

In earlier chapters, we saw that non-linear density evolution has a
large impact on weak gravitational lensing effects, e.g.~on the
magnification auto-correlation function $\xi_\mu(\phi)$. As mentioned
before, this is not the case for the deflection-angle auto-correlation
function $\xi_{\vec\alpha}(\phi)$ and the variance $\sigma(\phi)$
derived from it, because the filter function $F(l,\phi)$ relevant here
suppresses small-scale density fluctuations for which the effect of
non-linear evolution are strongest. Therefore, non-linear evolution is
expected to have less impact here. Only on small angular scales
$\phi$, the filter function extends into the sufficiently non-linear
regime. The curves in Fig.~\ref{fig:9.2} confirm and quantify this
expectation. Only on scales of $\phi\lesssim10'$, the non-linear
evolution does have some effect. Obviously, non-linear evolution
increases the deflection-angle variance in a manner quite independent
of cosmology. At angular scales $\phi\approx1'$, the increase amounts
to roughly a factor of two above the linear results.

\subsection{\label{sc:9.6}Change of CMB Temperature Fluctuations}

\subsubsection{\label{sc:9.6.1}Summary of Previous Results}

We are now ready to justify assumptions and approximations made
earlier, and to quantify the impact of weak gravitational lensing on
the Cosmic Microwave Background. The main assumptions were that (i)
the deflection-angle variance $\sigma(\phi)$ is small, and (ii)
$\sigma(\phi)\approx\epsilon\phi$, with $\epsilon$ a (small) constant
or a function slowly varying with $\phi$. The results obtained in the
previous section show that $\sigma(\phi)$ is typically about two
orders of magnitude smaller than $\phi$, confirming
$\epsilon\ll1$. Likewise, Fig.~\ref{fig:9.2} shows that the assumption
$\sigma(\phi)\propto\phi$ is valid on angular scales smaller than the
coherence scale for the deflection,
$\phi\lesssim\phi_\mathrm{g}\approx20'$. As we have seen, this
proportionality is a mere consequence of random coherent scattering of
neighbouring light rays in the fluctuating potential field. For angles
larger than $\phi_\mathrm{g}$, $\sigma(\phi)$ gradually levels off to
become constant, so that the ratio between $\sigma(\phi)$ and $\phi$
tends to zero while $\phi$ increases further beyond
$\phi_\mathrm{g}$. We can thus broadly summarise the numerical results
on the deflection-angle variance by
\begin{equation}
  \sigma(\phi) \approx \left\{\begin{array}{l@{\quad}l@{\quad}l}
    0.01\,\phi & \hbox{for} & \phi\lesssim20' \\
    0.7' & \hbox{for} & \phi\gg20' \\
  \end{array}\right.\;,
\label{eq:9.44}
\end{equation}
which is valid for cluster-normalised CDM quite independent of the
cosmological model; in particular,
$\sigma(\phi)<1'\approx3\times10^{-4}\,$radians for all $\phi$.

\subsubsection{\label{sc:9.6.2}Simplifications}

Accordingly, the argument of the exponential in eq.~(\ref{eq:9.16}) is
a truly small number. Even for large $l\approx10^3$,
$l^2\sigma^2(\phi)\ll1$. We can thus safely expand the exponential
into a power series, keeping only the lowest-order terms. Then,
eq.~(\ref{eq:9.16}) simplifies to
\begin{equation}
  \xi'_\mathrm{T}(\phi) = \xi_\mathrm{T}(\phi) -
  \sigma^2(\phi)\,\int_0^\infty\frac{l^3\d l}{4\pi}\,
  P_\mathrm{T}(l)\,\J(l\phi)\;,
\label{eq:9.45}
\end{equation}
where we have used that the auto-correlation function
$\xi_\mathrm{T}(\phi)$ is the Fourier transform of the power spectrum
$P_\mathrm{T}(l)$. Employing again the approximate relation
$\J''(x)\approx-\J(x)/2$ which holds for small $x$, we notice that
\begin{equation}
  \int_0^\infty\frac{l^3\d l}{4\pi}\,
  P_\mathrm{T}(l)\,\J(l\phi) \approx
  -\frac{\partial^2\xi_\mathrm{T}(\phi)}{\partial\phi^2}\;.
\label{eq:9.46}
\end{equation}
We can introduce a typical angular scale $\phi_\mathrm{c}$ for the CMB
temperature fluctuations in the same manner as for light deflection in
eq.~(\ref{eq:9.35}). We define $\phi_\mathrm{c}$ by
\begin{equation}
  \phi_\mathrm{c}^{-2} \equiv -\frac{1}{\xi_\mathrm{T}(0)}\,
  \left.
    \frac{\partial^2\xi_\mathrm{T}(\phi)}{\partial\phi^2}
  \right|_{\phi=0}\;,
\label{eq:9.47}
\end{equation}
so that, up to second order in $\phi$, eq.~(\ref{eq:9.45}) can be
approximated as
\begin{equation}
  \xi'_\mathrm{T}(\phi) \approx \xi_\mathrm{T}(\phi) -
  \frac{\sigma^2(\phi)}{\phi_\mathrm{c}^2}\,\xi_\mathrm{T}(0)\;.
\label{eq:9.48}
\end{equation}
We saw earlier that $\sigma(\phi)\approx\epsilon\phi$ for
$\phi\lesssim\phi_\mathrm{g}$. Equation~(\ref{eq:9.48}) can then
further be simplified to read
\begin{equation}
  \xi'_\mathrm{T}(\phi) \approx \xi_\mathrm{T}(\phi) -
  \epsilon^2\,\xi_\mathrm{T}(0)\,\frac{\phi^2}{\phi_\mathrm{c}^2}\;.
\label{eq:9.49}
\end{equation}

In analogy to eq.~(\ref{eq:9.26}), we can write the mean-square
temperature fluctuations of the CMB between two beams separated by an
angle $\phi$ as
\begin{equation}
  \sigma_\mathrm{T}^2(\phi) = \left\langle
    [\tau(\vec\theta) - \tau(\vec\theta + \vec\phi)]^2
  \right\rangle =
  2\,\left[\xi_\mathrm{T}(0) - \xi_\mathrm{T}(\phi)\right]\;.
\label{eq:9.50}
\end{equation}
Weak gravitational lensing changes this relative variance to
\begin{equation}
  \sigma_\mathrm{T}^{\prime\,2} = 2\,\left[
    \xi'_\mathrm{T}(0) - \xi'_\mathrm{T}(\phi)
  \right]\;.
\label{eq:9.51}
\end{equation}
Using eq.~(\ref{eq:9.49}), we see that the relative variance is {\em
increased\/} by the amount
\begin{equation}
  \Delta\sigma^2_\mathrm{T}(\phi) =
  \sigma^{\prime\,2}_\mathrm{T}(\phi) -
  \sigma^2_\mathrm{T}(\phi) \approx \epsilon^2\,\xi_\mathrm{T}(0)\,
  \frac{\phi^2}{\phi_\mathrm{c}^2}\;.
\label{eq:9.52}
\end{equation}
Now, the auto-correlation function at zero lag, $\xi_\mathrm{T}(0)$,
is the temperature-fluctuation variance, $\sigma_\mathrm{T}^2$.
Hence, we have for the {\em rms\/} change in the temperature variation
\begin{equation}
  \left[\Delta\sigma^2_\mathrm{T}(\phi)\right]^{1/2} =
  \epsilon\,\sigma_\mathrm{T}\,\frac{\phi}{\phi_\mathrm{c}}\;.
\label{eq:9.53}
\end{equation}
Weak gravitational lensing thus changes the CMB temperature
fluctuations only by a very small amount, of order
$\epsilon\approx10^{-2}$ for $\phi\approx\phi_\mathrm{c}$.

\subsubsection{\label{sc:9.6.3}The Lensed CMB Power Spectrum}

\begin{figure}[ht]
  \includegraphics[width=\hsize]{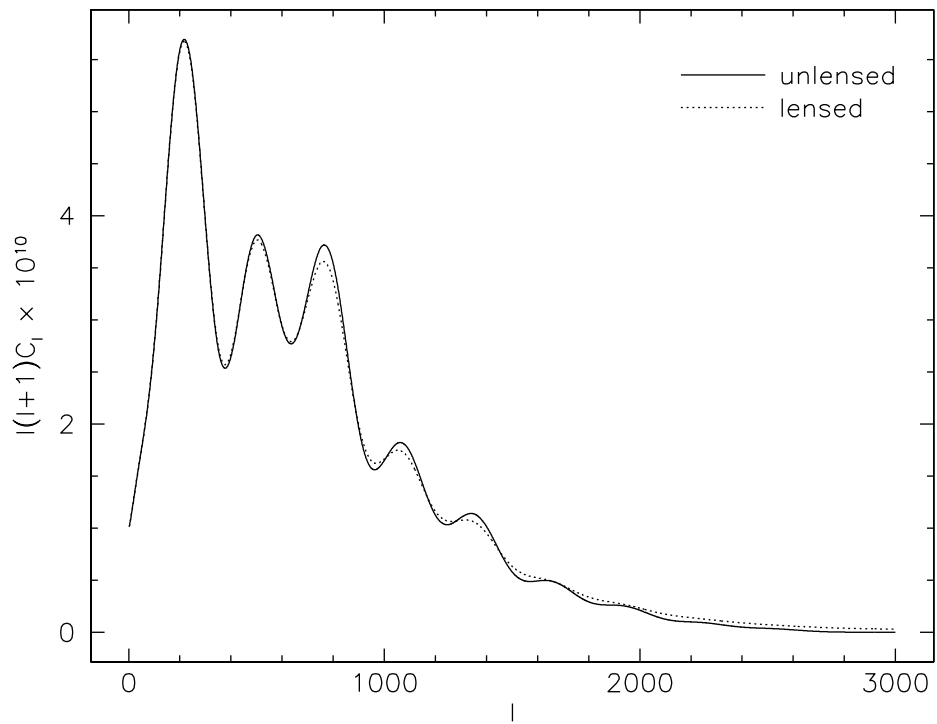}
\caption{The CMB power spectrum coefficients $l(l+1)C_l$ are shown as
a function of $l$. The solid line displays the intrinsic power
spectrum, the dotted line the lensed power spectrum for an Einstein-de
Sitter universe filled with cold dark matter. Evidently, lensing
smoothes the spectrum at small angular scales (large $l$), while it
has no visible effect on larger scales. The curves were produced with
the \texttt{CMBfast} code, see \protect\cite{zas98a}.}
\label{fig:9.3}
\end{figure}

However, we saw in eq.~(\ref{eq:9.23}) that the gravitationally lensed
CMB power spectrum is smoothed compared to the intrinsic power
spectrum. Modes on an angular scale $\phi$ are mixed with modes on
angular scales $\phi\pm\sigma(\phi)$, i.e.~the relative broadening
$\delta\phi/\phi$ is of order $2\sigma(\phi)/\phi$. For
$\phi\lesssim\phi_\mathrm{g}\approx20'$, this relative broadening is
of order $2\epsilon\approx2\times10^{-2}$, while it becomes negligible
for substantially larger scales because $\sigma(\phi)$ becomes
constant. This effect is illustrated in Fig.~\ref{fig:9.3}, where we
show the unlensed and lensed CMB power spectra for CDM in an
Einstein-de Sitter universe.

The figure clearly shows that lensing smoothes the CMB power spectrum
on small angular scales (large $l$), while it leaves large angular
scales unaffected. Lensing effects become visible at $l\gtrsim500$,
corresponding to an angular scale of
$\phi\lesssim(\pi/500)\,\mathrm{rad}\approx20'$, corresponding to the
scale where coherent gravitational light deflection sets in. An
important effect of lensing is seen at the high-$l$ tail of the power
spectra, where the lensed power spectrum falls systematically above
the unlensed one (\cite*{ME97.1}). This happens because the Gaussian
convolution kernel in eq.~(\ref{eq:9.23}) becomes very broad for very
large $l$, so that the lensed power spectrum at $l'$ can be
substantially increased by intrinsic power from significantly smaller
$l$. In other words, lensing mixes power from larger angular scales
into the otherwise feature-less damping tail of $P_\mathrm{T}(l)$.

\subsection{\label{sc:9.7}Discussion}

Several different approximations entered the preceding
derivations. Firstly, the deflection-angle variance $\sigma(\phi)$ was
generally assumed to be small, and for some expressions to be
proportional to $\phi$ with a small constant of proportionality
$\epsilon$. The numerical results showed that the first assumption is
very well satisfied, and the second assumption is valid for
$\phi\lesssim\phi_\mathrm{g}$, the latter being the coherence scale of
gravitational light deflection.

We further assumed the deflection-angle field to be a Gaussian random
field, the justification being that the deflecting matter distribution
is also a Gaussian random field. While this fails to be exactly true
at late stages of the cosmic evolution, we have seen that the
resulting expression can also be obtained when $\sigma(\phi)$ is small
and $\vec\alpha$ is not a Gaussian random field; hence, in practice
this assumption is not a limitation of validity.

A final approximation consists in the Born approximation. This should
also be a reasonable assumption at least in the case considered here,
where we focus on {\em statistical\/} properties of light
propagation. Even if the light rays would be bent considerably, the
statistical properties of the potential gradient along their true
trajectories are the same as along the approximated unperturbed rays.

Having found all the assumptions made well justifiable, we can
conclude that the random walk of light rays towards the surface of
recombination leads to smoothing of small-scale features in the CMB,
while large-scale features remain unaffected. The border line between
small and large angular scales is determined by the angular coherence
scale of gravitational light deflection by large-scale matter
distributions, which we found to be of order
$\phi_\mathrm{g}\approx20'$, corresponding to
$l_\mathrm{g}=2\pi\phi_\mathrm{g}^{-1}\approx1,000$. For the smallest
angular scales, well into the damping tail of the intrinsic CMB power
spectrum, this smoothing leads to a substantial re-distribution of
power, which causes the lensed CMB power spectrum to fall
systematically above the unlensed one at $l\gtrsim2000$, or
$\phi\lesssim2\pi l^{-1}\approx10'$. Future space-bound CMB
observations, e.g.~by the Planck Surveyor satellite, will achieve
angular resolutions of order $\gtrsim5'$, so that the lensed regime of
the CMB power spectrum will be well accessible. Highly accurate
analyses of the data of such missions will therefore need to take
lensing effects by large-scale structures into account.

One of the foremost goals of CMB observations is to derive
cosmological parameters from the angular CMB power spectrum
$C_l$. Unfortunately, there exists a parameter degeneracy in the sense
that for any given set of cosmological parameters fitting a given CMB
spectrum, a whole family of cosmological models can be found that will
fit the spectrum (almost) equally well (\cite*{zss97}). \cite{ME98.1}
showed that the rise in the damping-tail amplitude due to
gravitational lensing of the CMB can be used to break this degeneracy
once CMB observations with sufficiently high angular resolution become
available.

We discussed in Sect.~\ref{sc:4.2} how shapes of galaxy images can be
quantified with the tensor $Q_{ij}$ of second surface-brightness
moments. Techniques for the reconstruction of the intervening
projected matter distribution are then based on (complex)
ellipticities constructed from $Q_{ij}$, e.g.~the quantity $\chi$
defined in (\ref{eq:4.4}). Similar reconstruction techniques can be
developed by constructing quantities comparable to $\chi$ from the CMB
temperature fluctuations $\tau(\vec\theta)$. Two such quantities were
suggested in the literature, namely
\begin{equation}
  \tau_{,1}^2-\tau_{,2}^2+2\,\mathrm{i}\,\tau_{,1}\tau_{,2}
\label{eq:9.54}
\end{equation}
(\cite*{zas98}) and
\begin{equation}
  \tau_{,11}-\tau_{,22}+2\,\mathrm{i}\,\tau_{,12}
\label{eq:9.55}
\end{equation}
(\cite*{BE97.7}). As usual, comma-preceded indices $i$ denote
differentiation with respect to $\theta_i$.

Finally, it is worth noting that gravitational lensing mixes different
types of CMB polarisation (the ``electric'' and ``magnetic'', or $E$
and $B$ modes, respectively) and can thus create $B$-type polarisation
even when only $E$-type polarisation is intrinsically present
(\cite*{zas98a}). This effect, however, is fairly small in typical
cosmological models and will only marginally affect future CMB
polarisation measurements.

  % -*- LaTeX -*-

\section{\label{sc:10}Summary and Outlook}

We have summarised the basic ideas, theoretical developments, and
first applications of weak gravitational lensing. In particular, we
showed how the projected mass distribution of clusters can be
reconstructed from the image distortion of background galaxies, using
parameter-free methods, how the statistical mass distribution of
galaxies can be obtained from galaxy-galaxy lensing, and how the
larger-scale mass distribution in the Universe affects observations of
galaxy shapes and fluxes of background sources, as well as the
statistical properties of the CMB. Furthermore, weak lensing can be
used to construct a mass-selected sample of clusters of galaxies,
making use only of their tidal gravitational field which leaves an
imprint on the image shapes of background galaxies. We have also
discussed how the redshift distribution of these faint and distant
galaxies can be derived from lensing itself, well beyond the magnitude
limit which is currently available through spectroscopy.

Given that the first coherent image alignment of faint galaxies around
foreground clusters was discovered only a decade ago (\cite*{FO88.1};
\cite*{TY90.1}), the field of weak lensing has undergone a rapid
evolution in the last few years, for three main reasons: (i)
Theoreticians have recognised the potential power of this new tool for
observational cosmology, and have developed specific statistical
methods for extracting astrophysically and cosmologically relevant
information from astronomical images. (ii) Parallel to that effort,
observers have developed new observing strategies and image analysis
software in order to minimise the influence of instrumental artefacts
on the measured properties of faint images, and to control as much as
possible the point-spread function of the resulting image. It is
interesting to note that several image analysis methods, particularly
aimed at shape measurements of very faint galaxies for weak
gravitational lensing, have been developed by a coherent effort of
theoreticians and observers (\cite*{BO95.1}; \cite*{KA95.4};
\cite*{LU97.1}; \cite*{VA97.1}; \cite*{kai99}; \cite*{rrg99};
\cite*{kui99}), indicating the need for a close interaction between
these two groups which is imposed by the research subject.

(iii) The third and perhaps major reason for the rapid evolution is
the instrumental development that we are witnessing. Most spectacular
was the refurbishment of the Hubble Space Telescope (HST) in
Dec.~1993, after which this telescope produced astronomical images of
angular resolution unprecedented in optical astronomy. These images
have not only been of extreme importance for studying multiple images
of galaxy-scale lens systems (where the angular separation is of order
one arc second) and for detailed investigations of giant arcs and
multiple galaxy images in clusters of galaxies, but also for several
of the most interesting results of weak lensing. Owing to the lack of
atmospheric smearing and the reduced sky background from space, the
shape of fainter and smaller galaxy images can be measured on HST
images, increasing the useful number density of background galaxies,
and thus reducing the noise due to the intrinsic ellipticity
distribution. Two of the most detailed mass maps of clusters have been
derived from HST data (\cite*{SE96.2}; \cite*{hfk98}), and all but one
published results on galaxy-galaxy lensing are based on data taken
with the HST. In parallel to this, the development of astronomical
detectors has progressed quickly. The first weak-lensing observations
were carried out with CCD detectors of $\sim1,000^2$ pixels, covering
a fairly small field-of-view. A few years ago, the first
$(8\,\mathrm{K})^2$ camera was used for astronomical imaging. Its
$30'\times30'$ field can be used to map the mass distribution of
clusters at large cluster-centric radii, to investigate the potential
presence of filaments between neighbouring clusters (\cite*{kwl98}),
or simply to obtain high-quality data on a large area. Such data will
be useful for galaxy-galaxy lensing, the search for haloes using their
lensing properties only, for the investigation of cosmic shear, and
for homogeneous galaxy number counts on large fields, needed to obtain
a better quantification of the statistical association of AGNs with
foreground galaxies.

It is easy to foresee that the instrumental developments will remain
the driving force for this research field. By now, several
large-format CCD cameras are either being built or already installed,
including three cameras with a one square degree field-of-view and
adequate sampling of the PSF (MEGAPRIME at CFHT, MEGACAM at the
refurbished MMT, and OMEGACAM at the newly built VLT Support Telescope
at Paranal; see the recent account of wide-field imaging instruments
in \cite*{acm98}). Within a few years, more than a dozen 8- to
10-meter telescopes will be operating, and many of them will be
extremely useful for obtaining high-quality astronomical images, due
to their sensitivity, their imaging properties and the high quality of
the astronomical site. In fact, at least one of them (SUBARU on Mauna
Kea) will be equipped with a large-format CCD camera. One might
hypothesise that weak gravitational lensing is one of the main science
drivers to shift the emphasis of optical astronomers more towards
imaging, in contrast to spectroscopy. For example, the VLT Support
Telescope will be fully dedicated to imaging, and the fraction of time
for wide-field imaging on several other major telescopes will be
substantial. The Advanced Camera for Surveys (ACS) is planned to be
installed on the HST in 2001. Its larger field-of-view, better
sampling, and higher quantum efficiency -- compared to the current
imaging camera WFPC2 -- promises to be particularly useful for weak
lensing observations.

Even more ambitious ground-based imaging projects are currently under
discussion. Funding has been secured for the VISTA
project\footnote{see
\texttt{http://www-star.qmw.ac.uk/$\sim$jpe/vista/}} of a 4~m
telescope in Chile with a field-of-view of at least one square
degree. Another 4~m {\em Dark Matter Telescope\/} with a substantially
larger field-of-view (nine square degrees) is being discussed
specifically for weak lensing. \cite{ktl99} proposed a new strategy
for deep, wide-field optical imaging at high angular resolution, based
on an array of relatively small ($D\sim1.5$~m) telescopes with fast
guiding capacity and a ``rubber'' focal plane.

Associated with this instrumental progress is the evolution of
data-analysis capabilities. Whereas a small-format CCD image can be
reduced and analysed `by hand', this is no longer true for the
large-format CCD images. Semi-automatic data-reduction pipelines will
become necessary to keep up with the data flow. These pipelines, once
properly developed and tested, can lead to a more `objective' data
analysis. In addition, specialised software, such as for the
measurement of shapes of faint galaxies, can be implemented, together
with tools which allow a correction for PSF anisotropies and smearing.

Staying with instrumental developments for one more moment, the two
planned CMB satellite missions (MAP and Planck Surveyor) will provide
maps of the CMB at an angular resolution and a signal-to-noise ratio
which will most likely lead to the detection of lensing by the
large-scale structure on the CMB, as described in
Sect.~\ref{sc:9}. Last but not least, the currently planned Next
Generation Space Telescope (NGST, \cite*{kal99}), with a projected
launch date of 2008, will provide a giant step in many fields of
observational astronomy, not the least for weak lensing. It combines a
large aperture (of order eight meters) with a position far from Earth
to reduce sky background and with large-format imaging cameras. Even a
relatively short exposure with the NGST, which will be optimised for
observations in the near-infrared, will return images with a number
density of several hundred background galaxies per square arc minute,
for which a shape can be reliably measured; more accurate estimates
are presently not feasible due to the large extrapolation into unknown
territory. Comparing this number with the currently achievable number
density in ground-based observations of about 30 per square arc
minute, NGST will revolutionise this field.\footnote{Whereas with the
8- to 10-meter class ground-based telescopes deeper images can be
obtained, this does not drastically affect the `useful' number density
of faint galaxy images. Since fainter galaxies also tend to become
smaller, and since a reliable shape estimate of a galaxy is feasible
only if its size is not much smaller than the size of the seeing disk,
very much deeper images from the ground will not yield much larger
number densities of galaxy images which can be used for weak lensing.}
In addition, the corresponding galaxies will be at much higher mean
redshift than currently observable galaxy samples. Taken together,
these two facts imply that one can detect massive haloes at medium
redshifts with only half the velocity dispersion currently necessary
to detect them with ground-based data, or that the investigations of
the mass distribution of haloes can be extended to much higher
redshifts than currently possible (see \cite*{sck98}). The ACS on
board HST will provide an encouraging hint of the increase in
capabilities that NGST has to offer.

Progress may also come from somewhat unexpected directions. Whereas
the Sloan Digital Sky Survey (SDSS; e.g.~\cite*{sza98}) will be very
shallow compared to more standard weak-lensing observations, its huge
angular coverage may compensate for it (\cite*{ste96}). The VLA-FIRST
survey of radio sources (\cite*{wbh97}) suffers from the sparsely
populated radio sky, but this is also compensated by the huge sky
coverage (\cite*{rbk98}). The use of both surveys for weak lensing
will depend critically on the level down to which the systematics of
the instrumental image distortion can be understood and compensated
for.

Gravitational lensing has developed from a stand-alone research field
into a versatile tool for observational cosmology, and this also
applies to weak lensing. But, whereas the usefulness of strong lensing
is widely accepted by the astronomical community, weak lensing is only
beginning to reach that level of wide appreciation. Part of this
difference in attitude may be due to the fact that strong-lensing
effects, such as multiple images and giant arcs, can easily be seen on
CCD images, and their interpretation can readily be explained also to
the non-expert. In contrast, weak lensing effects are revealed only
through thorough statistical analysis of the data. Furthermore, the
number of people working on weak lensing on the level of data analysis
is still quite small, and the methods used to extract shear from CCD
data are rather intricate. However, the analysis of CMB data is
certainly more complicated than weak lensing analyses, but there are
more people in the latter field, who checked and cross-checked their
results; also, more people implies that much more development has gone
into this field. Therefore, what is needed in weak lensing is a
detailed comparison of methods, preferably by several independent
groups, analysing the same data sets, together with extensive work on
simulated data to investigate down to which level a very weak shear
can be extracted from them. Up to now, no show-stopper has been
identified which prohibits the detection of shear at the sub-percent
level.

Weak-lensing results and techniques will increasingly be combined with
other methods. A few examples may suffice to illustrate this point.
The analysis of galaxy clusters with (weak) lensing will be combined
with results from X--ray measurements of the clusters and their
Sunyaev-Zel'dovich decrement. Once these methods are better
understood, in particular in terms of their systematics, the question
will no longer be, ``Are the masses derived with these methods in
agreement?'', but rather, ``What can we learn from their comparison?''
For instance, while lensing is insensitive to the distribution of
matter along the line-of-sight, the X--ray emission is, and thus their
combination provides information on the depth of the cluster (see,
e.g., \cite*{ZA98.1}). One might expect that clusters will continue
for some time to be main targets for weak-lensing studies. In addition
to clusters selected by their emission, mass concentrations selected
only by their weak-lensing properties shall be investigated in great
detail, both with deeper images to obtain a more accurate measurement
of the shear, and by X-ray, IR, sub-mm, and optical/IR multi-colour
techniques. It would be spectacular, and of great cosmological
significance, to find mass concentrations of exceedingly high
mass-to-light ratio (well in excess of $1,000$ in solar units), and it
is important to understand the distribution of M/L for clusters. A
first example may have been found by \cite{ewm99}.

As mentioned before, weak lensing is able to constrain the redshift
distribution of very faint objects which do not allow spectroscopic
investigation. Thus, lensing can constrain extrapolations of the
$z$-distribution, and the models for the redshift estimates obtained
from multi-colour photometry (`photometric redshifts'). On the other
hand, photometric redshifts will play an increasingly important role
for weak lensing, as they will allow to increase the signal-to-noise
ratio of local shear measurements. Furthermore, if source galaxies at
increasingly higher redshifts are considered (as will be the case with
the upcoming giant telescopes, cf.~\cite*{CL98.1}), the probability
increases that more than one deflector lies between us and this
distant screen of sources. To disentangle the corresponding projection
effects, the dependence of the lensing strength on the lens and source
redshift can be employed. Lenses at different redshifts cause
different source-redshift dependences of the measured shear. Hence,
photometric redshifts will play an increasingly important role for
weak lensing. Whereas a fully three-dimensional mass distribution will
probably be difficult to obtain using this relatively weak redshift
dependence, a separation of the mass distribution into a small number
of lens planes appears feasible.

Combining results from cosmic-shear measurements with the power
spectrum of the cosmic density fluctuations as measured from the CMB
will allow a sensitive test of the gravitational instability picture
for structure formation. As was pointed out by \cite{hut98},
cosmic-shear measurements can substantially improve the accuracy of
the determination of cosmological parameters from CMB experiments, in
particular by breaking the degeneracies inherent in the latter (see
also \cite*{ME98.1}). The comparison between observed cosmic shear and
theory will at least partly involve the increasingly detailed
numerical simulations of cosmic structure evolution, from which
predictions for lensing observations can directly be obtained. For
example, if the dark matter haloes in the numerical simulations are
populated with galaxies, e.g., by using semi-empirical theories of
galaxy evolution (\cite*{kns97}), detailed prediction for
galaxy-galaxy lensing can be derived and compared with observations,
thus constraining these theories. The same numerical results will
predict the relation between the measured shear and the galaxy
distribution on larger scales, which can be compared with the
observable correlation between these quantities to investigate the
scale- and redshift dependence of the bias factor.

The range of applications of weak lensing will grow in parallel to the
new instrumental developments. Keeping in mind that many discoveries
in gravitational lensing were not really expected (like the existence
of Einstein rings, or giant luminous arcs), it seems likely that the
introduction and extensive use of wide-field cameras and giant
telescopes will give rise to real surprises.

\section*{Acknowledgements}

We are deeply indebted to Lindsay King and Shude Mao for their very
careful reading of the manuscript and their numerous constructive
remarks.

\bibliography{LensBook_R,WeakLens_R}
\bibliographystyle{WeakLens}

\end{document}